\documentclass[letterpaper,fleqn,usenatbib]{mnras}

\usepackage[T1]{fontenc}
\usepackage{ae,aecompl}

\usepackage{graphicx}
\usepackage{amsmath}
\usepackage{amssymb}
\usepackage{multirow}

\title[UV and X-ray variability of AGN with Swift]{Ultraviolet and X-ray variability of active galactic nuclei with \it{Swift}}

\author[D. J. K. Buisson et al.]{
D. J. K. Buisson,$^{1}$\thanks{E-mail: djkb2@ast.cam.ac.uk}
A. M. Lohfink,$^{1}$
W. N. Alston$^{1}$ and 
A. C. Fabian.$^{1}$
\\
$^{1}$Institute of Astronomy, Madingley Road, Cambridge CB3 0HA\\
}

\date{Accepted 2016 September 27. Received 2016 September 19; in original form 2016 August 05}

\pubyear{2016}

\begin{document}
\label{firstpage}
\pagerange{\pageref{firstpage}--\pageref{lastpage}}
\maketitle

\begin{abstract}
We analyse a sample of 21 active galactic nuclei (AGN) using data from the Swift satellite to study the variability properties of the population in the X-ray, UV and optical band.
We find that the variable part of the UV-optical emission has a spectrum consistent with a powerlaw, with an average index of $-2.21\pm0.13$, as would be expected from central illumination of a thin disc (index of $-7/3$).
We also calculate the slope of a powerlaw from UV to X-ray variable emission, $\alpha_{\rm OX,Var}$; the average for this sample is  $\alpha_{\rm OX,Var}=-1.06\pm0.04$. The anticorrelation of $\alpha_{\rm OX}$ with the UV luminosity, $L_{\rm UV}$, previously found in the average emission is also present in the variable part: $\alpha_{\rm OX,Var} = (-0.177\pm0.083)\log (L_{\nu,\rm Var}(2500\,\text{\AA})) + (3.88\pm2.33)$.
Correlated variability between the emission in X-rays and UV is detected significantly for 9 of the 21 sources. All these cases are consistent with the UV lagging the X-rays, as would be seen if the correlated UV variations were produced by the reprocessing of X-ray emission. The observed UV lags are tentatively longer than expected for a standard thin disc.

\end{abstract}


\section{Introduction}

The structure of AGN is difficult to determine in part because their small size means they cannot be resolved by current instruments. Fortunately, this small size implies a short light-crossing time and hence AGN emission can vary on observable time-scales.
Observations of variations in the emission from AGN have shown that they do indeed vary on all time-scales and at all wavelengths probed.
The nature of these variations may be used to infer properties of the structure of AGN.

Our fundamental picture of this structure is that the central regions of AGN comprise an accretion disc principally emitting thermally in UV \citep{pringleagndisc} and a central hot corona which Compton upscatters some of these photons to X-rays \citep{haardtagnxrays}. A fraction of the X-rays are then emitted back towards the disc, which heats it, increasing its UV emission \citep{lightman88}. In these two ways, the X-ray and UV emission is linked and studying the details of the interaction can retrieve information about the nature of the UV and X-ray emitting regions.

The variability of the emission is strongest and occurs over the shortest time-scales at high energies \citep{mushotzkyvar}, indicating that the hard X-rays are produced on the smallest scales. X-ray variability studies have become a field of their own, mapping the innermost regions around the black hole \citep[e.g.][]{fabianlag,vaughan11,mchardy13,uttley14,alston14,karaglobal}.
UV and optical measurements over longer time-scales form the basis for studies at longer wavelengths \citep[e.g.][]{cackettopticallags,cameronlags}, allowing a larger region of the accretion flow to be probed. The relation of the two bands is also studied \citep[e.g. review by][]{gaskell03}.

While variability studies do their part in enhancing our understanding of the innermost regions of AGN, studying the time-averaged emission in great detail also provides crucial information.
The time-averaged UV spectra of AGN \citep[e.g.][]{shull12,telfer02,vandenberk01,francis91,schneider91} have been measured for many different samples and wavelength ranges. When the continuum is fit with a powerlaw, $F_{\lambda}\propto\lambda^{\alpha}$, the quasars of SDSS have $\alpha=-1.56$ over $1300-5000$\,\AA\ \citep{vandenberk01}. This is softer than the spectrum of a thin accretion disc ($\alpha=-7/3$, \citet{shaksun73}), although the variable part of the spectrum  of NGC~7469 has been found to be consistent with that value \citep{collier99}.
The difference from the theoretical value may be influenced
by the strength of absorption in the UV band \citep[e.g.][]{melendez11}, the presence of strong emission lines \citep[e.g.][]{krolikkallman88} and host starlight \citep[e.g.][]{bentz06,bentz09}.
\citet{cackettopticallags} study optical AGN spectra and derive the reddening values necessary for the difference spectra to match a thin disc spectrum. Their reddening values match those from the flux-flux or Balmer decrement method, indicating that the variable spectrum is indeed shaped like that of a thin disc.
Softer spectra are found at shorter wavelengths: $\alpha$ is $-1.32\pm0.14$ over $1200-1750$\,\AA\ and $-0.59\pm0.21$ over $550-1000$\,\AA\ \citep[both][]{shull12}, suggesting the presence of a turn over at wavelengths probing the highest temperatures in the disc. The high-energy cut-off in the coolest sources may also redden the average spectral index at longer wavelengths.

When the mean UV emission is compared with that of the X-rays, the power is found to be tightly correlated: the X-ray luminosity scales as $L_{\rm X}\propto L_{\rm UV}^k$ with $k=0.5-0.8$ \citep[e.g.][]{lusso16,steffen06,vignali03sdss} which results in the UV (2500\,\AA) to X-ray (2\,keV) slope, $\alpha_{\rm OX}$, being anticorrelated with luminosity: $\alpha_{\rm OX}=a\log L_{\rm UV}+{\rm const}$, $-0.2\lesssim a\lesssim-0.1$ \citep{vagnetti10,just07,strateva05}. This relation suggests that the processes producing the UV and X-ray radiation are closely related, as would be expected for an accretion disc--corona system.

The link between the X-ray and different UV bands can also be studied by comparing their correlation for a given source across time. Lags between changes in each band are interpreted as being due to the light travel time between the regions responsible for the emission in the different bands and hence the distances between them can be inferred. 

Such lags have been sought in various sources \citep[e.g.][]{shemmer01,maoz00} and compared to the predictions for a steady state accretion disc \citep{shaksun73}. Where lags are found, they often disagree with thin disc theory, usually showing a longer lag than expected.

Before precision cosmology provided a largely unquestioned value of $H_0\simeq70$\,km\,s$^{-1}$\,Mpc$^{-1}$, the luminosity of a standard disc was used to provide a distance modulus and hence $H_0$ \citep{collier99}. However,  the disc sizes from the measured lags implied $H_0=42\pm9$\,km\,s$^{-1}$\,Mpc$^{-1}$ \citep{collier99} or $H_0=44\pm5$\,km\,s$^{-1}$\,Mpc$^{-1}$ \citep{cackettopticallags}, so the disc is not as bright as is expected for its size.
Other studies also find deviations from a standard disc. For example, studies of NGC~5548 by \citet{STORM2} and \citet{fausnaugh16} describe the disc as larger than expected for its mass and accretion rate.
Similarly, \citet{troyer16} and \citet{mcg6lira} find the best fitting accretion rate, $\dot{M}$, is unreasonably high for a standard disc model in NGC~6814 and MCG-6-30-15 respectively.
The longer lags being associated with a larger disc than expected is corroborated by quasar microlensing observations, which find emitting regions a factor of a few ($2-3$, \citet{chartas16}; $\sim4$, \citet{morgan10}) larger than predicted.

However, a larger emitting region may not be the whole answer, as longer lags are not always found: \citet{mchardy16} study the low mass AGN NGC~4395 ($3.6\times10^{5}M_{\odot}$) and find lags which are not markedly different from standard thin disc theory.

Despite the lags often being longer than expected, the lag-wavelength relation found by these studies in the UV to IR bands is usually consistent with the predicted $\tau\propto\lambda^{4/3}$ for a standard accretion disc.

Lags have also been sought in the short time-scale variability within an X-ray observation with simultaneous UV monitoring. \citet{smith07} analyse \textit{XMM} observations of 8 sources but find no significant correlations. \citet{mcg6arevalo05} find the UV emission leading the X-rays by $\sim160$\,ksec (1.9\,days) in a 430\,ksec observation, although this lag is a large fraction of the observation length.

These studies of lags in individual sources have shown that, at least for some sources, the reprocessing of X-ray radiation does not behave as expected for a centrally illuminated thin disc.
A study of many sources has the potential to show what proportion of sources has a longer lag and whether this correlates with other AGN properties.

The \textit{Swift} satellite \citep{Gehrels2004}, principally designed for the detection of GRBs, is ideal for such broadband variability analysis: it has detectors for X-rays \citep{XRT} and ultraviolet/optical emission from 1700--6000\,\AA\  \citep{UVOT}. Since it has been operating for more than a decade, many AGN lightcurves covering time-scales of several years are available.

Here, the amount of variability in the UV and X-rays and the time differences between them are analysed for a sample of AGN to determine properties of AGN as a population.
Emission from the \textit{V}-band to X-rays is included to consider a large extent of the accretion disc.

The choice of sources and observations and the reduction of data is described in Section~\ref{sec:odr}. The methods and results of the analysis are described in Section~\ref{sec:res}. In particular, the UV variability is considered in Section~\ref{sub:UVvar} and the X-ray in Section~\ref{sub:xrayvar}. The bands are compared in terms of power in Section~\ref{sub:power} and time lags are explored in Section~\ref{sec:lags}. These results are interpreted in Section~\ref{sec:discussion}. Comments on individual sources are given in Section~\ref{indsources}.

\section{Observations and Data Reduction}
\label{sec:odr}

\begin{table*}
	\centering
	\caption{List of sources in our sample, with number of observations, black hole mass and reddening values for each object}
	\label{tab:source_table}
	\begin{tabular}{lccccccc}
		\hline
			Source & Number of & UV bands with & ${\rm{log}}_{10}(M_{\rm{BH}})$ & Ref. & $E(B-V)$ & $E(B-V)$ & Ref. \\
			&  Observations & $\geq10$ Observations & & & (Galactic) & (Intrinsic) & \\
		\hline
NGC 5548        &  744 & 6 & $7.59_{-0.21}^{+0.24}$ & P & 0.0171 & 0.152 & G\\
MRK 335         &  339 & 6 & $7.23_{-0.04}^{+0.04}$ & B & 0.0307 & 0.00 & C\\
NGC 7469        &  224 & 6 & $6.96_{-0.05}^{+0.05}$ & B & 0.0591 & 0.09 & C\\
FAIRALL 9       &  168 & 6 & $8.41_{-0.09}^{+0.11}$ & P & 0.0223 & 0 & N\\
1H 0707-495     &  119 & 1 & $6.72_{-0.5}^{+0.5}$ & Pa & 0.0816 & 0 & N\\
MCG--6-30-15     &  105 & 1 & $6.7_{-0.2}^{+0.1}$ & P & 0.0521 & 0.54 & Wa\\
MRK 766         &  100 & 6 & $6.82_{-0.06}^{+0.05}$ & P & 0.0169 & 0.613 & G\\
ARK 120         & 	90 & 2 & $8.18_{-0.06}^{+0.05}$ & P & 0.1094 & 0.04 & C\\
IRAS 13224--3809 & 	70 & 1 & $6.76_{-0.5}^{+0.5}$ & Z & 0.0601 & 0.628 & Po\\
PG 1211+143     & 	68 & 6 & $8.17_{-0.15}^{+0.11}$ & P & 0.0293 & 0 & G\\
NGC 4051        & 	59 & 1 & $6.13_{-0.16}^{+0.12}$ & P & 0.011 & 0.12 & Wi\\
NGC 3516        & 	58 & 1 & $7.40_{-0.06}^{+0.04}$ & B & 0.0359 & 0.15 & C\\
MRK 1383        & 	35 & 2 & $9.01_{-0.07}^{+0.11}$ & B & 0.0275 & 0 & N\\
PG 1247+267     & 	33 & 1 & $8.91_{-0.17}^{+0.15}$ & T & 0.0112 & 0 & N\\
H 0557--385      & 	28 & 1 & $7_{-1}^{+1}$ & A & 0.0375 & 0.511 & Ki\\
MRK 509         & 	27 & 3 & $8.05_{-0.04}^{+0.04}$ & B & 0.0493 & 0.11 & C\\
MRK 841         & 	26 & 6 & $7.88_{-0.1}^{+0.1}$ & V & 0.0255 & 0 & Wi\\
PDS 456         & 	22 & 2 & $8.91_{-0.5}^{+0.5}$ & Z & 0.4450 & 0 & N\\
IC 4329A        & 	20 & 6 & $8.34_{-0.3}^{+0.3}$ & M & 0.0501 & 0.98 & M\\
3C 120          &   71 & 5 & $7.75_{-0.04}^{+0.04}$ & B & 0.2558 & 0.05 & H \\
Zw 229--15       &   71 & 1 & $6.91_{-0.12}^{+0.08}$ & B & 0.0615 & 0 & N \\
		\hline
		\smallskip
	\end{tabular}
	
	A: \citet{ashton06}, B: \citet{bentzmbh}, M: \citet{markowitz09}, P: \citet{peterson04}, Pa: \citet{pan16}, T: \citet{trevese14}, V: \citet{vestergaard02}, Z: \citet{zhou05}.

	C: \citet{cackettopticallags}, G: \citet{grupe10}, H: \citet{hjorth95}, Ki: \citet{kishimoto11}, M: \citet{marziani92}, N: No value found in literature -- $E(B-V)=0$ assumed, Po: \citet{Polletta99}, Wi: \citet{winter10}, Wa: \citet{ward87}
	
\end{table*}

Our sample consists of all those AGN from the CAIXA sample \citep{CAIXAdesc} with at least 20 \textit{Swift} observations by September 2015. CAIXA comprises all unobscured radio-quiet AGN with public targeted \textit{XMM-Newton} observations as of March 2007. The radio-loud AGN 3C~120 was also included to test whether radio-loud AGN have grossly different properties and the Seyfert 1 galaxy Zw~229--15 was included as a large number of \textit{Swift} observations was present in the archive. This provides 21 sources, shown in Table~\ref{tab:source_table} along with their mass and reddening. 

Our analysis uses the data from both telescopes on board of \textit{Swift} \citep{Gehrels2004}: the UV \citep[UVOT,][]{UVOT} as well as the X-ray \cite[XRT,][]{XRT} telescope.

The X-ray light curves used in this work were produced using the automatic pipeline of the UK Swift Science Data Centre \citep{Evans2007,Evans2009}. Light curves were produced with a resolution of one bin per observation in 8 fine energy bands (0.3--0.6\,keV, 0.6--0.9\,keV, 0.9--1.2\,keV, 1.2--1.5\,keV, 1.5--3.0\,keV, 3.0--5.0\,keV, 5.0--7.0\,keV, 7.0--10\,keV)and the broader bands 0.3--10\,keV, 0.3--1.5\,keV and 1.5--10\,keV. For our analysis we used the default grade selection. When converting counts to flux, we correct for Galactic absorption using values from the LAB survey \citep{labnh}.

The UVOT observations were taken in different filters (\textit{V}:~5440~\AA, \textit{B}:~4390~\AA, \textit{U}:~3450~\AA, \textit{W1}:~2510~\AA, \textit{M2}:~2170~\AA, \textit{W2}:~1880~\AA) and we reduce each filter individually for each observation. We started our UVOT reduction from the level II image files, performing photometry with the tool \texttt{uvotsource}. To obtain the source counts, we assumed a circular source region with a 5 arcsec radius and a, also circular, background region with a 15 arcsec radius.
Some areas of the UVOT detector have shown spuriously low fluxes \citep{STORM2}, so we excluded observations where the source region overlaps these bad areas.
Count rates were converted to fluxes for each source using calibration factors from \citet{swiftctstoflux}.

Where necessary, the UV emission for each source was dereddened using the idl tools \texttt{ccmunred} \citep{ccm,odonnell94} for Galactic dust and \texttt{mszdst}\footnote{https://heasarc.gsfc.nasa.gov/xanadu/xspec/models/dust.html} for host absorption, with the extinction law from the LMC \citep{pei}.
 
UV/optical fluxes include a significant contribution from the host galaxy. Removing this component is difficult and requires high resolution images of the host galaxy. Here, we avoid the problem by using methods which are not affected by the additional constant host flux.

The luminosity distance was calculated with $H_0=70\,\mathrm{km s^{-1}\,Mpc^{-1}}$. The relative luminosities are not sensitive to cosmological parameters since all sources apart from PG1247+267 ($z=2.038$) are at low redshift ($z\ll1$).

\section{Results}
\label{sec:res}

\subsection{UV variability}
\label{sub:UVvar}
We initially consider the variability of the 6 \textit{Swift} UV/optical bands.

\subsubsection{Presence of variability}

To characterise the variability of the luminosity in each band, the measured standard deviation ($\sigma$) and mean square error ($\bar{\epsilon^2}$) were calculated and the intrinsic variability was estimated as the error-corrected rms variability, $L_{\rm Var}=\sqrt{\sigma^2-\bar{\epsilon^2}}$ \citep{nandraexvar,edelsonfracvar}. We only consider UV bands with at least 10 data points to remove lightcurves where the uncertainty would be excessively dominated by the stochastic variations due to sampling.

X-ray studies usually consider the fractional (rms) variablity, $F_{\rm Var}=L_{\rm Var}/\bar{L}$ (although this is usually calculated directly from the measured count rates), where $\bar{L}$ is the mean of the measured lightcurve, to avoid the effects of absorption. Absorption is a multiplicative effect which changes $L_{\rm Var}$ and $\bar{L}$ by the same factor, so their ratio is unchanged. In the UV, host galaxy contamination is a more significant problem, so $\bar{L}$ from the AGN alone is not precisely known. This constant addition to the measured $\bar{L}$ has no effect on the variance of the lightcurve. Hence, we prefer the absolute luminosity variance. The standard deviation (rms variability) is preferred to the variance because of its more familiar dimension.

We define a source as variable if the variations in measured flux are larger than would be expected due to measurement errors alone. The approximate error in the lightcurve variability is given by equation~(\ref{eq:exvarerr}) \citep[][eq. 11]{exvarerr}, where $N$ is the number of datapoints.
\begin{equation}
\mathrm{err}(L_{\rm Var}^2)=\frac{1}{\sqrt{N}}\sqrt{\left(\sqrt{2}\bar{\epsilon^2}\right)^2+\left(2\sqrt{\bar{\epsilon^2}}L_{\rm Var}\right)^2}
\label{eq:exvarerr}
\end{equation}
We assess the variability for each source and band separately. In four bands out of 90 with sufficient data ($\geq10$ points) to calculate variability (IC~4329A \textit{B}, \textit{W2}; MRK~766 \textit{M2}; PDS~456 \textit{W1}), the error was greater than the measured variability. Since the measured variability is subject to random variations depending on when a source is observed, a varying source may by chance be observed when it varies little. Since there is no reason to expect this small fraction of bands to not be varying, these bands were not excluded from further analysis to avoid biasing results towards observations which happened to catch higher variability.

\subsubsection{UV variability spectra}
\label{sub:exvar}

For 12 sources, $L_{\lambda,{\rm Var}}$ measurements exist in at least 2 wavebands; for these, we produced a spectrum of the variable component of the emission. The variable UV spectrum of Fairall~9, which is typical of the sample, is shown in Fig.~\ref{fig:exvarspec}, fitted with a powerlaw. Spectra of the whole sample are shown in Appendix~\ref{app:exvarspec}.

\begin{figure}
\includegraphics[width=\columnwidth]{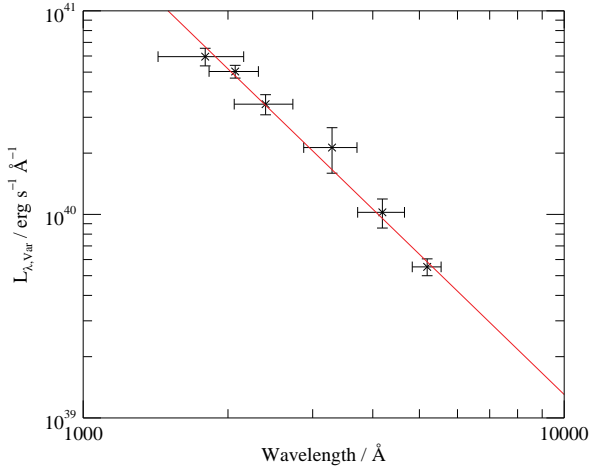}
\caption{The spectrum of the variable UV emission of Fairall~9, along with the best-fitting powerlaw, $\alpha=-2.29\pm0.10$.}
\label{fig:exvarspec}
\end{figure}

To characterise the shape of the spectrum, the spectrum of variable luminosity was fitted with a powerlaw, $L_{\lambda,{\rm Var}}\propto\lambda^\alpha$. We performed a chi-square minimisation with \texttt{mpfit} in IDL in the logarithmic domain. The wavelengths of each band were taken as the nominal central value. Errors of the luminosity variability were estimated from equation~(\ref{eq:exvarerr}). Values for $\alpha$ are given in Table~\ref{tab:exvarind}.

In most sources, the powerlaw provides a good fit; however, for 3C~120, Mrk~766 and PG~1211+143 the reduced $\chi^2$ value is unacceptably high. Exploring the reasons for this, we find that in Mrk~766, the \textit{W2} measurement is much higher than would be expected from extrapolating a powerlaw fitted to the remaining values. Since for Mrk~766 the \textit{W2} band has a large number of samples from 2012-13 which are not taken in the other filters, the W2 variability measurement may be increased by the difference in variability in the  different epochs. In PG1211+143, the \textit{U}-band is somewhat higher than expected and the \textit{M2} much lower. In 3C~120, the \textit{U} and \textit{M2}-bands are both higher than the fit. Since these sources show scatter rather than curvature, it is possible that the errors have been underestimated.

\begin{table}
	\centering
	\caption{Index of powerlaw fit to variable part of UV spectrum, with $1\sigma$ errors. Most sources are consistent with $-2>\alpha>-2.33$, as predicted for a thin accretion disc}
	\label{tab:exvarind}
	\begin{tabular}{lccc}
		\hline
			Source & Index, $\alpha$ & $\chi^2_{\rm red}$ & d.o.f.\\
		\hline
 3C 120 & $-2.21\pm0.10$ &  5.10 & 3\\
ARK 120 & $-2.41\pm0.83$ &  - & 0\\
 Fairall 9 & $-2.29\pm0.10$ &  0.94 & 4\\
IC 4329A & $-3.14\pm0.27$ &  1.78 & 1\\
MRK 335 & $-2.27\pm0.11$ &  2.14 & 4\\
MRK 509 & $-1.96\pm2.41$ &  0.04 & 1\\
MRK 766 & $-3.57\pm0.09$ & 34.31 & 4\\
MRK 841 & $-2.11\pm0.16$ &  0.59 & 4\\
MRK 1383 & $-2.81\pm0.88$ &  - & 0\\
NGC 5548 & $-2.71\pm0.07$ &  0.67 & 4\\
NGC 7469 & $-3.13\pm0.39$ &  1.65 & 3\\
PG 1211+143 & $-2.01\pm0.15$ &  9.90 & 4\\
                \hline
        \end{tabular}
\end{table}

From the fits, we find an average slope of $\alpha=-2.6\pm0.8$. For all sources apart from IC~4329A, MRK~766, NGC~5548 and NGC~7469, the index
is consistent with $\alpha=-2$ to $-2.33$, as predicted by \citet{davis} for a thin accretion disc. The variable spectrum of NGC~5548 has also been measured by \citet{STORM2} with a subset of the data used here, finding $\alpha=-1.98\pm0.20$ assuming no intrinsic reddening (this is consistent with our value before dereddening). For IC~4329A and Mrk~766, the intrinsic reddening is strong (Table~\ref{tab:source_table}), so the uncertainty in the reddening may allow their indices to be consistent with $\alpha=-2$ to $-2.33$. \citet{collier99} also measured the variable spectrum of NGC~7469, finding an index consistent with $\alpha=-2.33$.

For the sources with $E(B-V)\leq0.05$ and excluding those sources with very large uncertainties ($\Delta\alpha>0.75$), the average index is $\alpha=-2.21\pm0.13$. We consider this our best estimate of the slope.

\begin{figure}
\includegraphics[width=\columnwidth]{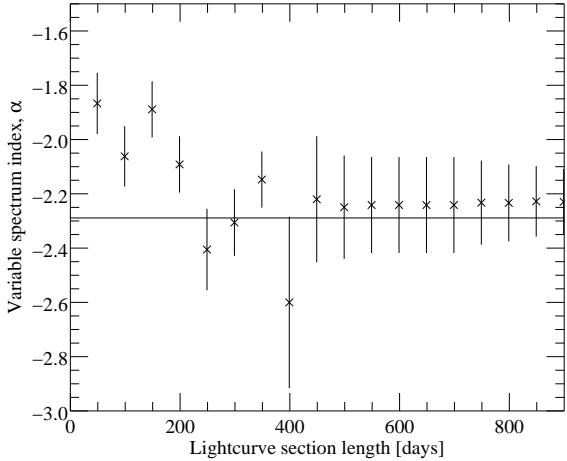}
\caption{Plot of index of variable spectrum of Fairall~9 against duration of lightcurve used to calculate variability. The index shown at each timescale is the average over all sections of given length. The horizontal line shows the value for the full lightcurve. The index appears to converge to the value measured from the full lightcurve as section length increases.
}
\label{fig:secind}
\end{figure}

In order to investigate the nature of the variability on different time-scales and ensure that the non-uniformly sampled lightcurves do not bias the results based on the different time-scales sampled in different observations, the lightcurves were also split into sections of different lengths. The index of the variable spectrum was calculated in the same way as for the full lightcurve. For each section length, we calculate the average index of all sections. For Fairall~9, which has enough data to split on many time-scales, Fig.~\ref{fig:secind} shows the spectral index as a function of section length. The index converges towards the value calculated from the full lightcurve as the section length increases, which justifies the use of the full lightcurve.

\begin{figure}
	\includegraphics[width=\columnwidth]{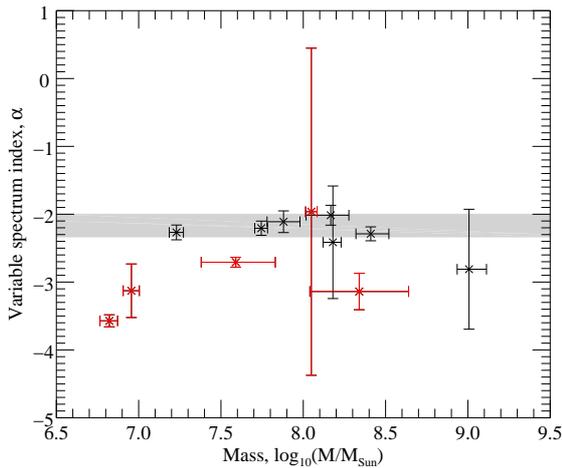}
    \caption{Plot of spectral index of variable emission against black hole mass. Most sources are consistent with an index of  $-2>\alpha>-2.33$ (indicated by the shaded region), as predicted for a thin accretion disc. Those sources with a steeper spectrum tend to have been the most dereddened: sources with $E(B-V)_{\rm int}>0.05$ are shown in red. No correlation with mass is apparent ($r=0.36$).}
    \label{fig:exvarind}
\end{figure}

\begin{figure*}
\includegraphics[width=\textwidth]{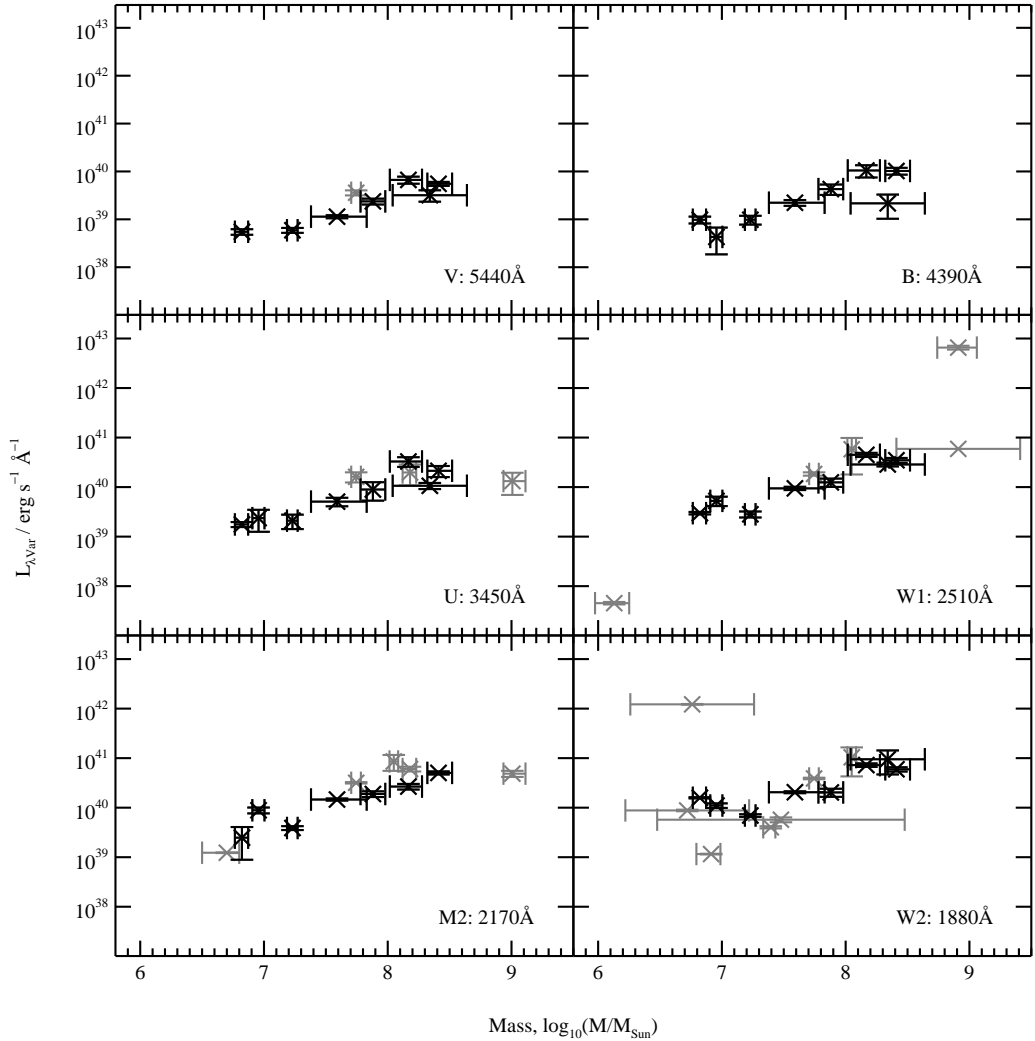}
\caption{Plot of variable luminosity, $L_{\rm Var}$, against black hole mass in each waveband. Sources in black are observed in all bands, other sources are in grey.}
\label{fig:msigma}
\end{figure*}

Finally, we consider how the spectra of variable power depend on black hole mass.
Plotting the spectral indices, $\alpha$, of the variable part of the spectrum against mass (Fig.~\ref{fig:exvarind}) shows no clear correlation ($r=0.36$).
Fig.~\ref{fig:msigma} shows the variable luminosity in each UV band against black hole mass, $M_{\rm BH}$. Sources with observations in all 6 bands are shown in black and the remaining sources in grey. It is apparent that $L_{\rm Var}$ increases with mass. While the black points show a tight correlation, this is a selection effect: the power is also dependent on the Eddington ratio and, for this subsample, the Eddington ratio smoothly decreases with mass (Table~\ref{tab:edd}). This flattens the observed correlation so the slope of the relationship is not meaningful. The effect of scattered Eddington ratios can be seen in the greater scatter when including the remaining sources (grey points).
 The correlations between $M_{\rm BH}$ and $L_{\rm Var}$ seen in the black points do show that the scaling with mass is consistent between the different bands, as would be expected if the spectral shape is constant.

\subsection{X-ray Variability}
\label{sub:xrayvar}

To compare the variability of the thermal disc emission with that of the coronal  emission, we also calculated the excess variance in the 8 X-ray bands.
We estimated the error in the excess variance due to measurement error from simulated lightcurves. To do this, we simulated lightcurves by adding values from a Gaussian with the same width as the error on each observation to the measured lightcurve and used the distribution of excess variance from 20000 realisations to estimate 1$\sigma$ and 2$\sigma$ confidence intervals. As the error is comparable to the measured value in some sources, we converted the confidence limits from variance to rms directly, as in \citet{poutanenexvarerr}, rather than using differential error propagation.

The variable X-ray luminosity spectra this produces are shown alongside the UV variable spectra in Appendix~\ref{app:exvarspec}. The X-ray variable spectra look superficially similar in shape to the mean spectra of their respective sources. 

\begin{figure}
\includegraphics[width=\columnwidth]{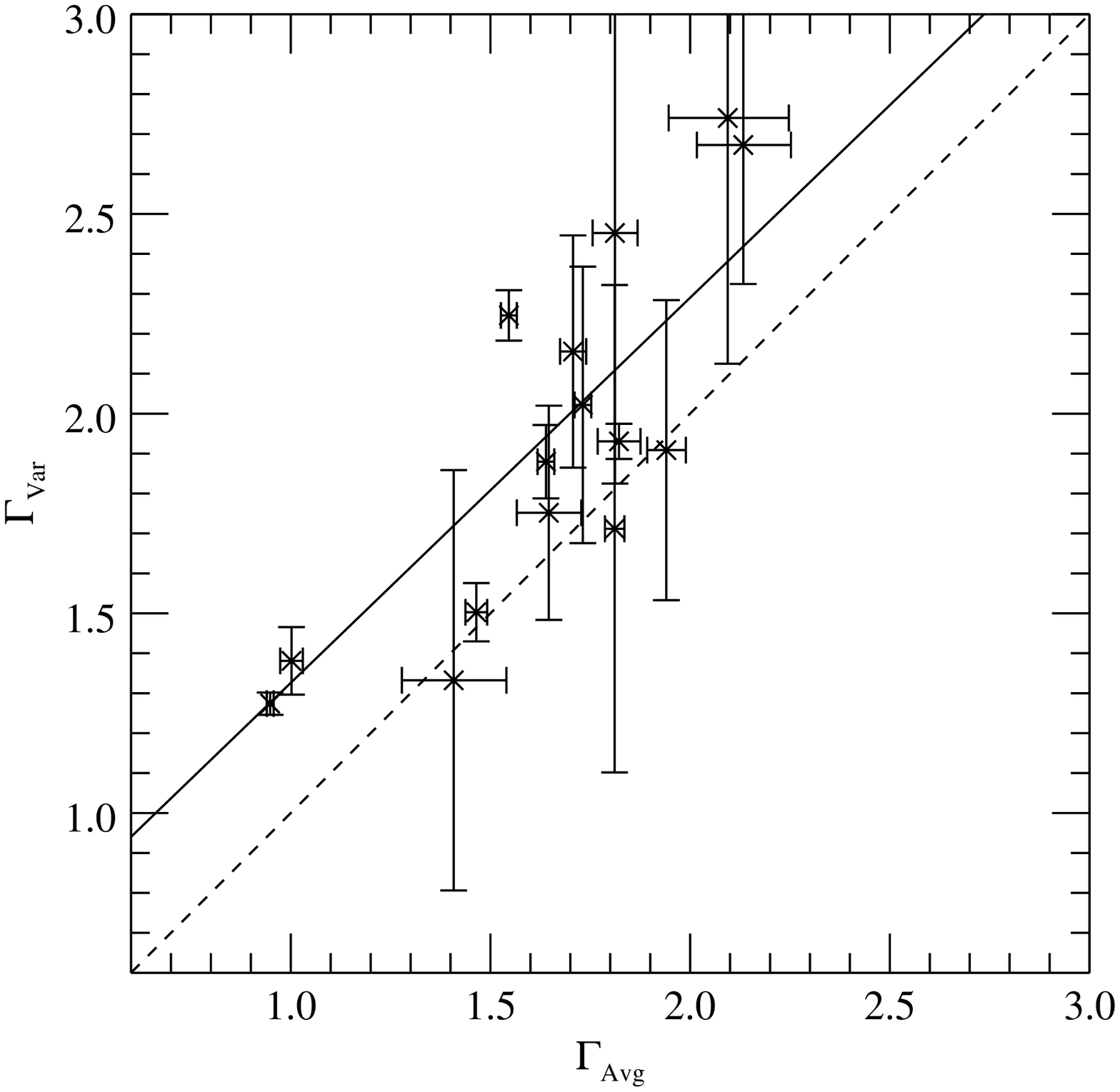}
\caption{Photon index ($\Gamma$) of powerlaw fit to average and variable parts of 1.5-10\,keV X-ray spectra. Solid line: best fit; dashed line: equal variable and average indices. The variable part is usually softer than the average spectrum, with an average $\Delta\Gamma=0.28\pm0.02$ ($1\sigma$ error).}
\label{fig:xrayvarind}
\end{figure}

To quantify the differences between the mean and variable spectra, we fitted powerlaws to the hard (1.5-10\,keV) X-ray band for each source. For consistency with other X-ray measurements, we use the photon index defined as $N(E)\propto E^{-\Gamma}$ (note that this converts from the wavelength spectral index as $\Gamma=\alpha_{\rm X}+3$). These photon indices, $\Gamma_{\rm{Var}}$, are compared to the indices of the mean spectra over the same band, $\Gamma_{\rm{Avg}}$, in Fig.~\ref{fig:xrayvarind}. This shows that the two indices are well correlated ($r=0.81$) and the variable spectra are softer than the average spectra over the 1.5-10\,keV band. We estimate the relation between the variable and average indices with a linear function; the best fitting is $\Gamma_{\rm{Var}}=(0.97\pm0.07)\Gamma_{\rm{Avg}}+(0.36\pm0.08)$, compatible with a constant offset. Fitting for a constant offset gives $\Delta\Gamma = 0.28\pm 0.02$ between average and variable index.

\subsection{Comparison of UV and X-ray variable power}
\label{sub:power}

To help understand the interactions between the disc and corona, we compared the power of the variability in the X-ray ($L_{\rm X,Var}$) and UV ($L_{\rm UV,Var}$).

\begin{figure}
\includegraphics[width=\columnwidth]{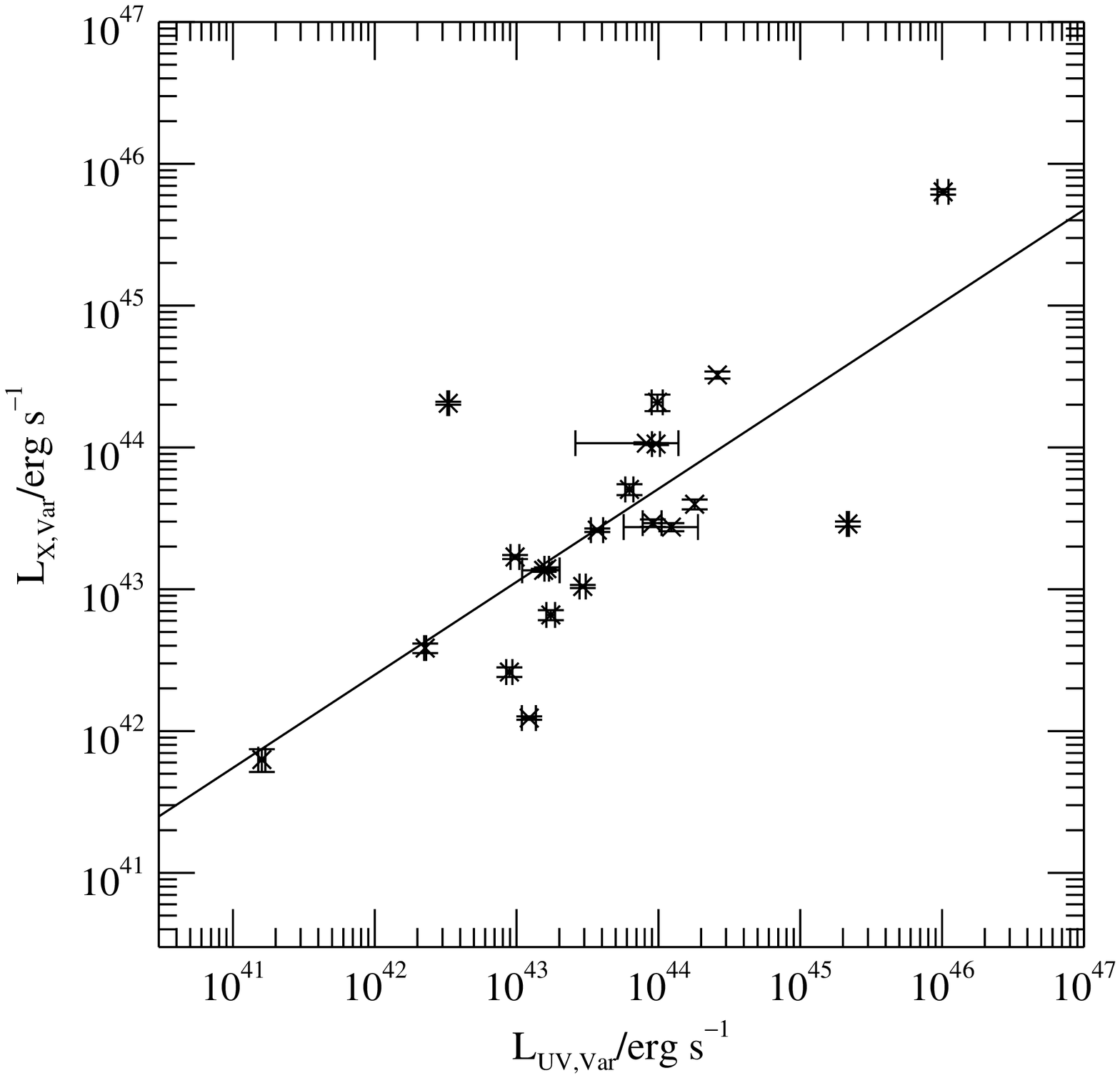}
\caption{Plot of total UV against X-Ray variable power
in measured bands: UV: $1500-5815$\AA; X-ray: $0.3-10$\,keV. The line shows the best-fitting powerlaw, $L_{\rm X,Var} \propto L_{\rm UV,Var}^{0.66\pm0.22}$}
\label{fig:uvxpow}
\end{figure}

We first considered the variable power in the directly measured energy ranges, $0.3-10$\,keV (X-ray) and $1500-5815$\,\AA\ (UV). The X-ray variability power was calculated from a direct sum over the power in each energy bin. Since the UV bands do not fully and evenly cover their overall wavelength range, we used a powerlaw fit to the datapoints integrated over the total range of the \textit{Swift} filters. The variable power in the X-ray and UV is plotted in Fig.~\ref{fig:uvxpow}; this shows that the UV and X-ray power is broadly comparable. There is a strong correlation ($r=0.74$) between the two quantities with the UV power increasing somewhat faster than the X-ray power. Approximating the relation with a powerlaw $L_{\rm X,Var} \propto L_{\rm UV,Var}^{\beta}$ gives a best fit of $\beta=0.66\pm0.22$.
This is consistent with the relationship between the average luminosities, for which $\beta = 0.72\pm0.01\ {\rm or}\ 0.75\pm0.06$ have been found by \citet{steffen06} and \citet{vignali03sdss} respectively.

We also consider the relative specific luminosity, using the definition of $\alpha_{\rm OX}$ applied to the variable part of the emission. $\alpha_{\rm OX}$ is the index of a powerlaw between the specific luminosity at 2500\,\AA\ and 2\,keV \citep[e.g.][]{vagnetti10}:

\begin{equation}
\begin{split}
\alpha_{\rm OX} & = \frac{\log(L_\nu(2\,{\rm keV})/L_\nu(2500\,{\text{\AA}}))}{\log(\nu_{\rm 2\,keV}/\nu_{2500\,\text{\AA}})}\\
 & =0.3838\log\left(\frac{L_\nu(2\,{\rm keV})}{L_\nu(2500\,\text{\AA})}\right)
\end{split}
\end{equation}

\begin{figure}
\includegraphics[width=\columnwidth]{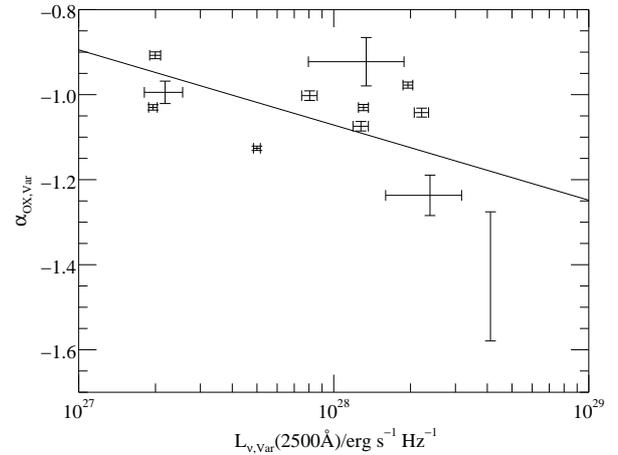}
\caption{Plot of $\alpha_{\rm OX,Var}$ against $L_{\nu, {\rm Var}}(2500\,\text{\AA})$. The solid line shows the least squares fit, $\alpha_{\rm OX,Var}=(-0.177\pm0.083)\log L_{\nu, {\rm Var}}(2500\,\text{\AA})+(3.88\pm2.33)$.}
\label{fig:alphaox}
\end{figure}

Here, we measure $\alpha_{\rm OX,Var}$ from the variable specific luminosities. The average is $\alpha_{\rm OX,Var}=-1.06\pm0.04$, which is flatter than some measurements of $\alpha_{\rm OX}$ measured from mean spectra of similarly bright sources (e.g. $-1.32\pm0.03$, \citealt{steffen06}) but similar to values found in \citet{xu11}, which includes some sources which are in our sample. $\alpha_{\rm OX,Var}$ for each source is shown against specific variable UV luminosity at 2500\,\AA\ in Fig.~\ref{fig:alphaox}. 
These are anticorrelated, $r=-0.73$, and the least-squares fit is
\begin{equation}
\alpha_{\rm OX,Var}=(-0.177\pm0.083)\log L_{\nu, {\rm Var}}(2500\,\text{\AA})+(3.88\pm2.33)
\end{equation}

This relation has a slope within those found by previous authors for the mean spectrum, such as that by \citet{gibson08}:
\begin{equation}
\alpha_{\rm OX}=(-0.217\pm0.036)\log L_{2500\,\text{\AA}}+(5.075\pm1.118)
\end{equation}
or the flatter relation found by \citet{just07}:
\begin{equation}
\alpha_{\rm OX}=(-0.140\pm0.007)\log L_{2500\,\text{\AA}}+(2.705\pm0.212)
\end{equation}
\citet{xu11} studied low-luminosity AGN and found: 
\begin{equation}
\alpha_{\rm OX}=(-0.134\pm0.031)\log L_{2500\,\text{\AA}}+(2.406\pm0.785)
\end{equation}
\citet{grupe10} also analysed a sample of local ($z<0.35$) sources with \textit{Swift}, so provide the most similar reference sample:
\begin{equation}
\alpha_{\rm OX}=(-0.114\pm0.014)\log L_{2500\,\text{\AA}}+(1.177\pm0.305)
\end{equation}
However, \citet{vagnetti13} corrected for the galactic dilution in this sample and found:
\begin{equation}
\alpha_{\rm OX}=(-0.135\pm0.015)\log L_{2500\,\text{\AA}}+(2.645\pm0.446)
\end{equation}
While our value is within the range found by previous authors for the mean spectrum, the spread in indices from measurements of mean spectra is much larger than the error for a given sample and therefore a reliable comparison between the $\alpha_{\rm OX} - \log L_{2500\,\text{\AA}}$ relations for the mean and variable parts of the spectra would require that both be calculated from the same sample and dataset.

\subsection{Interband Lags}
\label{sec:lags}

So far, we have only considered the amount of variability in each band. To investigate how the variability in different bands is related in time, we calculated the correlations between the different bands.

To look for correlated variability between the X-rays (0.3-10\,keV) and each of the UV bands for each source, we used the discrete cross-correlation function (DCF) \citep{dccf}. The X-rays were chosen as the reference band since this is the only band which is measured for every source.
Since the number and frequency of observations differs widely between sources, we grouped the DCF into lag bins by number of observation pairs rather than a fixed lag width. We chose 100 points as the minimum compatible with little apparent noise in the well-sampled sources.
Before calculating the DCF, a 30 day half-width boxcar running mean was subtracted from the lightcurves to remove the effects of long-term variations and highlight the expected lags of a few days \citep{welsh99}.

To determine the significance of any correlations, we used the distribution of the DCFs from 10000 uncorrelated simulated lightcurves in each band.
Using the method of \citet{powerlawnoise}, we produced lightcurves with the same power spectra as the real lightcurves and a resolution equal to the average observation length. We used X-ray power spectra from \citet{psdbend} and estimated UV power spectra by fitting a powerlaw to the periodogram of the \textit{Swift} data. We added the rms-flux relation by taking the exponential of the lightcurves \citep{rmsflux}. From these regular lightcurves, we extracted count measurements at points corresponding to the actual observation times and simulated observational noise by drawing the final simulated data from a Poisson distribution with mean equal to these count measurements. We calculated the DCF for each UV/X-ray pair and used the distribution of the DCFs at each lag value to estimate the 95\% and 99\% confidence intervals outside which a correlation is unlikely to be produced by random noise.
This found significant (>99\%) correlations in 9 sources out of the sample of 21. These sources either have more datapoints or a clear peak in the lightcurve, so the non-detection in the remaining sources is likely due to a lack of data quality rather than less intrinsic correlation. MCG--6-30-15 notably has many datapoints but no significant lag detection; see Section~\ref{sub:mcg6} for a detailed discussion.

Where a significant correlation was found, we estimated the potential lag between bands using the centroid of the DCF peak. To generate enough points to produce a smooth distribution to centroid, the X-ray lightcurve was linearly interpolated onto a finer grid ($\Delta t=0.1$\,days) and the ICCF of each UV band was measured against it \citep{gaskellsparke,gaskellpeterson}. We used the centroid of the region with a correlation coefficient of at least 0.8 times the maximum value \citep[e.g.][]{troyer16}.

Errors on the lag values obtained were estimated using subset selection/flux randomisation \citep{peterson98,peterson04}. From 2000 realisations, we estimated 1$\sigma$ errors by percentile.

Lag values are shown in Table~\ref{tab:lags}. In all 9 sources, the UV bands are consistent with a lag behind the X-rays. For the two best measured sources, NGC~5548 and Fairall~9, the lower energy bands have a longer delay. For the remaining sources, the lags are not sufficiently well constrained to determine differences in lag between the different UV bands.

\begin{table}
\centering
\caption{Lags of each UV band behind X-Rays, with $1\sigma$ errors.}
\label{tab:lags}
	\begin{tabular}{lcc}
		\hline
			Source & Band & Lag/days\\
		\hline
\multirow{2}{*}{ 3C 120} & U & $0.4\pm 4.5$\\
 & W2 & $-0.2\pm 2.3$\\
		\hline
\multirow{1}{*}{ARK 120} & U & $2.4\pm 2.3$\\
		\hline
\multirow{6}{*}{ Fairall 9} & V & $4.2\pm 2.8$\\
 & B & $2.7\pm 1.9$\\
 & U & $3.1\pm 1.5$\\
 & W1 & $2.2\pm 1.0$\\
 & M2 & $1.7\pm 1.0$\\
 & W2 & $1.7\pm 0.8$\\
		\hline
\multirow{1}{*}{ IRAS 13224--3809} & W2 & $6.4\pm 3.7$\\
		\hline
\multirow{1}{*}{MRK 335} & W2 & $0.0\pm 2.9$\\
		\hline
\multirow{1}{*}{MRK 1383} & M2 & $4.3\pm 8.5$\\
		\hline
\multirow{1}{*}{NGC 3516} & W2 & $1.6\pm 1.5$\\
		\hline
\multirow{6}{*}{NGC 5548} & V & $2.0\pm 1.1$\\
 & B & $1.5\pm 0.8$\\
 & U & $1.4\pm 0.7$\\
 & W1 & $1.0\pm 0.7$\\
 & M2 & $0.8\pm 0.7$\\
 & W2 & $0.7\pm 0.5$\\
		\hline
\multirow{4}{*}{NGC 7469} & U & $1.1\pm 1.0$\\
 & W1 & $-0.3\pm 1.2$\\
 & M2 & $1.3\pm 1.7$\\
 & W2 & $0.8\pm 0.7$\\
		\hline
	\end{tabular}
\end{table}

We compared these lags with the theoretical lags for a thin accretion disc around a black hole of the appropriate mass and accretion rate \citep{shaksun73}.
To do so, we calculated the accretion rate in terms of the Eddington ratio, $\dot{m}$, derived from the average hard (2-10\,keV) X-ray luminosity during the \textit{Swift} monitoring. We used correction factors, $\kappa$, to convert from X-ray to bolometric luminosity based on the most recent measurement in \citet{vasudevan10, vasudevanf09, vasudevan07}. Where a source does not have a value for $\kappa$ in any of these papers, we estimated $\dot{m}$ from the average X-ray spectral index, $\Gamma_{\rm Avg}$, using the relation from \citet{shemmer08}. The resulting Eddington ratios are given in Table~\ref{tab:edd}.

\begin{table}
\caption{X-ray luminosities and Eddington ratios for each source in the sample. $\dot{m}$ is calculated by converting ${\rm log}_{10}L_{\rm 2-10\,keV}$ to a bolometric luminosity where $\kappa$ is available. For the remaining sources, $\dot{m}$ is estimated from $\Gamma_{\rm Avg}$.}
\label{tab:edd}
	\begin{tabular}{lccc}
		\hline
			Source & ${\rm log}_{10}L_{\rm 2-10\,keV}$/erg\,s$^{-1}$ & $\kappa$ & $\dot{m}$ \\
		\hline
 1H 0707--495 & $41.90\pm  0.01$ & $ $ & $ 0.37$\\
 3C 120 & $44.09\pm  0.01$ & $8.29 $ & $ 0.14$ \\
ARK 120 & $43.78\pm  0.02$ & $25.0 $ & $ 0.08$ \\
 Fairall 9 & $43.98\pm  0.01$ & $10.5 $ & $ 0.03$ \\
 H 0557--385 & $42.76\pm  0.02$ & $ $ & $ 0.02$\\
IC 4329A & $43.83\pm  0.01$ & $14.8 $ & $ 0.04$ \\
 IRAS 13224--3809 & $42.34\pm  0.06$ & $ $ & $ 0.68$\\
IRAS 13349+2438 & $43.64\pm  0.05$ & $ $ & $ 0.07$\\
 MCG--6-30-15 & $42.65\pm  0.01$ & $22.2 $ & $ 0.16$\\
MRK 335 & $42.79\pm  0.01$ & $102 $ & $ 0.30$ \\
MRK 509 & $43.97\pm  0.01$ & $12.5 $ & $ 0.08$ \\
MRK 766 & $42.66\pm  0.01$ & $70.5 $ & $ 0.39$ \\
MRK 841 & $43.48\pm  0.01$ & $27.4 $ & $ 0.09$ \\
MRK 1383 & $44.07\pm  0.01$ & $33.5 $ & $ 0.03$ \\
NGC 3516 & $42.58\pm  0.01$ & $17.7 $ & $ 0.02$ \\
NGC 4051 & $41.03\pm  0.01$ & $16.5 $ & $ 0.01$ \\
NGC 5548 & $43.25\pm  0.00$ & $18.8 $ & $ 0.07$ \\
NGC 7469 & $43.06\pm  0.01$ & $38.7 $ & $ 0.39$ \\
PDS 456 & $44.56\pm  0.01$ & $ $ & $ 0.24$\\
PG 1211+143 & $43.55\pm  0.02$ & $92 $ & $ 0.18$ \\
PG 1247+267 & $45.82\pm  0.04$ & $ $ & $ 0.22$\\
 Zw 229-15 & $42.56\pm  0.02$ & $ $ & $ 0.12$\\
		\hline
	\end{tabular}
\end{table}

For each source, the lags and theoretical predictions (black lines) are plotted in Appendix~\ref{app:lags}. The red lines indicate the $1\sigma$ uncertainties due to the mass and luminosity uncertainties but do not include uncertainties in $\kappa$.

The two best-measured sources, NGC~5548 and Fairall~9, both have lags slightly longer than the predicted values but each individual value is still consistent. The remaining sources do not have enough points of sufficient quality to determine inconsistencies as individuals.

To explore any potential global deviation from the lags expected for a thin accretion disc, the lags for each source were scaled to represent the accretion disc around a black hole with a common mass, $10^{8}M_{\odot}$, and Eddington fraction, $\dot{m}=0.1$. The lags were scaled by $\tau\sim R/c\sim \dot{m}^{1/3}M^{2/3}$ \citep{shaksun73}.

\begin{figure}
	\includegraphics[width=\columnwidth]{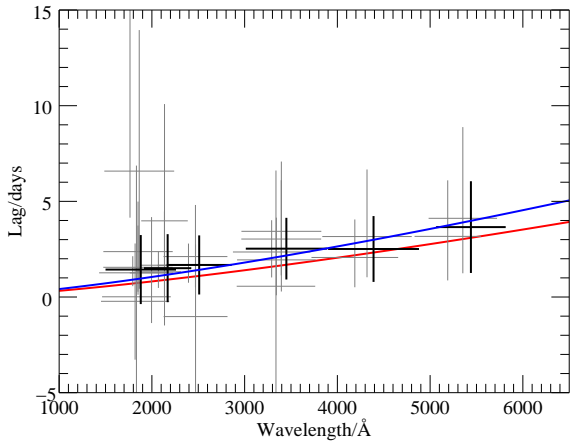}
    \caption{Lags relative to X-rays for each source scaled to a black hole mass of $10^8 M_\odot$ and Eddington rate $\dot{m}=0.1$. The red (lower) line shows the lags expected for a thin disc; the blue (upper) line shows the best fitting scaled lags.}
    \label{fig:normlags}
\end{figure}

These scaled lags are plotted in Fig.~\ref{fig:normlags}, along with the expected lags for a thin accretion disc (red) and the best fitting size for the measured lags (blue). In each band, the average lag (corrected to the rest wavelength of the band) is shown in black. For every band, the measured lag is greater than the theoretical by an average factor of 1.3. However, the overall deviation is only $1.5\sigma$.

\section{Discussion}
\label{sec:discussion}

\subsection{Summary of results}

From our study of 21 AGN monitored by \textit{Swift} over time-scales of several years in bands from optical through to X-ray, we find that:
\begin{itemize}

\item The UV variable spectra are consistent with a powerlaw. The average index is $\alpha=-2.21\pm0.13$.

\item The amount of variable power increases with mass at the same rate for all UV bands.

\item The UV variable luminosity and the index from UV to X-rays, $\alpha_{\rm OX}$, are anticorrelated:
$\alpha_{\rm OX}=\left(-0.177\pm0.083\right)\log L_{\rm UV,Var}+(3.88\pm2.33)$

\item The variable X-ray spectra are softer than the average spectra of their respective sources by $\Delta\Gamma=0.28\pm0.06$.

\item Significant correlations between X-ray and UV variability are detectable in 9 sources. The remaining sources are generally less well sampled and any correlated variability is too weak to detect with the current data.

\item Every lag measurement is consistent with variations in the UV lagging behind the X-rays.
\end{itemize}

\subsection{Comparison with average spectra}

In the unified AGN model \citep{antonucci93,urry95}, the UV emission of AGN is principally from an accretion disc \citep{lyndenbell69} surrounding the central SMBH. 

We can compare the indices of the variable spectra we measure with those of the average spectra, e.g. those by \citep{shull12} from \textit{HST-COS} or \citet{vandenberk01}, who study quasar spectra over a more similar wavelength range ($1300-5000$\,\AA) to \textit{Swift}. We find harder indices than either of these measurements. This may be due to our individual dereddening of the sources or there being more fractional variability at shorter wavelengths.
An additional constant component which is cooler than the disc would make the total average nuclear spectrum appear redder than the variable part.
The consistency of the variable spectra with a flat disc suggests that variable clouds obscuring the disc are unlikely to play a major role at the radii ($\sim 50-1000\,r_g$) probed here.

Contrastingly, we find that the variable X-ray spectra are softer than the average spectra.
The X-ray variability is thought to come from fluctuations in the accretion rate in the innermost regions of the disc \citep[e.g.][]{lyubarskii97,king04,zdziarski05,kelly11}. 
The softer variable spectra are probably due to the shortest time-scale variations in the harder emission being averaged out across an observation.
There may also be a contribution from the continuum varying more than the reflected part, which blurs out variations because it is produced over a larger region. Since the continuum is softer than the reflection it produces, the variable spectrum is softer.

Various groups \citep[e.g.][]{just07,gibson08,grupe10,vagnetti10,vagnetti13} study the relationship between the UV and X-ray average luminosity and find an $\alpha_{\rm OX}$-$L_{\rm UV}$ anticorrelation: brighter UV sources have a lower $L_{\rm X,Avg}/L_{\rm UV,Avg}$. We find that this is also the case for the variable part of the emission. This means that the fractional variability of UV and X-ray emission behaves similarly between sources, suggesting a link between the two, as expected for an accretion disc and corona system.

\subsection{Sources of variability}

We find that our variable UV spectra are described well by a simple powerlaw. The indices are consistent with those expected for the emission from a disc heated by internal dissipation in the accretion process or by illumination by a central source.

To determine whether the variable UV spectral slope is the result of illumination or a variable accretion disc, we consider the expected time-scales of the two potential variability mechanisms.
The shortest time-scales occur towards the centre of the disc, which are principally probed by the higher energy bands. For our fiducial $10^8\, M_{\odot}$ black hole, the shortest wavelength \textit{W2}-band has a half light radius of $\sim  50\, r_g$. The viscous time-scale at this radius is of the order of 1000\,years, so variations in accretion rate due to viscous processes would not occur over the duration of our observations.
However, the light-crossing time-scale, which governs variable illumination, is around 1\,day and therefore consistent with these observations.

\subsection{X-ray reprocessing}

The variable X-ray emitting corona above the accretion disc directs some of its radiation towards the disc; this is seen both in X-ray reflection spectra \citep[e.g.][]{mcg6tanaka95} and X-ray reverberation \citep[e.g.][]{fabianlag,karaglobal,alston14}. If some of this energy is absorbed by the disc, it will be re-emitted thermally in the UV \citep{lightman88}. This would suggest a connection between the UV and X-ray emission aside from the feeding of the corona with disc photons.

The correlation between X-ray and UV variability with the UV variations occurring after those in the X-rays in 9 of our sources strongly suggests that X-ray variations drive at least part of the UV variability in those objects. This could be explained by some of the UV variability being due to reprocessing of X-ray radiation.

We find that there is comparable variable power in the UV and X-ray bands that we measure. While this does not include all of the power in either band, it shows that the X-ray variations are sufficiently powerful to drive a significant component of the observed UV variability.

We also note that the X-rays have a level of steady emission: the minimum flux of a given source is typically around one third of the peak flux. If variable illumination is indeed producing notable changes in UV flux, the effect of steady illumination will also be a significant factor in the average flux of the disc.

Further support for the reprocessing scenario comes from the fact that of the 9 sources in which we detect correlated variability, 5 also have iron K reverberation lags \citep[][]{karaglobal,IRAS13224kara13,mrk335kara}.
Two of the sources which have measured UV but not iron K lags, Fairall~9 and Ark~120, have relatively high black hole masses ($M_{\rm BH}=8.41$ and $8.18$ respectively) so the expected iron K lags become difficult to detect due to the length of an \textit{XMM} orbit.
Alternatively, the detection of only UV lags may be because the coronal emission is less focussed to the central regions in sources without iron K lags, so more coronal variable power is delivered to the outer regions of the disc which respond in UV.
Equivalently, iron K reverberation is measured in 1H~0707--495 \citep{fabianlag,1h0707zoghbi10,1h0707karalags}, IC~4329A and PG~1211+143 \citep[both][]{karaglobal} but we do not find significant correlations, which may reflect a greater proportion of the coronal variability being focussed towards the central regions.

Some previous studies \citep[e.g.][]{STORM2,troyer16,shappee14} have found that the lags in the DCF are longer than the light travel time for a standard thin disc by a factor of a few. Our measurement of the average lag is not sufficiently precise to distinguish between a standard thin disc and these longer measurements.
Even where lag times are incompatible with a standard thin disc, there are various explanations for this which still allow for X-ray reprocessing to occur. For example, the perceived disc lags can also be increased by UV emission from emission lines and the Balmer continuum \citep{korista01}. If these come from the larger BLR, the lags will appear longer. \citet{dexter11} suggest that local fluctuations in the disc temperature allow hot regions to exist significantly further out than their average radius. These distant hot regions will increase the measured lags while maintaining a thin disc temperature profile on average.

However, the lightcurves are sometimes far from perfectly correlated (for example, the peak correlation coefficient for Fairall~9 is around 0.5), which suggests that at least one of the bands shows variations which do not affect the other.
For example, relativistic light bending \citep{miniutti04} allows the disc to see different variability from a distant observer, particularly if the coronal geometry is changing.
It is likely that there are also intrinsic fluctuations in the disc which add to the UV variable power. Variable reddening of the disc emission would also increase the observed UV variable power. The presence of some additional UV variability is supported by \citet{uttley03var}, who find more fractional variability in NGC~5548 in the optical (5100\,\AA) than X-rays in lightcurves binned on a 30 day time-scale.

We therefore conclude that X-ray reprocessing is likely to happen to some extent in all sources and that it is the origin of the observed UV/X-ray lags.
However, X-ray reprocessing may not be the only driver of UV variability. Since the viscous time-scale is so much longer than the time-scale of the observed variations, any additional fluctuations must be governed by other processes such as magnetism.

\section{UV/X-ray correlations in individual sources -- comparison with previous results}
\label{indsources}

\subsection{1H0707--495}

In agreement with our results, \citet{1h0707robertson15} did not find strong correlations between X-ray and W1 band variability in a 7\,day long \textit{XMM} observation, which they ascribe to a particularly compact corona. They found low significance ($\sim95$\%) UV leads, which we also find (at similarly low significance) in the longer lightcurves presented here. These could be due to upscattering of UV emission in the corona.
Seemingly at odds with these findings, X-ray studies have found strong evidence of iron K and L reverberation lags \citep{fabianlag,1h0707zoghbi10,1h0707karalags} and relativistic reflection \citep{1h0707karalow,1h0707dauser12} in this source.
However, the \textit{Swift} monitoring includes a period where 1H0707--495 is in a very low state, and the X-ray reflection is concentrated towards the centre-most region of the disc due to illumination from a low corona ($\sim2R_{\rm g}$) \citep{1h0707fabian12}. The most illuminated region ($\lesssim10R_{\rm g}$) is smaller than the region ($\sim500R_{\rm g}$) responsible for the bulk of the \textit{W2}-band emission. This would make UV--X-ray lags hard to detect.

\subsection{Fairall~9}

We detect UV lags in all 6 bands, which are largely consistent with the expectations from a thin disc.
\citet{f9recondo97} find variability of up to a factor of $33\pm4$ in \textit{IUE} data of the source, but the mean sampling interval of 96\,days is too long to detect lags, giving an upper limit on the lag of 4298\,\AA\ behind 1400\,\AA\ of 80\,days, consistent with our results of lag lengths of a few days.
\citet{f9lohfink14} study 2.5\,months of \textit{Swift} monitoring and an \textit{XMM} observation. They find correlated variability between UV-optical bands on all time-scales measured, down to the \textit{Swift} sampling time. The additional data now available allow us to show that the longer wavelengths lag the shorter ones. They also find rapid UV flares in the \textit{XMM} observation, with a lag behind the X-rays of $1-2$\,hours at 2$\sigma$ significance. We do not have sufficient sampling cadence to verify this result.

\subsection{MCG--6-30-15}
\label{sub:mcg6}
Despite the well-sampled \textit{M2}-band lightcurve, we are unable to detect lags in MCG--6-30-15.
\citet{mcg6lira} studied long-term lightcurves in X-rays and optical/near-IR; they found only a weak correlation between X-rays and the \textit{B}-band which does not put useful constraints on the lag. They do find correlations between different optical bands with longer wavelengths lagging shorter ones; these lags are consistent with a $\tau\propto\lambda^{4/3}$ relation but for a disc larger than expected by up to a factor of 4, indicating that weak reprocessing might be taking place.
\citet{mcg6arevalo05} studied a $\sim5$\,day long \textit{XMM} observation and found that the \textit{U}-band emission leads the X-rays by $1.9_{-0.8}^{+0.5}$\,days, as would be expected for X-rays which are produced by Compton upscattering of UV photons. We would expect to detect such a lag if it were present in the observations analysed here.

While a broad iron line attributed to disc reflection has been detected in the mean X-ray spectrum \citep{mcg6tanaka95,mcg6fabian03},
\citet{mcg6kara} were unable to find an iron K lag despite having enough counts and variability. They suggest that some variability may be due to geometrical changes in the corona, which would not cause correlated changes in the iron K emission. \citet{mcg6miller08} suggest that the red wing of the iron line may be caused by complex absorption, which would not show an iron K lag.
However, using high energy \textit{NuSTAR} data, \citet{mcg6marinucci14} favour the relativistic reflection interpretation.

Rapid changes of the geometry of the source could explain the difficulty of detecting X-ray reverberation and reprocessing in the source despite the existence of reflection features in the spectrum.

\subsection{Mrk~335}

We detect a correlation between X-rays and the \textit{W2}-band but only constrain the lag to $0\pm2.9$\,days. \citet{mrk335grupe12} studied the first half of the \textit{Swift} monitoring used here, finding strong variability but no significant correlation between the X-ray and UV flux. The additional \textit{W2} measurements collected after 2012 allow for the detection of a significant correlation now.
While we do not determine the direction of the lag, strong reflection \citep[e.g.][]{parker14} indicates that the X-rays illuminate the disc, so some reprocessing is likely to occur and may well be the cause of the correlated variability.

\subsection{Mrk~509}

We do not detect a significant lag in Mrk~509. Since we have only 27 data points for Mrk~509, we would not expect to do so.
\citet{mrk509marshall08} find the optical (\textit{R}-band) flux leads the X-rays by 15\,days in observations from \textit{RXTE} and ground-based measurements by the SMARTS consortium.
\citet{mehdipour11} find a correlation between disc and soft X-ray flux in \textit{XMM} and \textit{Swift} measurements, which they attribute to warm Comptonisation producing the soft excess.
This interpretation agrees with \citet{mrk509boissay14}, who study spectra from  a large \textit{XMM}/\textit{INTEGRAL} campaign.

\subsection{NGC~3516}

We find a $1.6\pm1.5$\,day lag of the \textit{W2}-band behind the X-rays.
\citet{edelson00} find no significant UV/X-ray correlation in a 3\,day observation with \textit{HST}, \textit{RXTE} and \textit{ASCA}. Such a short observation is unlikely to detect a lag of the length which we find.
\citet{maoz02} find a possible 100\,day lag of the X-ray relative to the \textit{R}-band but the correlation is not detected in a longer observation \citep{maoz02}. They suggest that this may be due to the initial section of the observation being at a higher flux level, in which the X-rays are dominated by a component of emission which does correlate with the UV but is less significant at lower fluxes.
\citet{noda16} studied observations from \textit{Suzaku} and Japanese ground-based telescopes. They found a correlation between the hard X-rays and the \textit{B}-band with the X-rays lagging by $2.0_{-0.6}^{+0.7}$\,days, which like our measurement is larger than expected for a thin disc.

\subsection{NGC~4051}

The \textit{Swift} data for NGC~4051 are insufficient in number and frequency to measure the expected lags. While we find a formally significant correlation at 15\,days lag, this corresponds to superposing the well sampled ends of the 35\,day lightcurves with the unsampled middle of the lightcurve.
\citet{peterson00} analysed 3\,years of \textit{RXTE} and ground-based observations and found that the long time-scale (>30\,day) variability is correlated between optical and X-rays with a lag range of $-106$ to 68\,days.
\citet{ngc4051shemmer03} found in 60 days of intensive \textit{RXTE} monitoring
that the DCF centroid showed a UV lead but that the peak may be at a lag, suggesting that both inward propagating fluctuations and X-ray reprocessing are responsible for some of the correlation.
\citet{ngc4051alstonuvx} studied the UV/X-ray variability on short time-scales with \textit{XMM} and found a 3\,ks \textit{W1}-band lag relative to the X-rays. The strength of correlation indicates that 25\% of the UV variance is caused by X-rays. The lag is somewhat shorter than the $\sim0.2$\,day lag detected by \citet{ngc4051mason02} with \textit{XMM} in the same band at only 85\% confidence.
\citet{breedt10} correlated 12\,years of \textit{RXTE} observations with \textit{u} to \textit{I}-band measurements, finding that a $\lambda^{4/3}$ relation fits well and that the scaling, subject to significant uncertainties, is consistent with thin disc predictions. As well as lags of a few days, they find lags of $\sim40$\,days which they suggest may be due to the dusty torus which surrounds the inner regions.

\subsection{NGC~5548}

We find lags in all bands, increasing with wavelength and always longer than predicted for a thin disc, but consistent within the uncertainties.
The \textit{STORM} campaign has provided an extensive dataset from X-rays to IR: \citet{STORM2} measure the UV/X-ray lags against the Hubble 1315\,\AA\ band; and \citep{fausnaugh16} extend the wavelength range to the \textit{z}-band ($\sim9160$\,\AA). This finds that the lags are broadly consistent with $\tau\propto\lambda^{4/3}$ but that the disc radius is around 3 times the thin-disc prediction. This is in agreement with our findings, which is expected as the majority of \textit{Swift} data for NGC~5548 is part of the \textit{STORM} campaign.

\subsection{NGC~7469}

We detect significant correlations in 4 UV bands. Our lag measurements are consistent with a thin disc but are poorly constrained.
Studying the UV only, 
\citet{ngc7469wanders97} find delays of UV lines and continuum behind the emission at 1315\,\AA . They find lags increasing with wavelength over 1315--1825\,\AA.
\citet{ngc7469collier} find lags between 1315\,\AA\ and 4865,6962\,\AA.
\citet{ngc7469kriss00} use a \textit{HST-FOS} spectrum over 1150--3300\,\AA\ to better extract spectral bands which are less contaminated by line emission.   These measurements of UV continuum lags find that the lags follow a $\lambda^{4/3}$ relation.
\citet{collier99} use these lags to determine $H_0$; their value for $H_0=42\pm9$\,km\,s$^{-1}$\,Mpc$^{-1}$ is lower than is now accepted, corresponding to the lags being longer than expected.
However, \citet{ngc7469nandra98} find a 4\,day UV (1315\,\AA) lead relative to the X-rays. This could indicate that UV upscattering and X-ray reprocessing are both responsible for some of the correlations.

\subsection{PG~1211+143}

We find peaks in the DCF at $0\pm5$\,days, consistent with previous findings, but these are not significant at the 99\% level.
\citet{bachev09} studied the first section of \textit{Swift} observations, along with ground-based photometry down to \textit{I}-band. They found lags compatible with a $\lambda^{4/3}$ relation at approximately twice the expected lag. \citet{papadakis16} also analyse the first \textit{Swift} section of this dataset but find that the UVOT measurements are consistent with constant flux.
\citet{lobban16} studied the second section of \textit{Swift} observations along with \textit{XMM-PN/OM} data, finding marginally significant X-ray/UV correlations with lags $\lesssim1$\,day.

\section{Conclusions}

We have presented a variability analysis of archival \textit{Swift} data from AGN monitoring.

We find that essentially all bands vary and that the variable part of the UV emission has a spectrum consistent with that of the thermal emission from dissipation in an accretion disc or central illumination of a flat disc. The time-scales of variability and lags of UV relative to X-ray variability show that the latter is principally responsible.

The variable power in sources with heavier black holes is higher.
The variable UV power increases faster than the variable X-ray power, as is the case for the average emission.

The X-ray and UV variations are significantly correlated in 9 sources; the data for the remaining sources are not sufficient to detect a correlation.
All measurements of correlated X-ray/UV variability are consistent with the UV lagging the X-rays.
We associate this with the reprocessing of X-rays on the accretion disc.

\section*{Acknowledgements}

We thank the referee for helpful comments. AL acknowledges useful discussions with Richard Mushotzky. ACF, AML and DJKB acknowledge support from the ERC Advanced Grant FEEDBACK 340442. WNA acknowledges support from the European Union Seventh Framework Programme (FP7/2013-2017) under grant agreement n.312789, StrongGravity. DB acknowledges an STFC studentship. This work made use of data supplied by the UK Swift Science Data Centre at the University of Leicester. 

\bibliographystyle{mnras}
\bibliography{paper}

\clearpage
\appendix
\section{Lightcurves}
Lightcurves for each source in our sample in each band. Rates are given in cts\,s$^{-1}$; X-rays are measured over 0.3-10\,keV
\begin{figure}
        \includegraphics[width=\columnwidth]{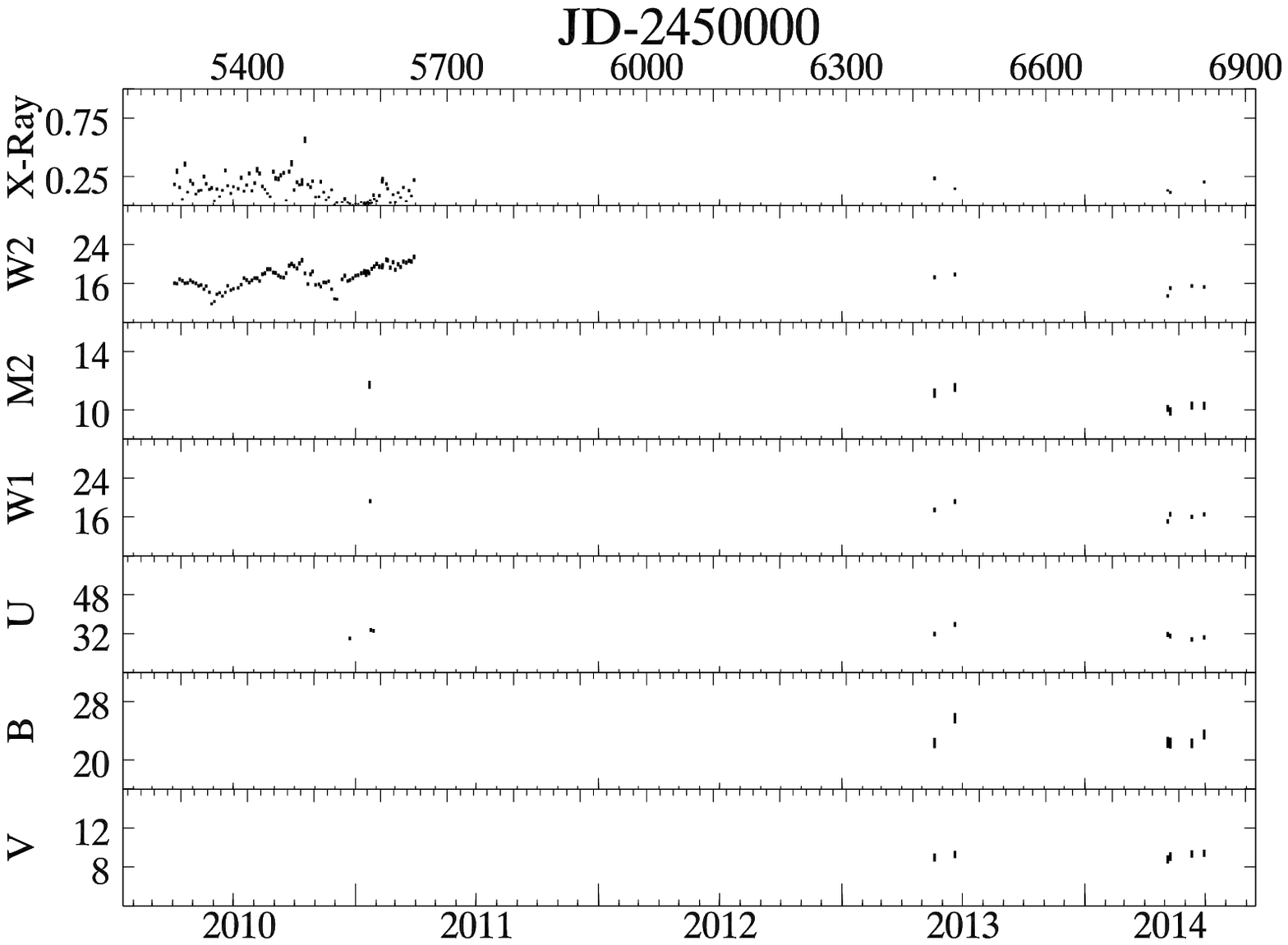}
    \caption{ 1H 0707--495}
    \label{fig:lc1h0707}
\end{figure}
\begin{figure}
        \includegraphics[width=\columnwidth]{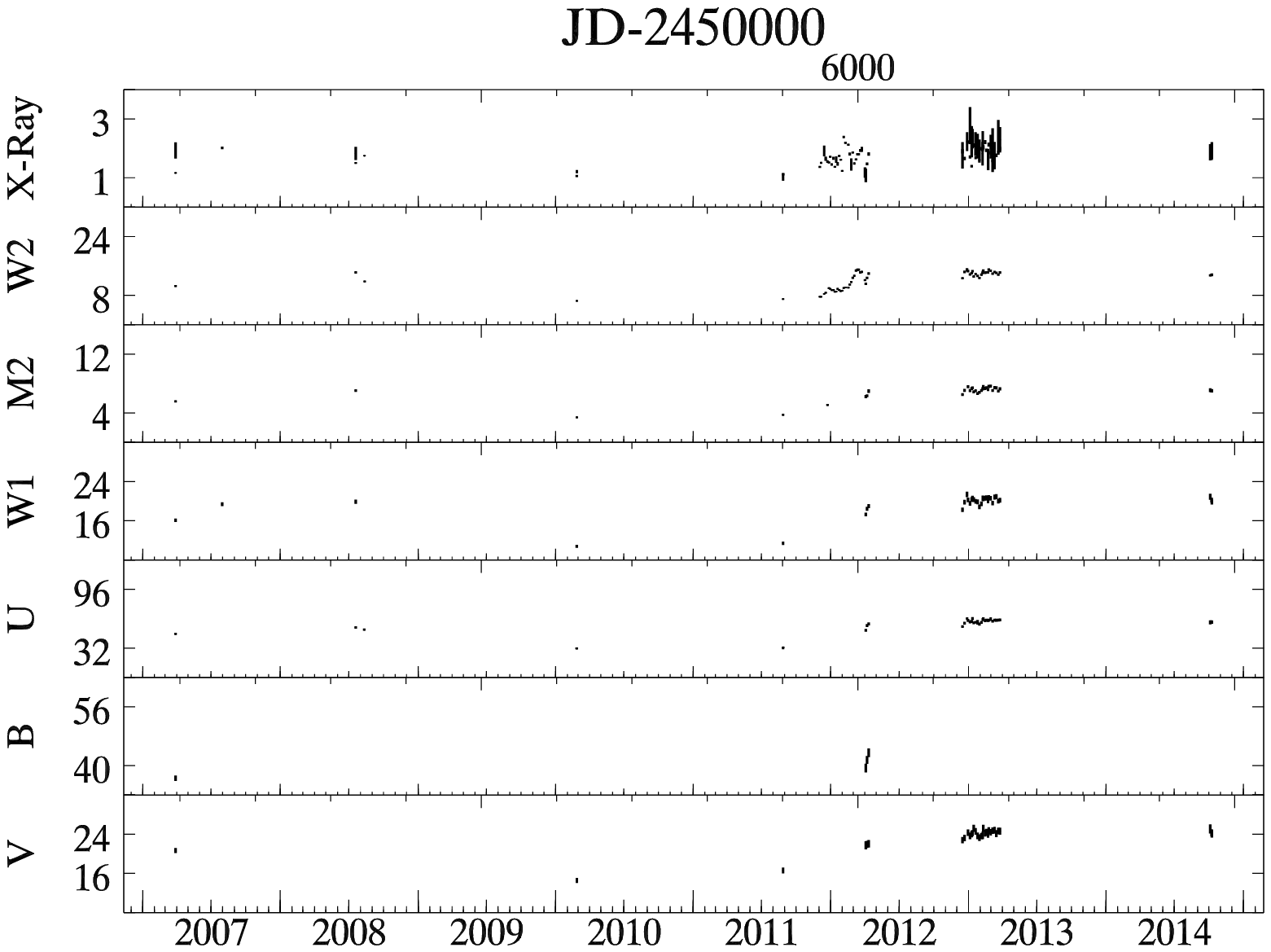}
    \caption{ 3C 120}
    \label{fig:lc3c120}
\end{figure}
\begin{figure}
        \includegraphics[width=\columnwidth]{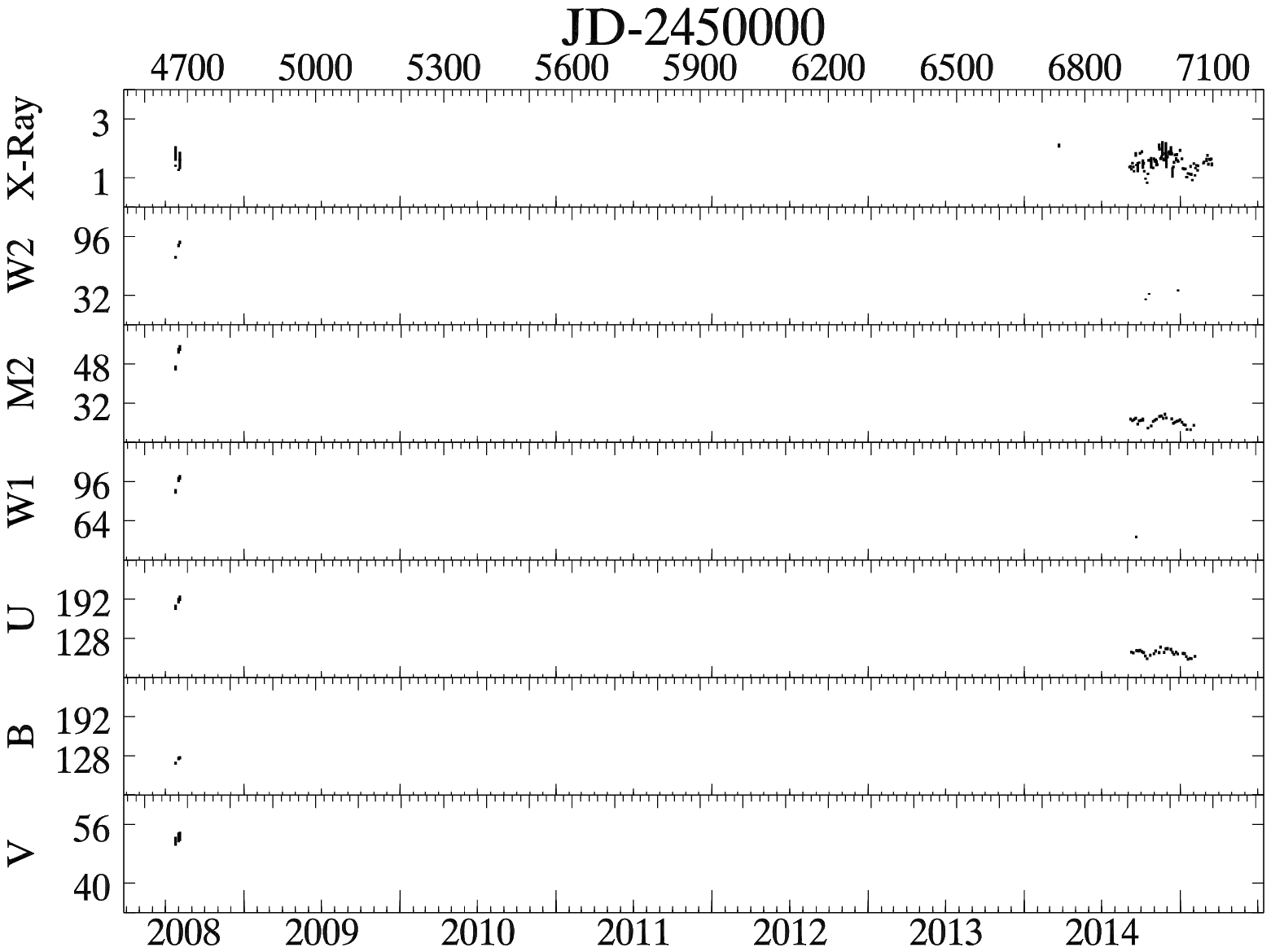}
    \caption{ARK 120}
    \label{fig:lcark120}
\end{figure}
\begin{figure}
        \includegraphics[width=\columnwidth]{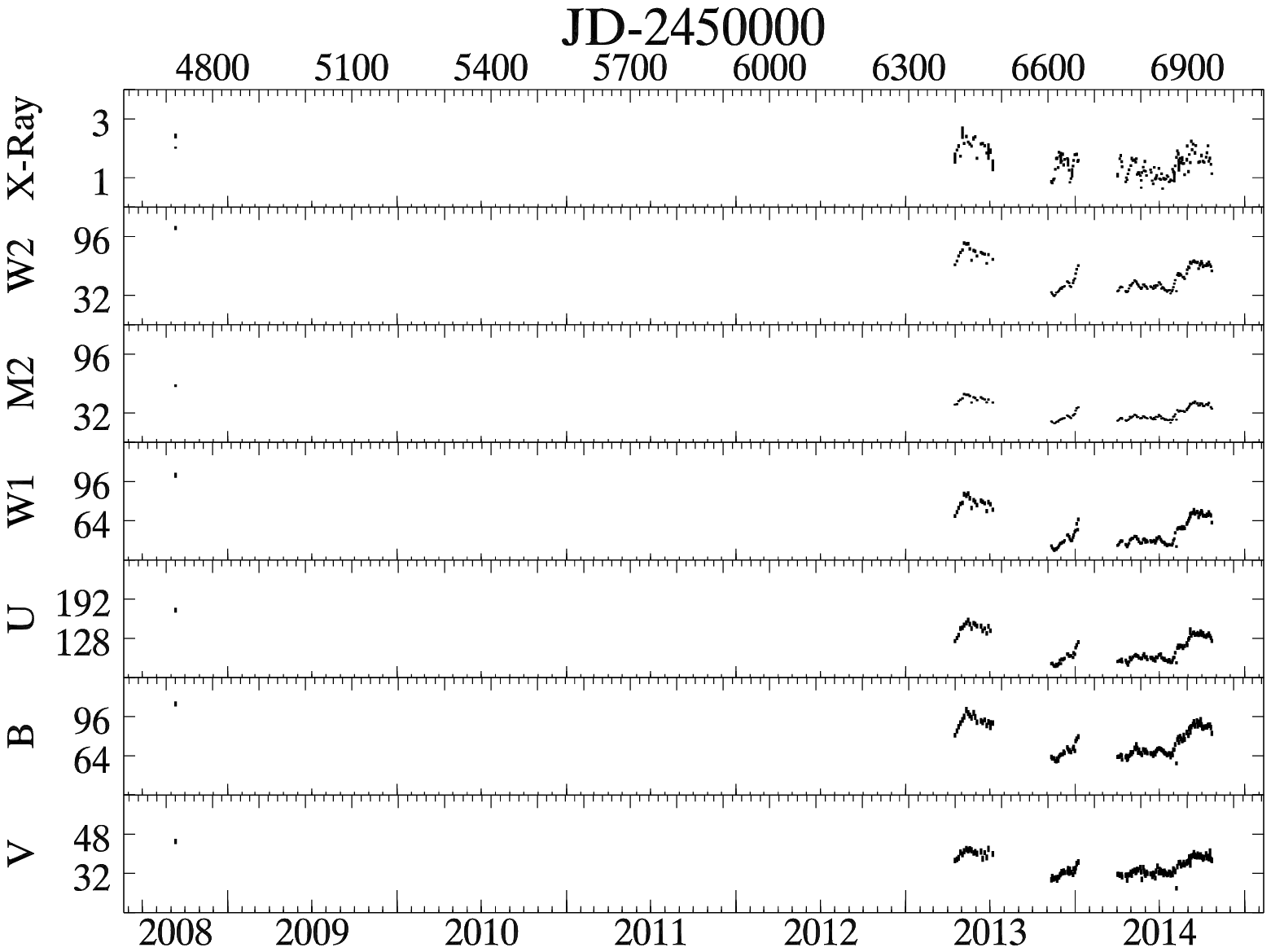}
    \caption{ Fairall 9}
    \label{fig:lcf9}
\end{figure}
\clearpage
\begin{figure}
        \includegraphics[width=\columnwidth]{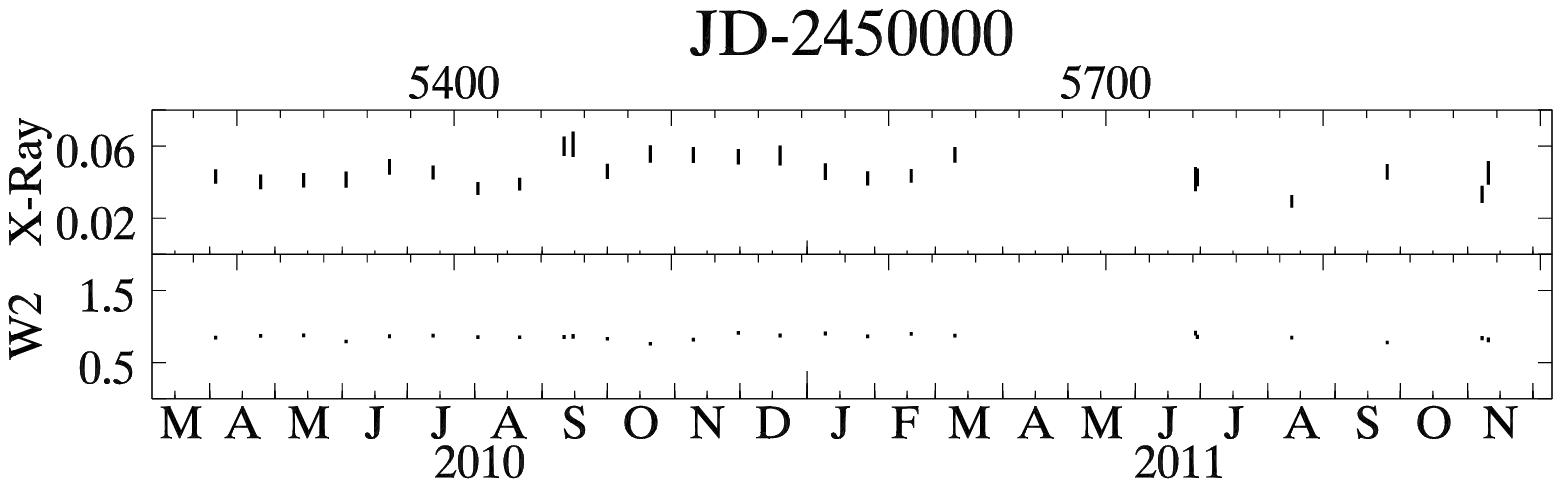}
    \caption{ H 0557--385}
    \label{fig:lch0557}
\end{figure}
\begin{figure}
        \includegraphics[width=\columnwidth]{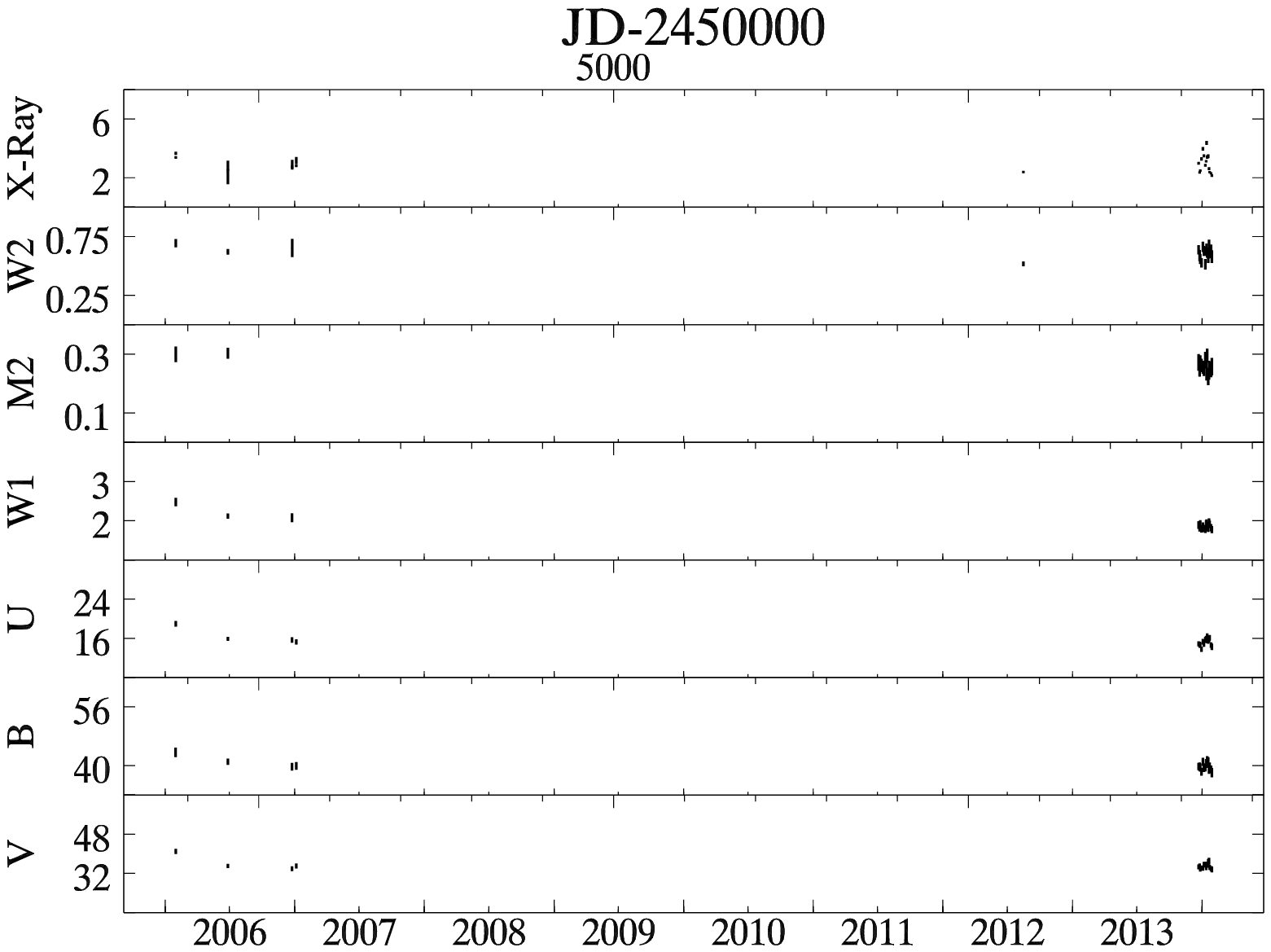}
    \caption{IC 4329A}
    \label{fig:lcic4329a}
\end{figure}
\begin{figure}
        \includegraphics[width=\columnwidth]{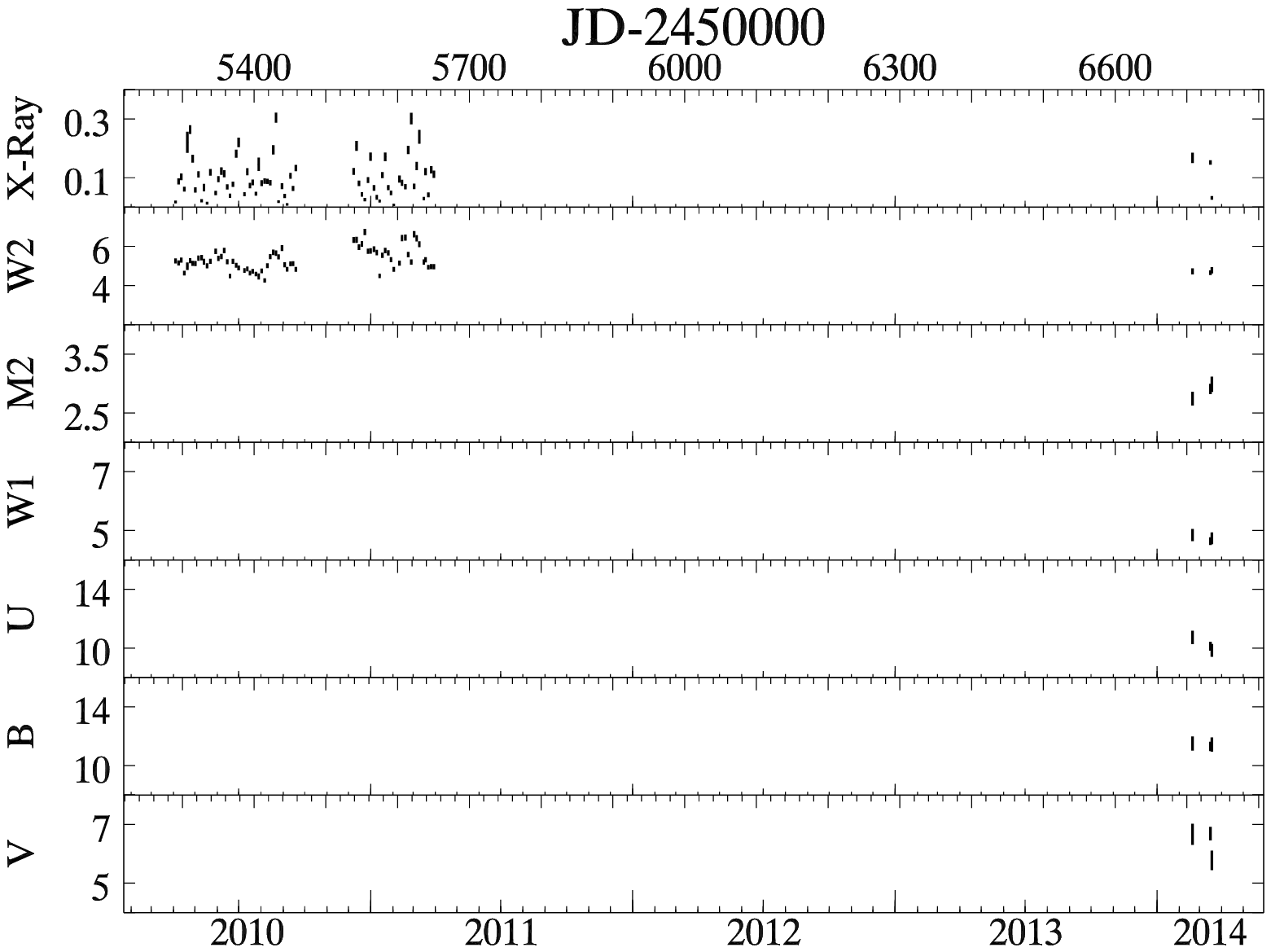}
    \caption{ IRAS 13224-3809}
    \label{fig:lcIRAS13224}
\end{figure}
\begin{figure}
        \includegraphics[width=\columnwidth]{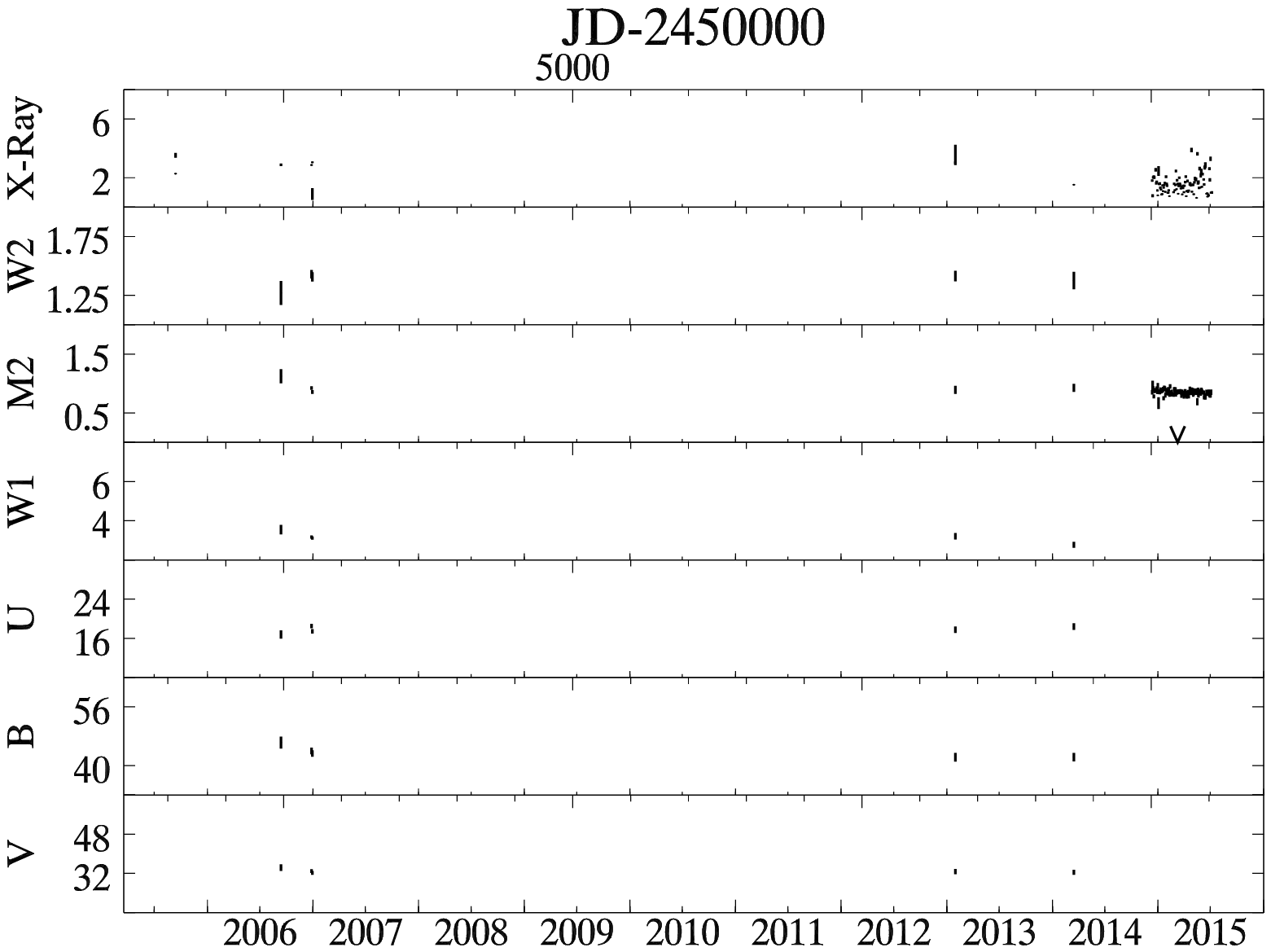}
    \caption{ MCG--6-30-15}
    \label{fig:lcmcg6}
\end{figure}
\begin{figure}
        \includegraphics[width=\columnwidth]{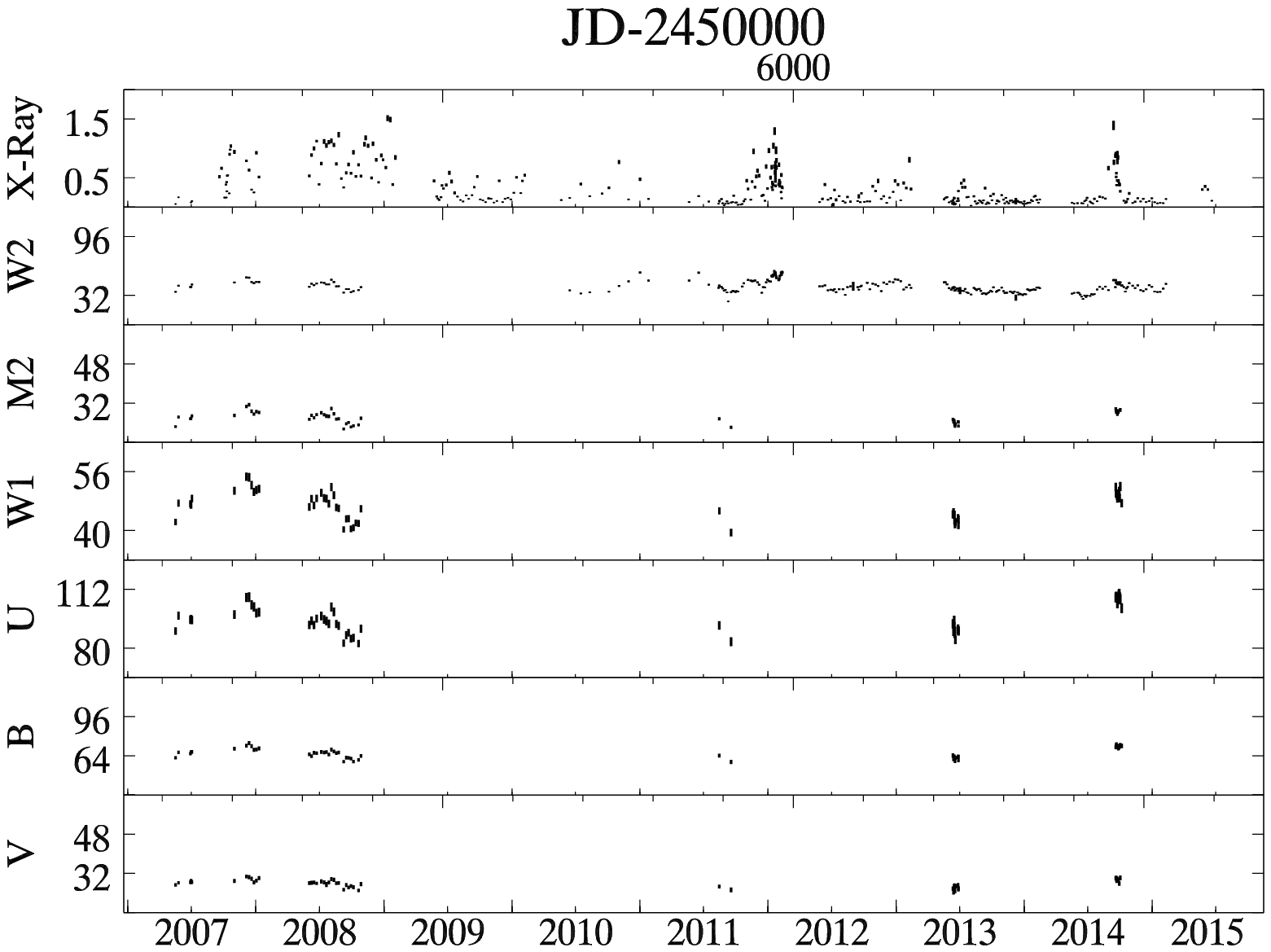}
    \caption{MRK 335}
    \label{fig:lcmrk335}
\end{figure}
\begin{figure}
        \includegraphics[width=\columnwidth]{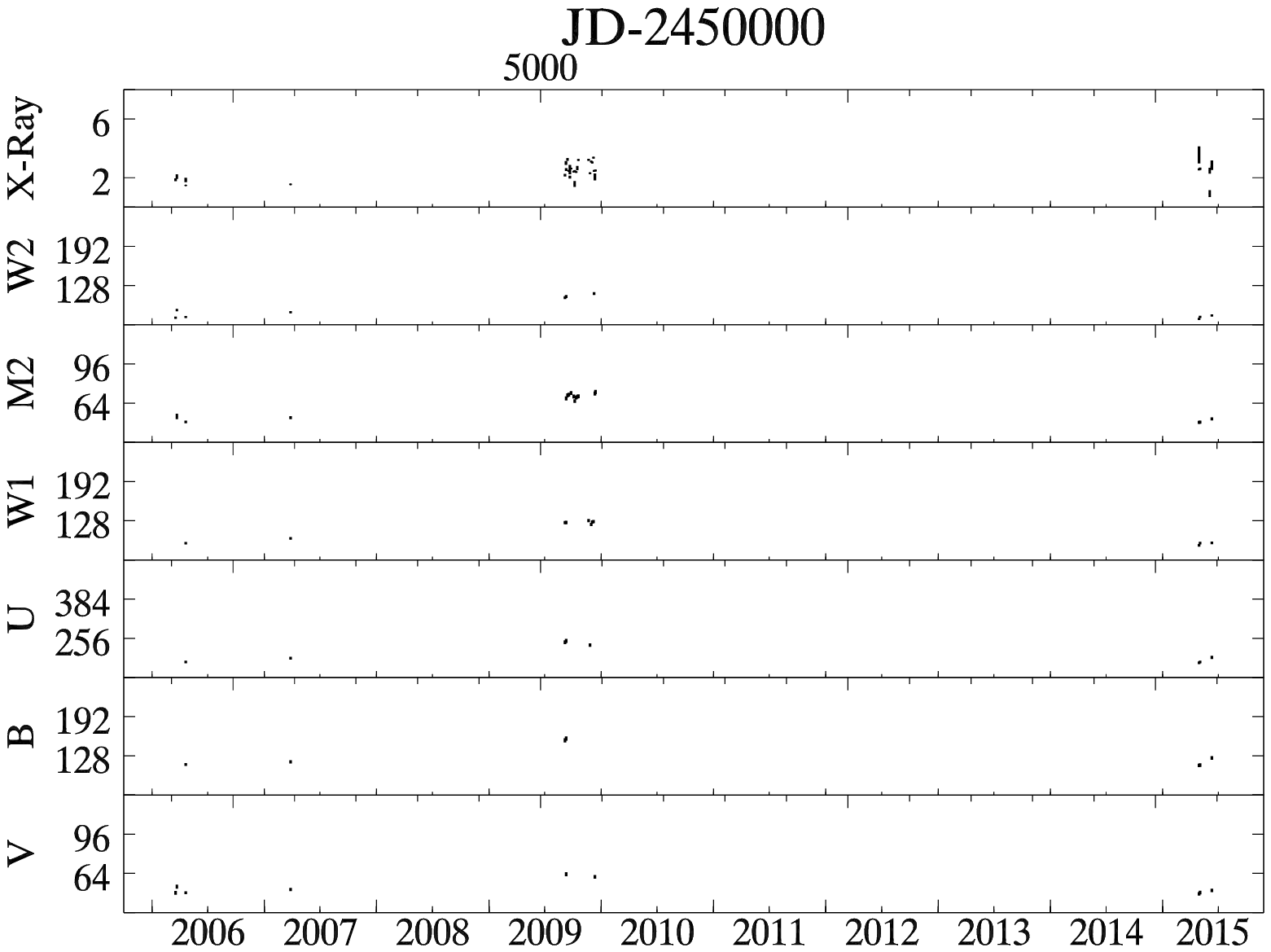}
    \caption{MRK 509}
    \label{fig:lcmrk509}
\end{figure}
\clearpage
\begin{figure}
        \includegraphics[width=\columnwidth]{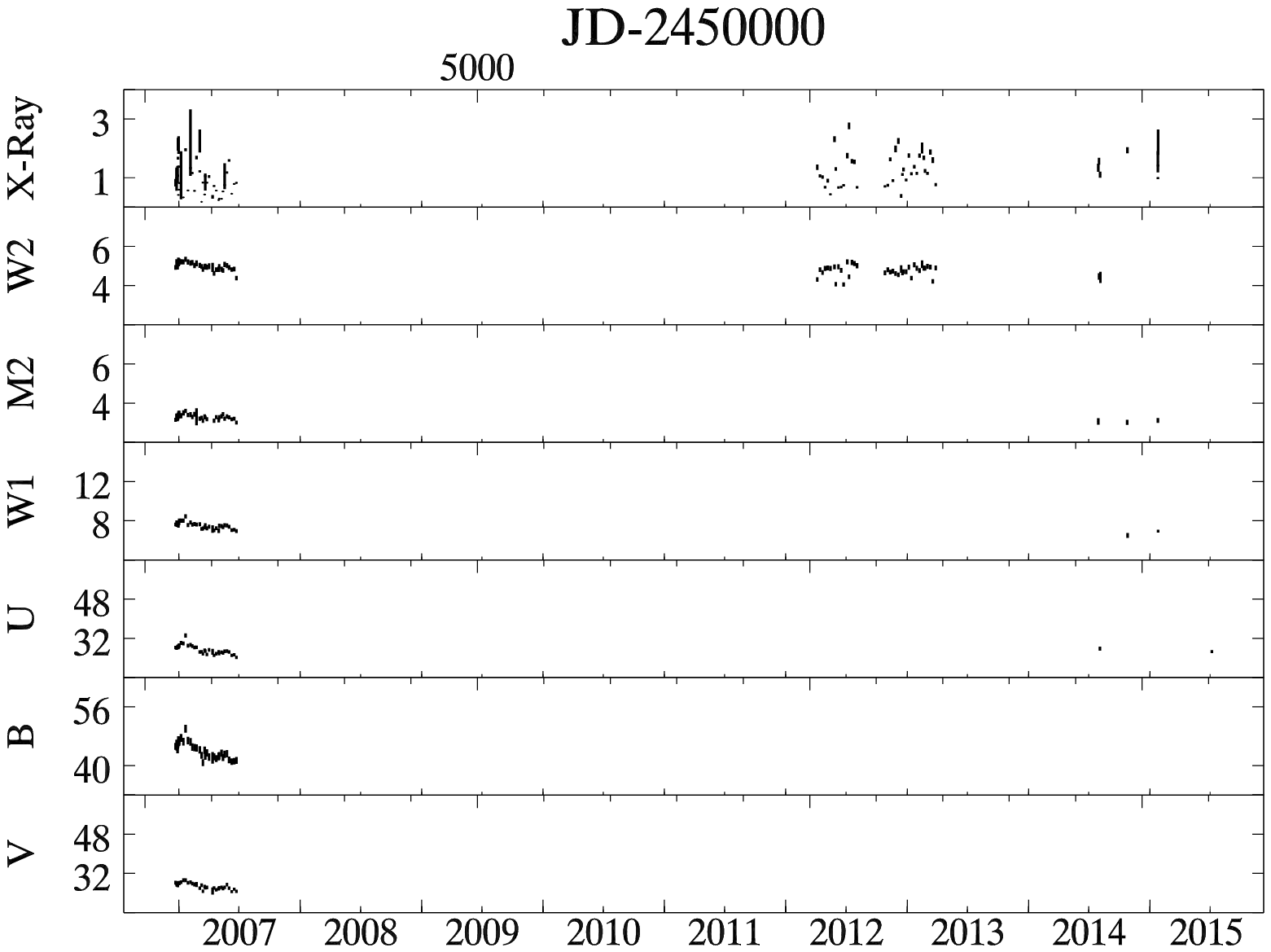}
    \caption{MRK 766}
    \label{fig:lcmrk766}
\end{figure}
\begin{figure}
        \includegraphics[width=\columnwidth]{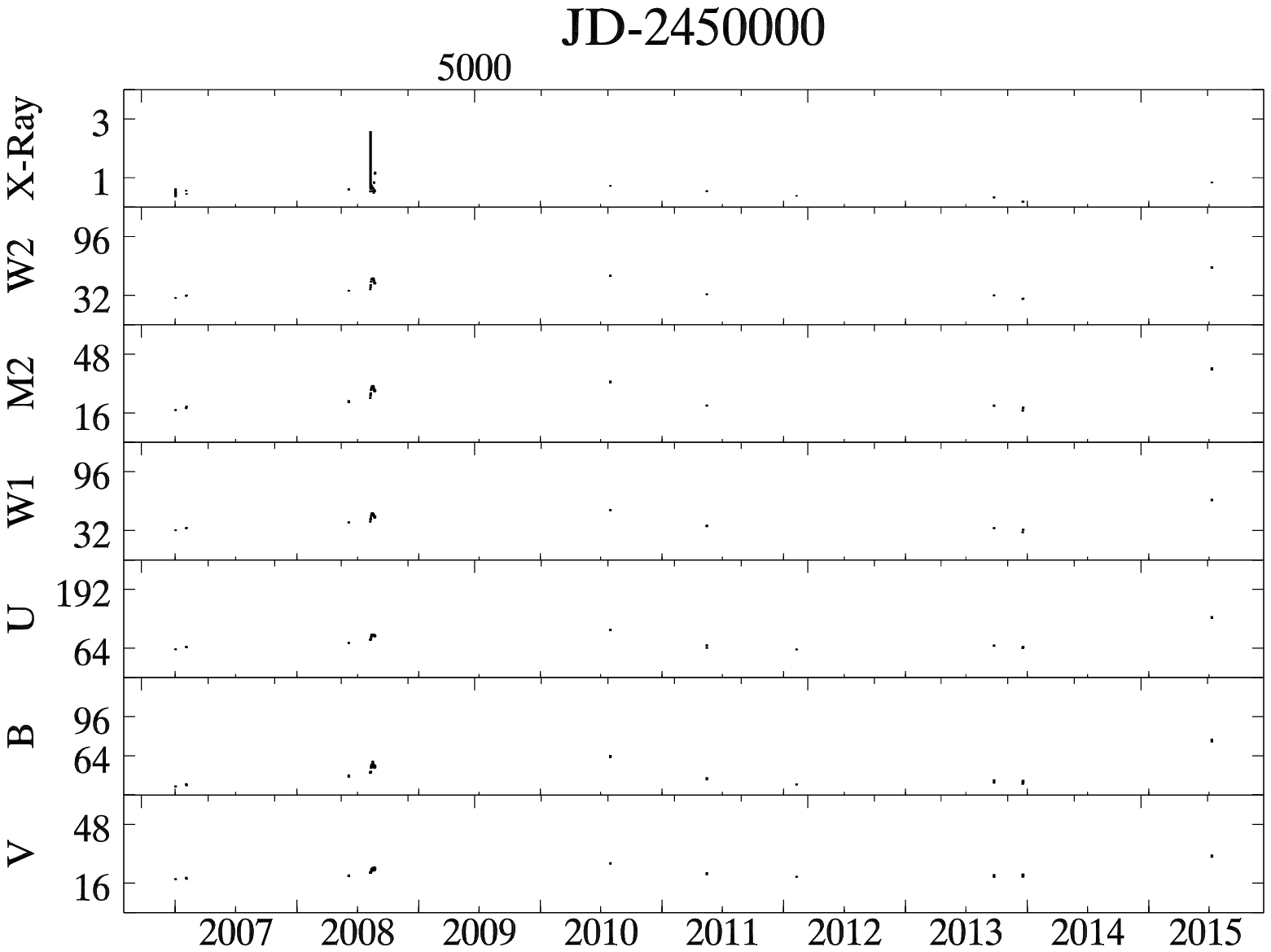}
    \caption{MRK 841}
    \label{fig:lcmrk841}
\end{figure}
\begin{figure}
        \includegraphics[width=\columnwidth]{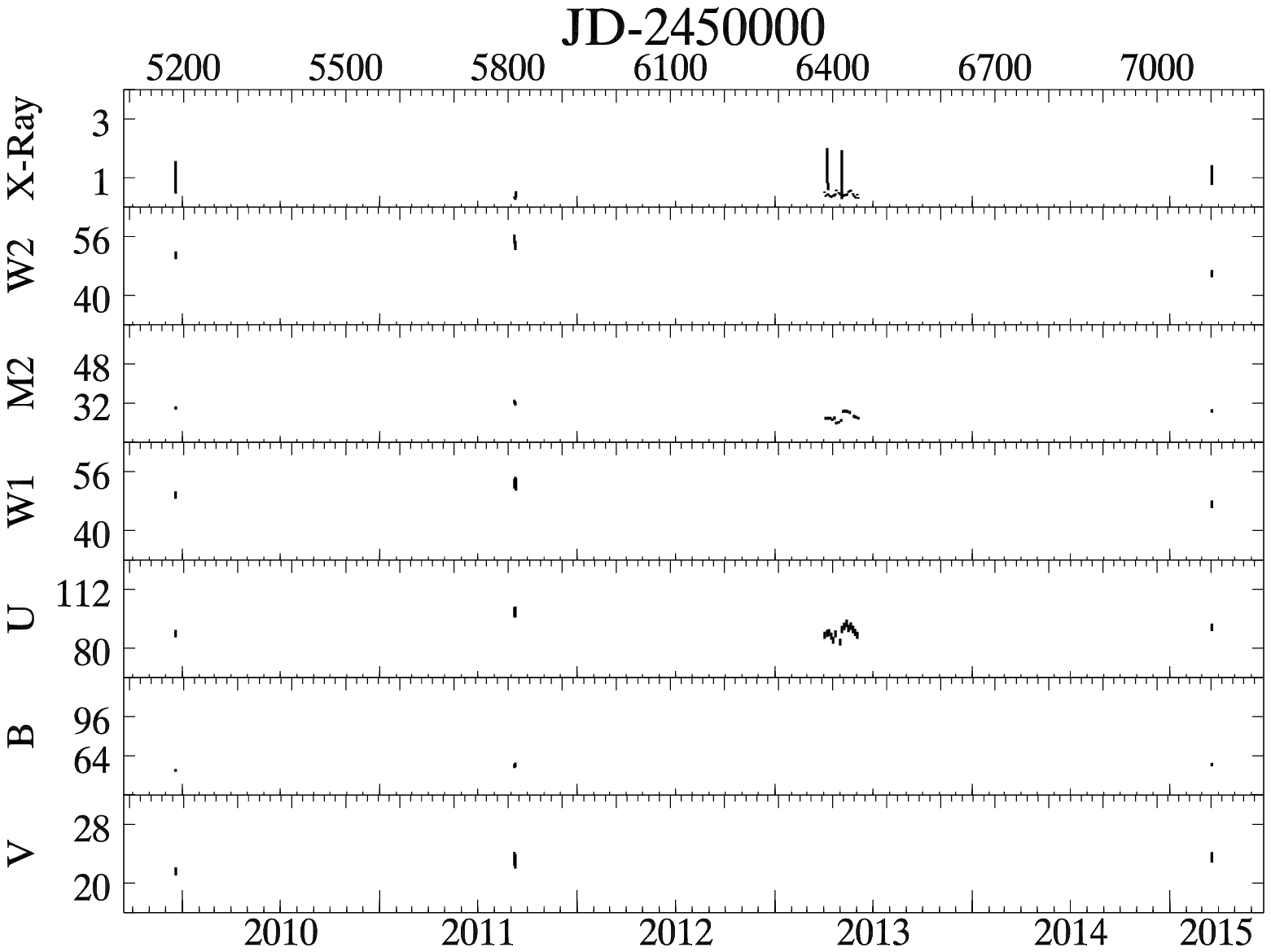}
    \caption{MRK 1383}
    \label{fig:lcmrk1383}
\end{figure}
\begin{figure}
        \includegraphics[width=\columnwidth]{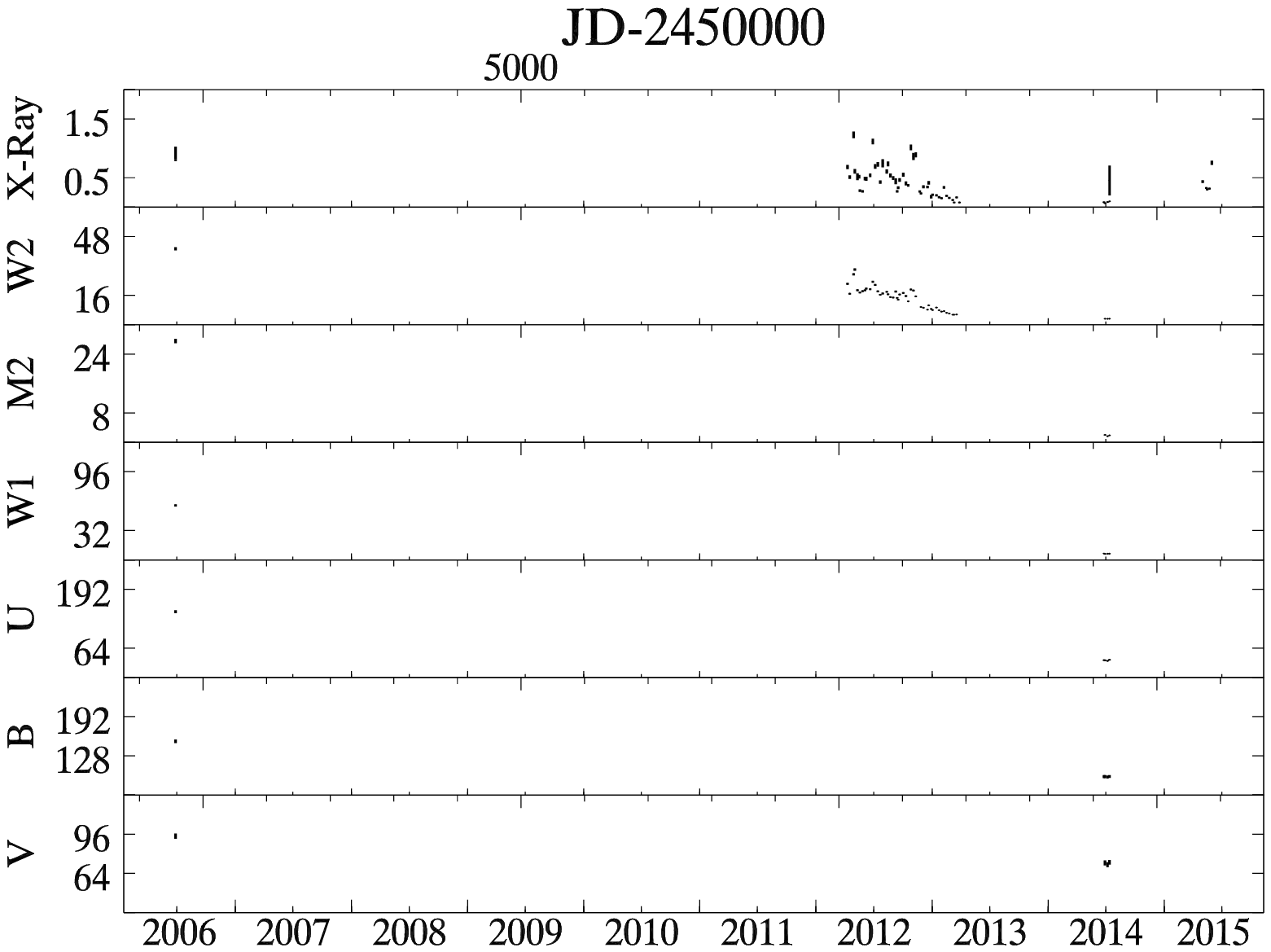}
    \caption{NGC 3516}
    \label{fig:lcngc3516}
\end{figure}
\begin{figure}
        \includegraphics[width=\columnwidth]{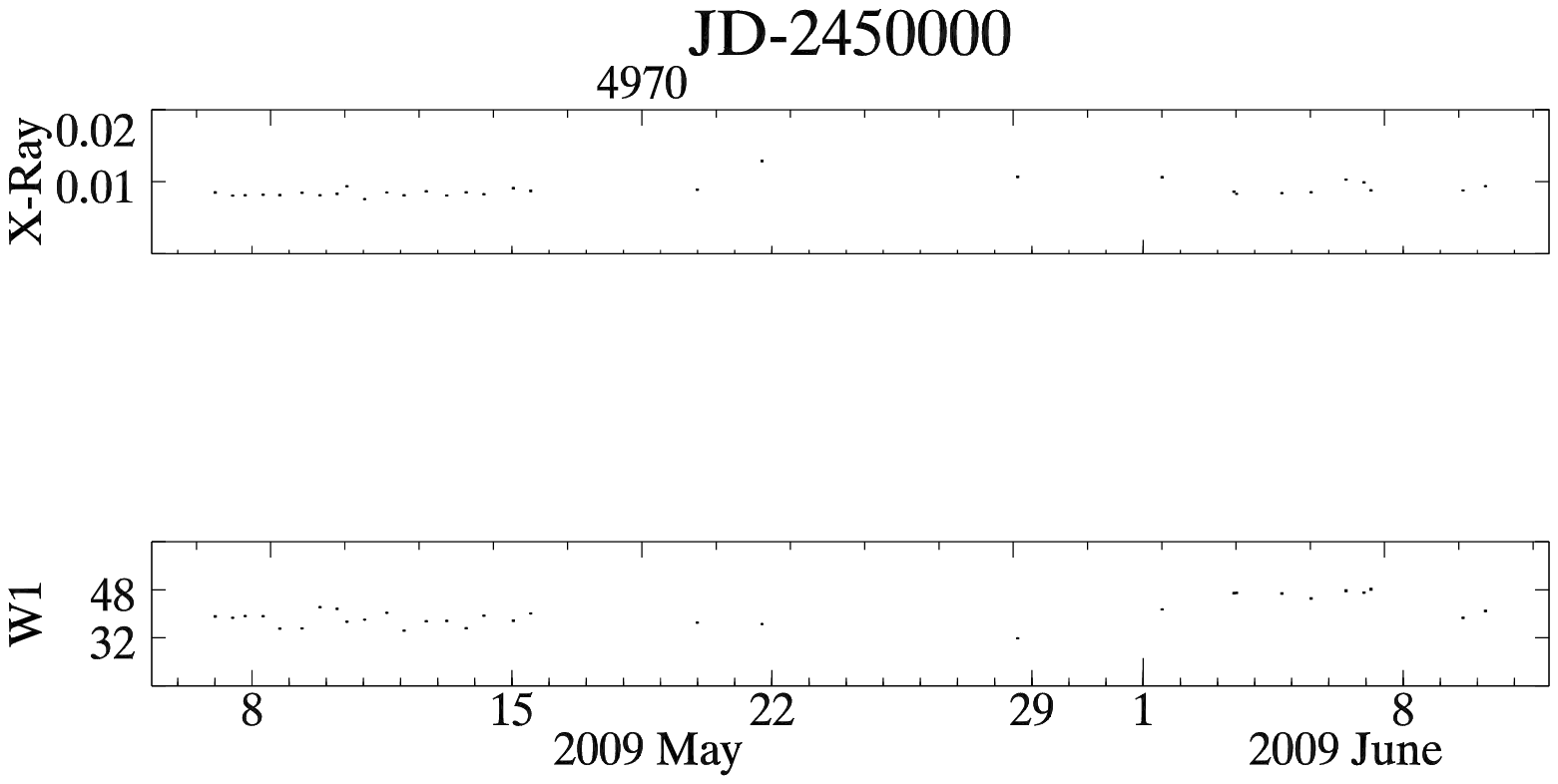}
    \caption{NGC 4051}
    \label{fig:lcngc4051}
\end{figure}
\begin{figure}
        \includegraphics[width=\columnwidth]{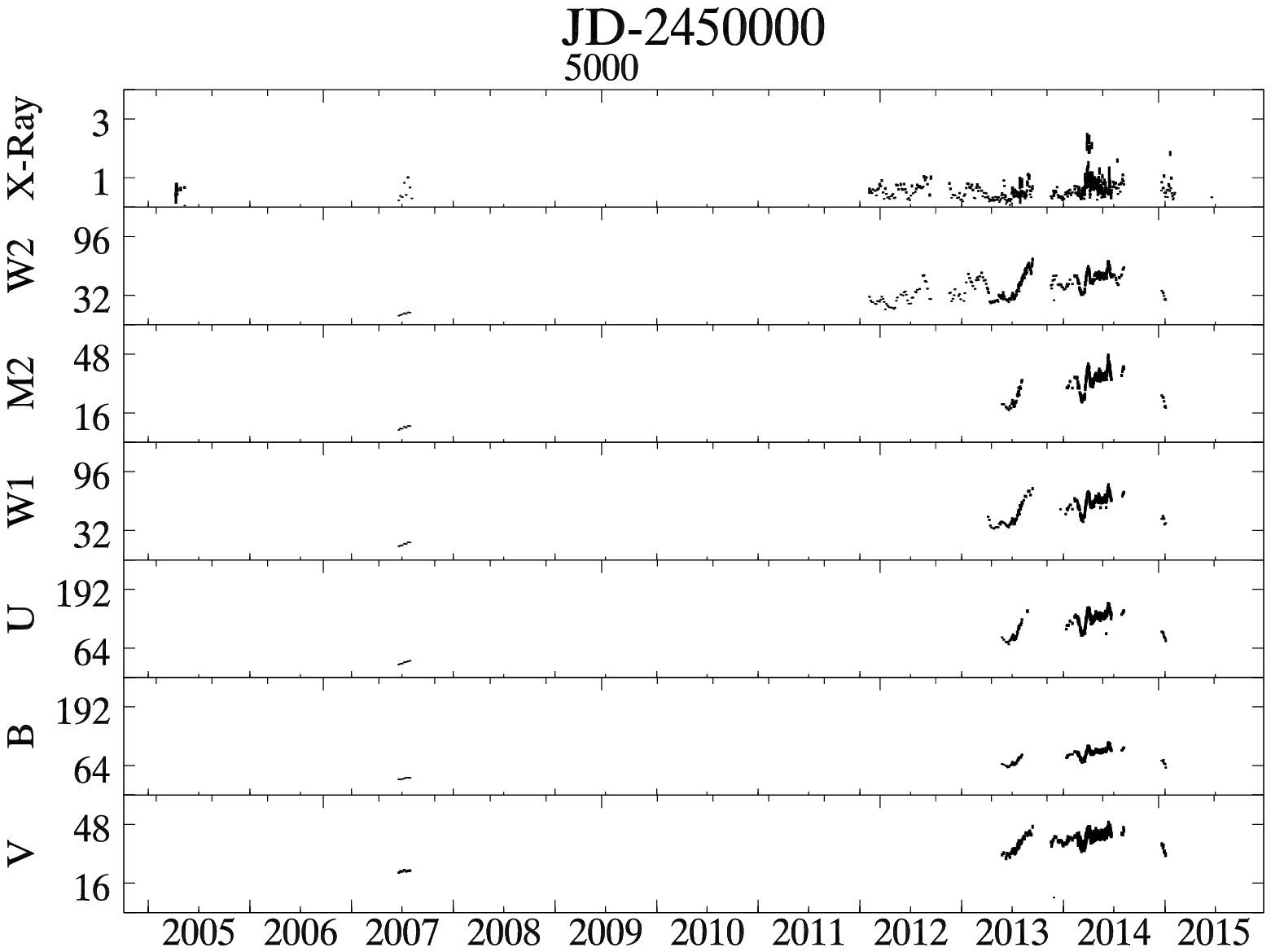}
    \caption{NGC 5548}
    \label{fig:lcngc5548}
\end{figure}
\clearpage
\begin{figure}
        \includegraphics[width=\columnwidth]{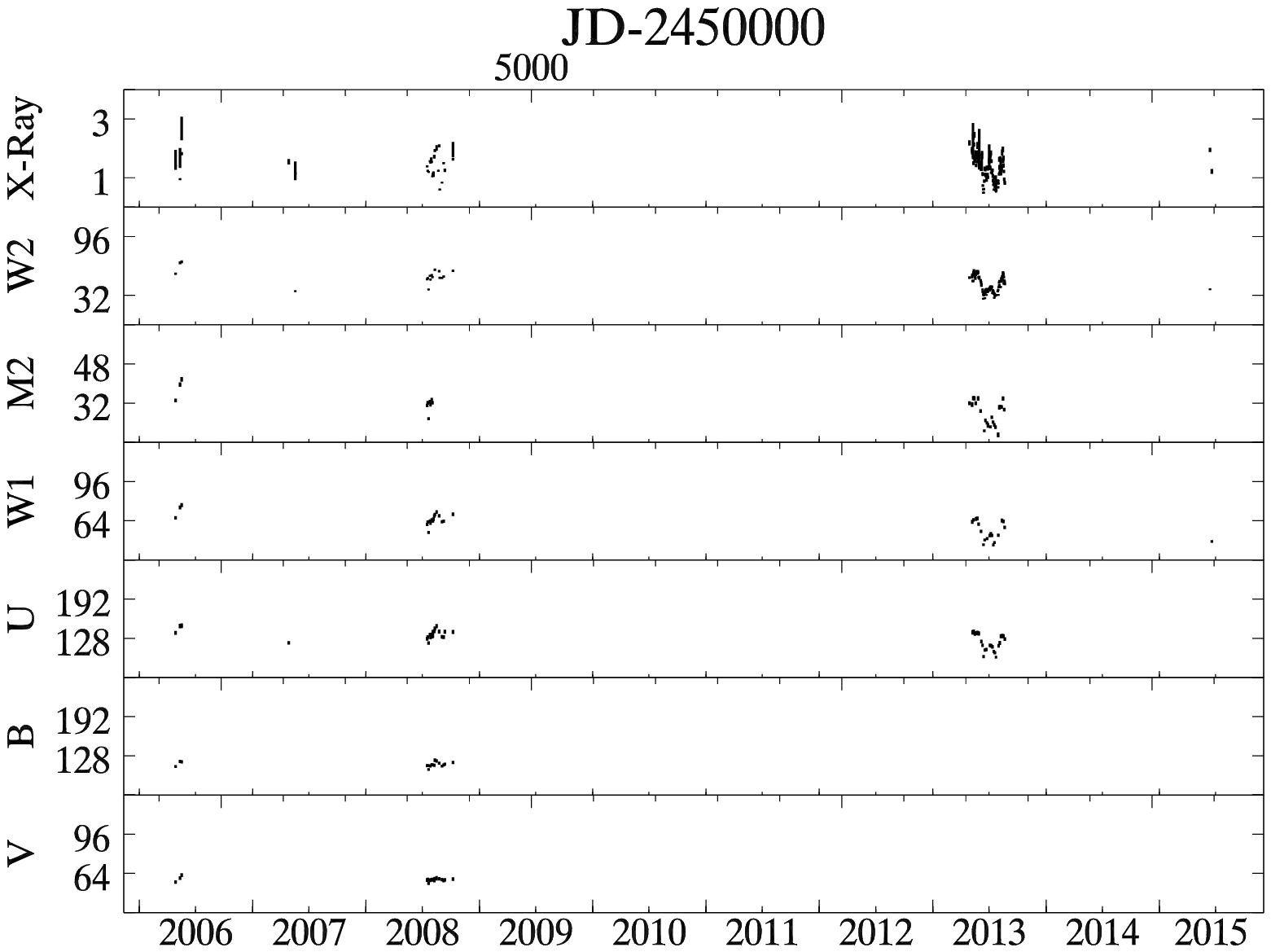}
    \caption{NGC 7469}
    \label{fig:lcngc7469}
\end{figure}
\begin{figure}
        \includegraphics[width=\columnwidth]{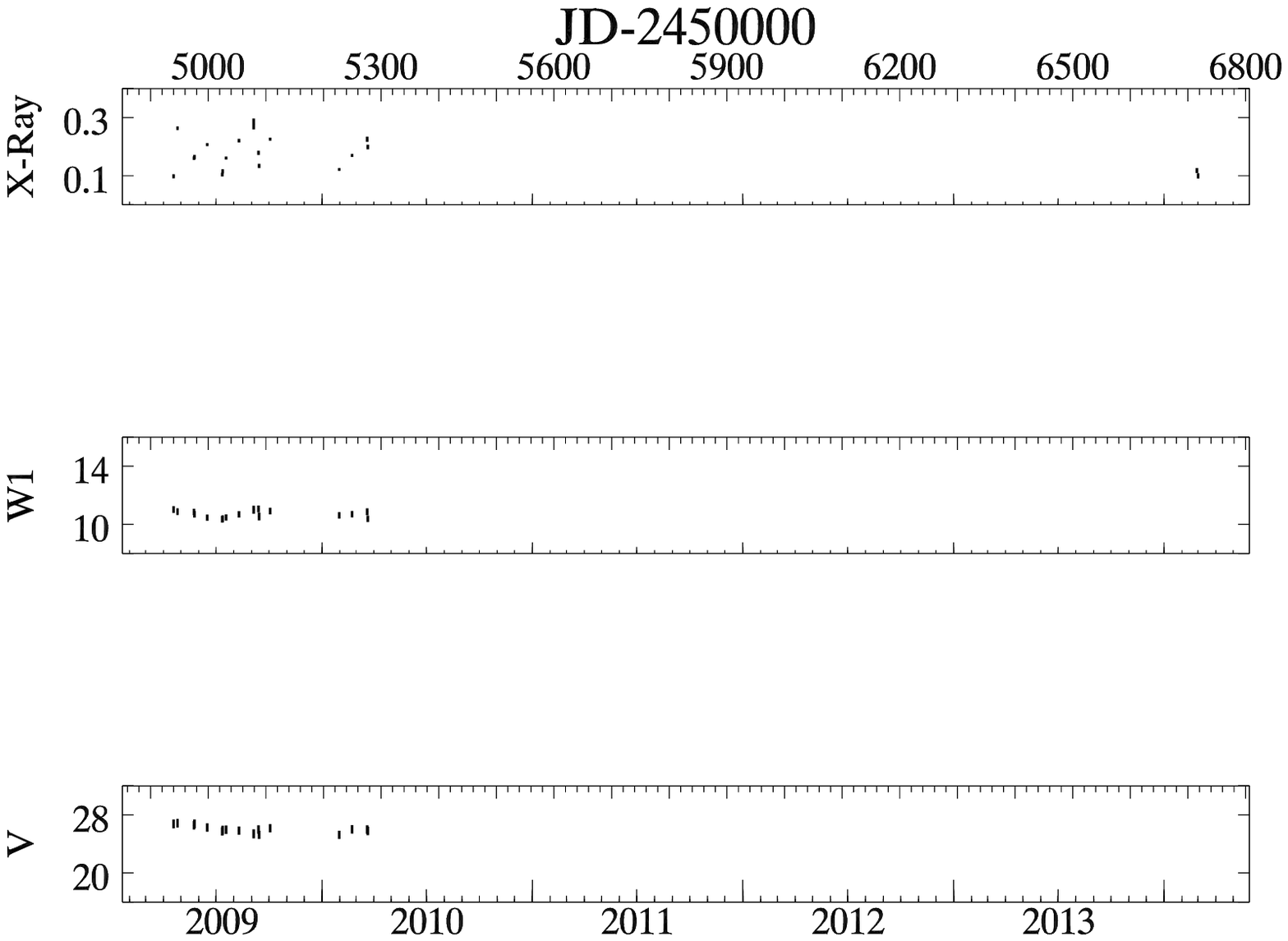}
    \caption{PDS 456}
    \label{fig:lcpds456}
\end{figure}
\begin{figure}
        \includegraphics[width=\columnwidth]{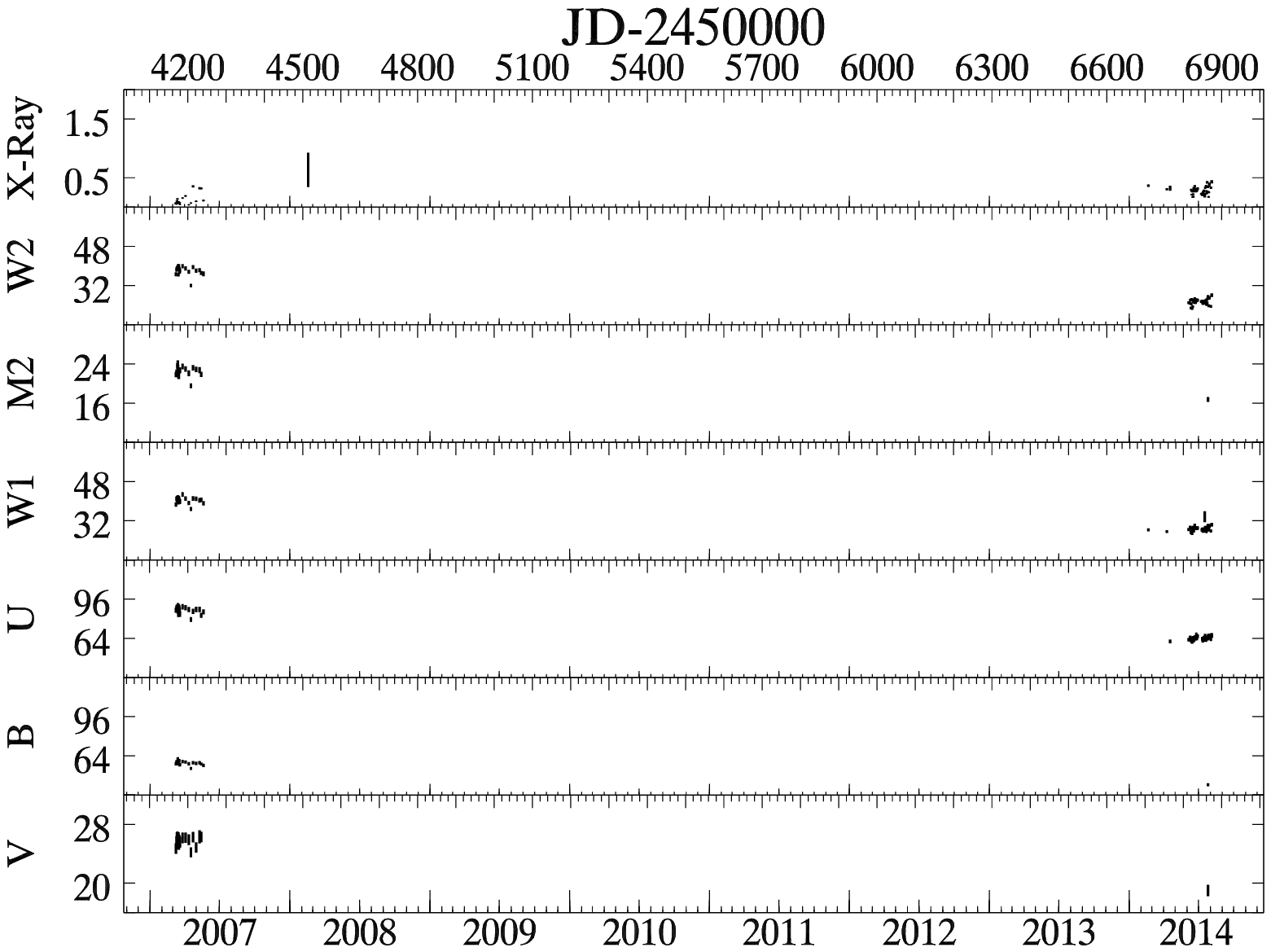}
    \caption{PG 1211+143}
    \label{fig:lcpg1211}
\end{figure}
\begin{figure}
        \includegraphics[width=\columnwidth]{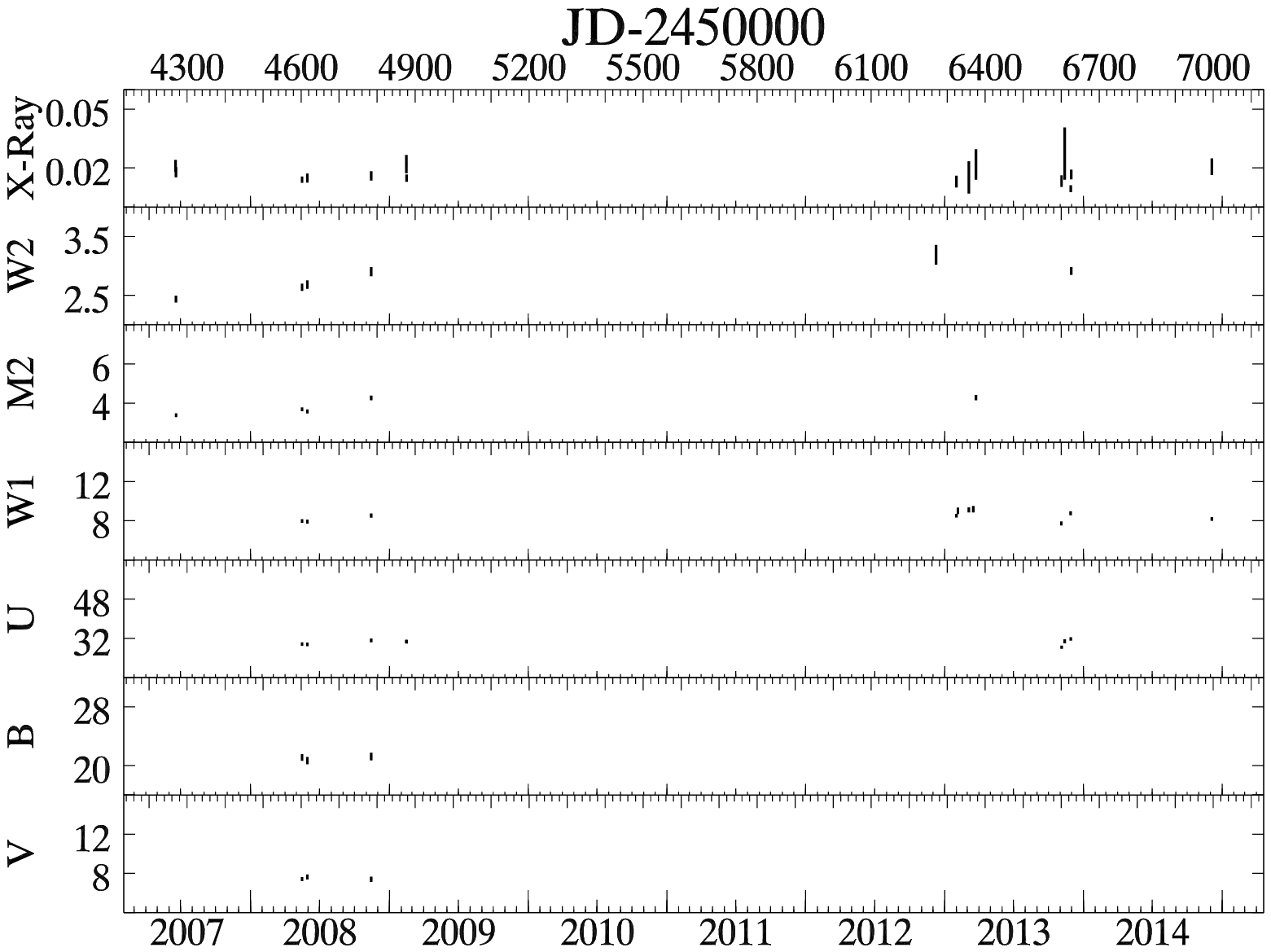}
    \caption{PG 1247+267}
    \label{fig:lcpg1247}
\end{figure}
\begin{figure}
        \includegraphics[width=\columnwidth]{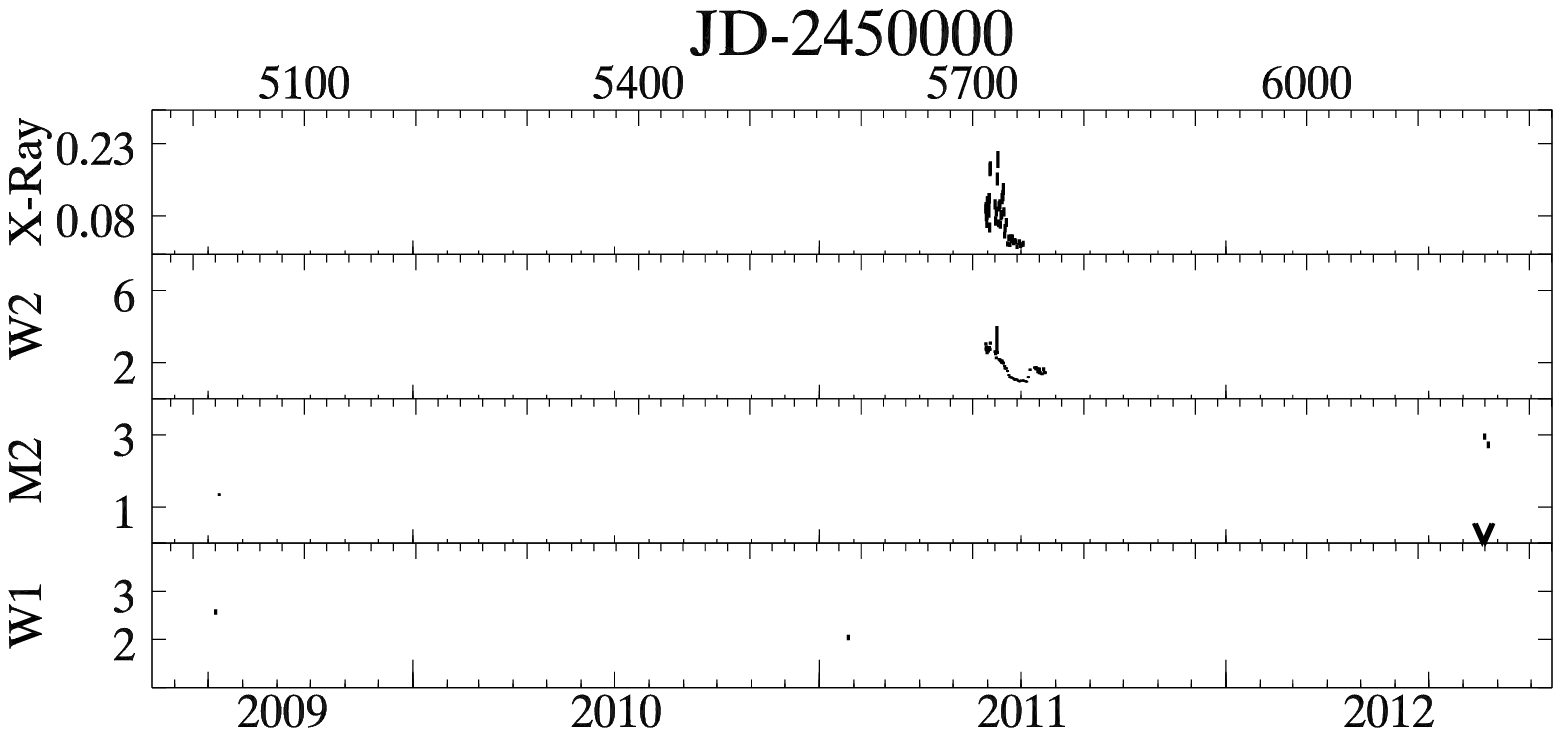}
    \caption{Zw229-15}
    \label{fig:lczw229}
\end{figure}

\clearpage

\section{Broadband Variability Spectra}
\label{app:exvarspec}

Energies are given in the source frame. UV errors are shown at 1-$\sigma$. 1,2 and 3-$\sigma$ X-ray errors are shown in maroon, red and pink respectively; triangles represent upper limits. Blue lines show powerlaw fits to UV and X-ray bands separately; those for X-rays only include points detected at 2-$\sigma$.

\begin{figure}
	\includegraphics[width=\columnwidth]{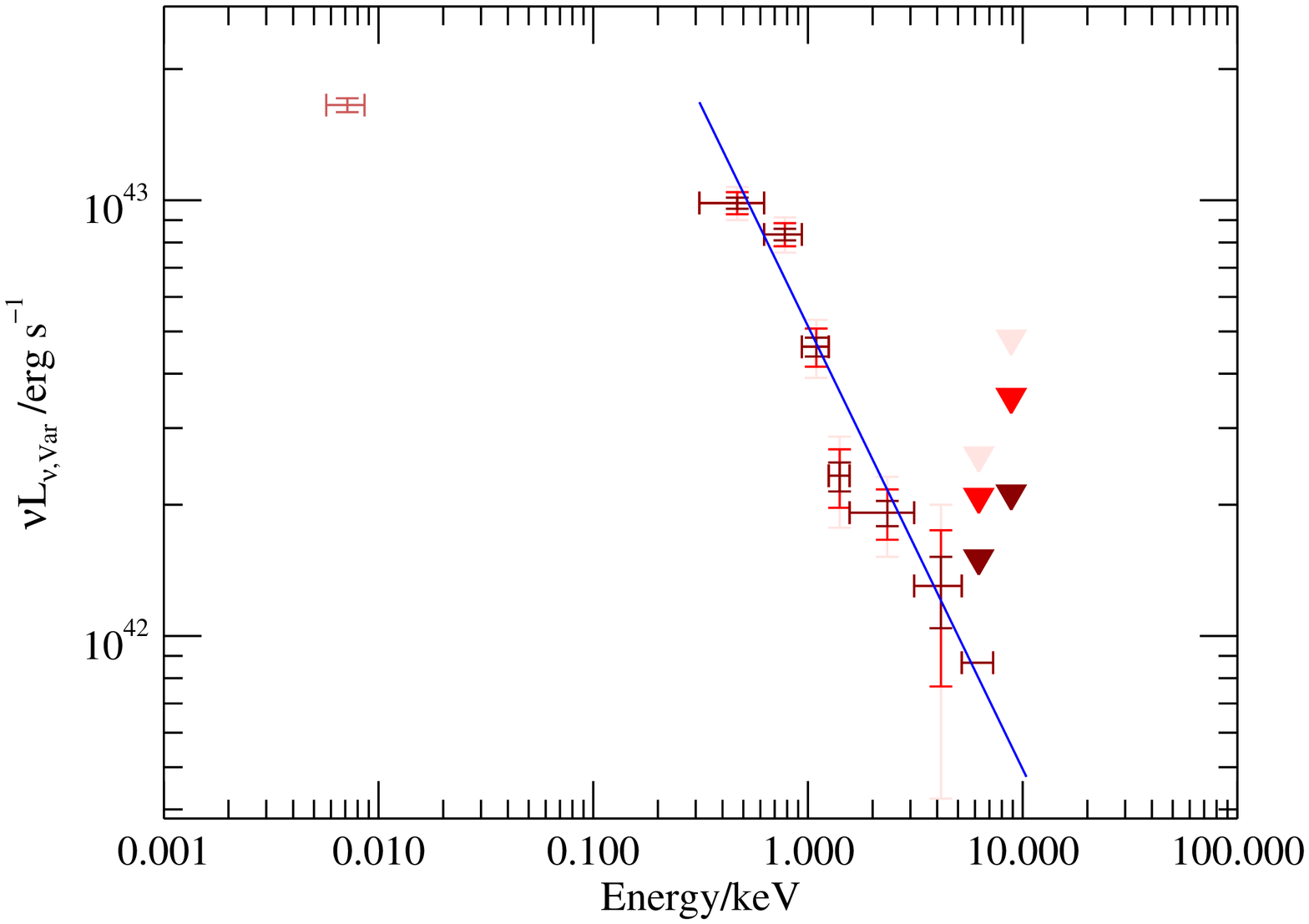}
    \caption{ 1H 0707--495}
    \label{fig:vs1h0707}
\end{figure}
\begin{figure}
	\includegraphics[width=\columnwidth]{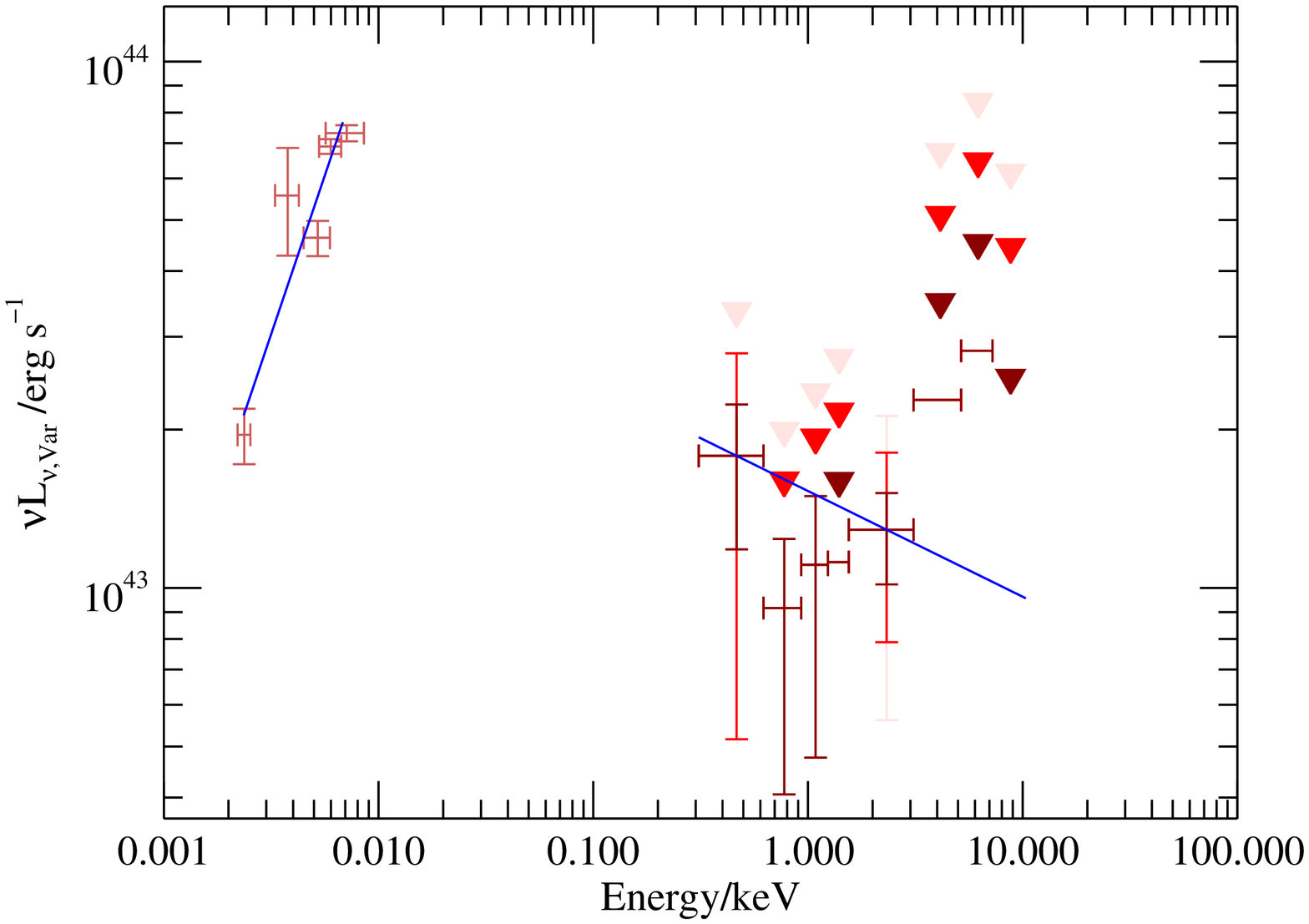}
    \caption{ 3C 120}
    \label{fig:vs3c120}
\end{figure}
\begin{figure}
	\includegraphics[width=\columnwidth]{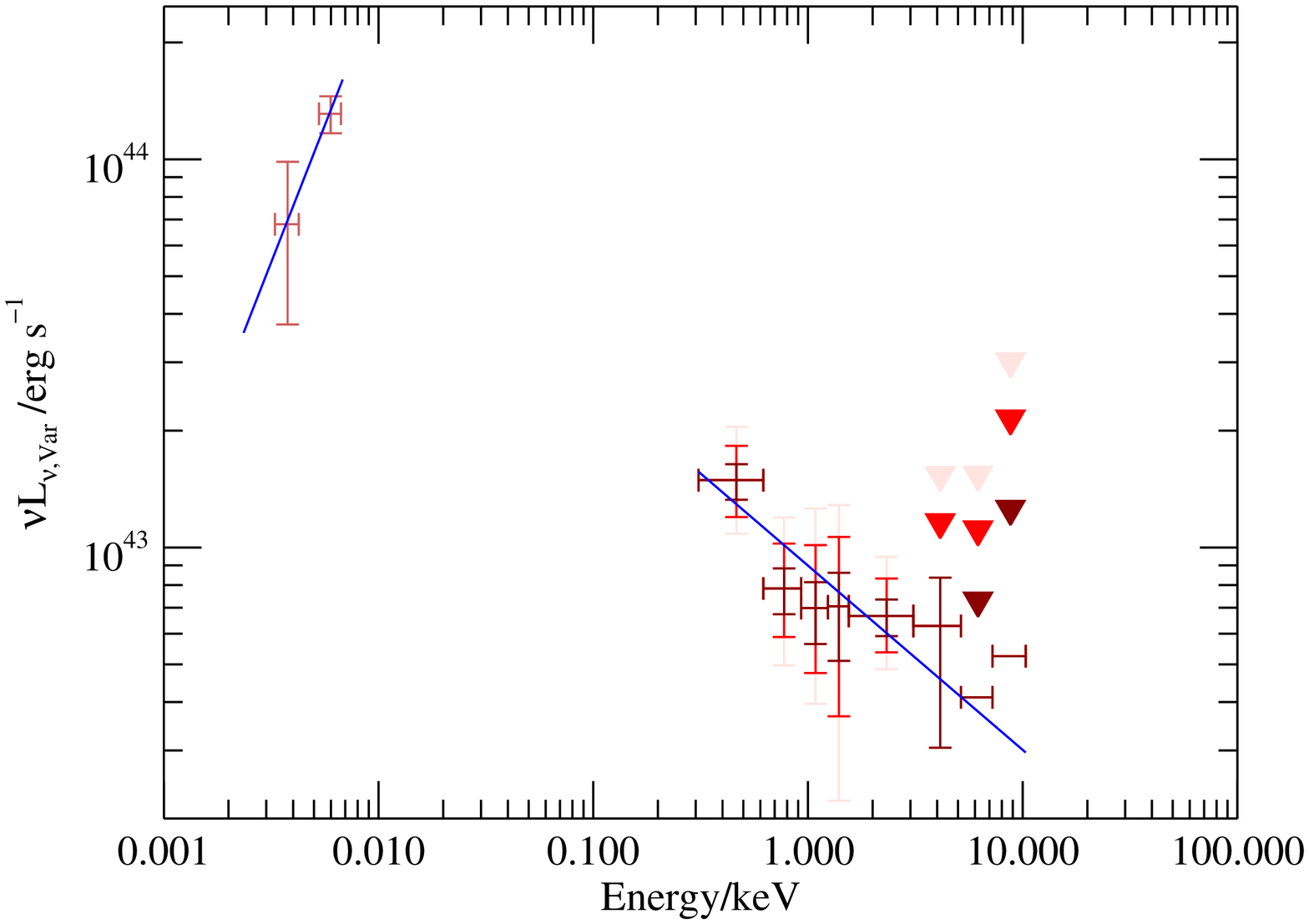}
    \caption{ARK 120}
    \label{fig:vsark120}
\end{figure}
\begin{figure}
	\includegraphics[width=\columnwidth]{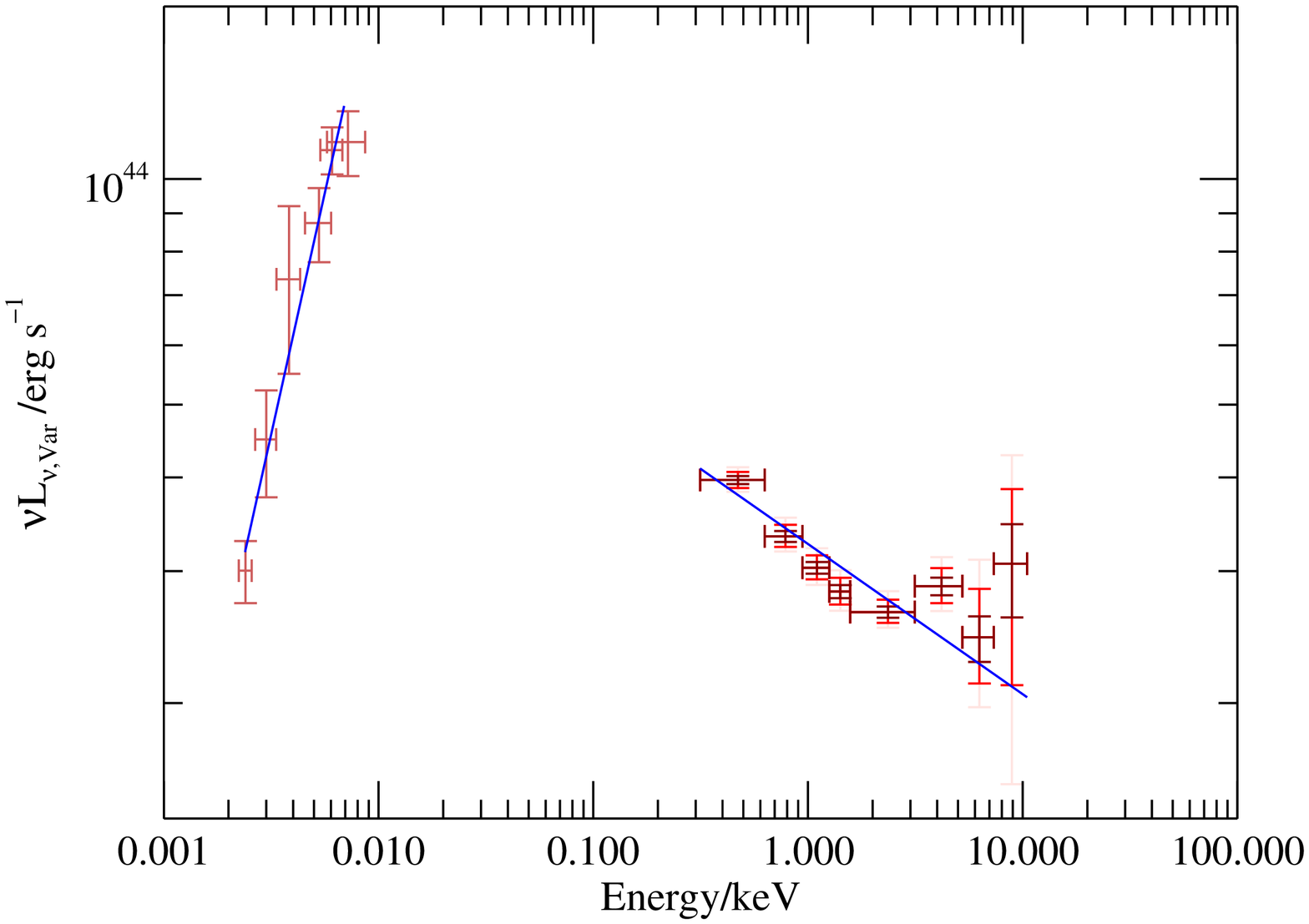}
    \caption{ Fairall 9}
    \label{fig:vsf9}
\end{figure}
\clearpage
\begin{figure}
	\includegraphics[width=\columnwidth]{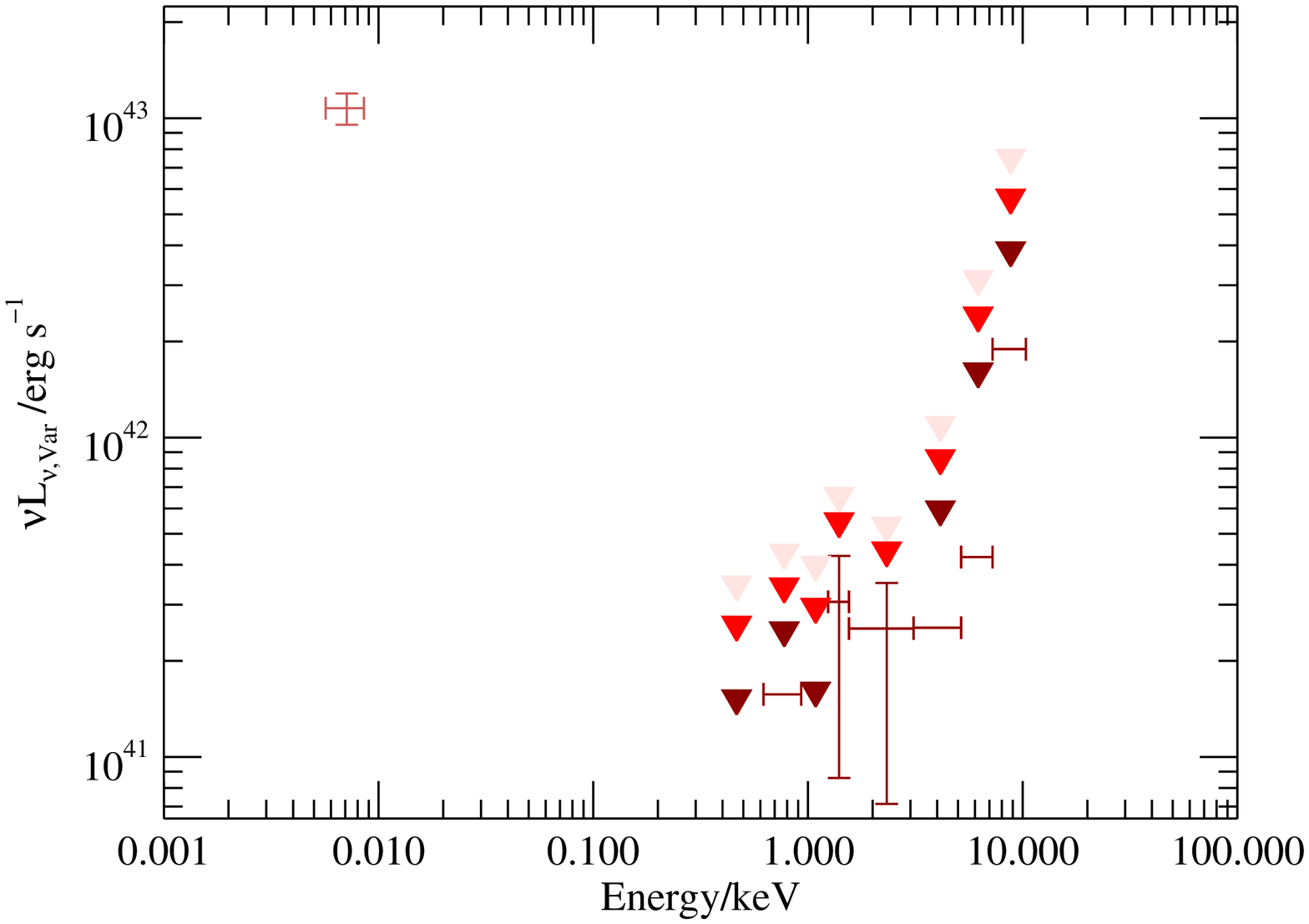}
    \caption{ H 0557--385}
    \label{fig:vsh0557}
\end{figure}
\begin{figure}
	\includegraphics[width=\columnwidth]{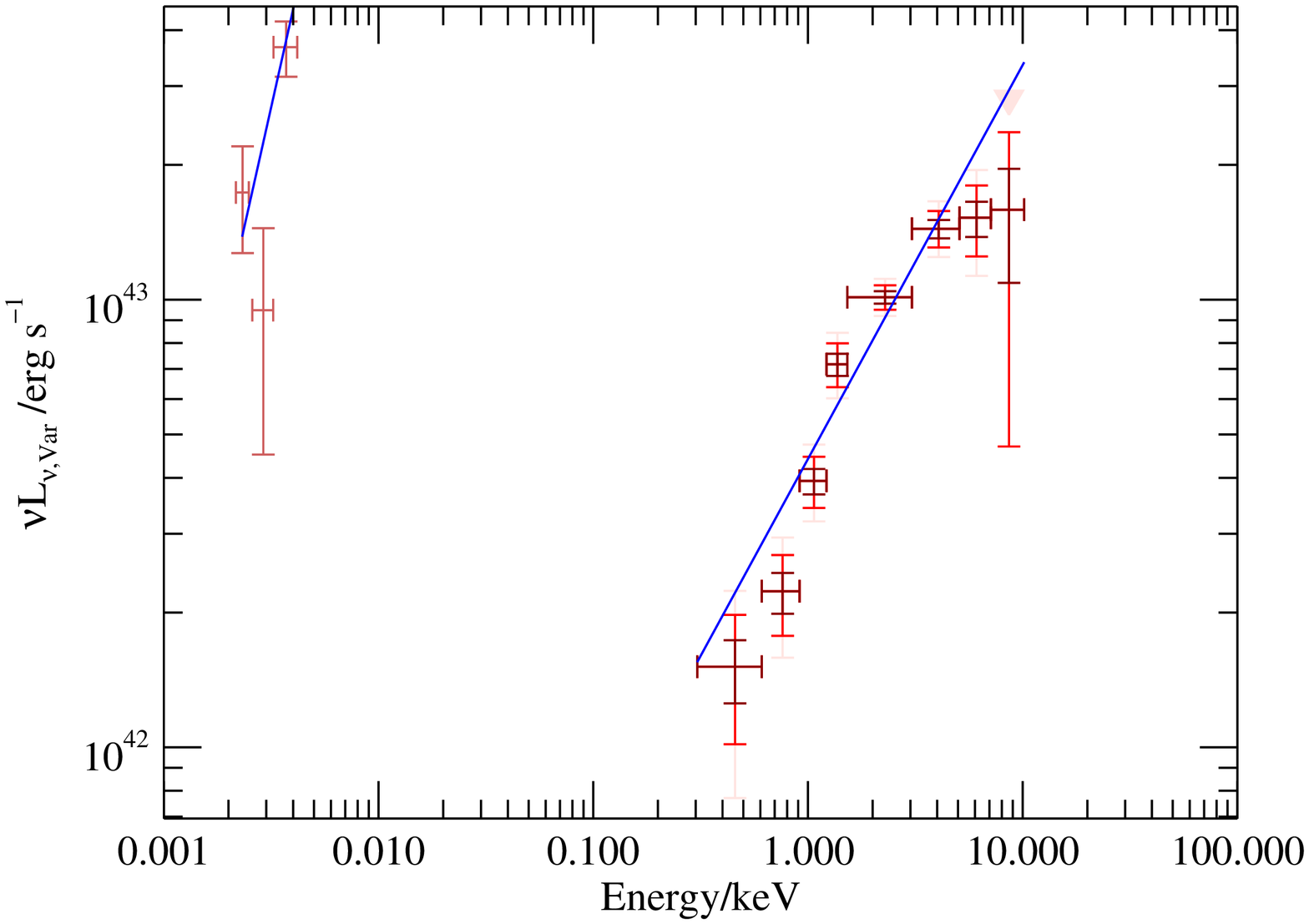}
    \caption{IC 4329A}
    \label{fig:vsic4329a}
\end{figure}
\begin{figure}
	\includegraphics[width=\columnwidth]{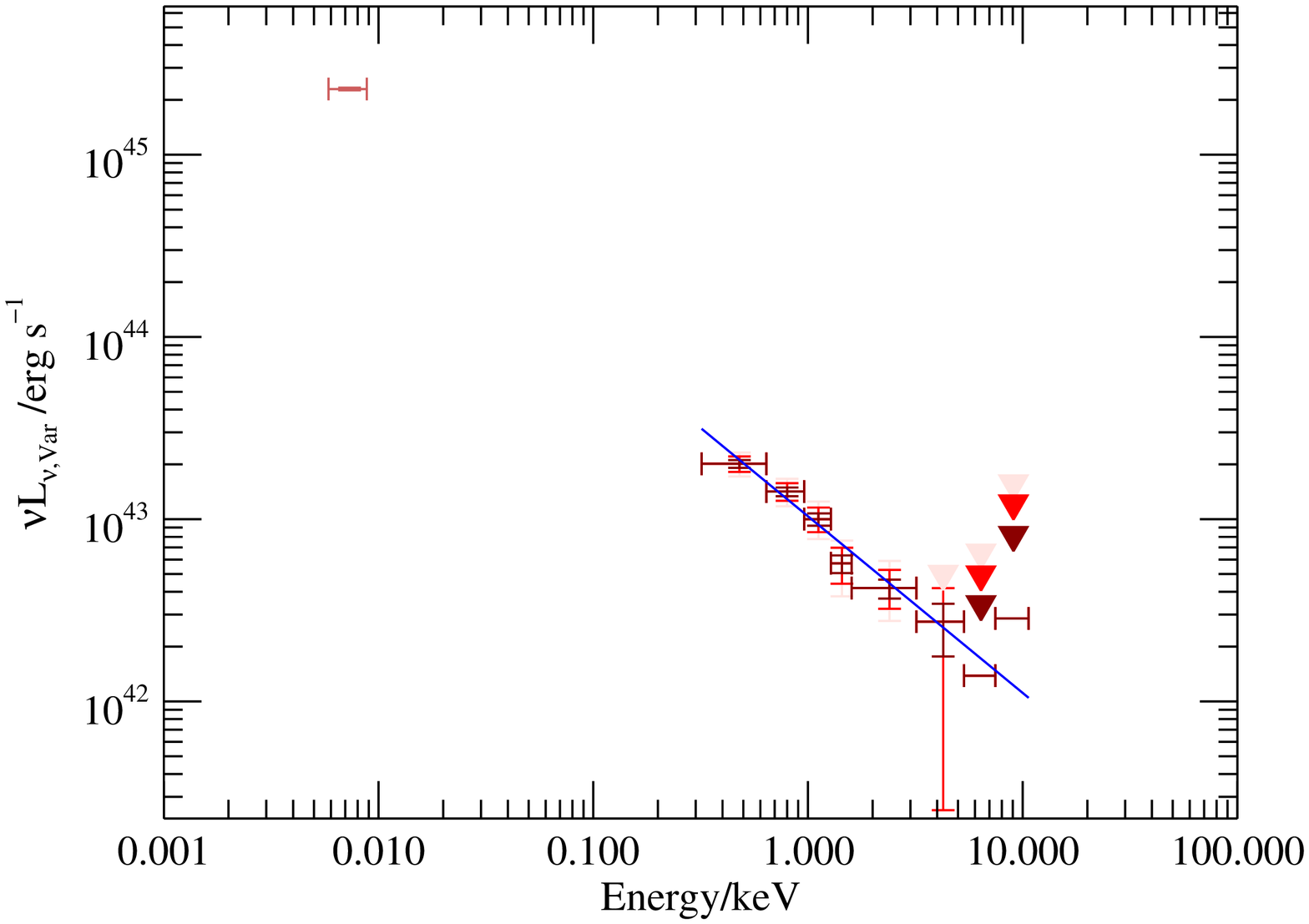}
    \caption{ IRAS 13224-3809}
    \label{fig:vsIRAS13224}
\end{figure}
\begin{figure}
	\includegraphics[width=\columnwidth]{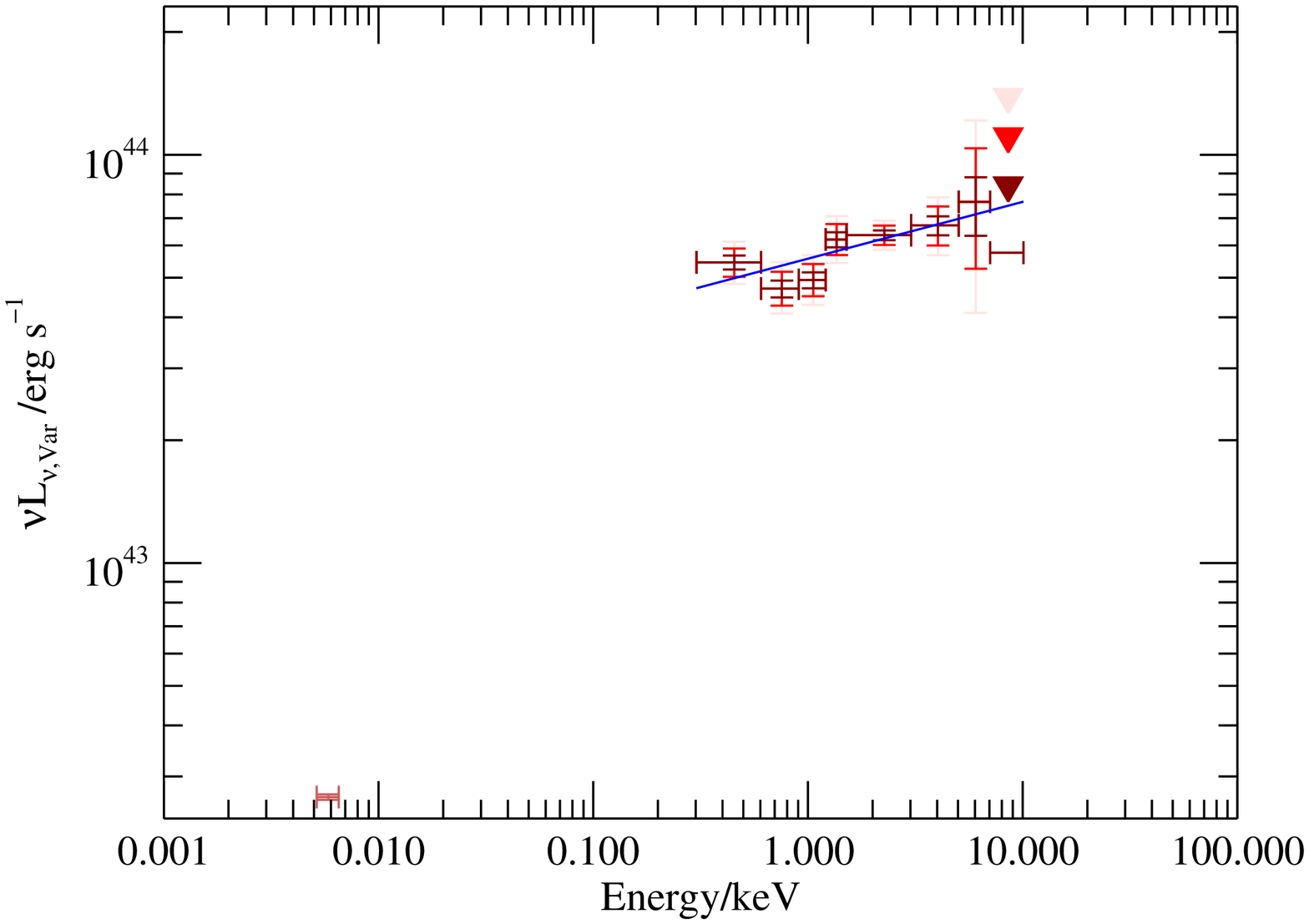}
    \caption{ MCG--6-30-15}
    \label{fig:vsmcg6}
\end{figure}
\begin{figure}
	\includegraphics[width=\columnwidth]{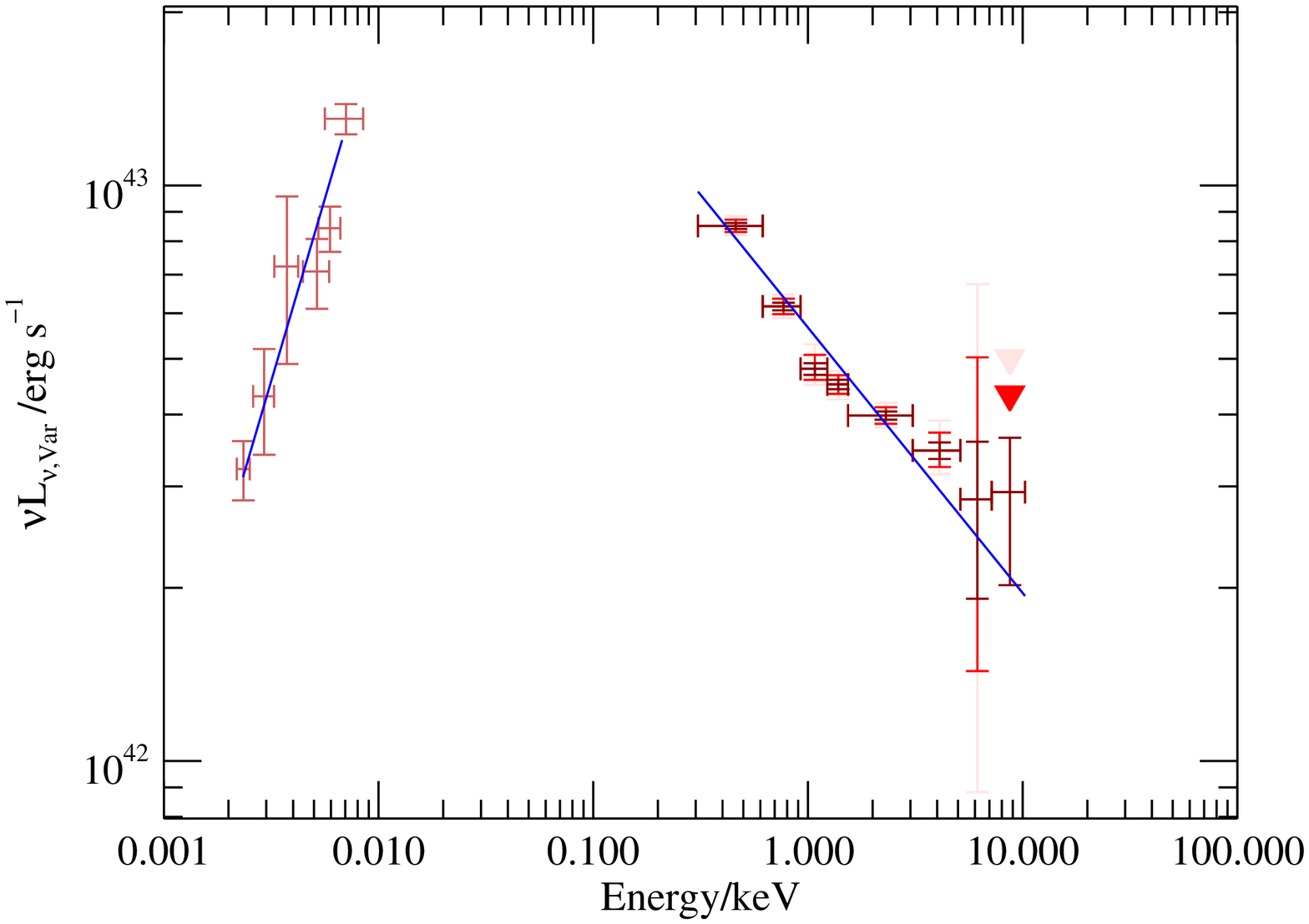}
    \caption{MRK 335}
    \label{fig:vsmrk335}
\end{figure}
\begin{figure}
	\includegraphics[width=\columnwidth]{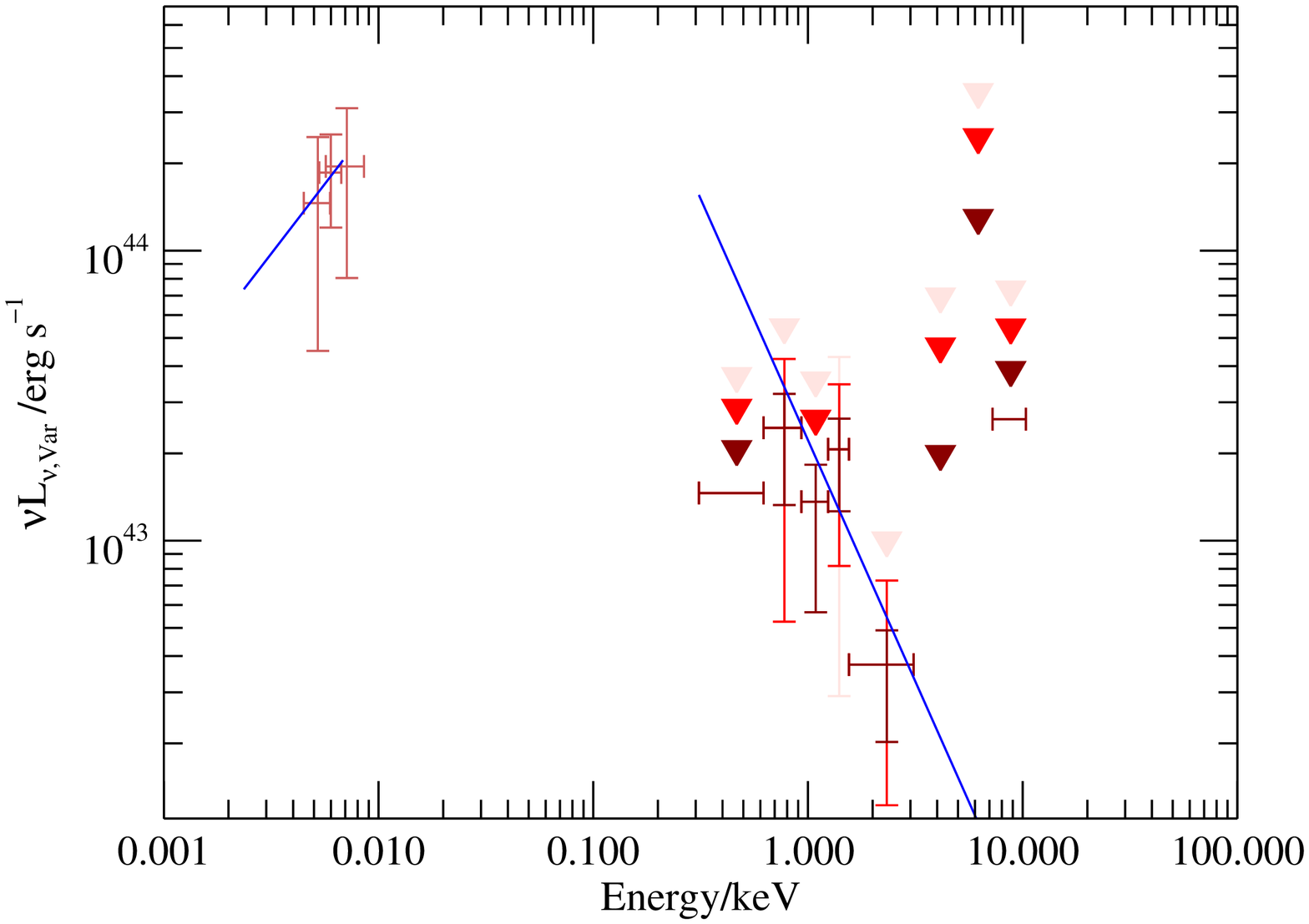}
    \caption{MRK 509}
    \label{fig:vsmrk509}
\end{figure}
\clearpage
\begin{figure}
	\includegraphics[width=\columnwidth]{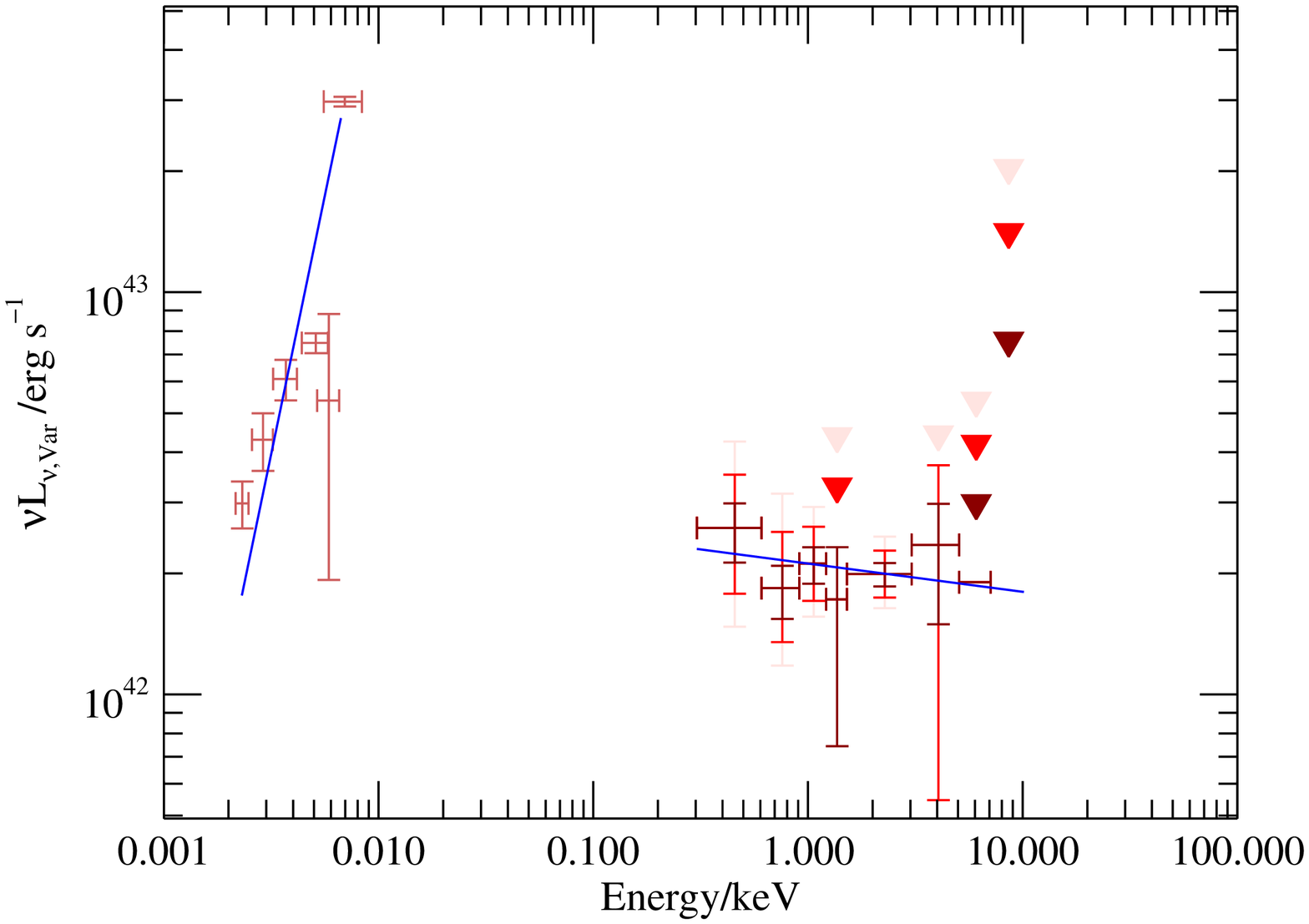}
    \caption{MRK 766}
    \label{fig:vsmrk766}
\end{figure}
\begin{figure}
	\includegraphics[width=\columnwidth]{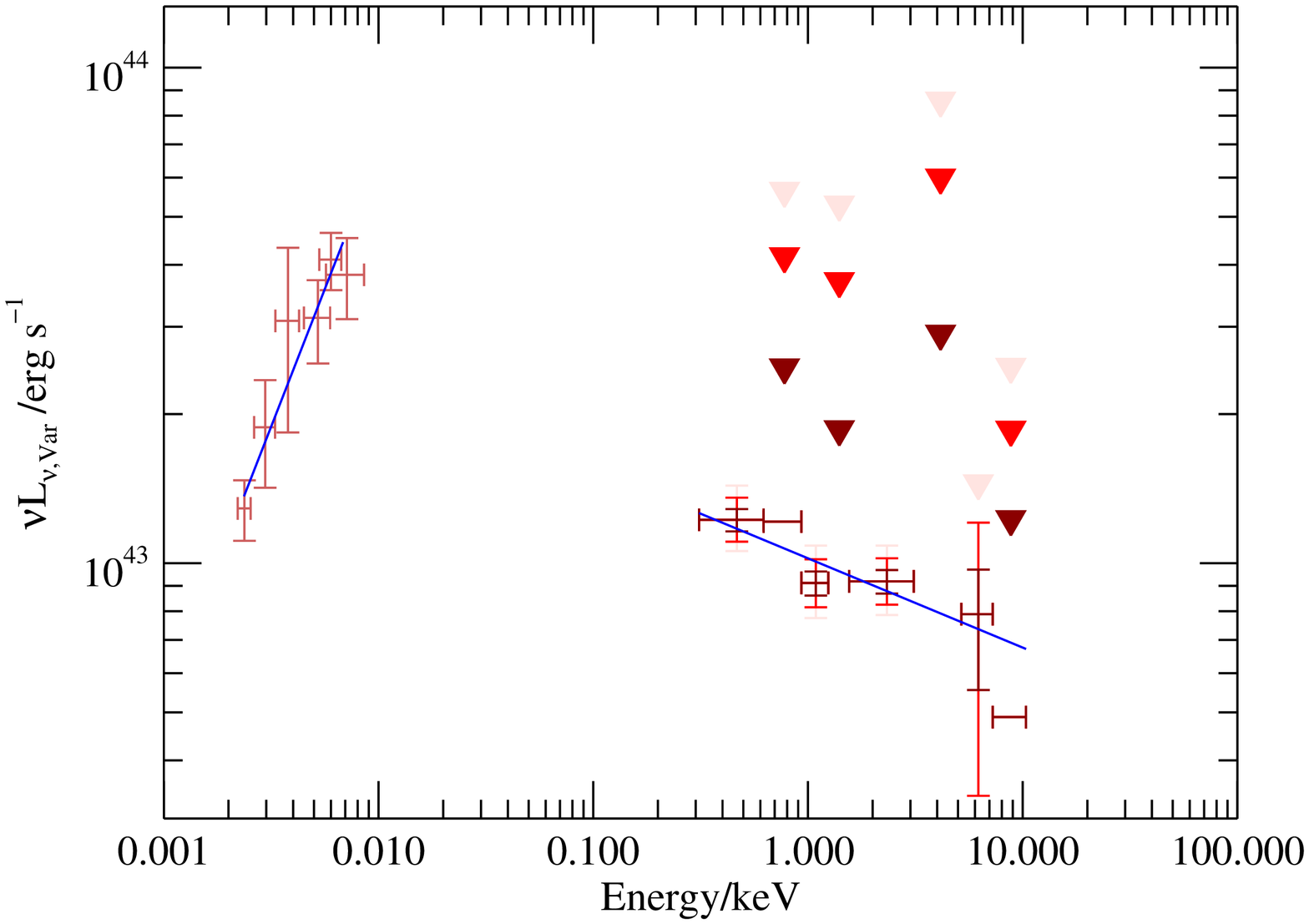}
    \caption{MRK 841}
    \label{fig:vsmrk841}
\end{figure}
\begin{figure}
	\includegraphics[width=\columnwidth]{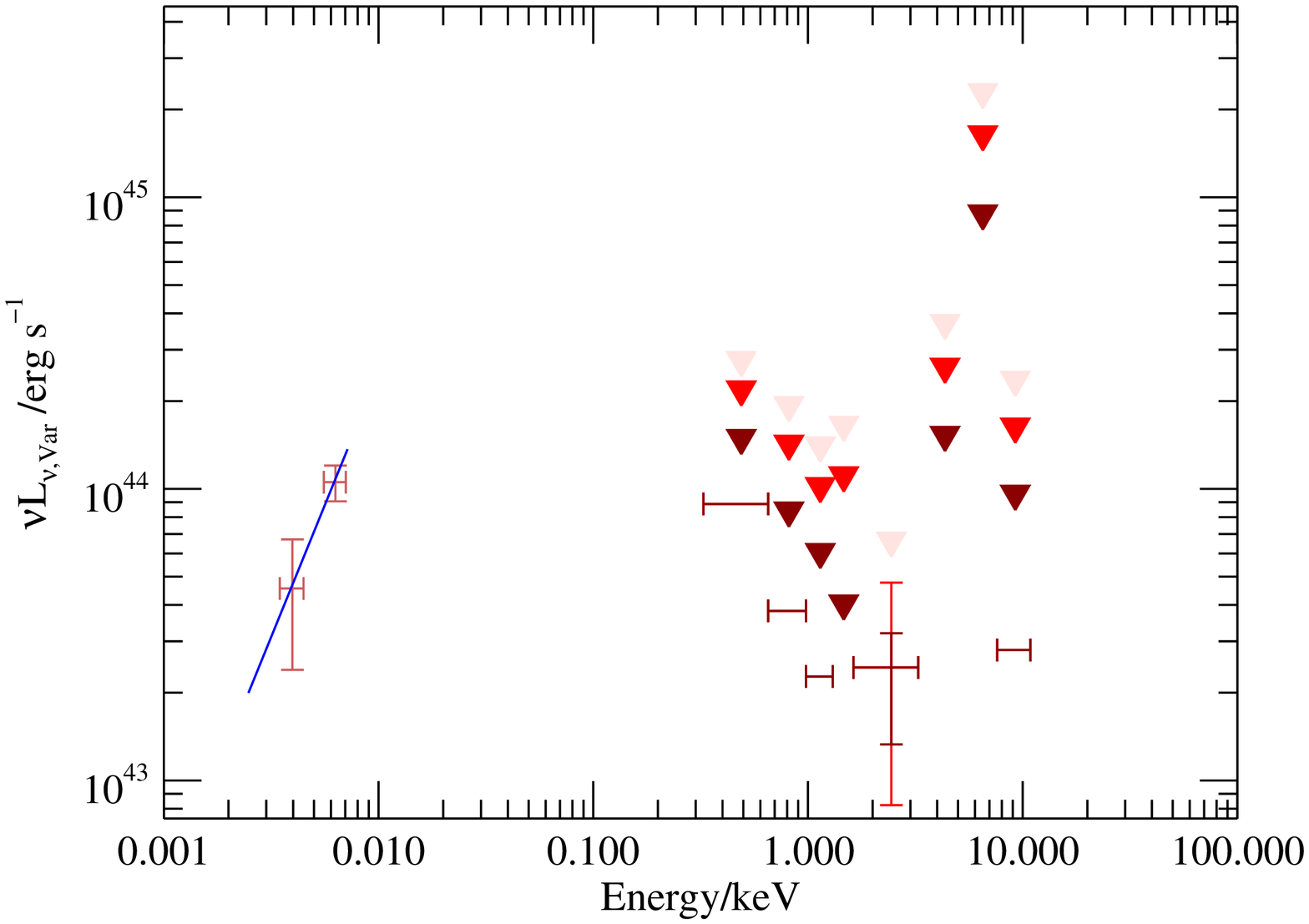}
    \caption{MRK 1383}
    \label{fig:vsmrk1383}
\end{figure}
\begin{figure}
	\includegraphics[width=\columnwidth]{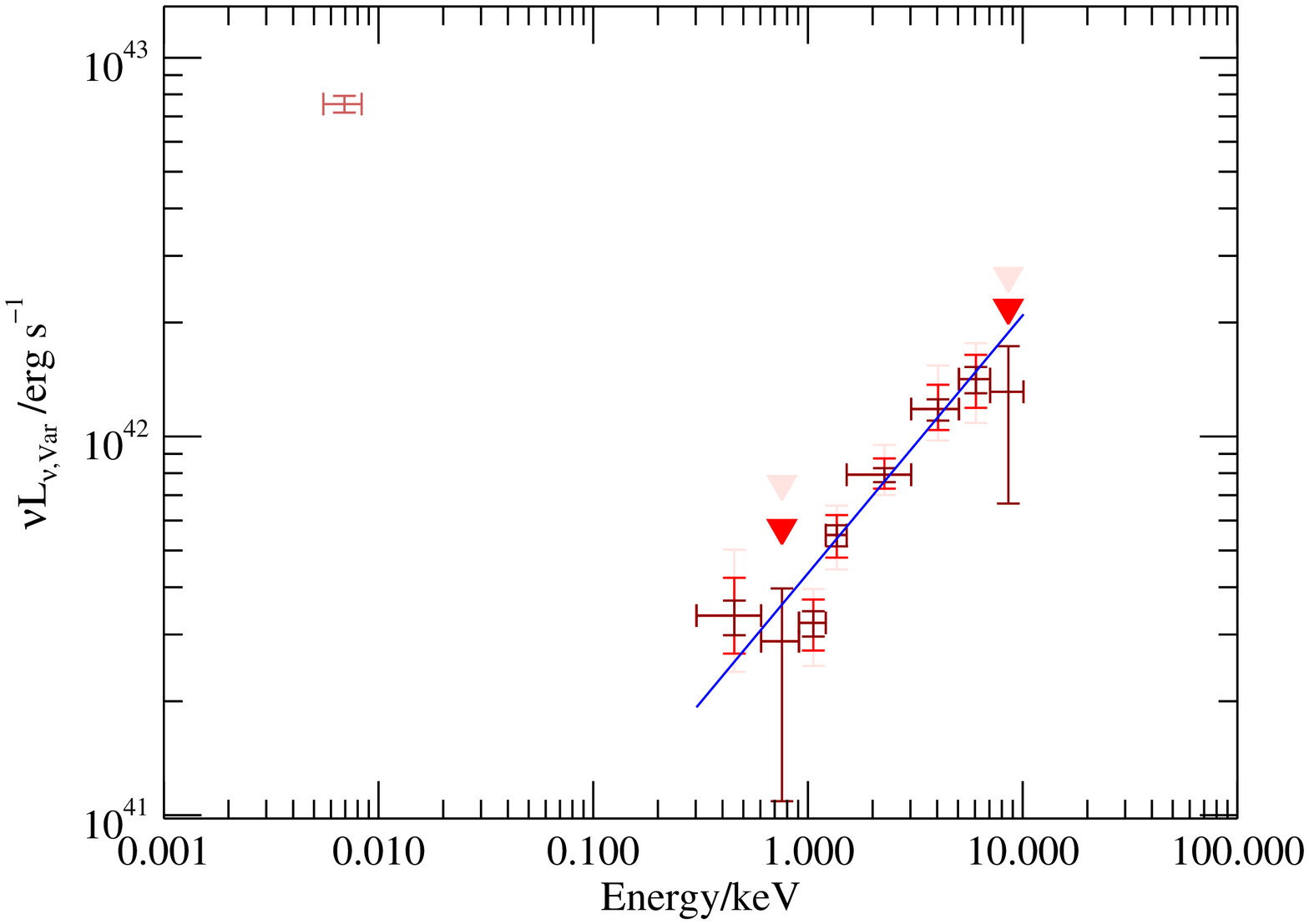}
    \caption{NGC 3516}
    \label{fig:vsngc3516}
\end{figure}
\begin{figure}
	\includegraphics[width=\columnwidth]{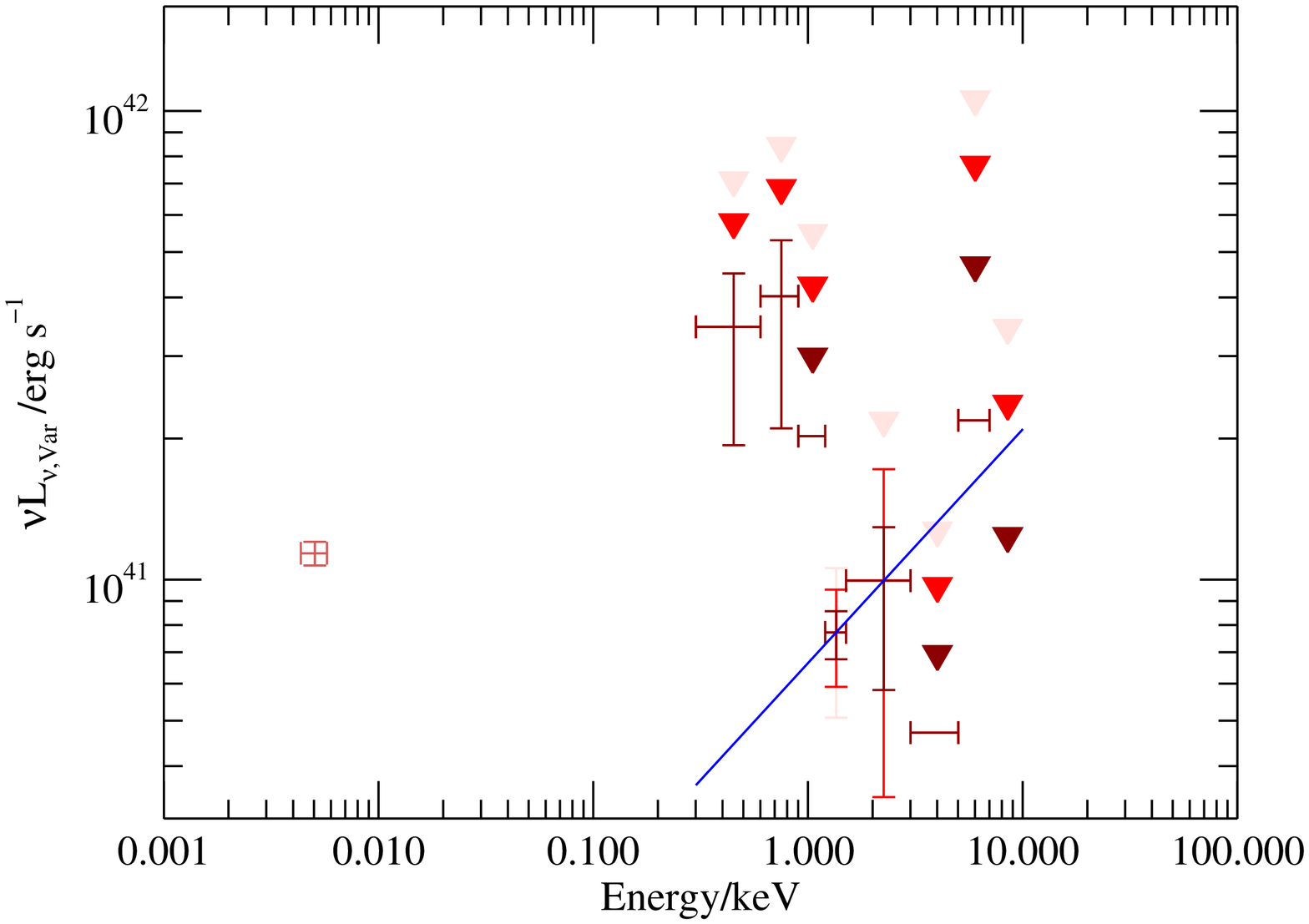}
    \caption{NGC 4051}
    \label{fig:vsngc4051}
\end{figure}
\begin{figure}
	\includegraphics[width=\columnwidth]{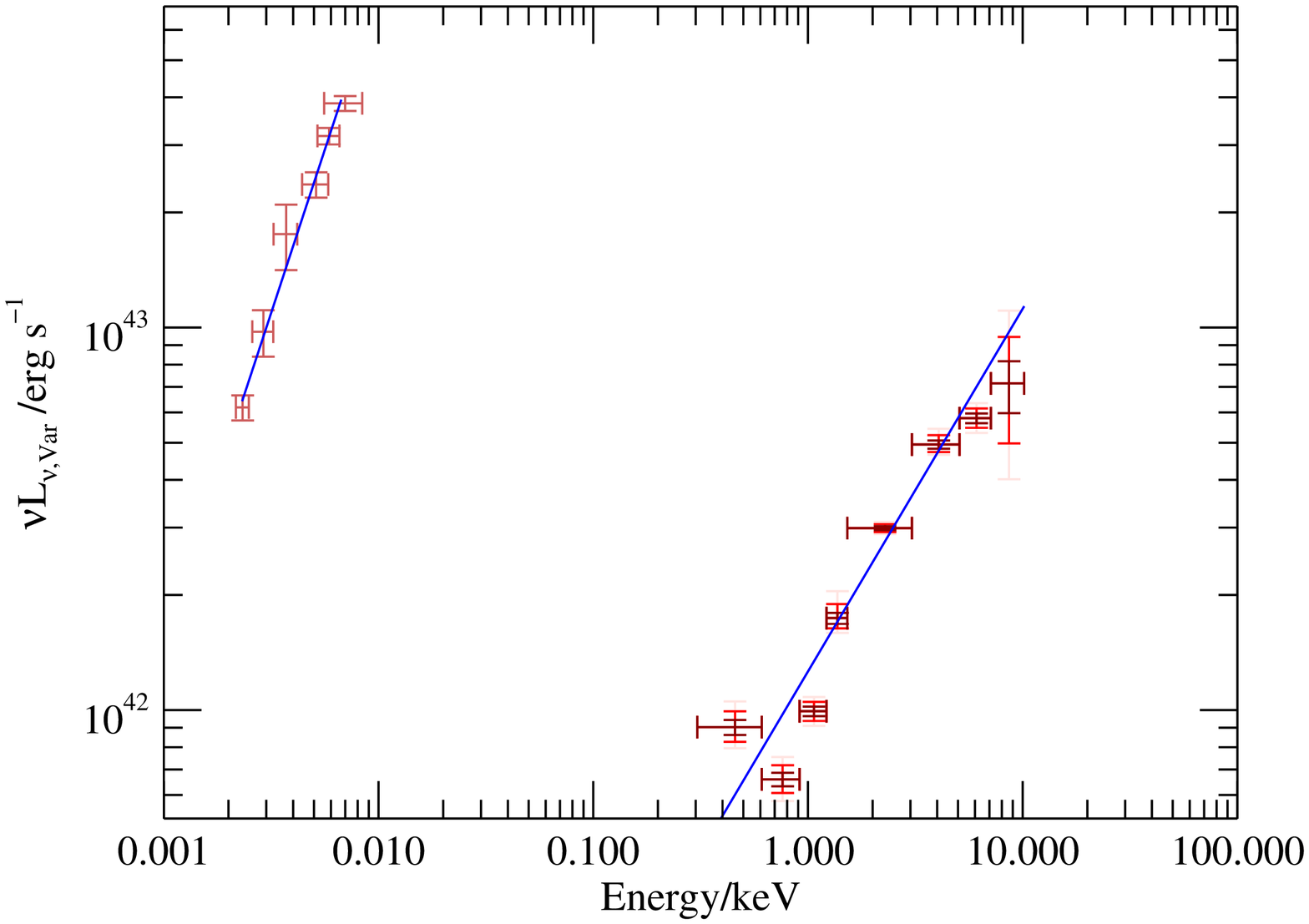}
    \caption{NGC 5548}
    \label{fig:vsngc5548}
\end{figure}
\clearpage
\begin{figure}
	\includegraphics[width=\columnwidth]{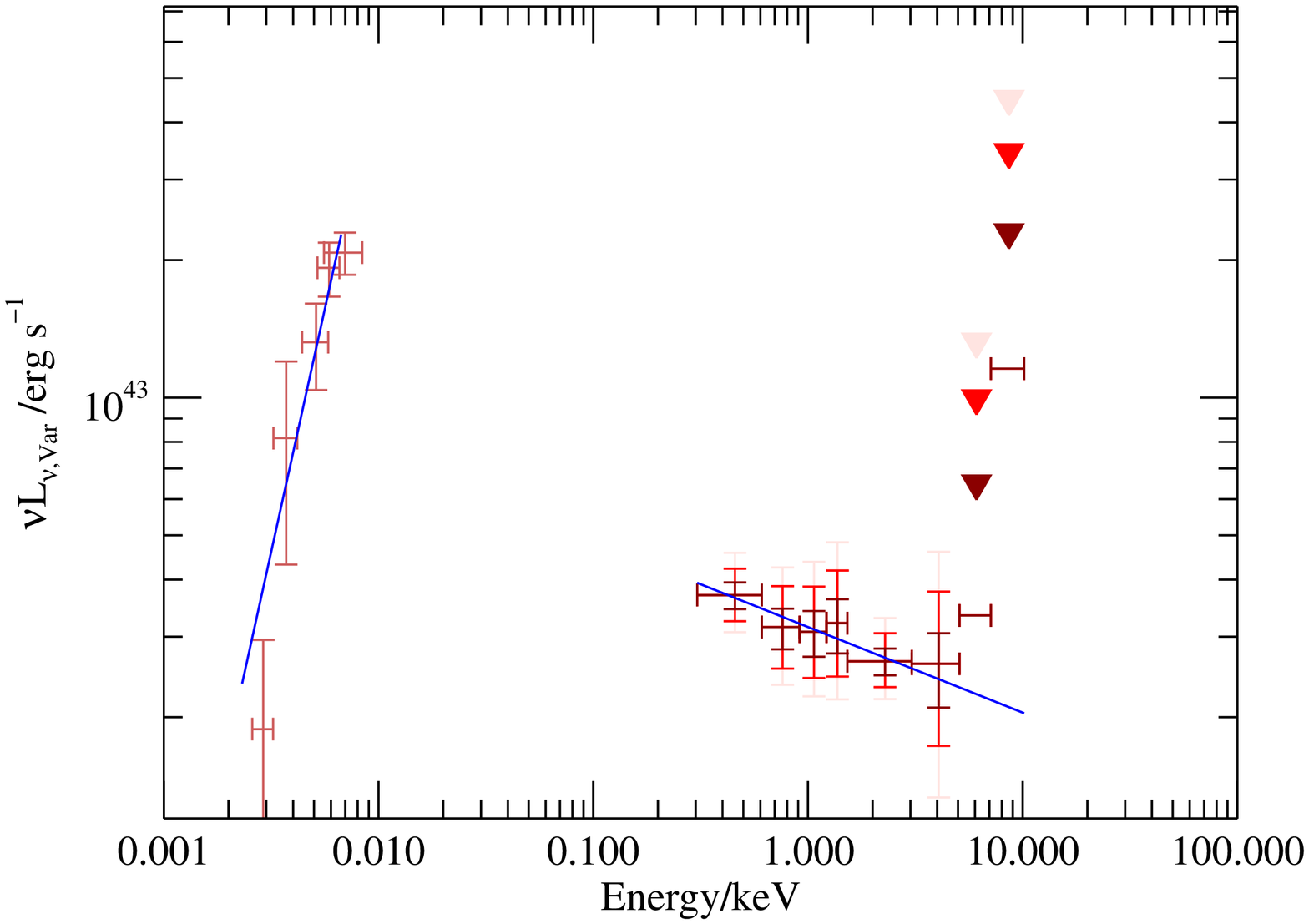}
    \caption{NGC 7469}
    \label{fig:vsngc7469}
\end{figure}
\begin{figure}
	\includegraphics[width=\columnwidth]{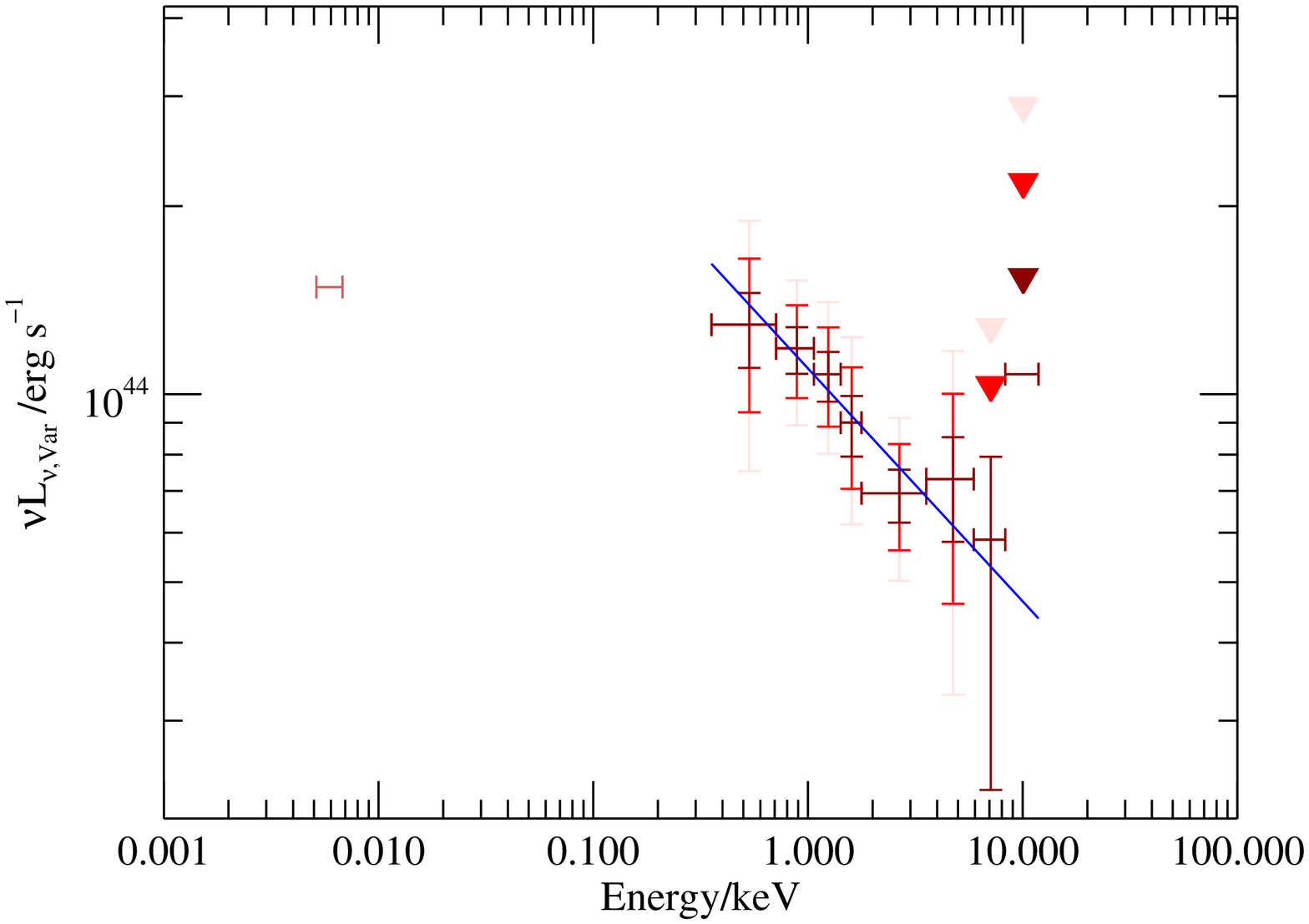}
    \caption{PDS 456}
    \label{fig:vspds456}
\end{figure}
\begin{figure}
	\includegraphics[width=\columnwidth]{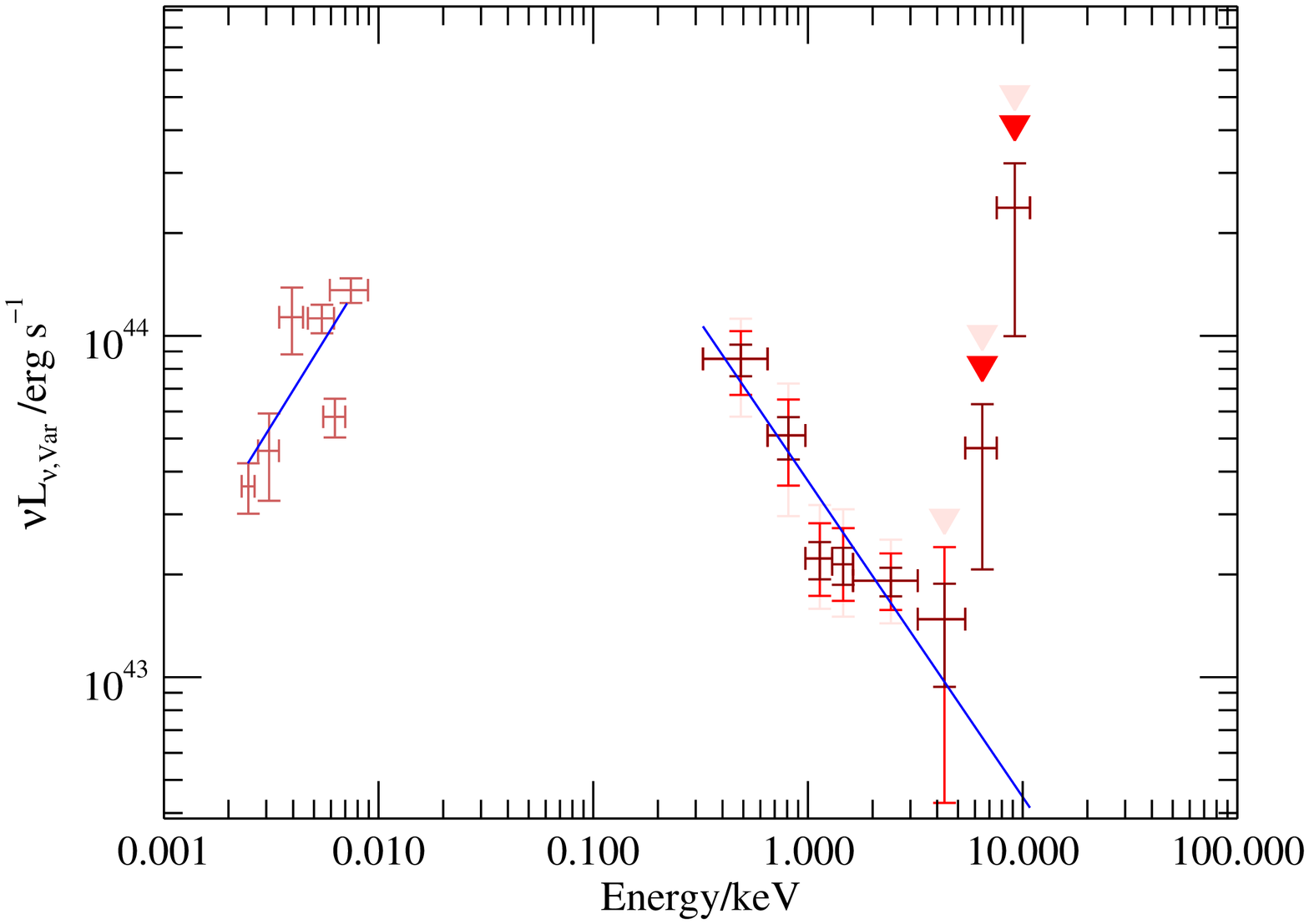}
    \caption{PG 1211+143}
    \label{fig:vspg1211}
\end{figure}
\begin{figure}
	\includegraphics[width=\columnwidth]{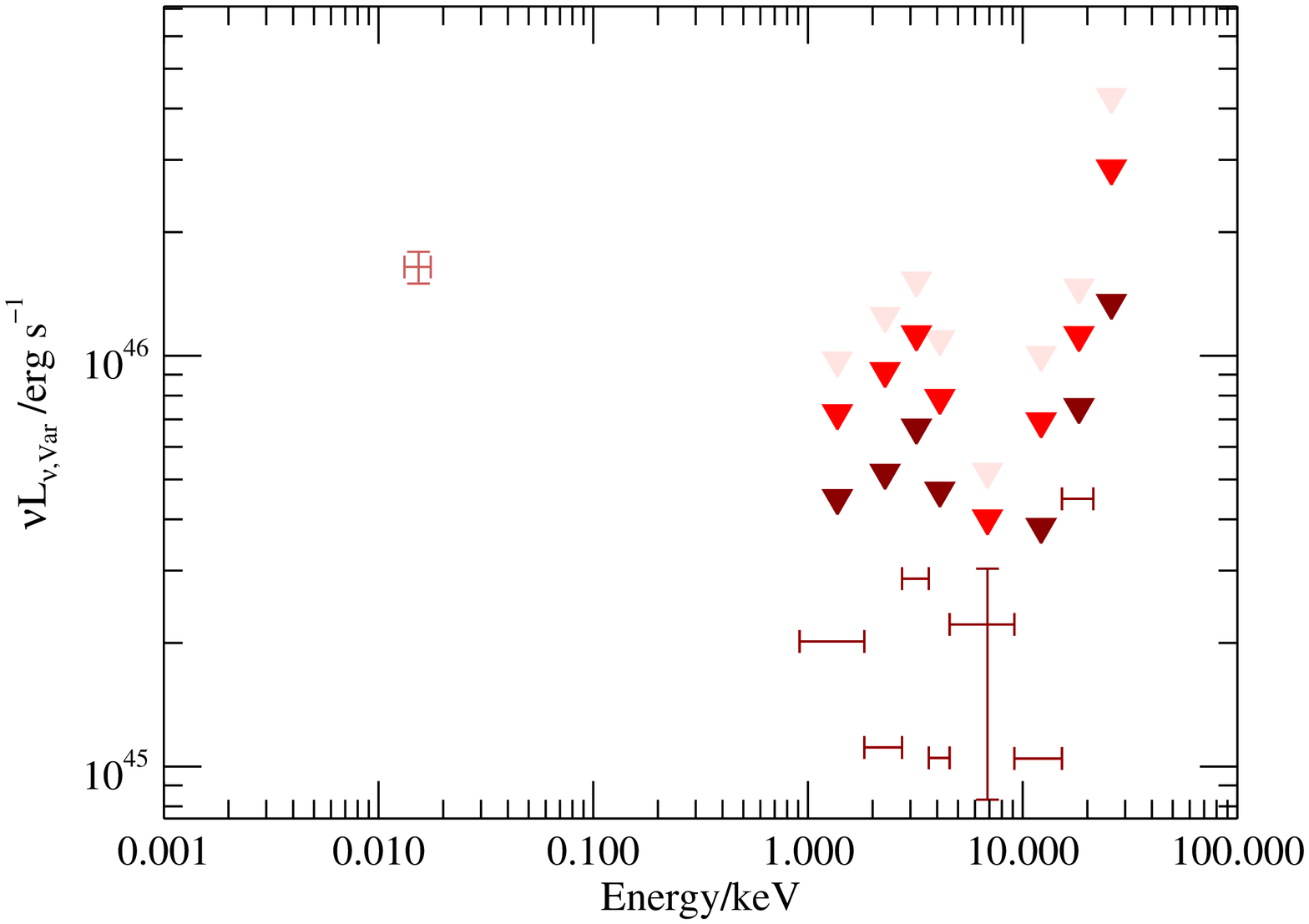}
    \caption{PG 1247+267}
    \label{fig:vspg1247}
\end{figure}
\begin{figure}
	\includegraphics[width=\columnwidth]{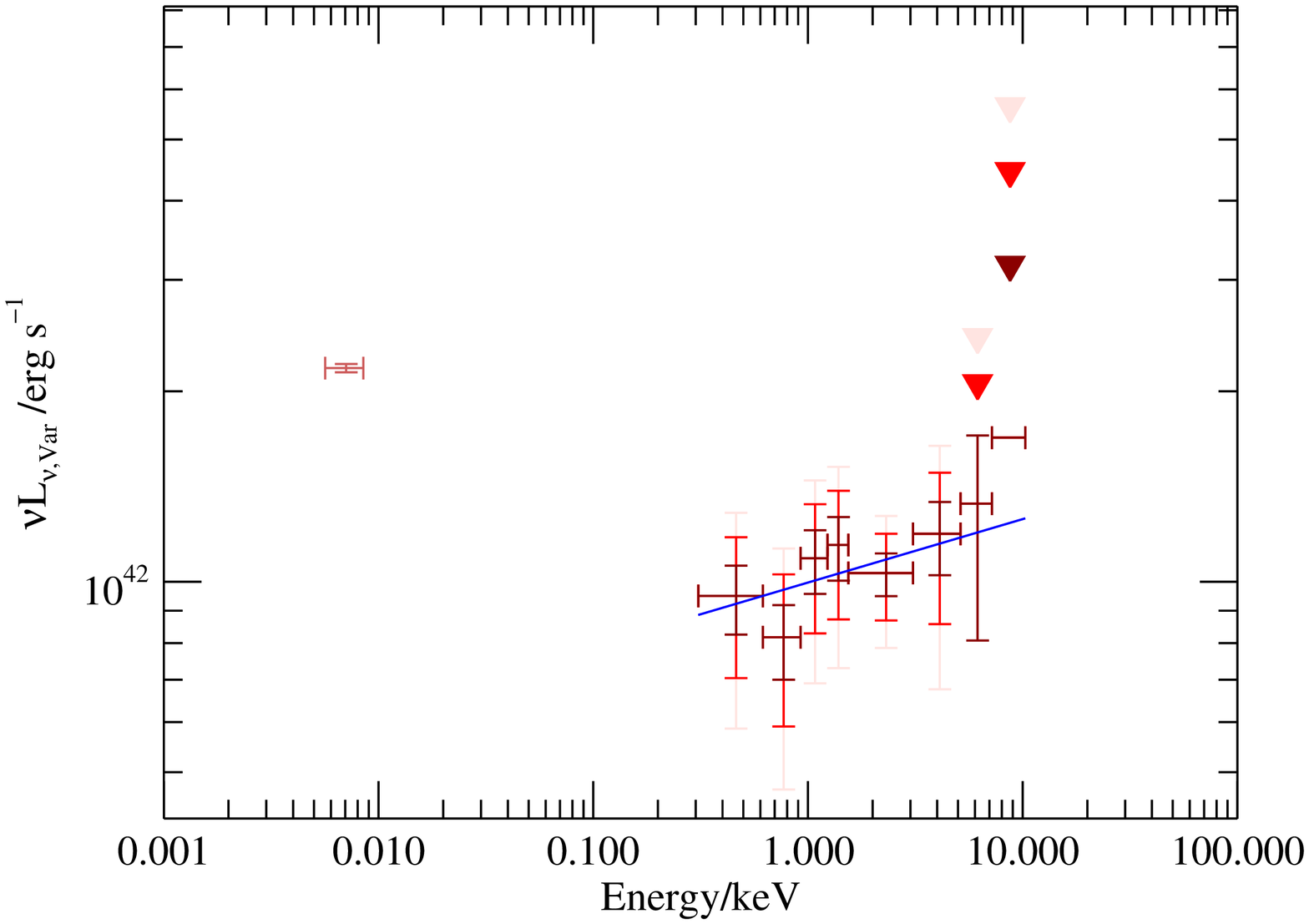}
    \caption{Zw 229-15}
    \label{fig:vszw229}
\end{figure}

\clearpage

\section{DCFs}
Discrete correlation function (black) of given band relative to X-rays. Positive lag indicates UV variations occurring after X-ray variations. 95 and 99\% confidence intervals are shown in blue and red respectively. The ACF of the X-rays and UV are shown in green dashed and purple dotted lines respectively.
Plots with fewer than 3 DCF bins are not shown.
\begin{figure*}
	\includegraphics[width=5.5cm]{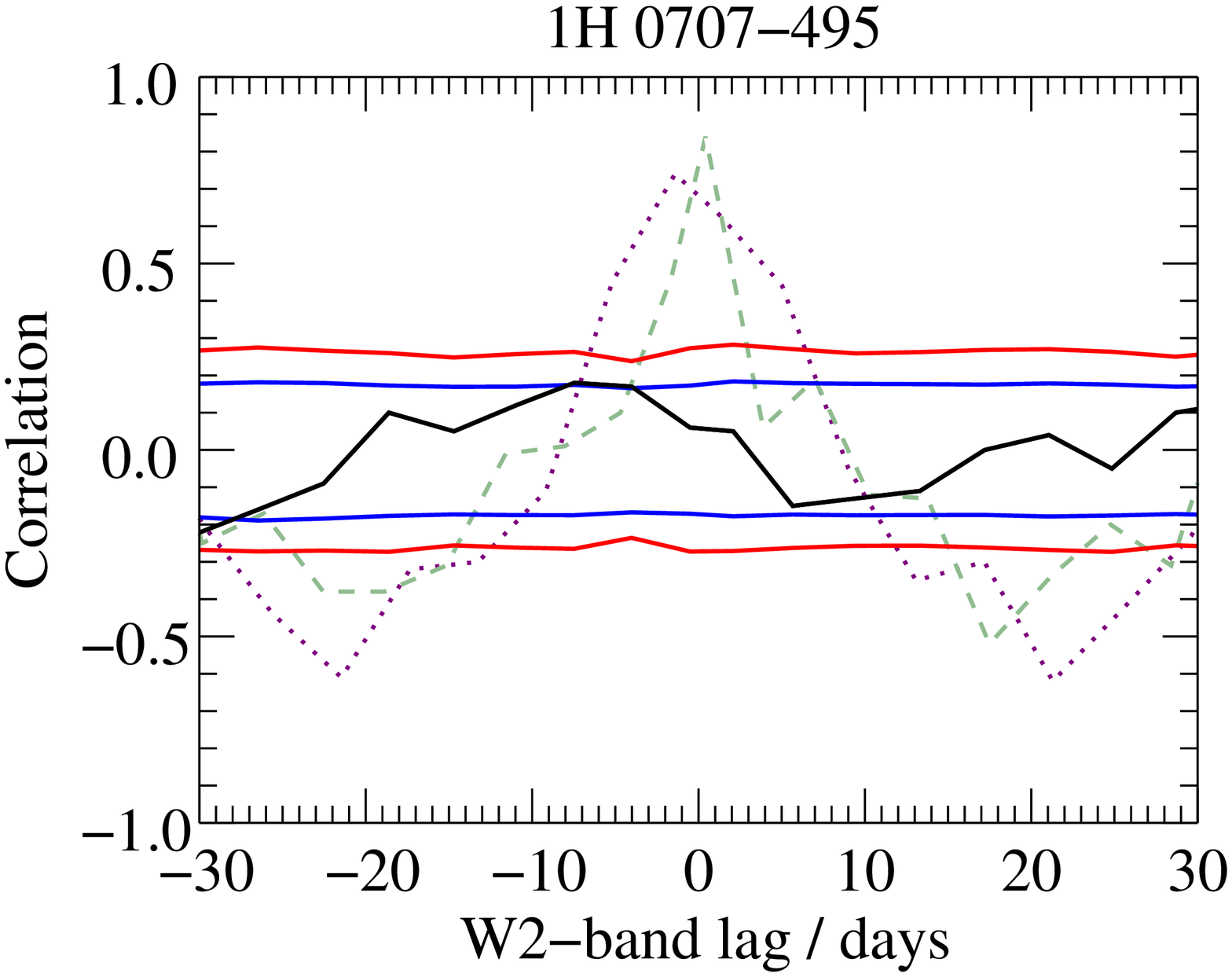}
	\includegraphics[width=5.5cm]{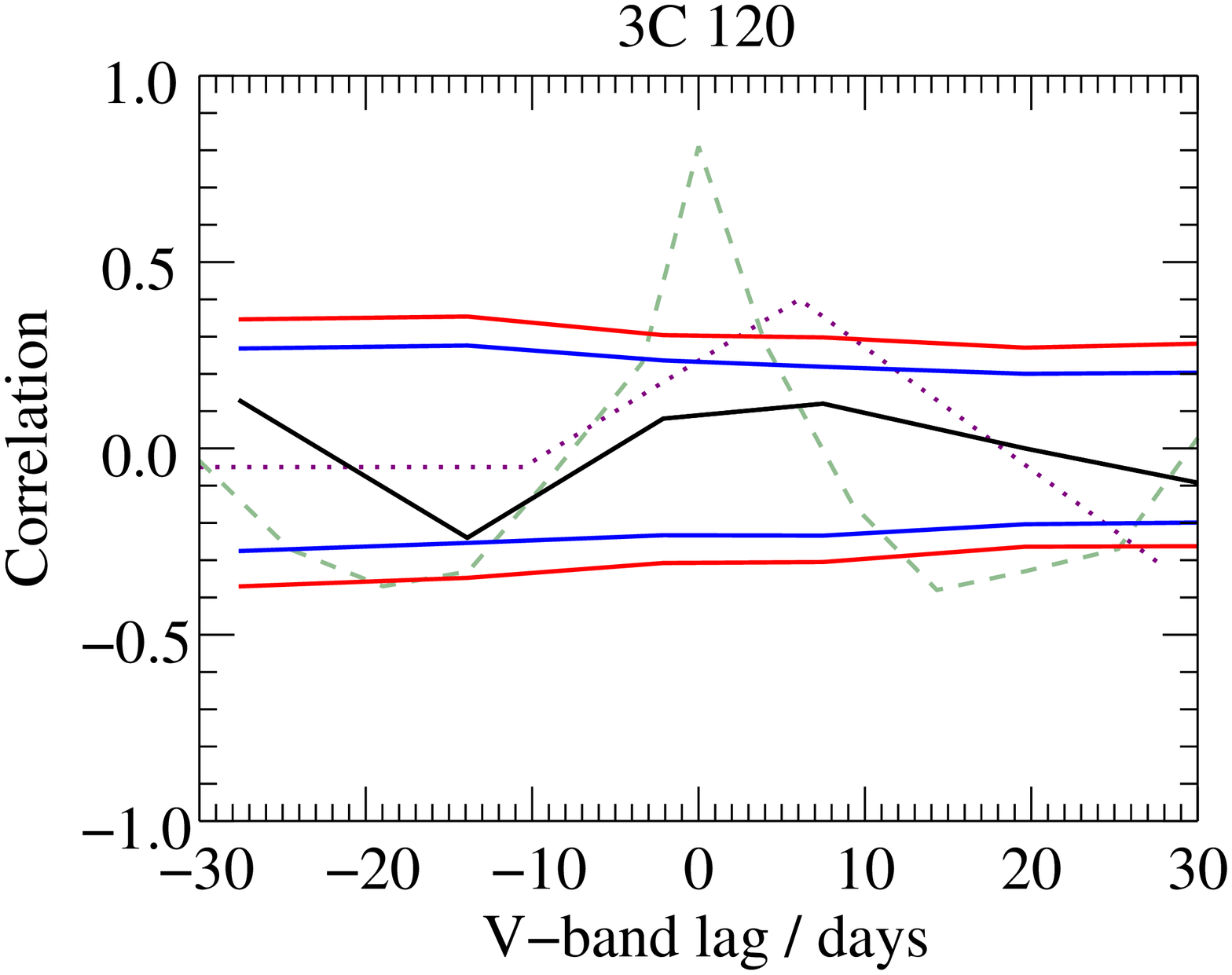}
	\includegraphics[width=5.5cm]{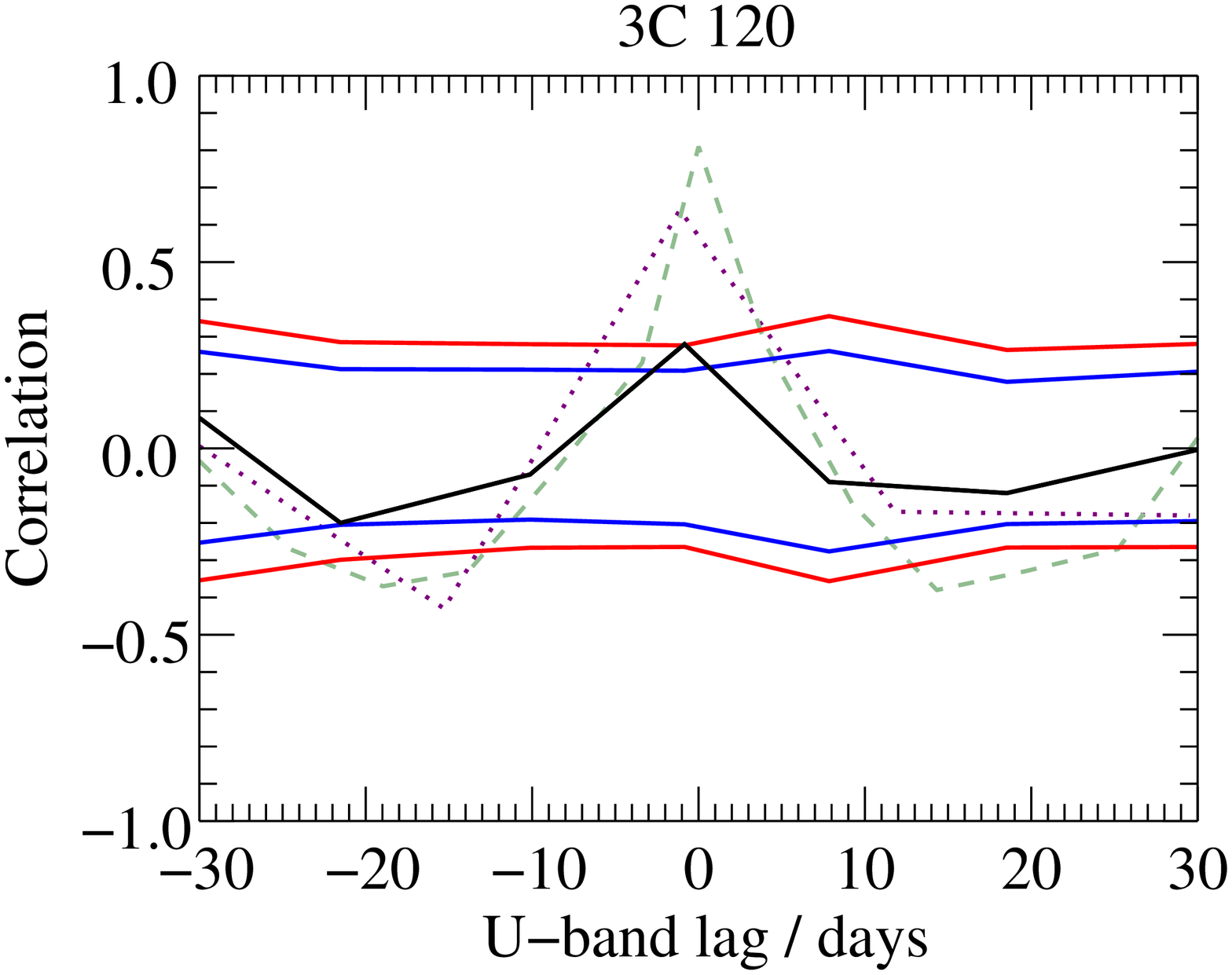}
	\includegraphics[width=5.5cm]{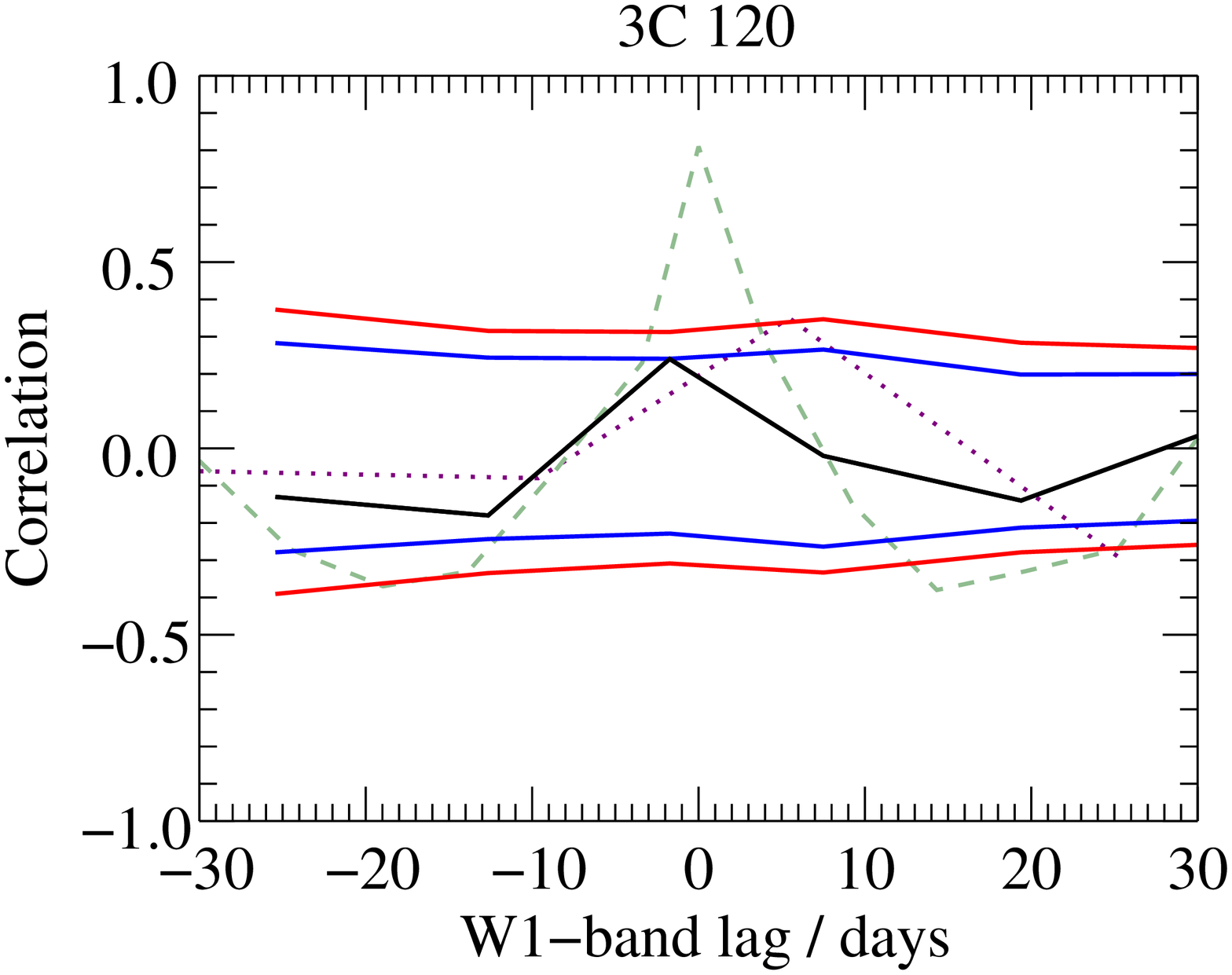}
	\includegraphics[width=5.5cm]{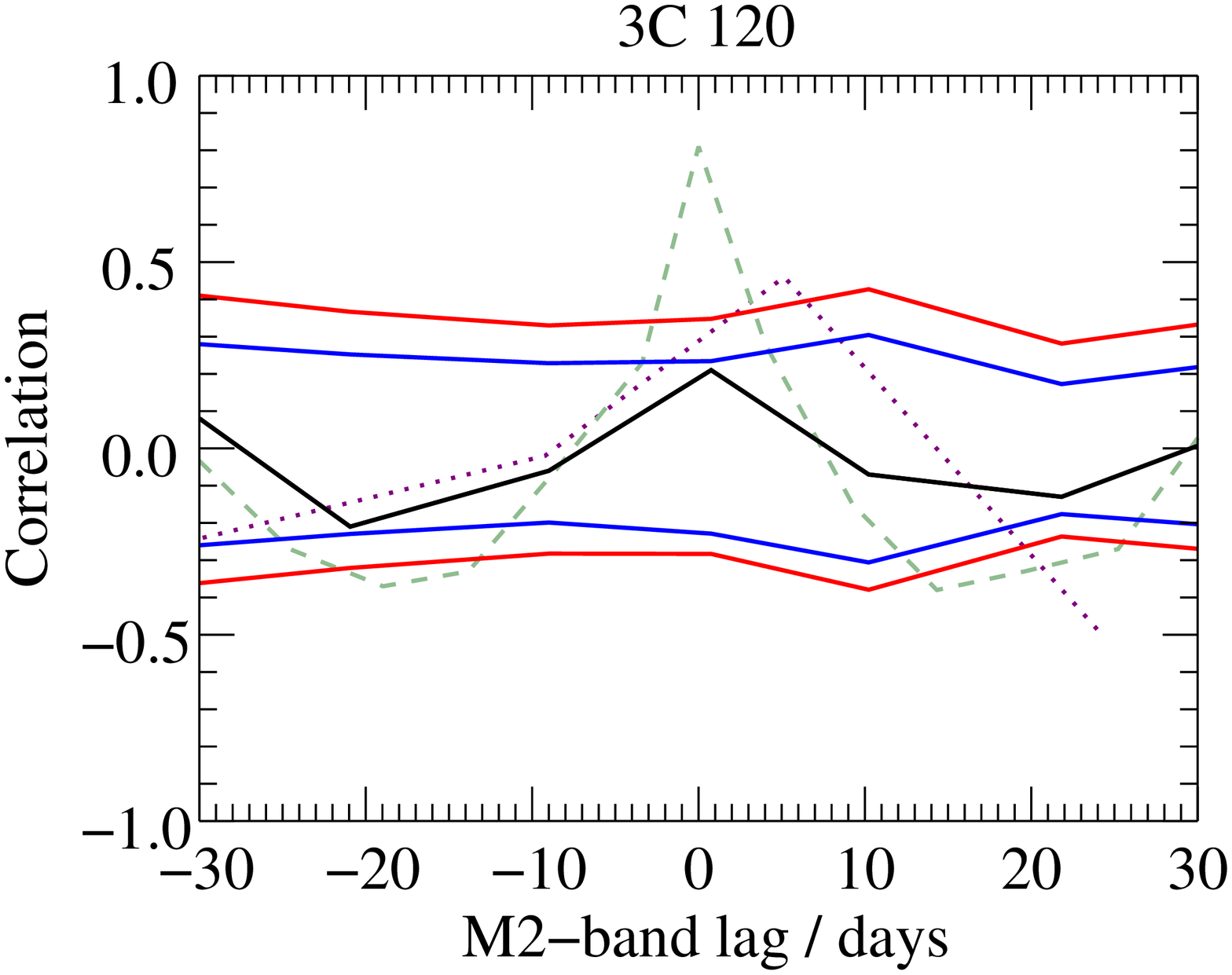}
	\includegraphics[width=5.5cm]{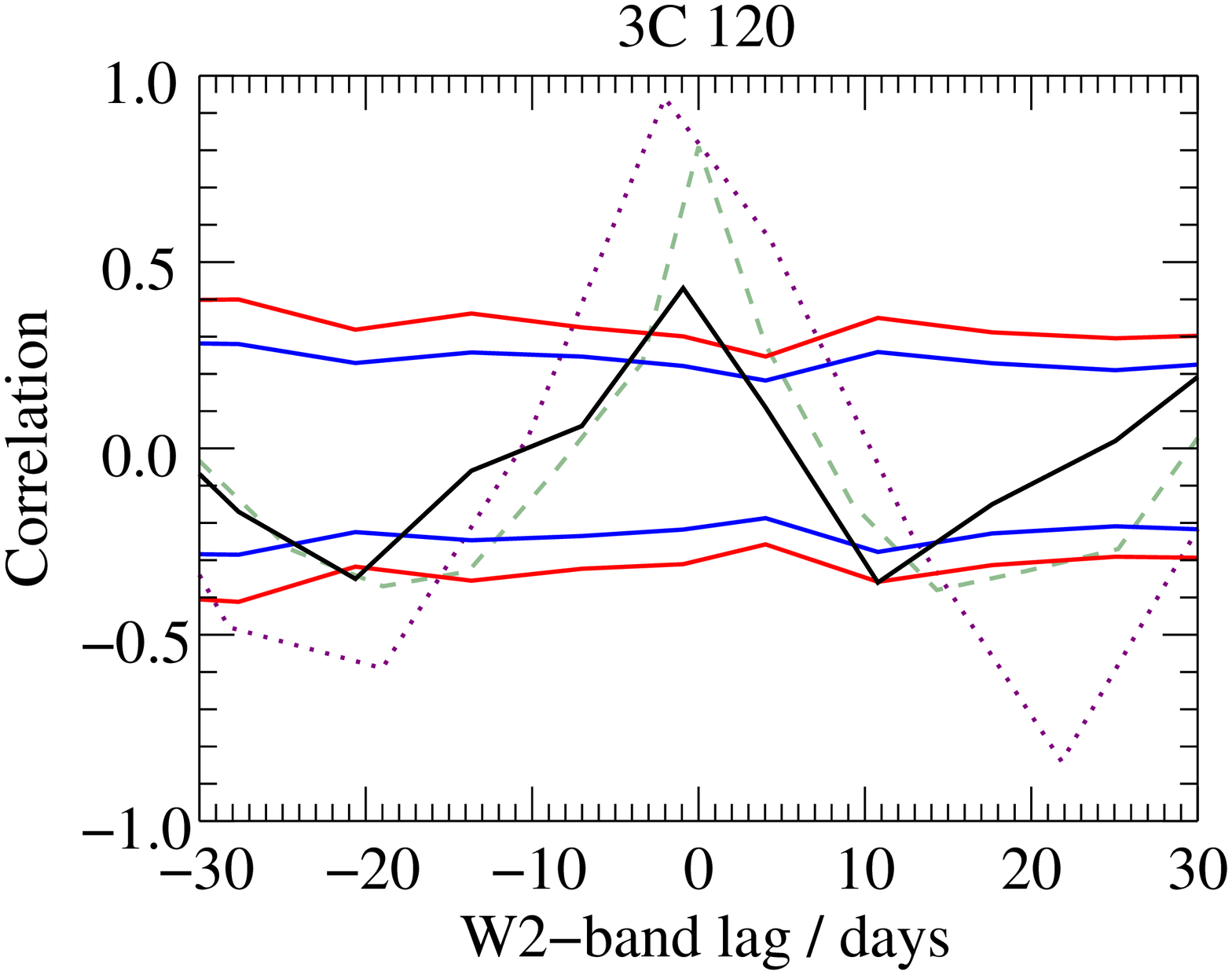}
	\includegraphics[width=5.5cm]{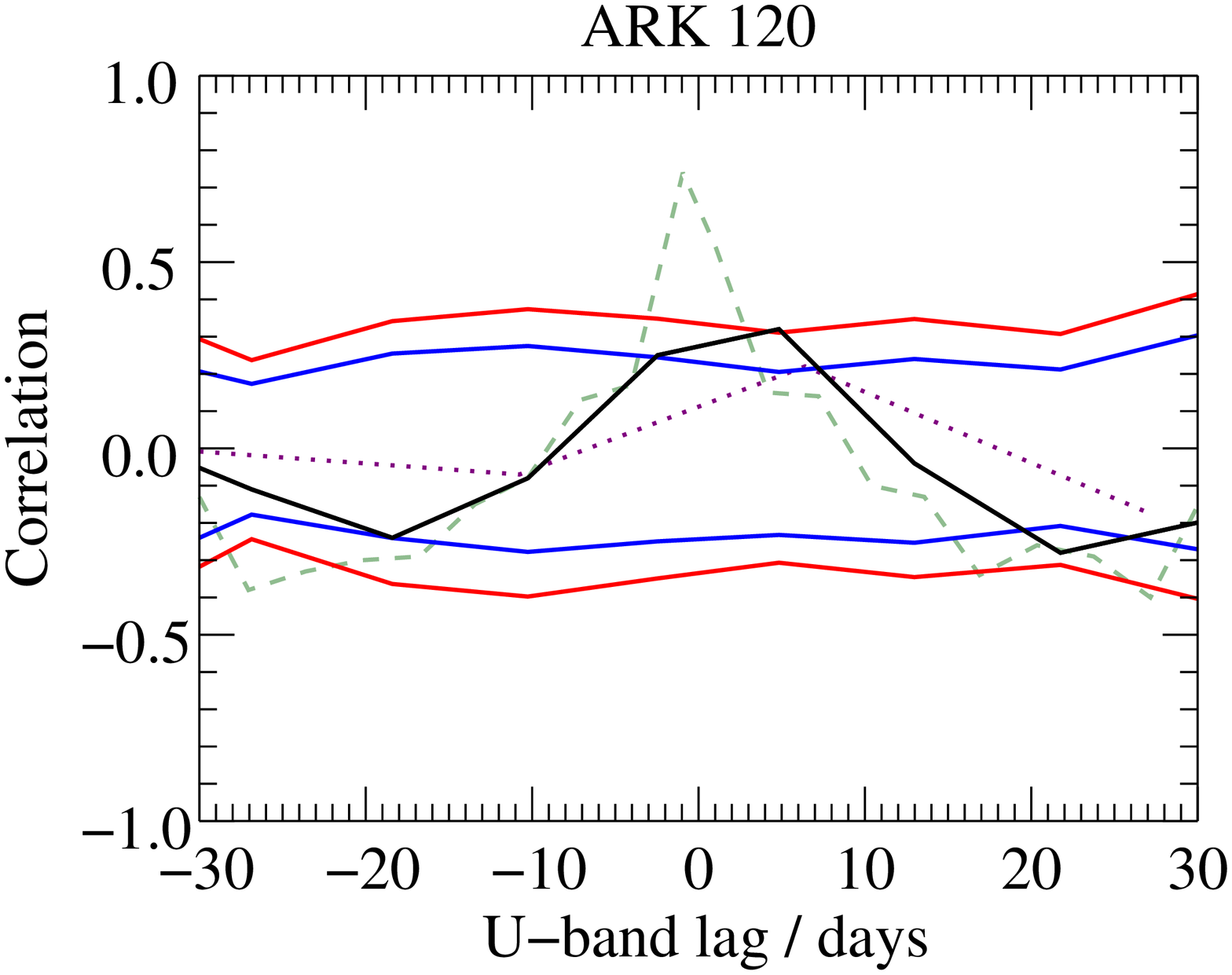}
	\includegraphics[width=5.5cm]{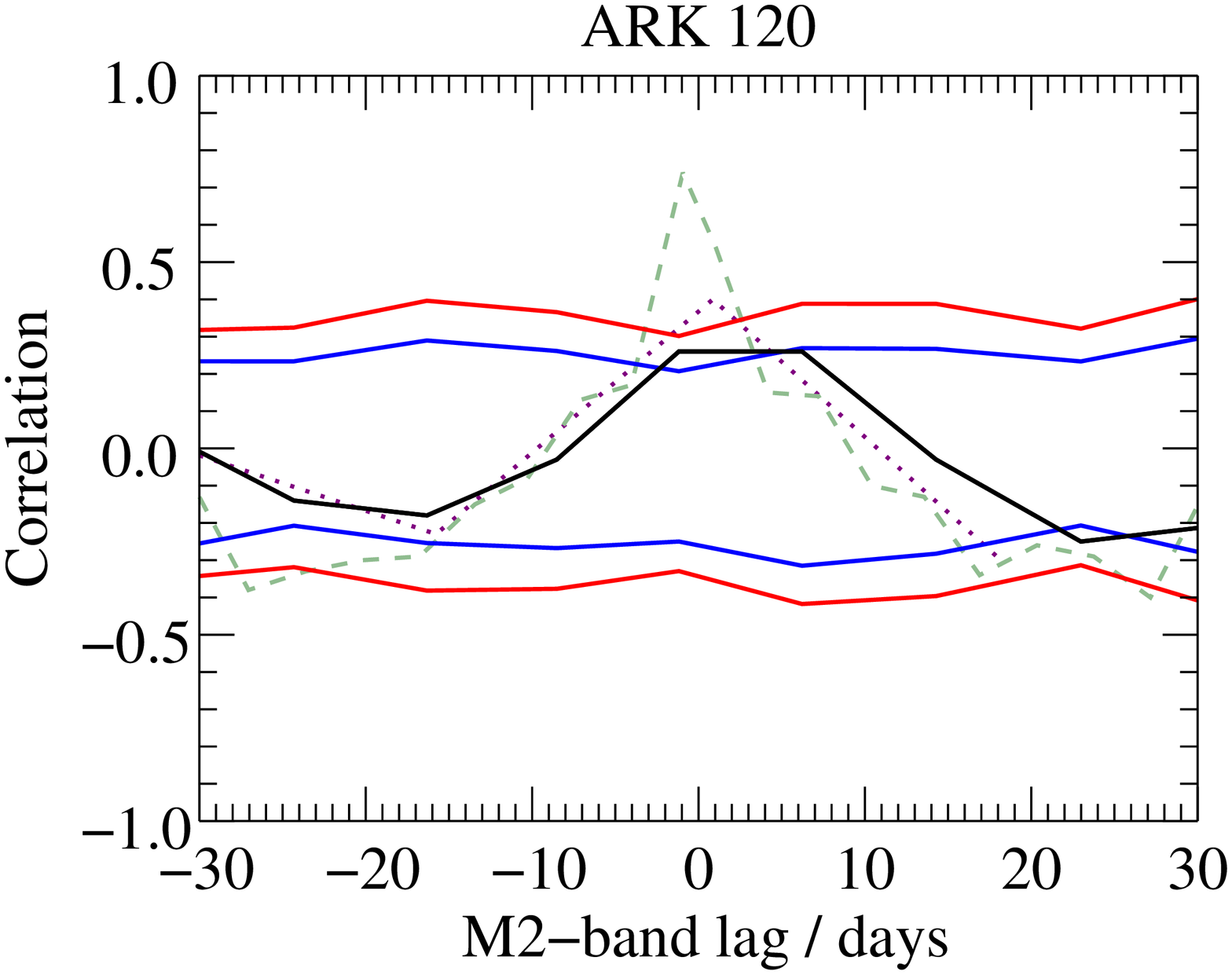}
	\includegraphics[width=5.5cm]{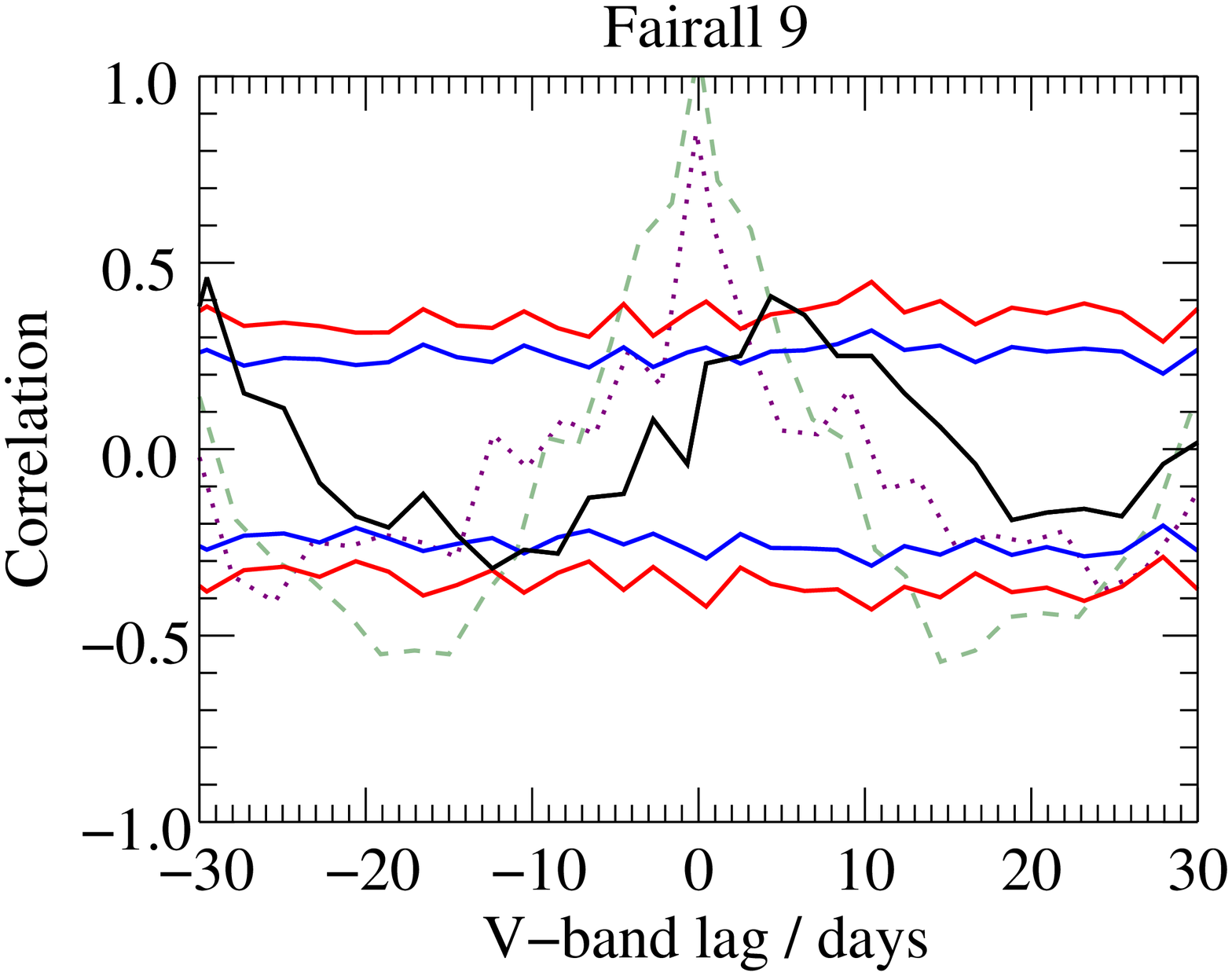}
\end{figure*}
\clearpage
\begin{figure*}
	\includegraphics[width=5.5cm]{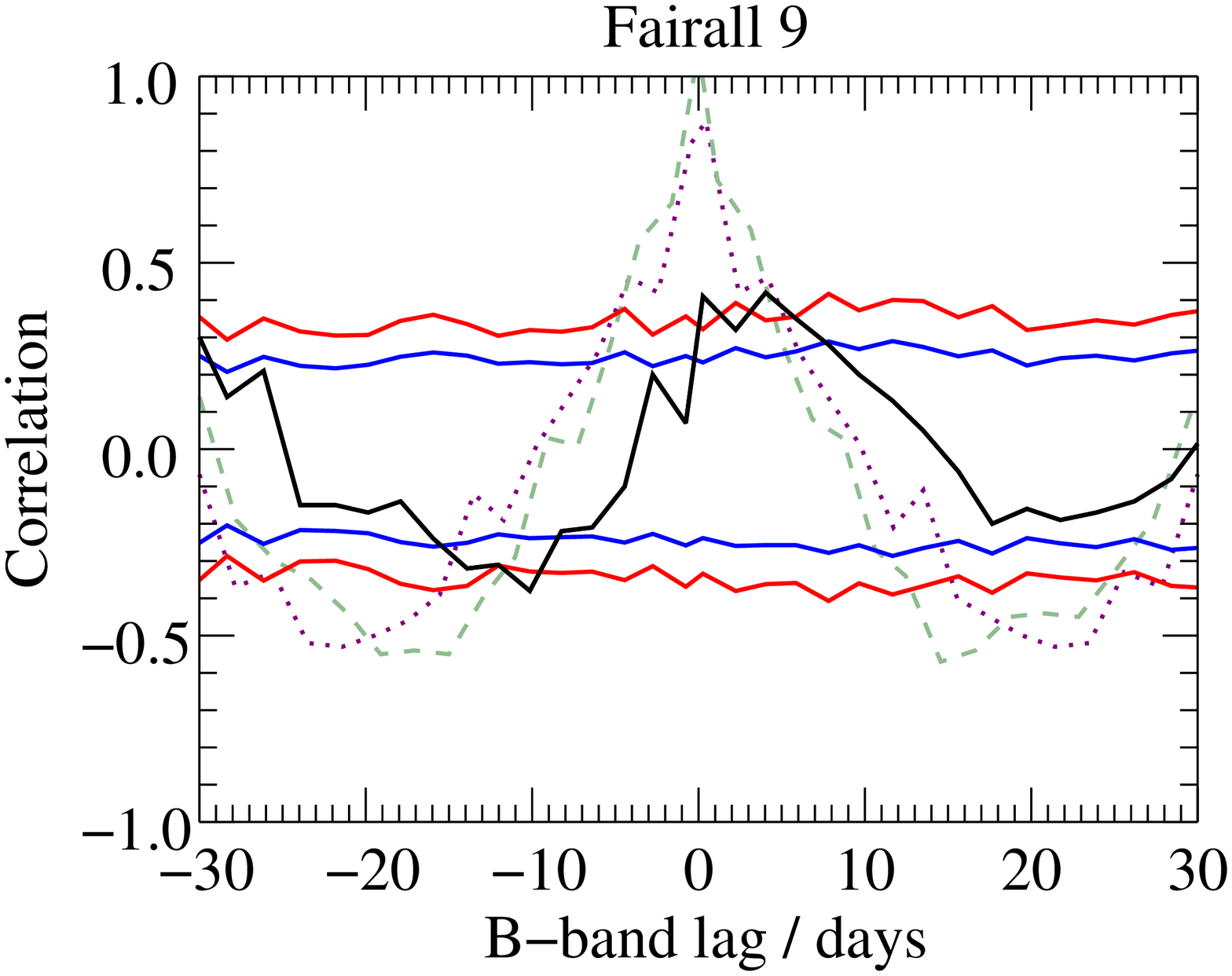}
	\includegraphics[width=5.5cm]{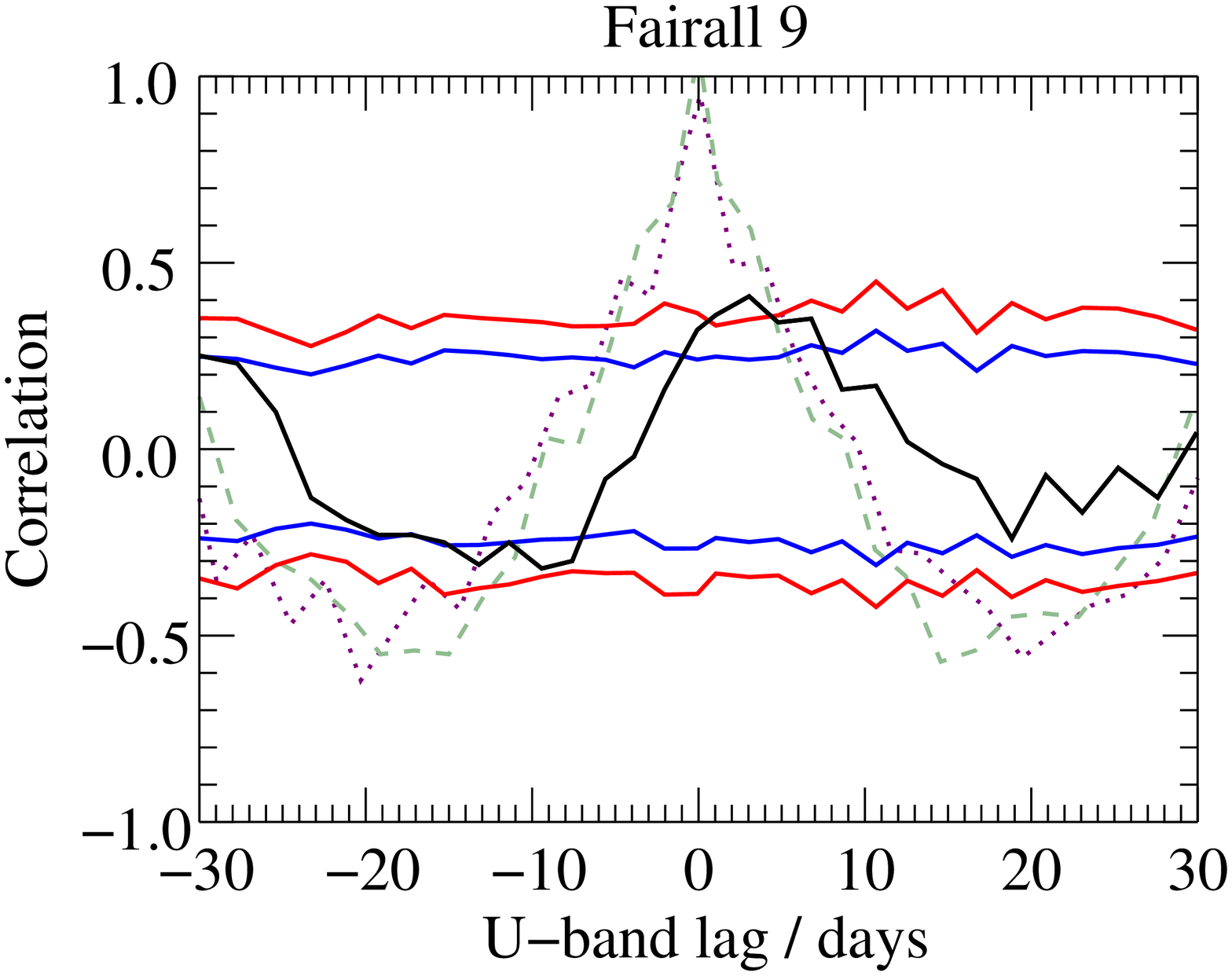}
	\includegraphics[width=5.5cm]{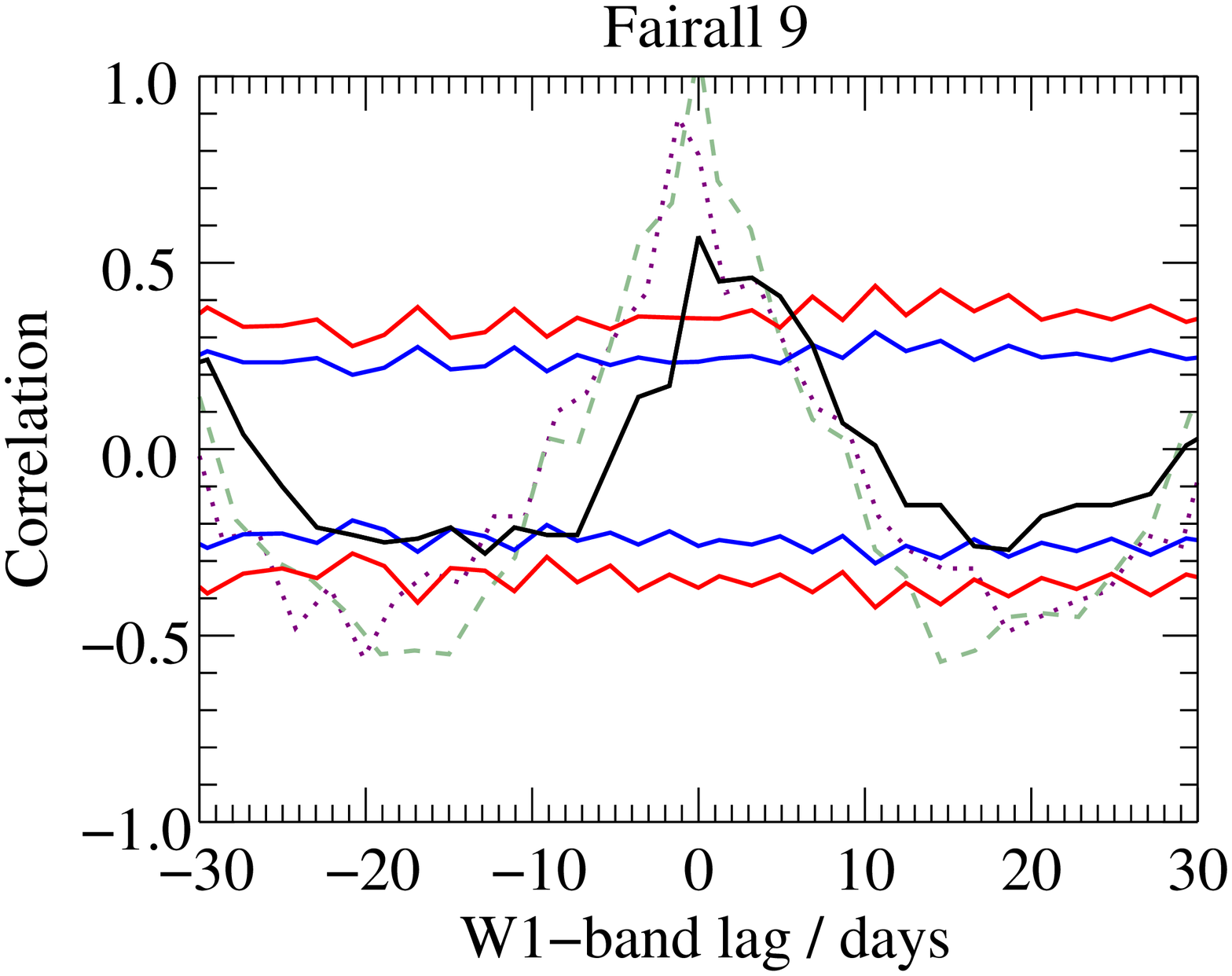}
	\includegraphics[width=5.5cm]{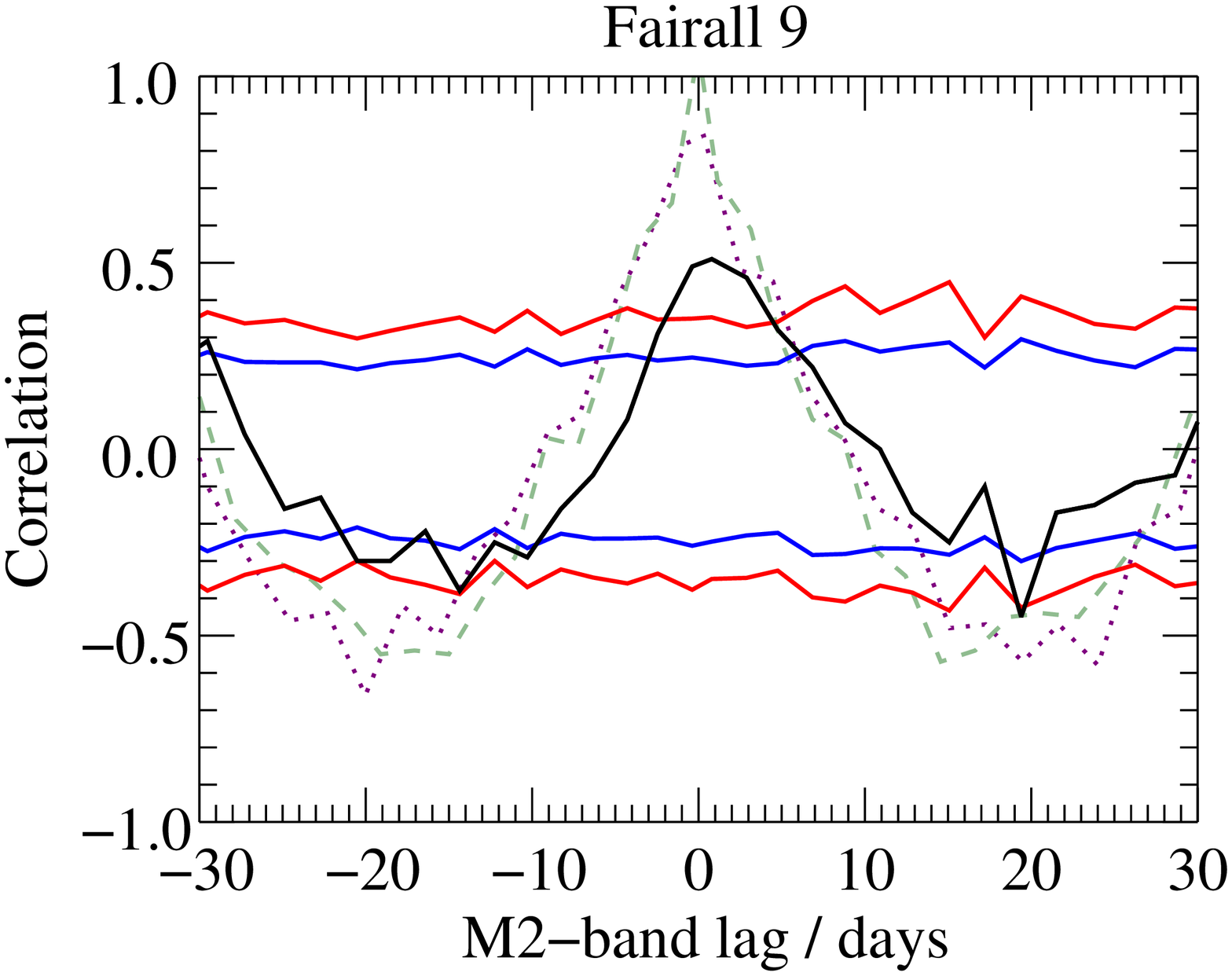}
	\includegraphics[width=5.5cm]{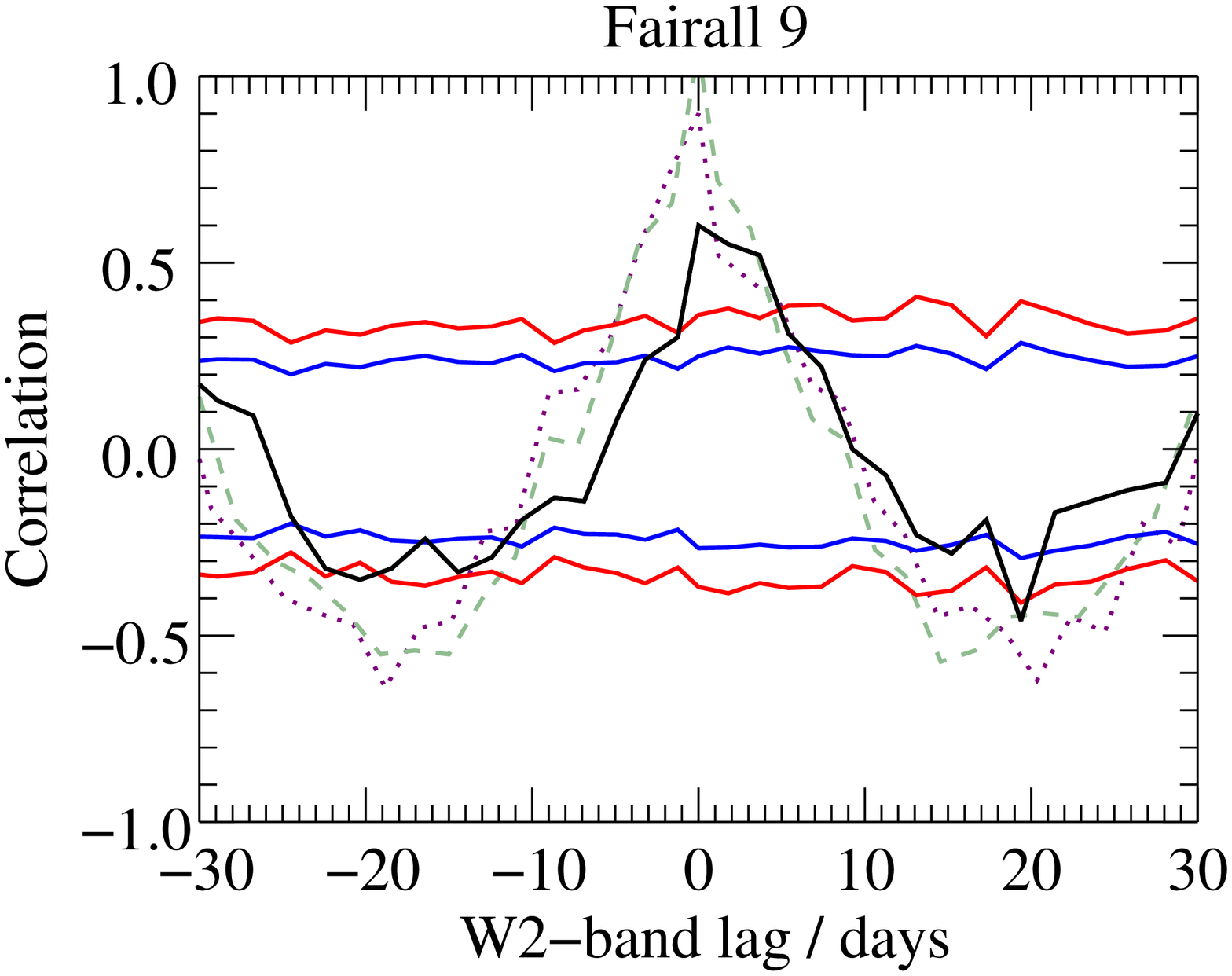}
	\includegraphics[width=5.5cm]{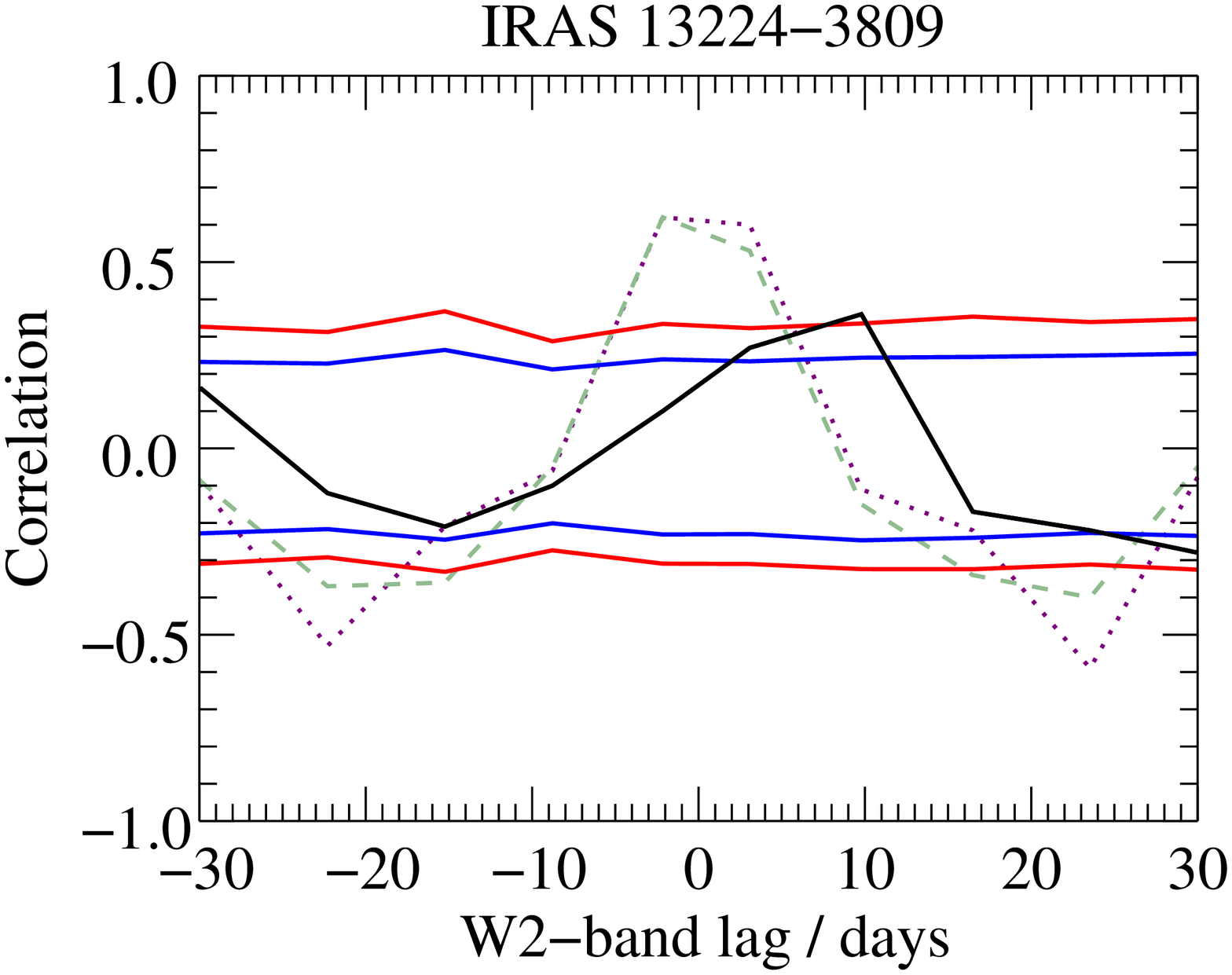}
	\includegraphics[width=5.5cm]{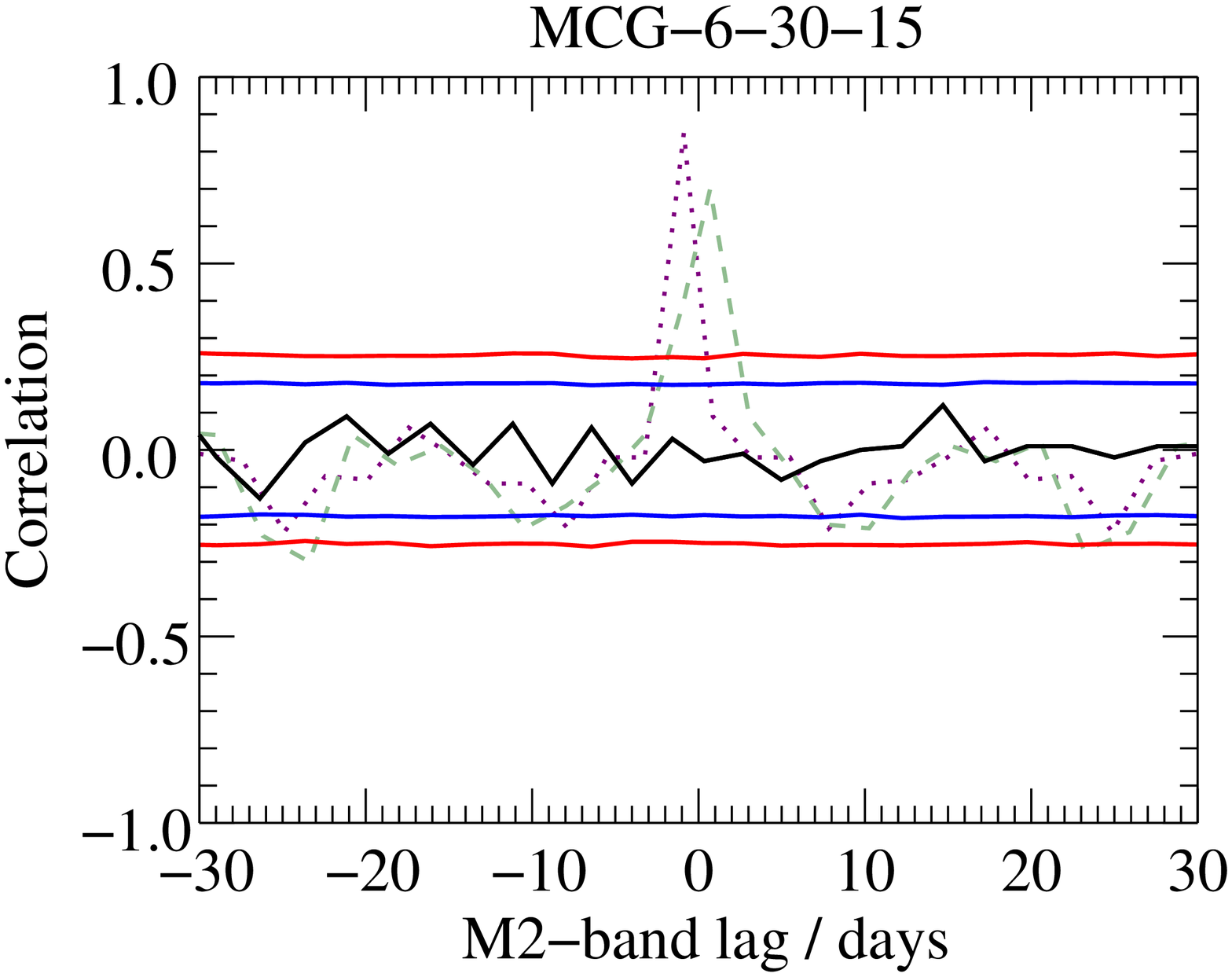}
	\includegraphics[width=5.5cm]{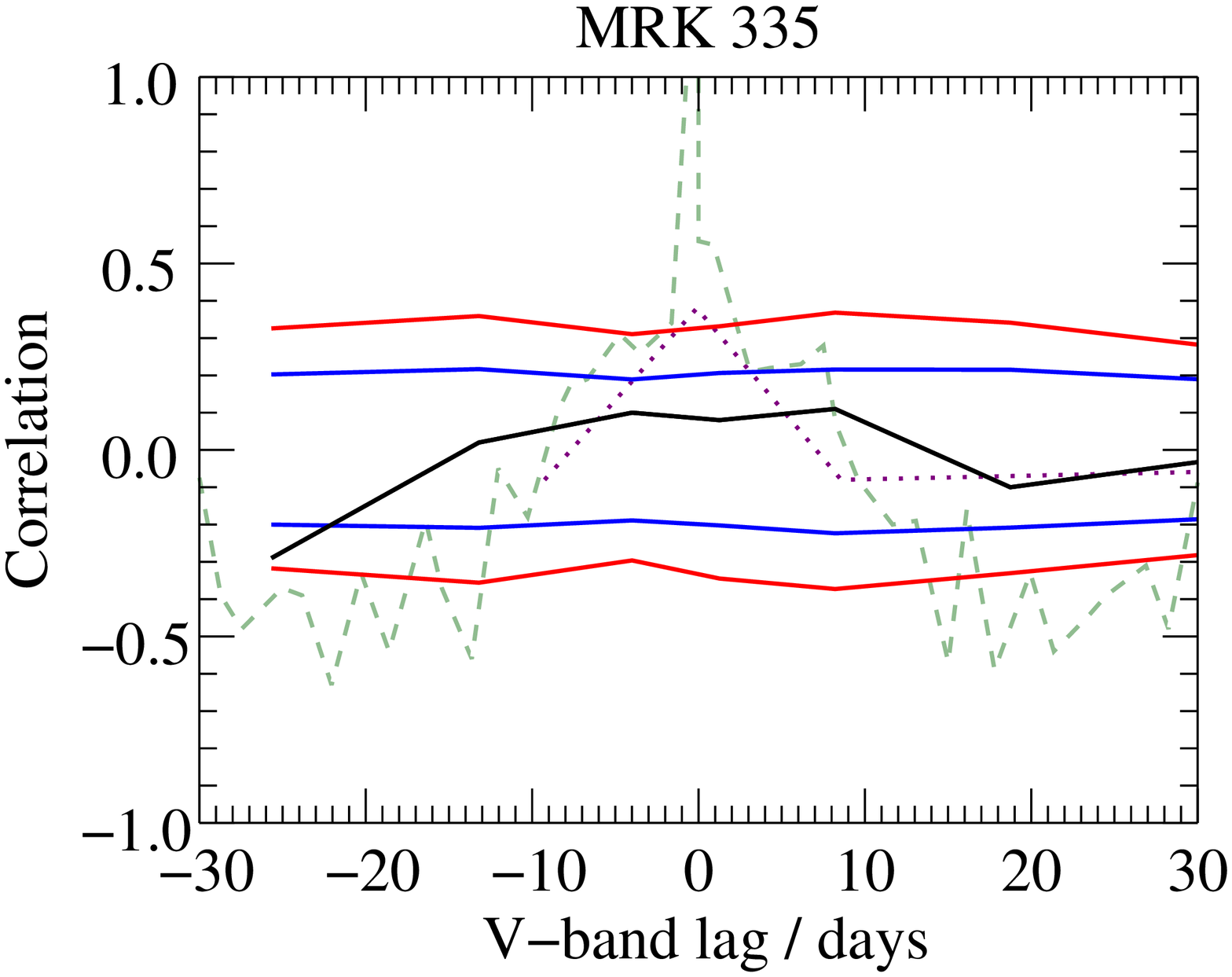}
	\includegraphics[width=5.5cm]{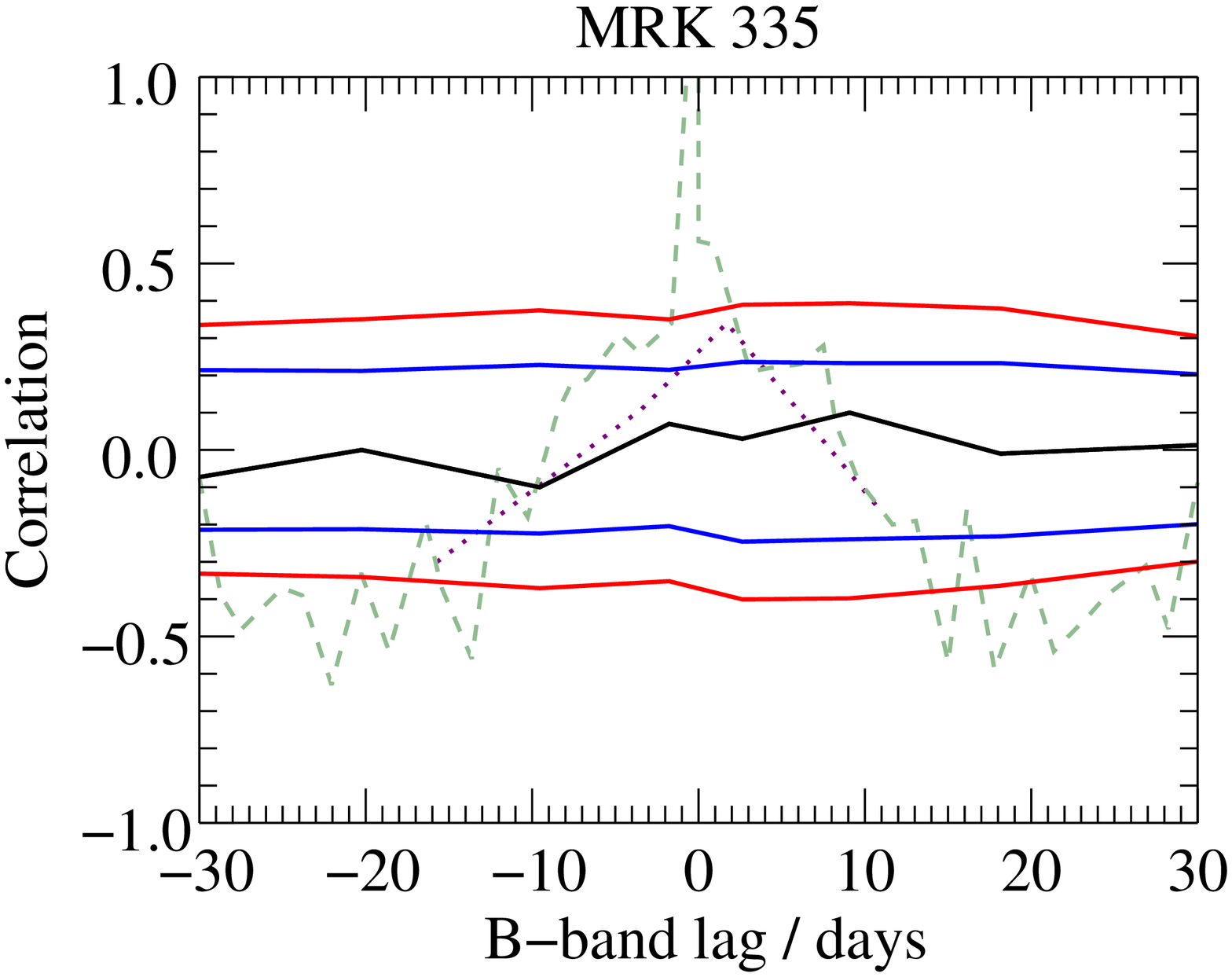}
	\includegraphics[width=5.5cm]{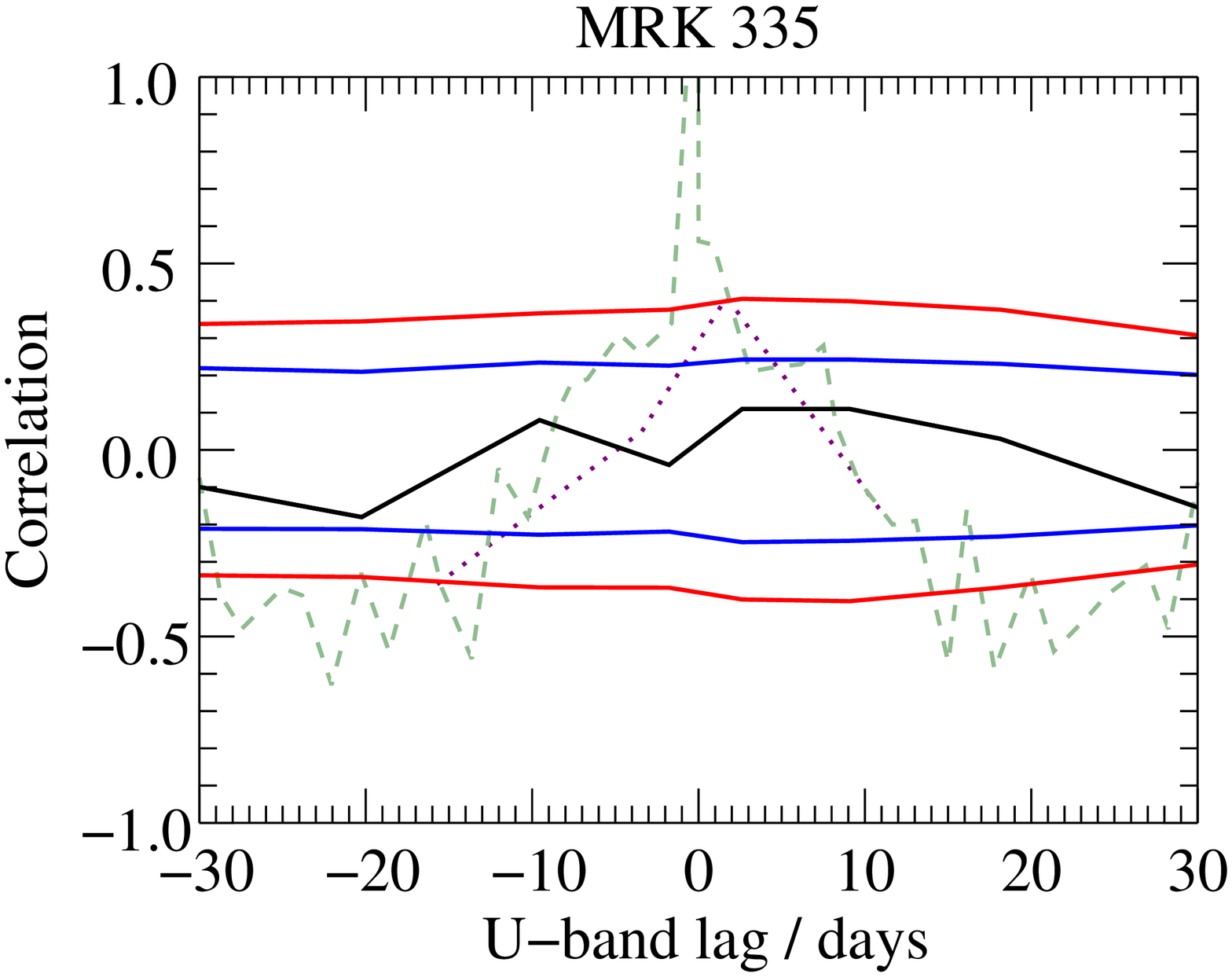}
	\includegraphics[width=5.5cm]{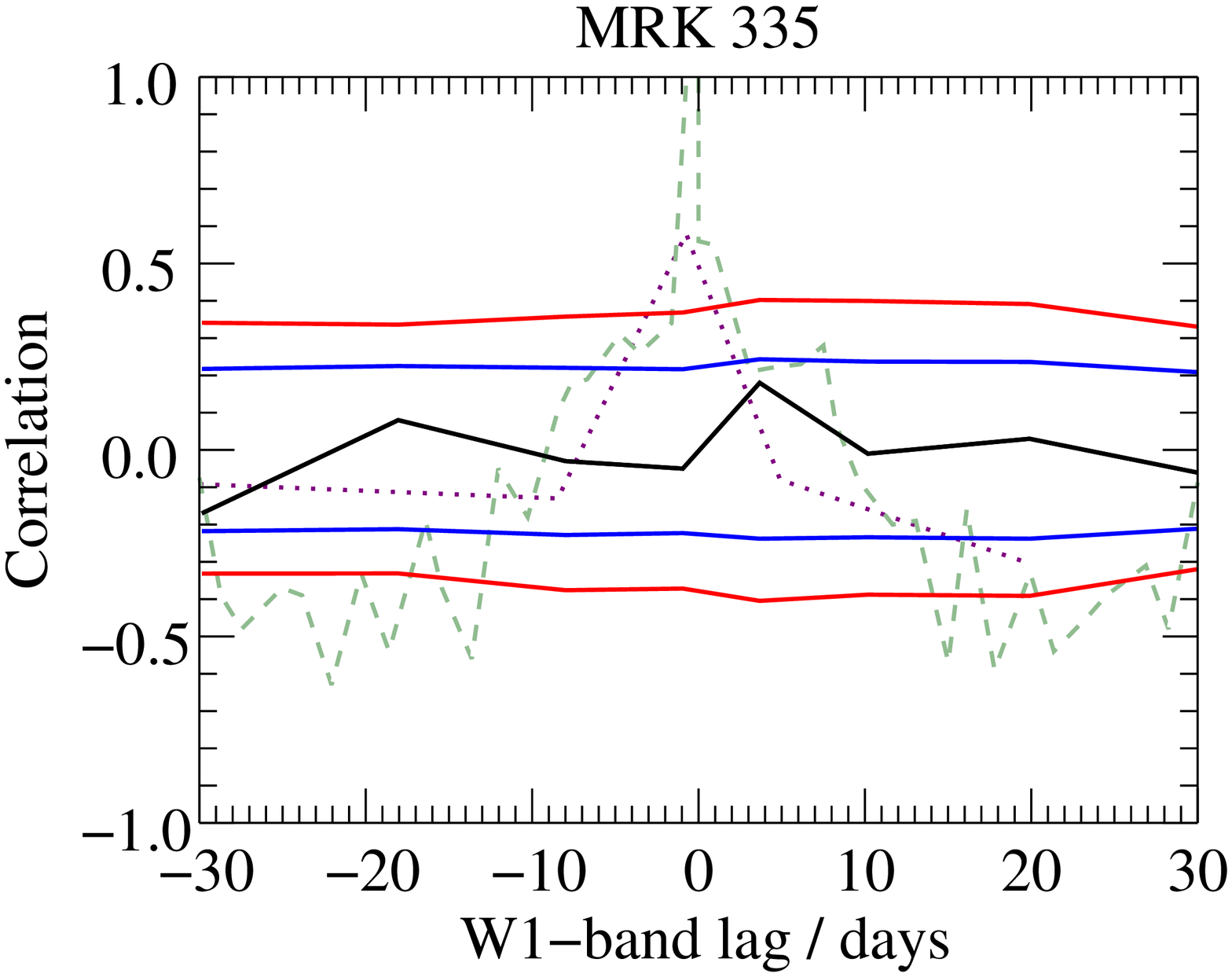}
	\includegraphics[width=5.5cm]{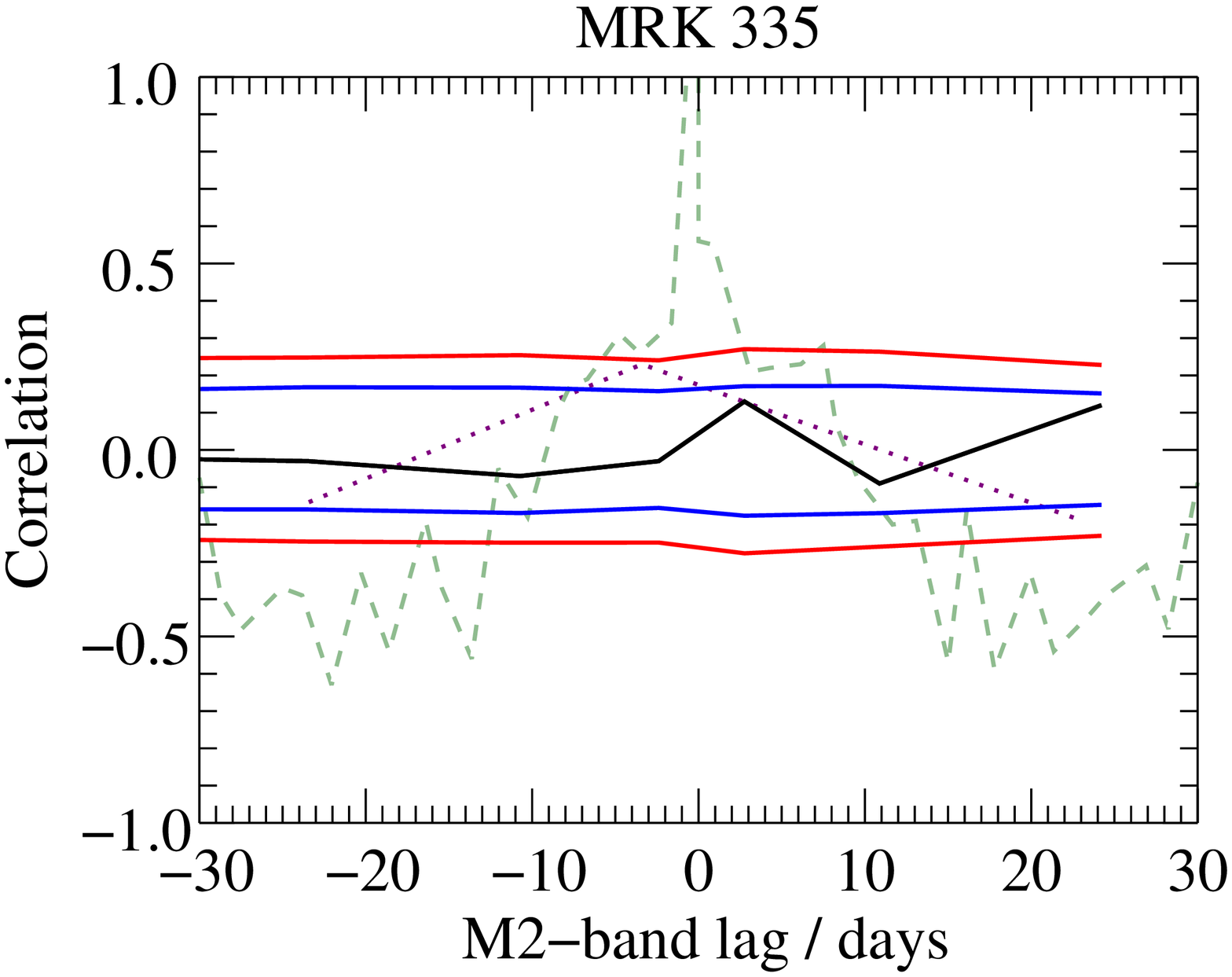}
	\includegraphics[width=5.5cm]{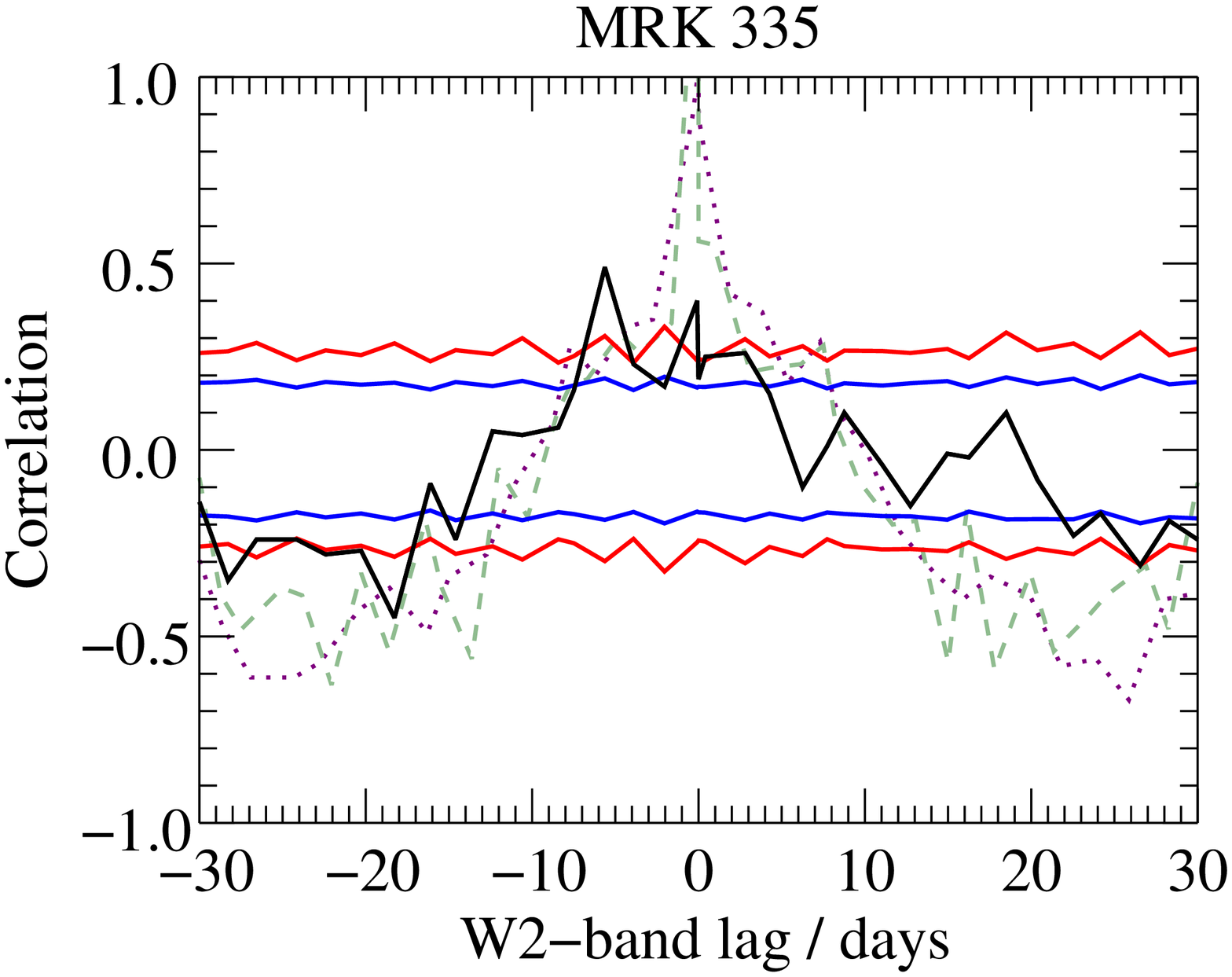}
	\includegraphics[width=5.5cm]{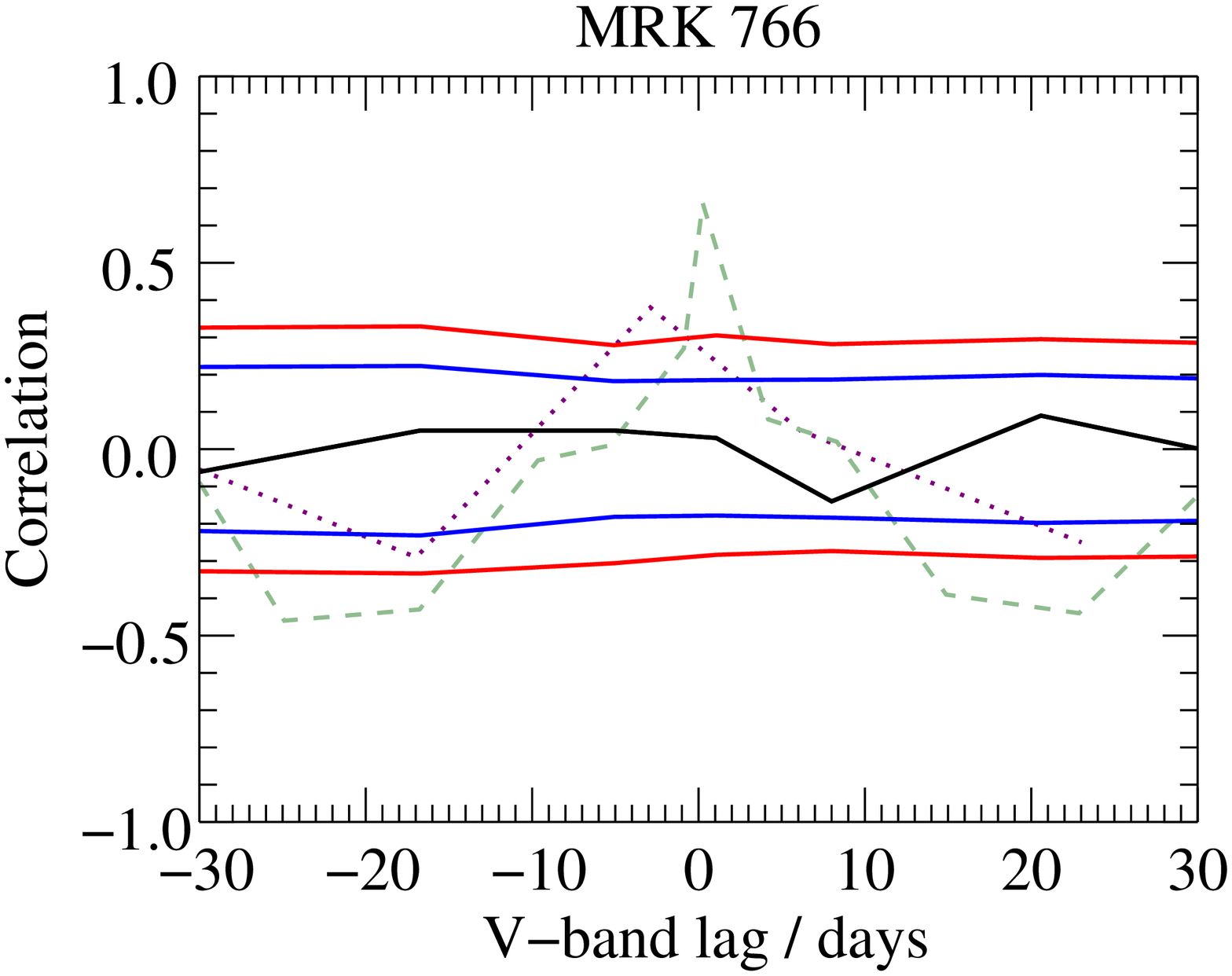}
	\includegraphics[width=5.5cm]{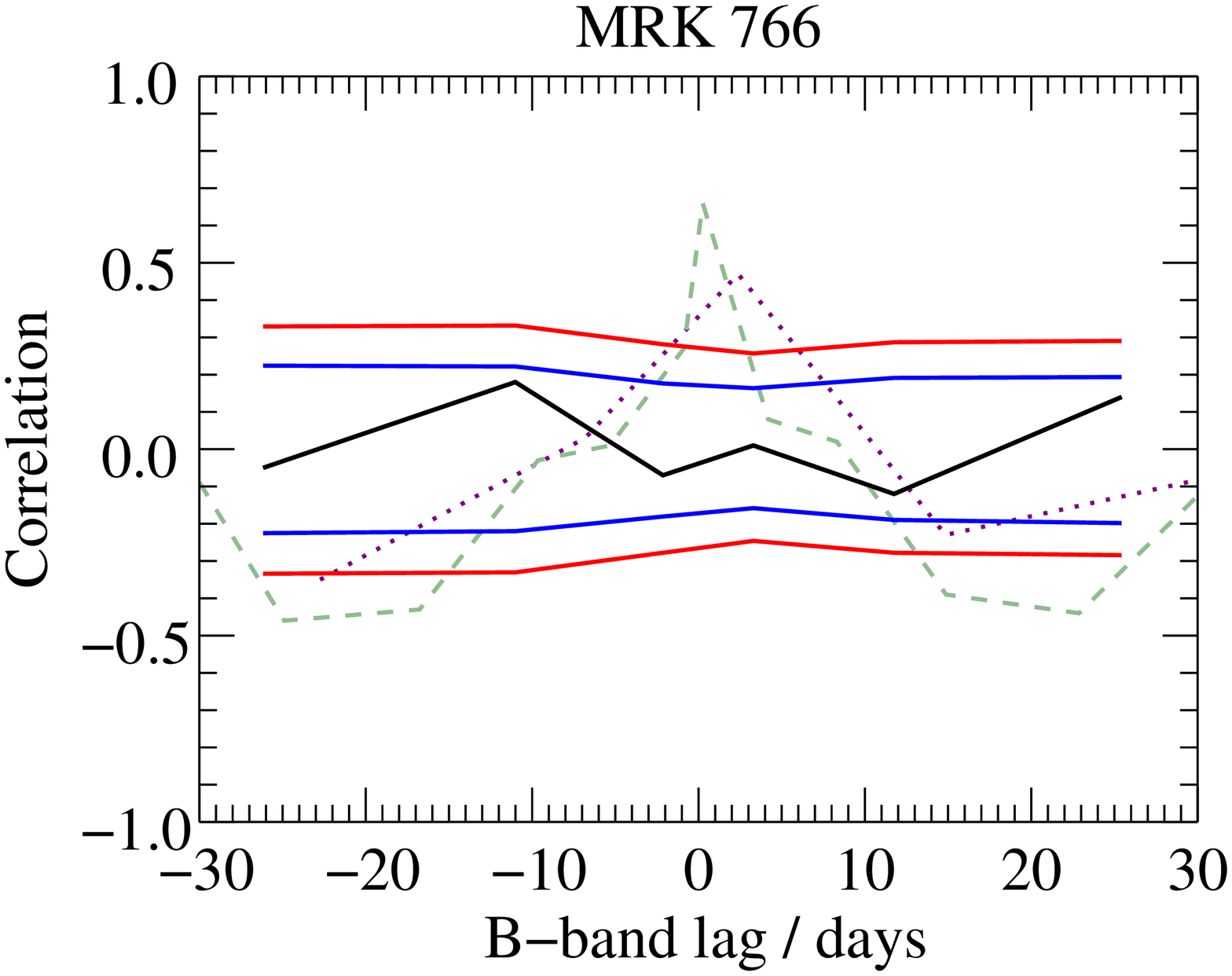}
\end{figure*}
\begin{figure*}
	\includegraphics[width=5.5cm]{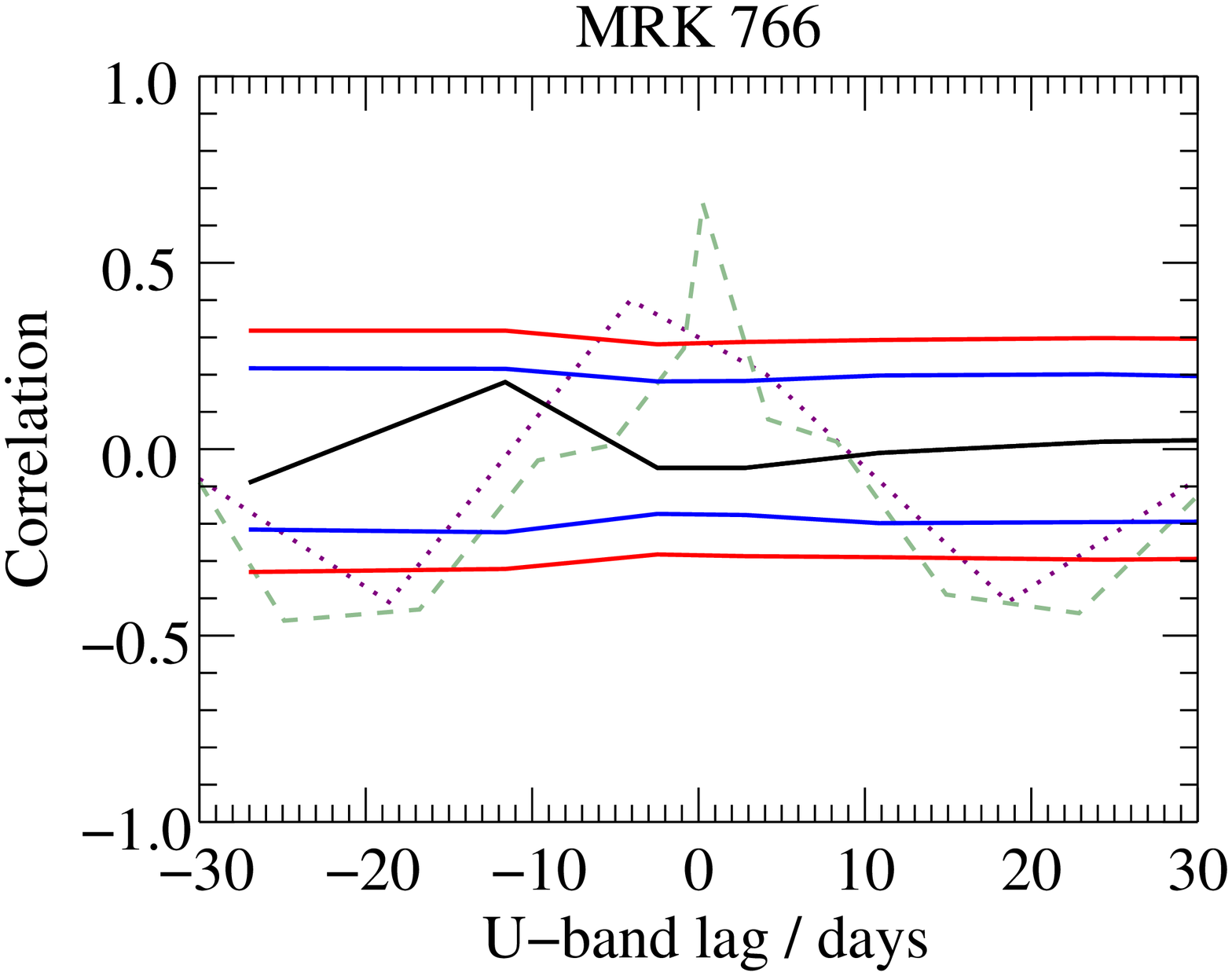}
	\includegraphics[width=5.5cm]{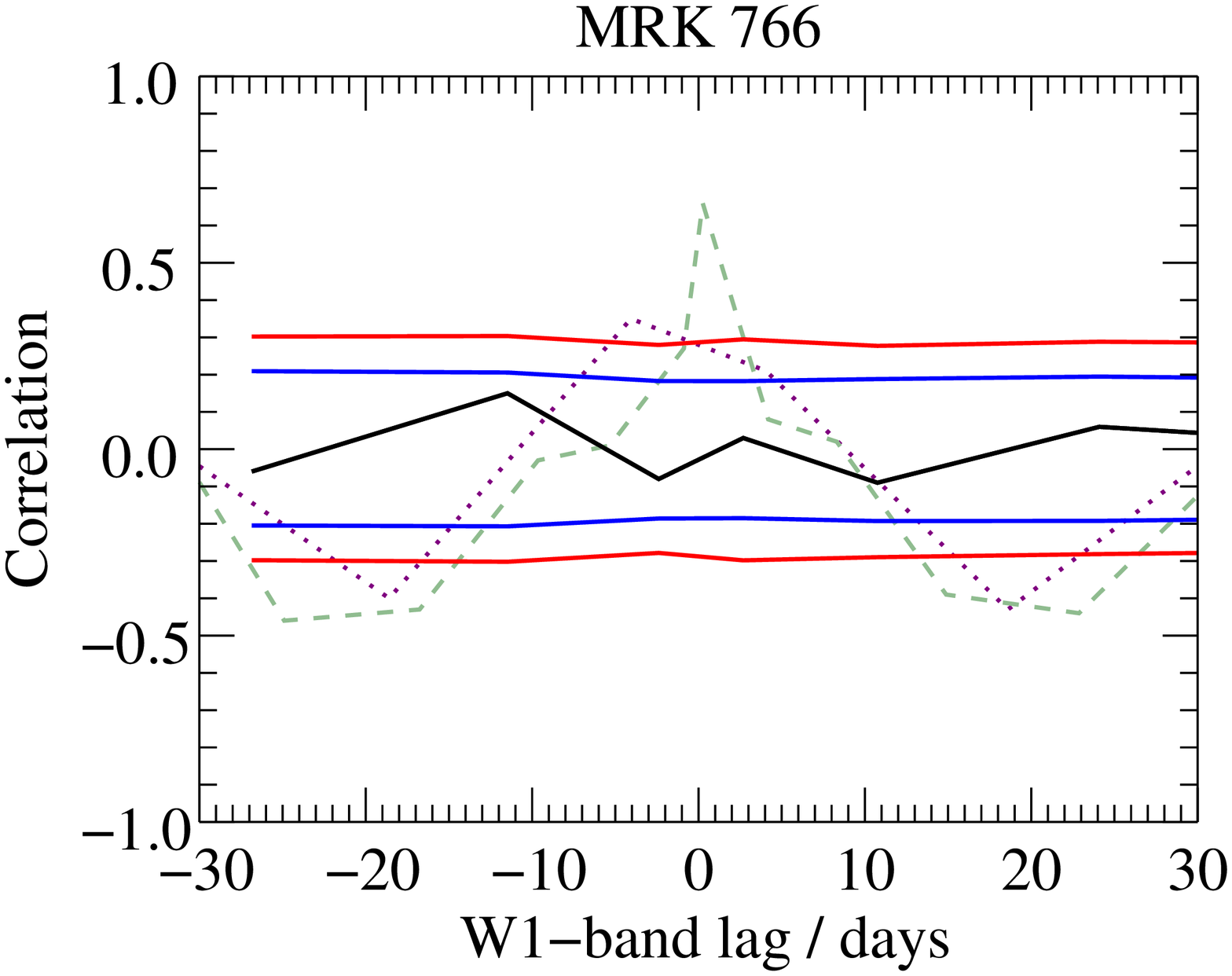}
	\includegraphics[width=5.5cm]{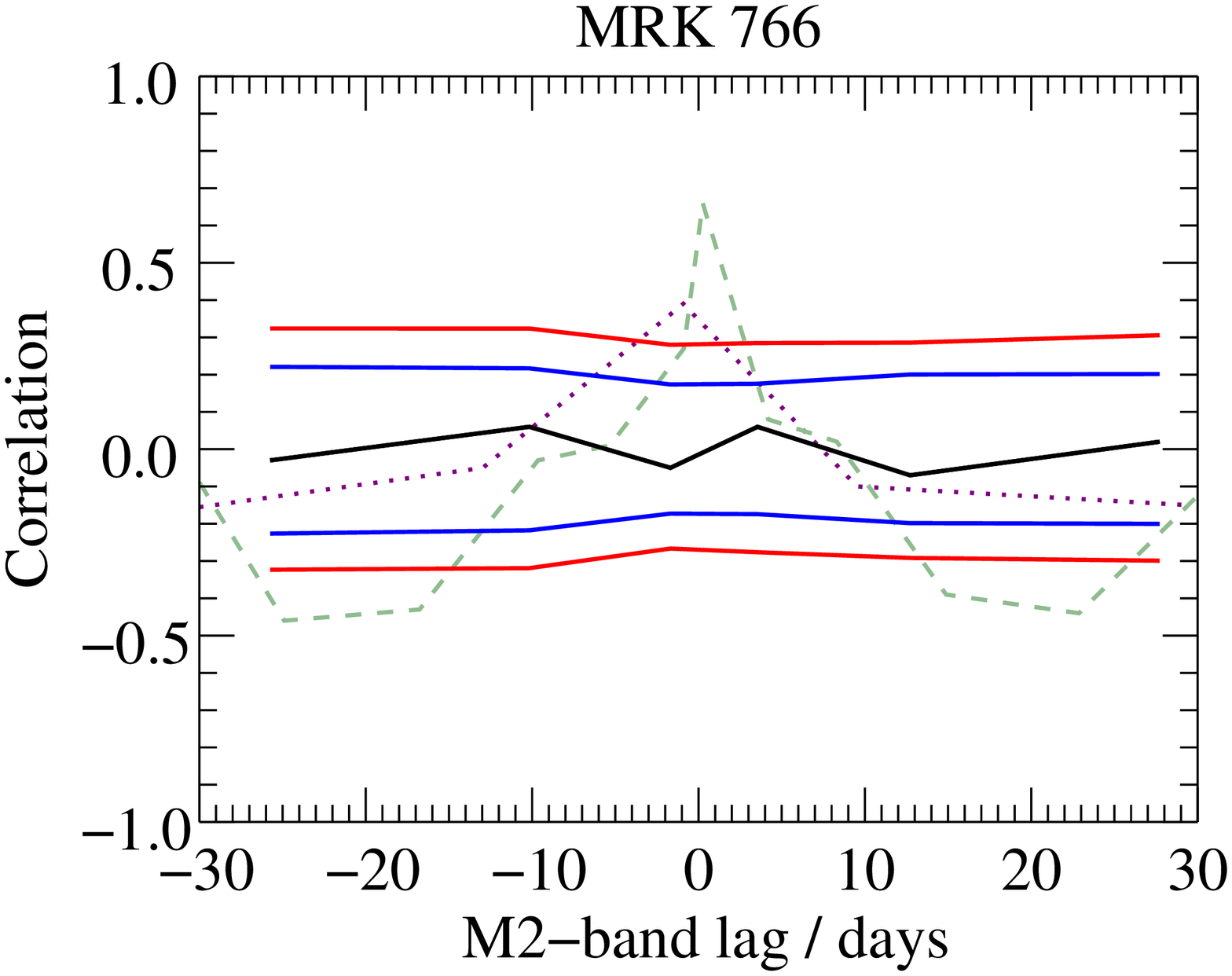}
	\includegraphics[width=5.5cm]{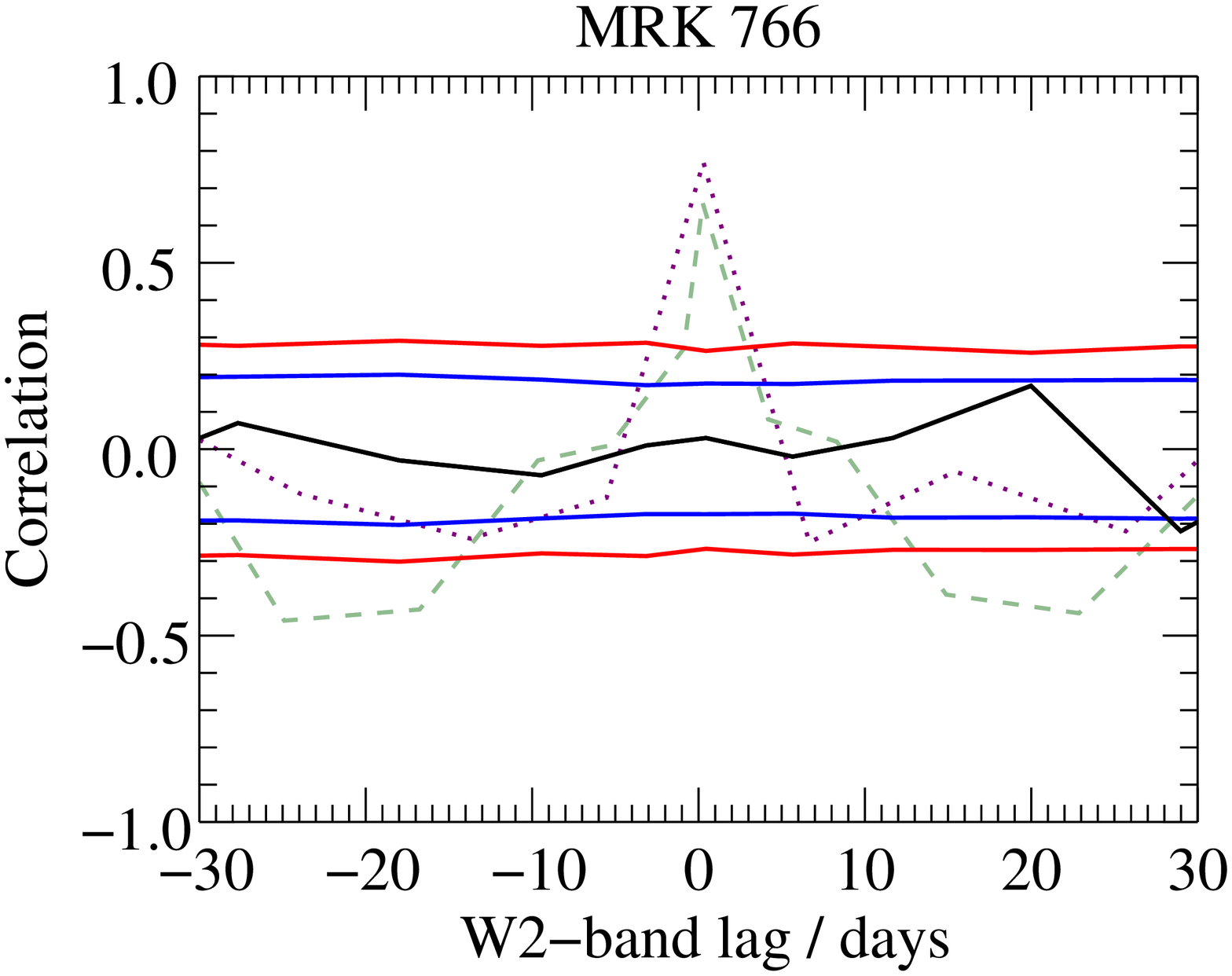}
	\includegraphics[width=5.5cm]{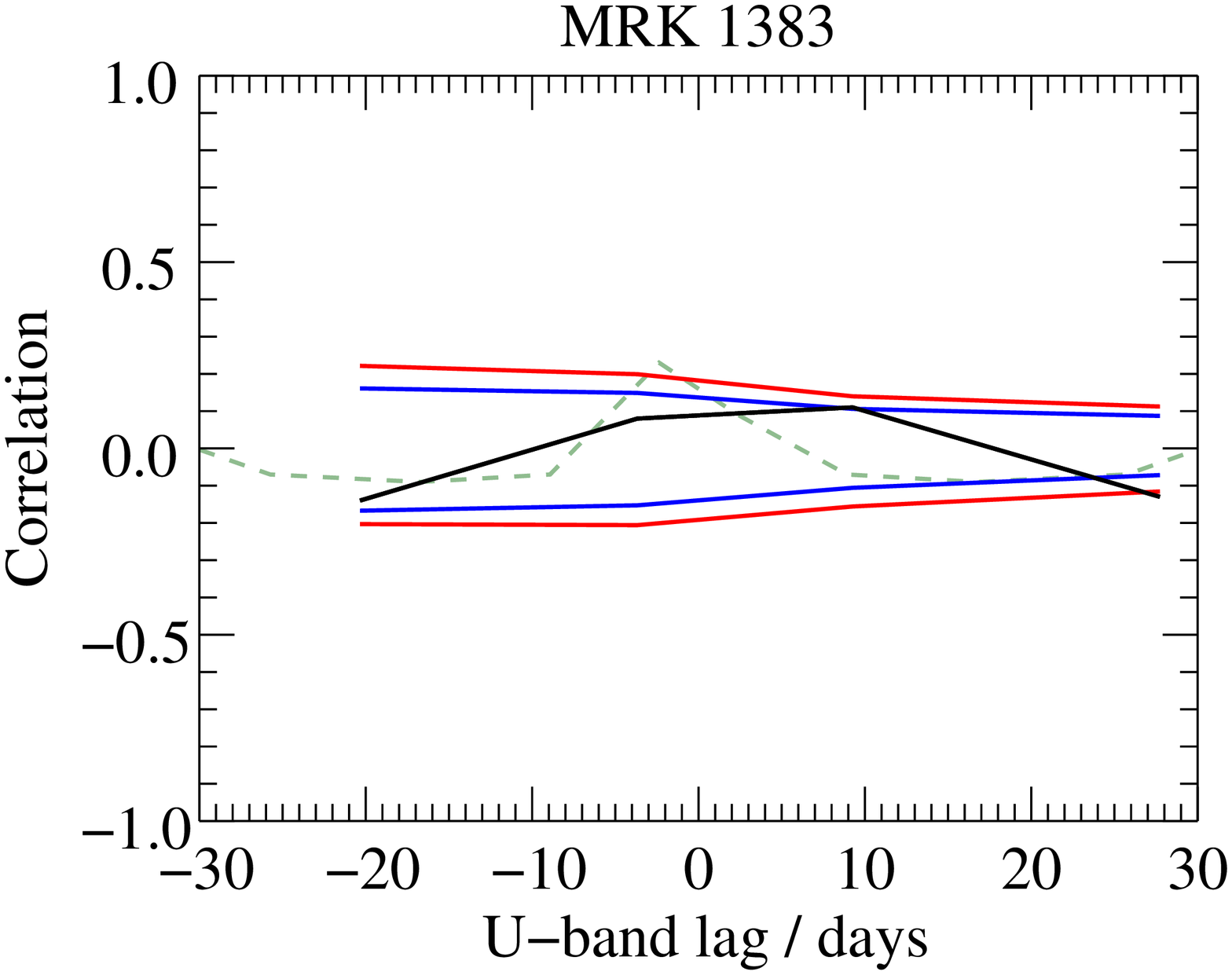}
	\includegraphics[width=5.5cm]{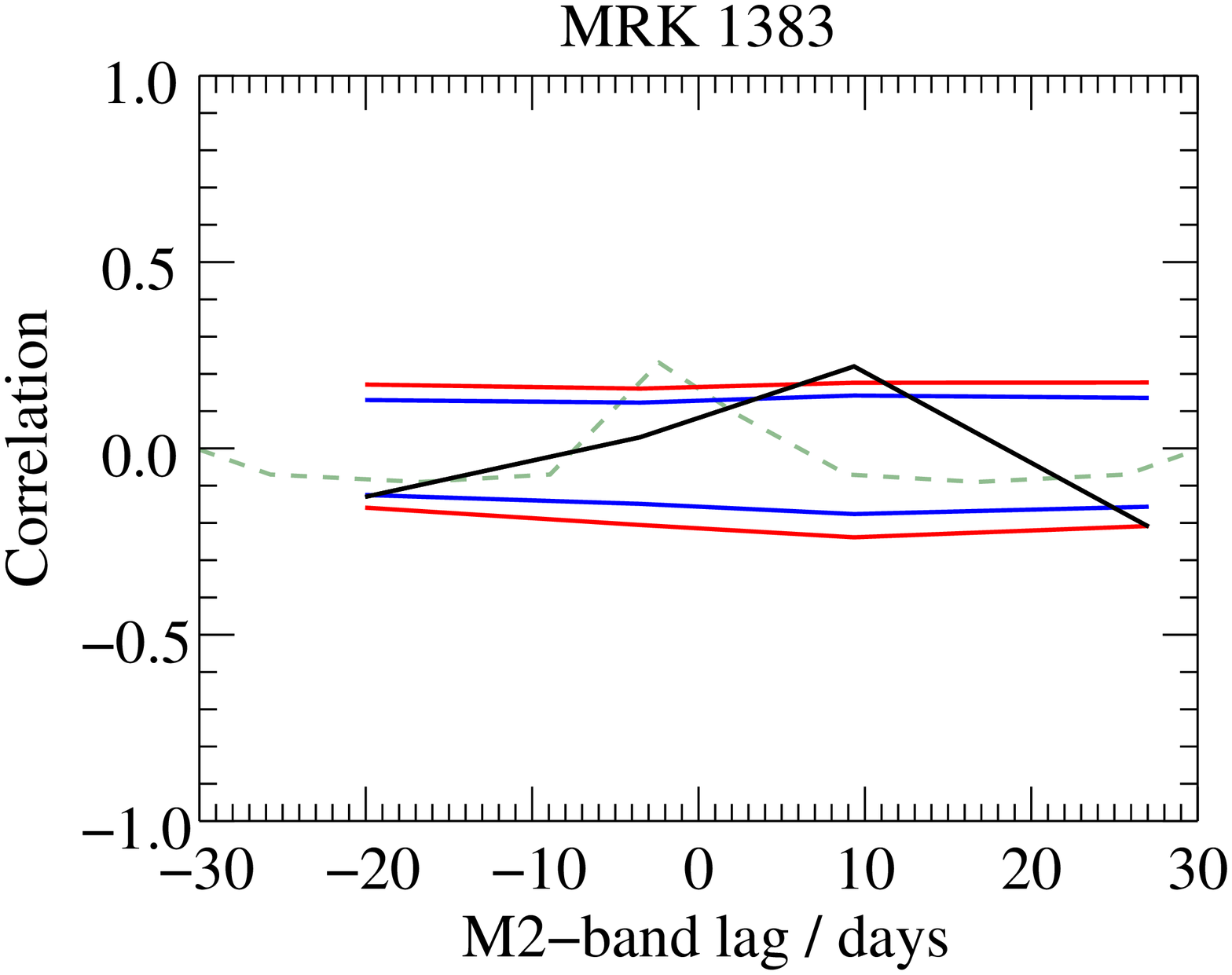}
	\includegraphics[width=5.5cm]{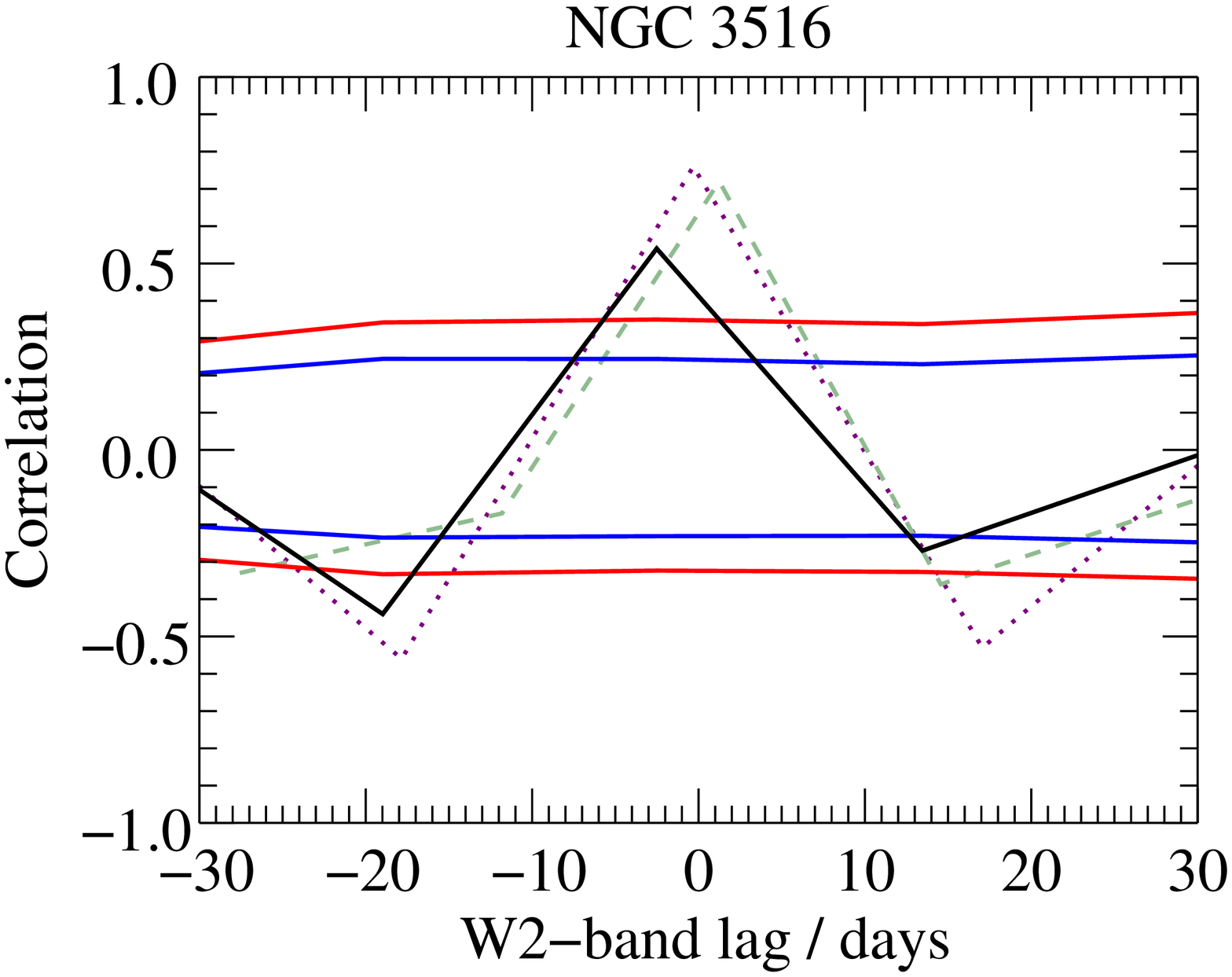}
	\includegraphics[width=5.5cm]{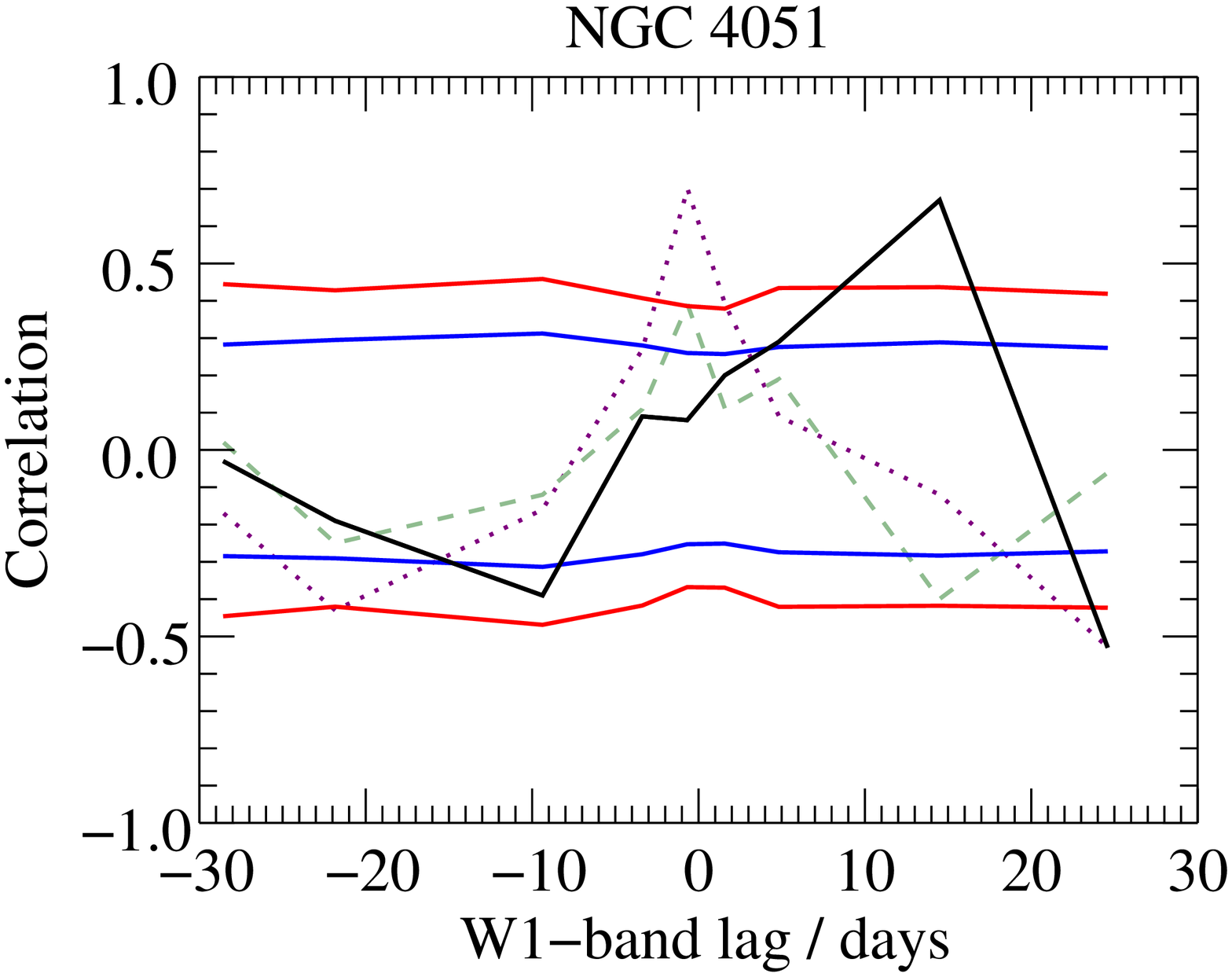}
	\includegraphics[width=5.5cm]{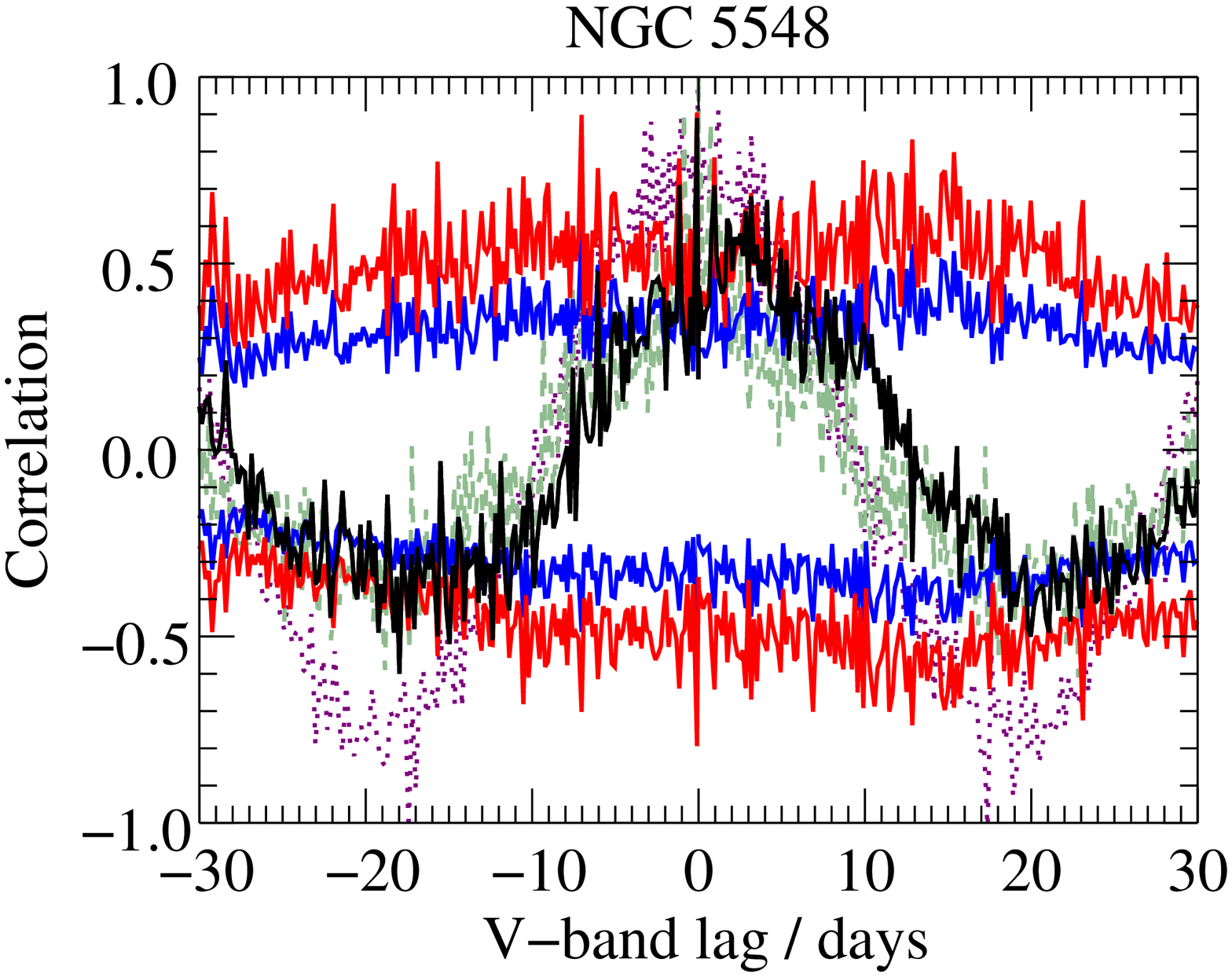}
	\includegraphics[width=5.5cm]{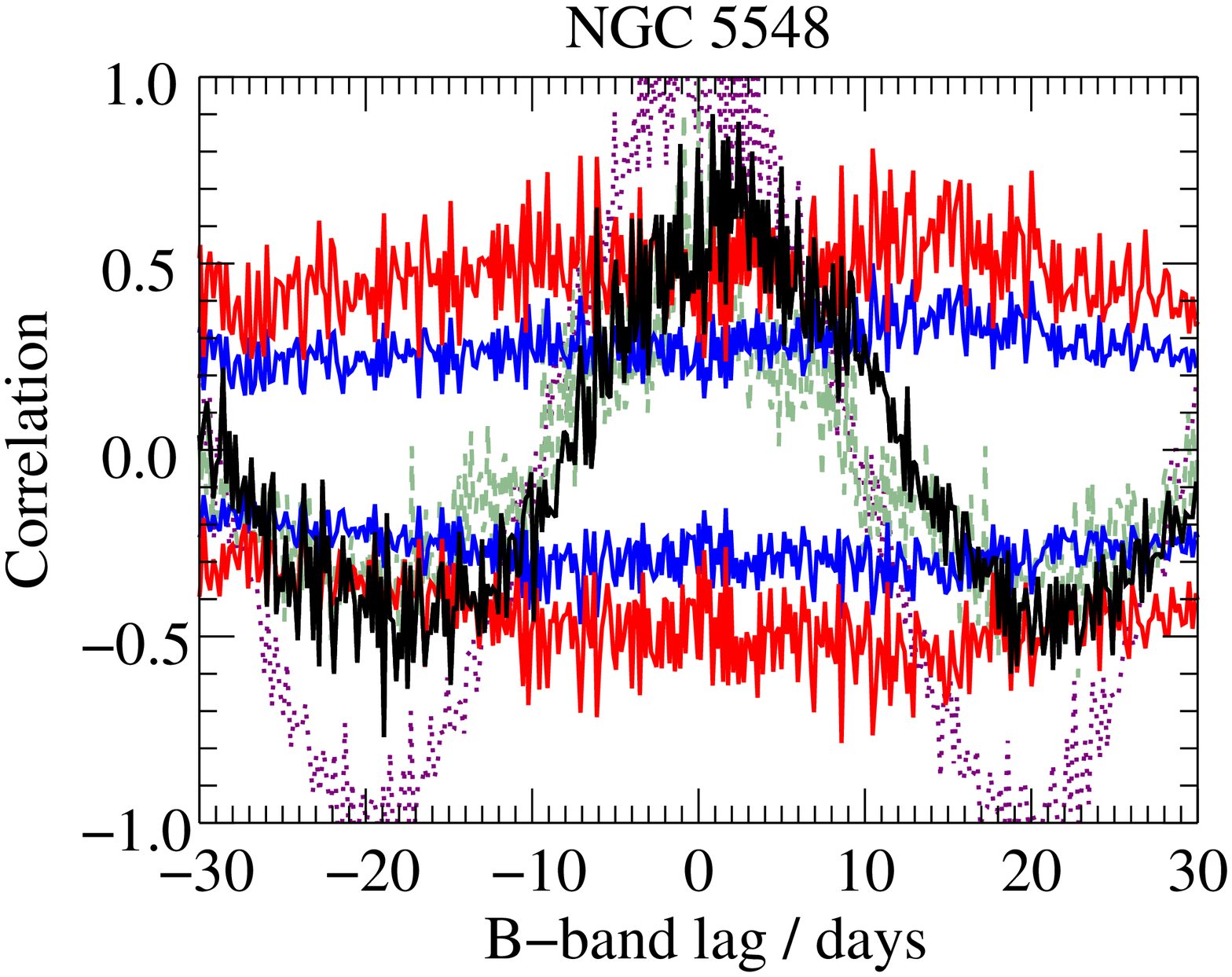}
	\includegraphics[width=5.5cm]{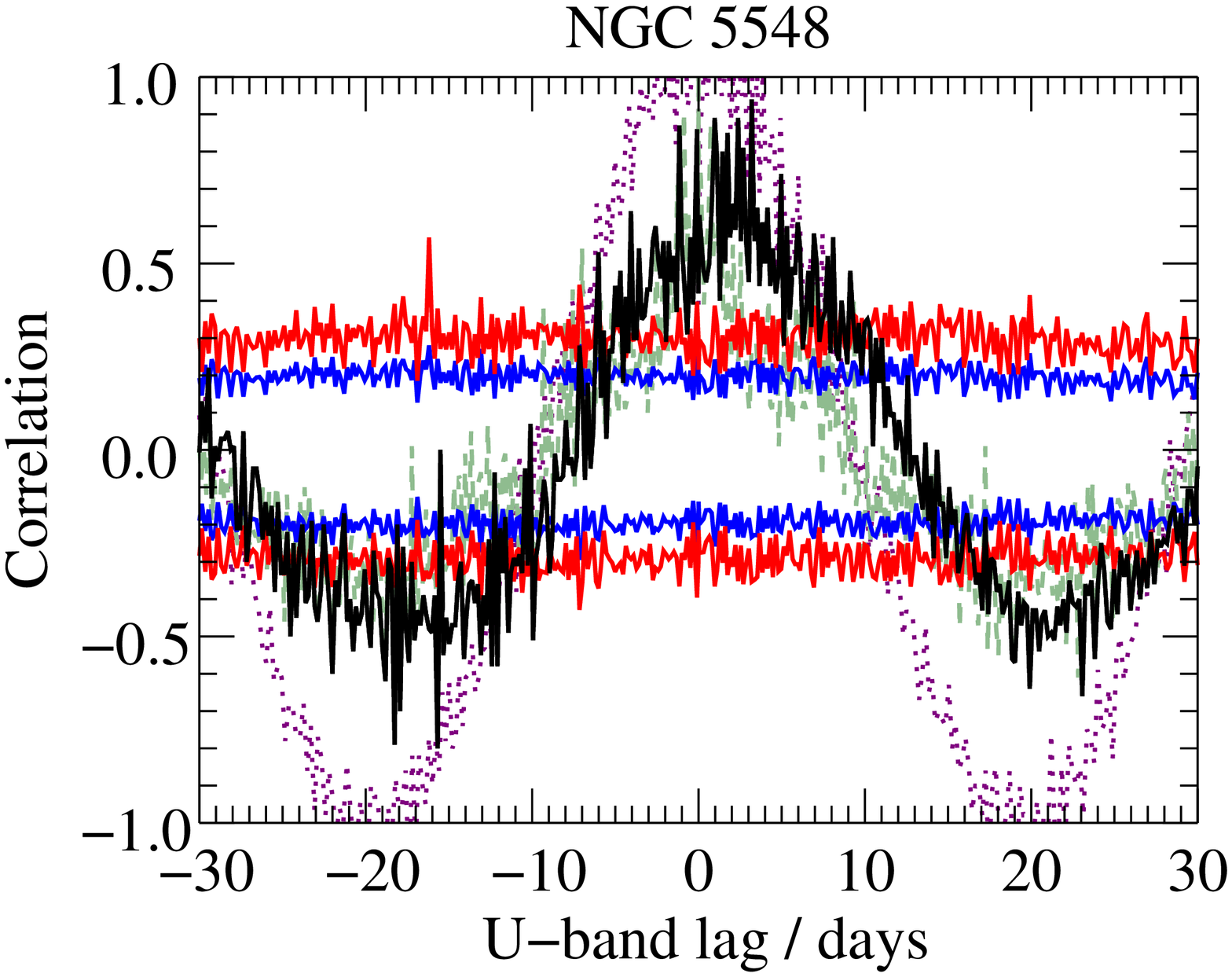}
	\includegraphics[width=5.5cm]{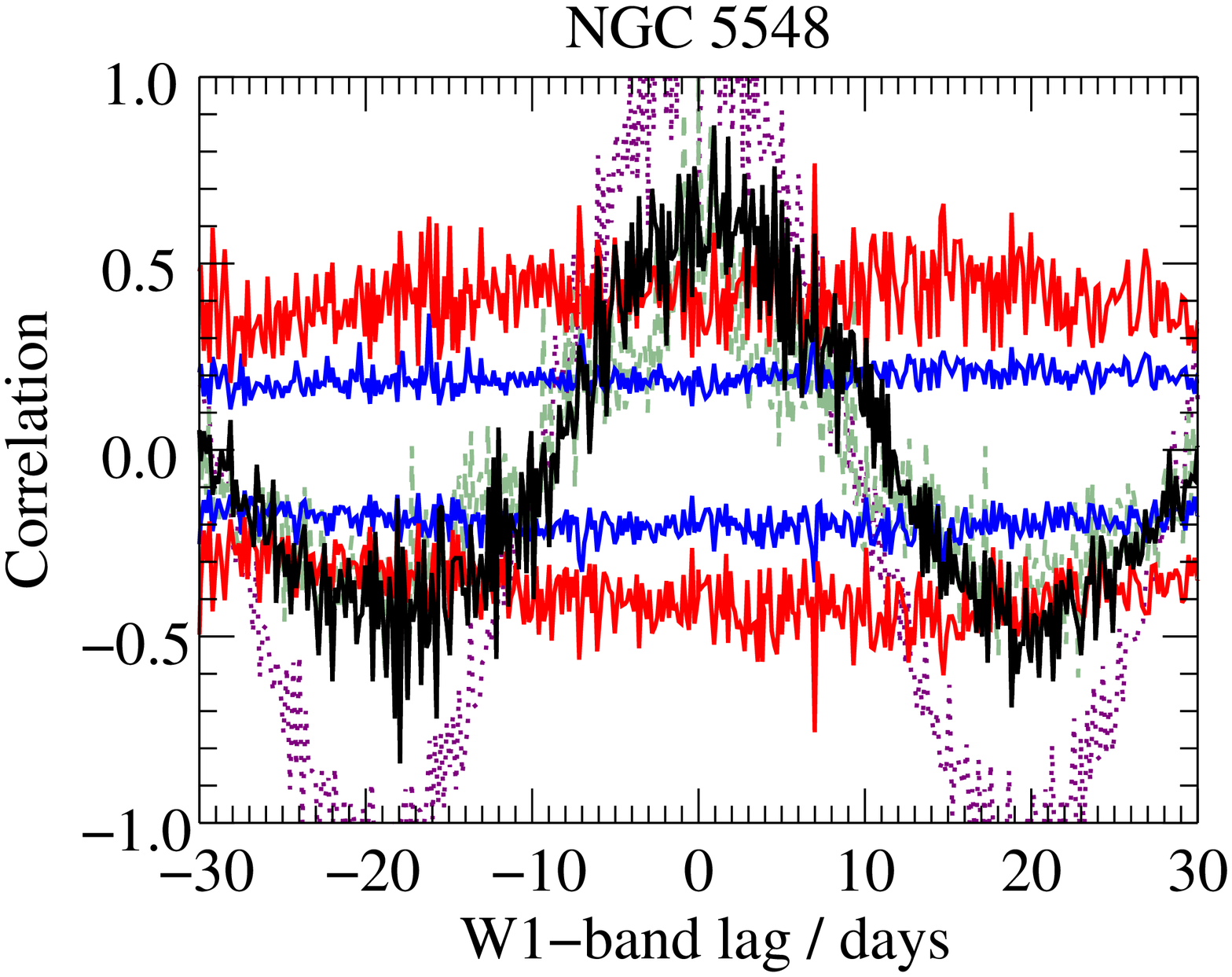}
	\includegraphics[width=5.5cm]{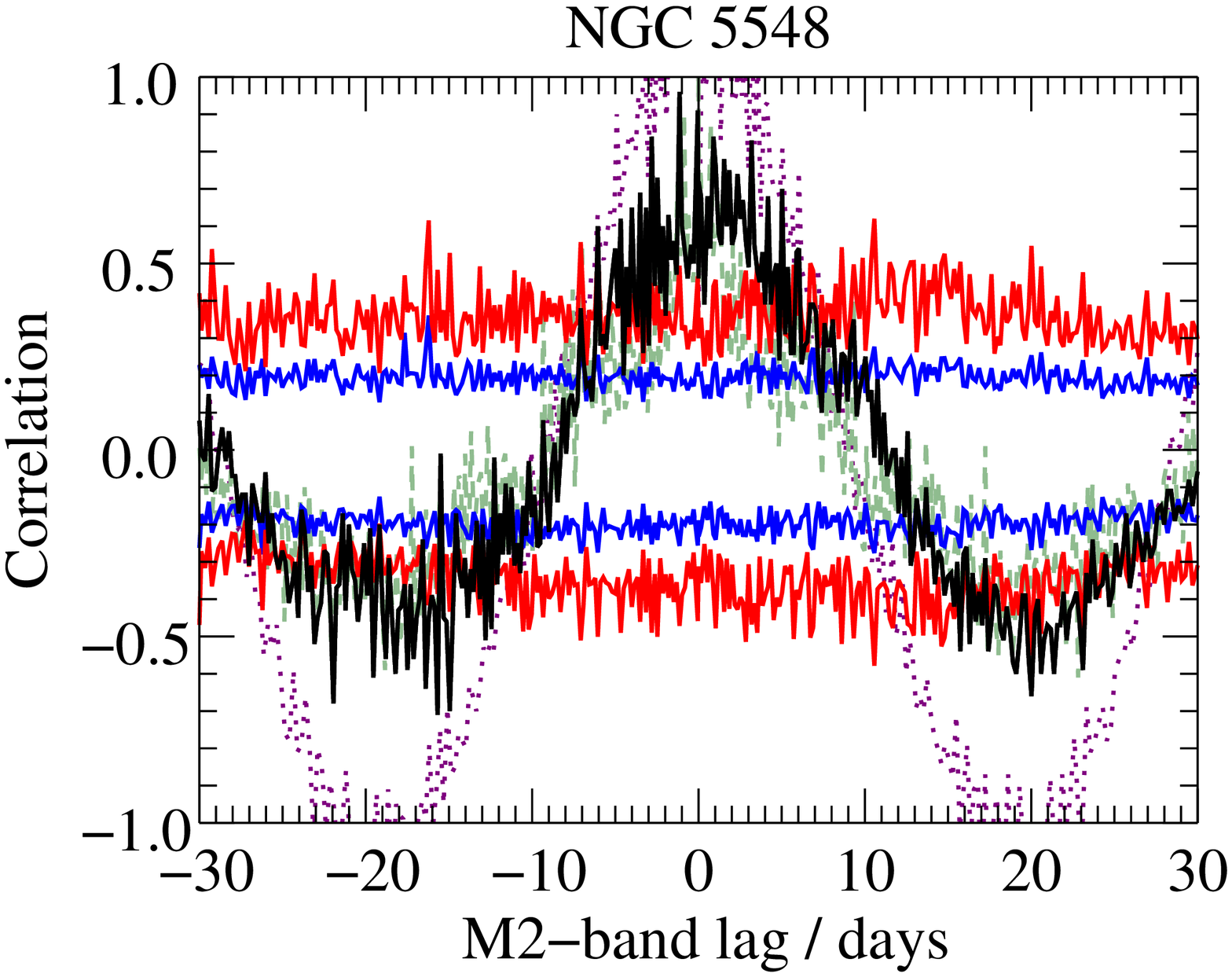}
	\includegraphics[width=5.5cm]{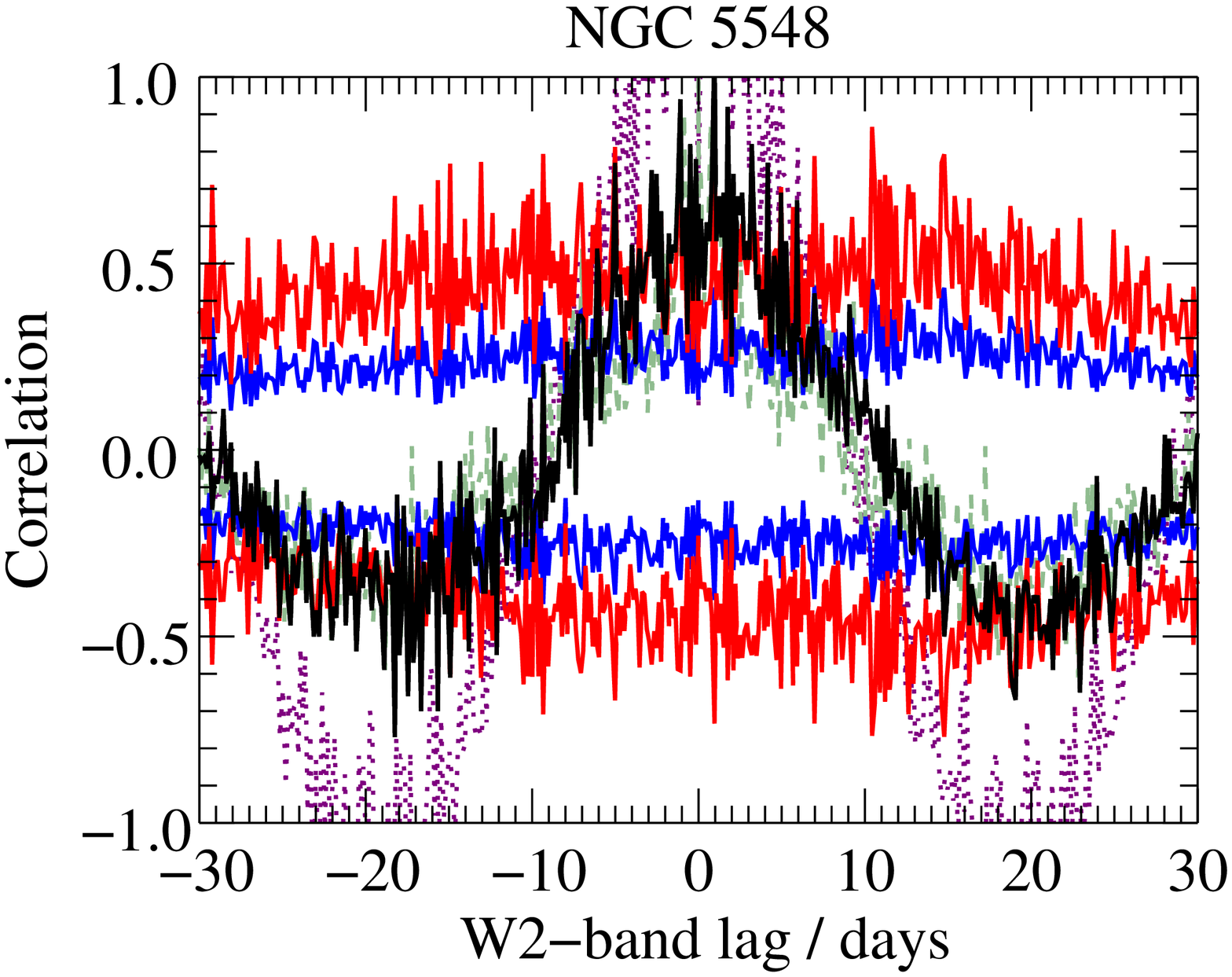}
	\includegraphics[width=5.5cm]{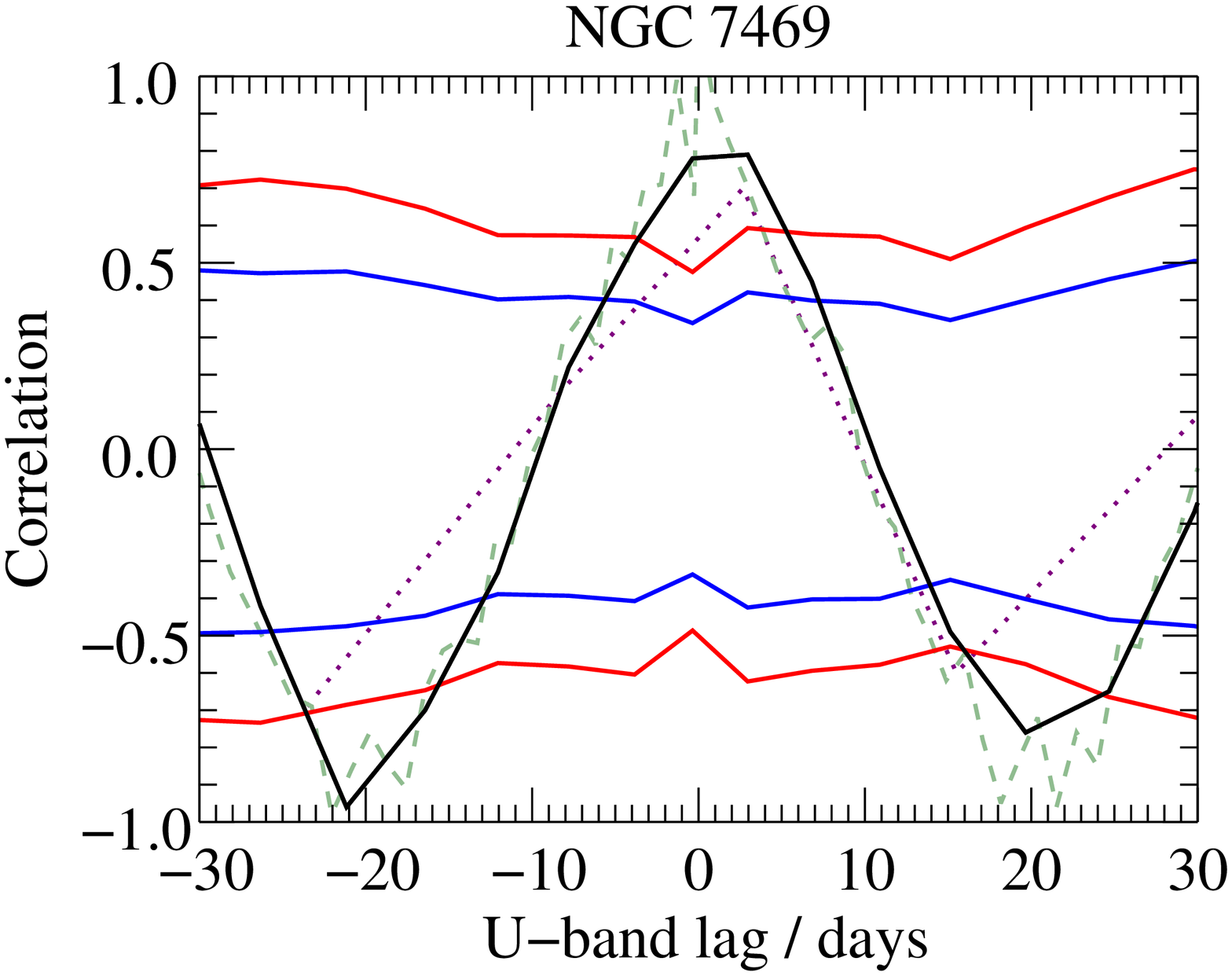}
\end{figure*}
\begin{figure*}
	\includegraphics[width=5.5cm]{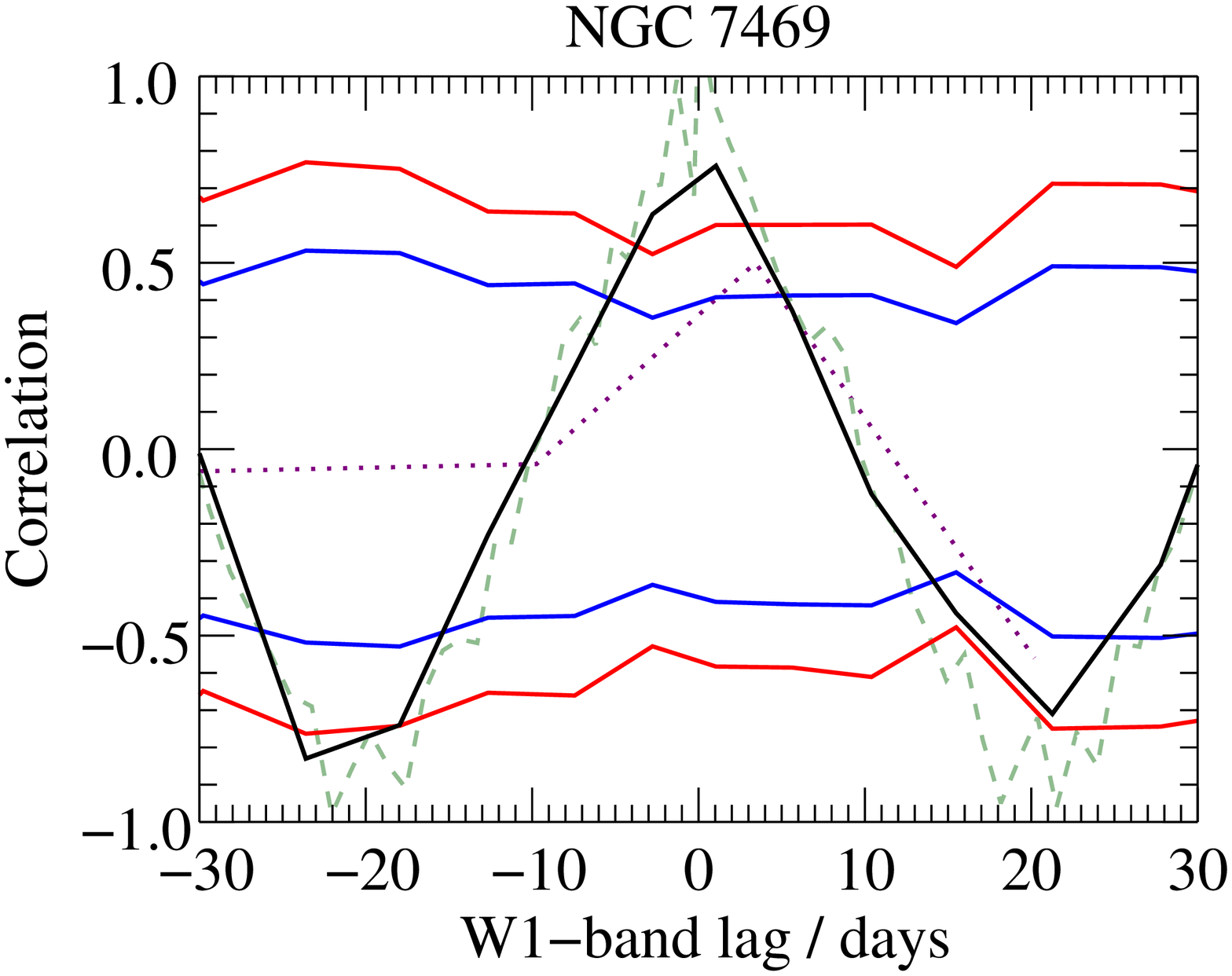}
	\includegraphics[width=5.5cm]{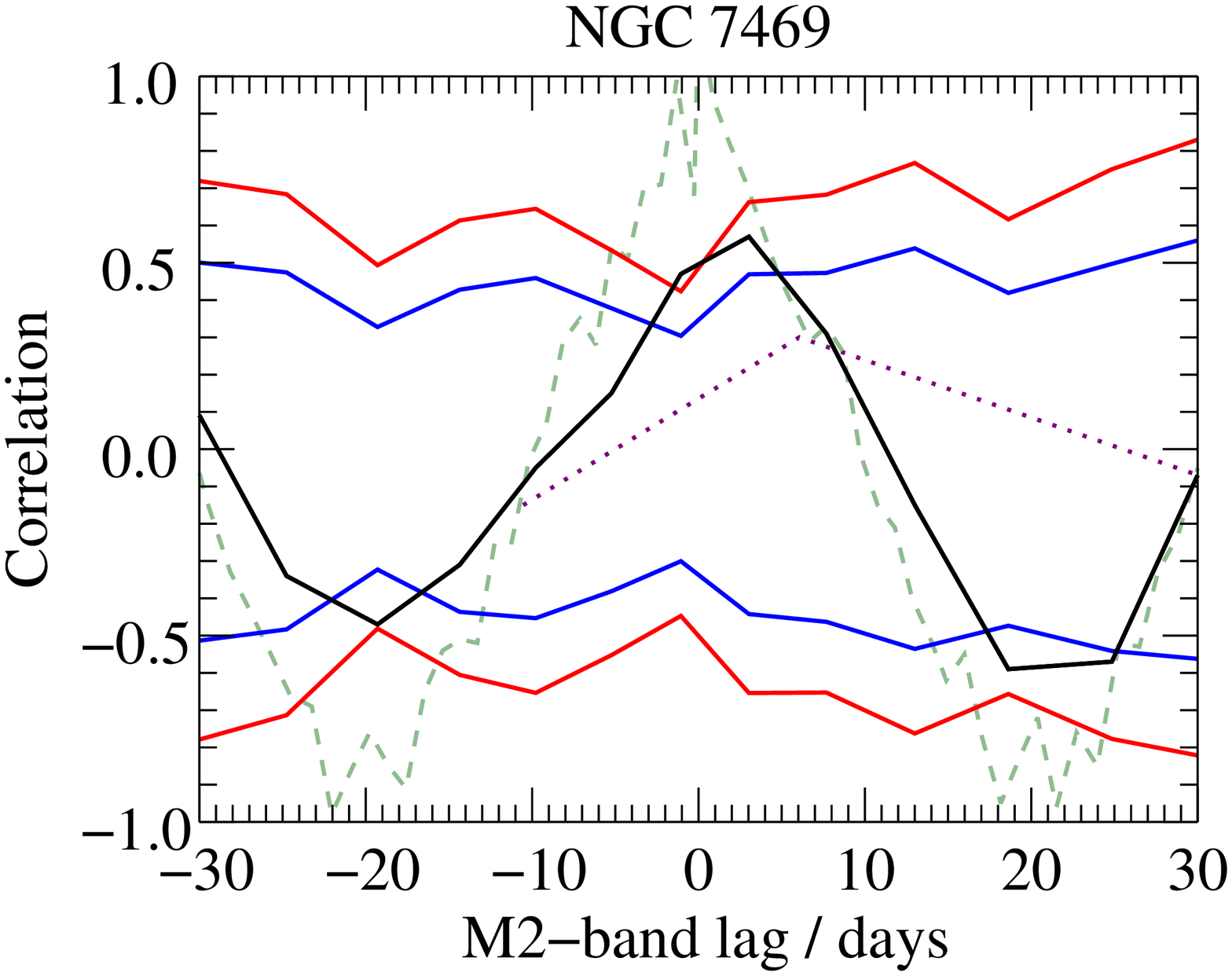}
	\includegraphics[width=5.5cm]{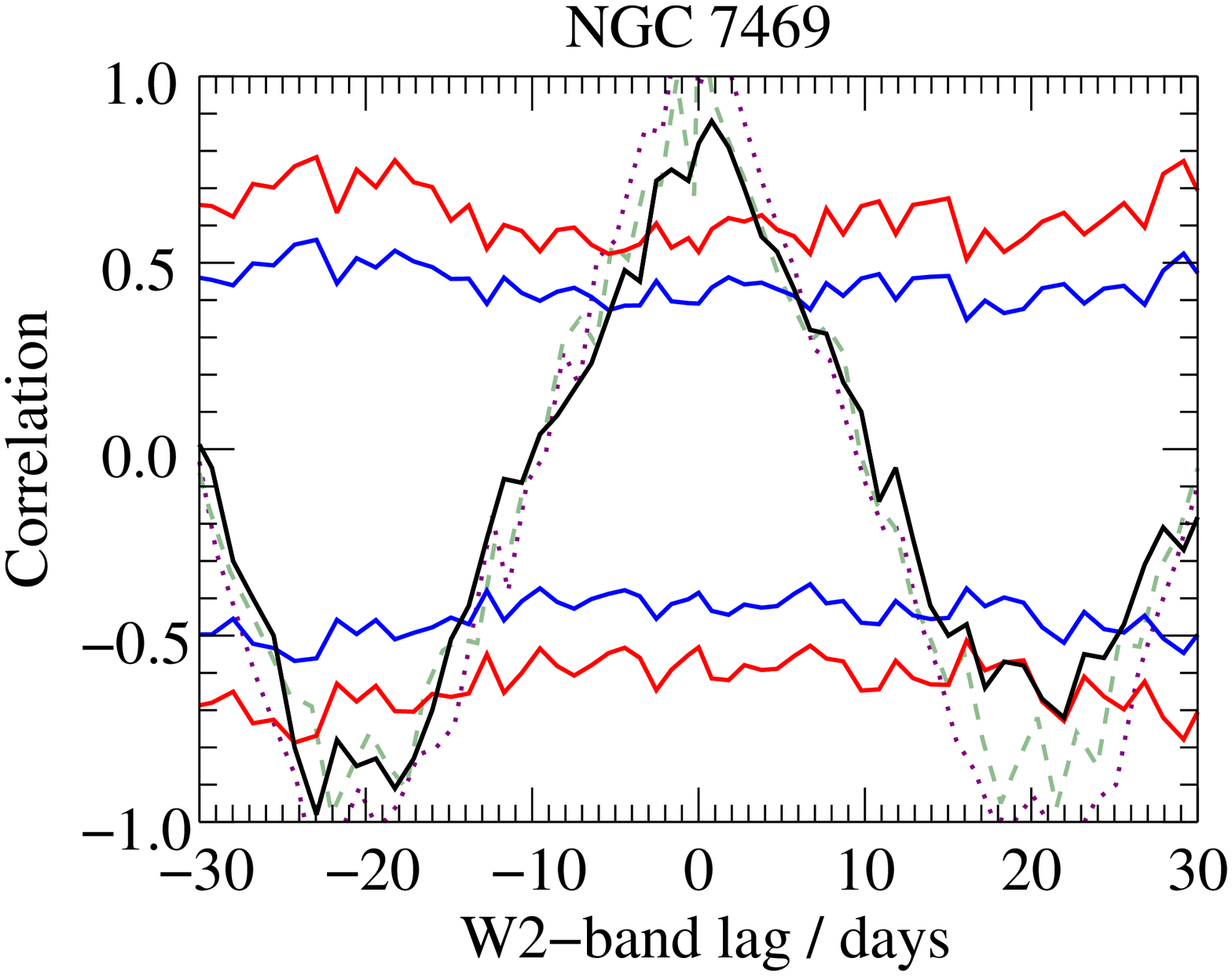}
	\includegraphics[width=5.5cm]{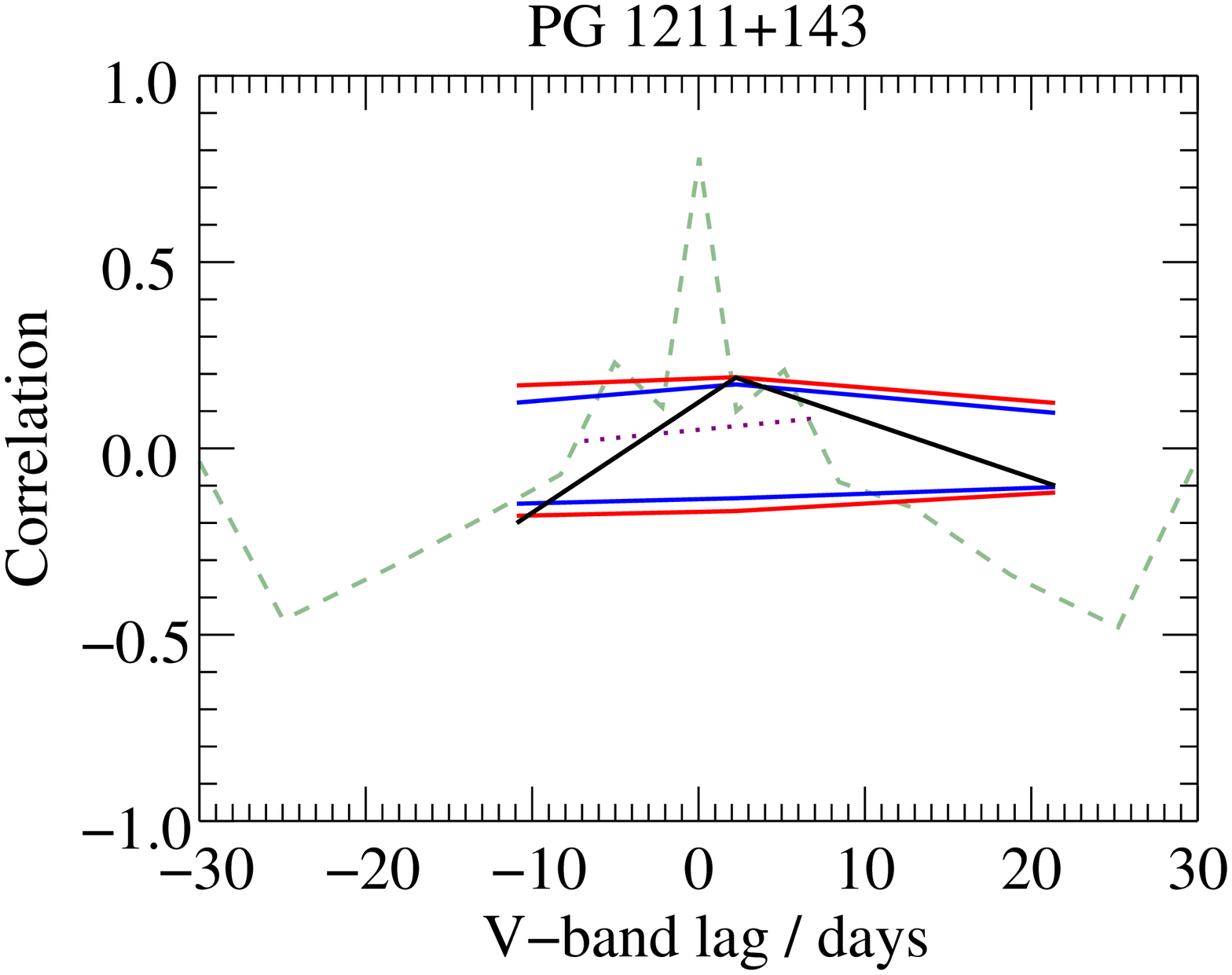}
	\includegraphics[width=5.5cm]{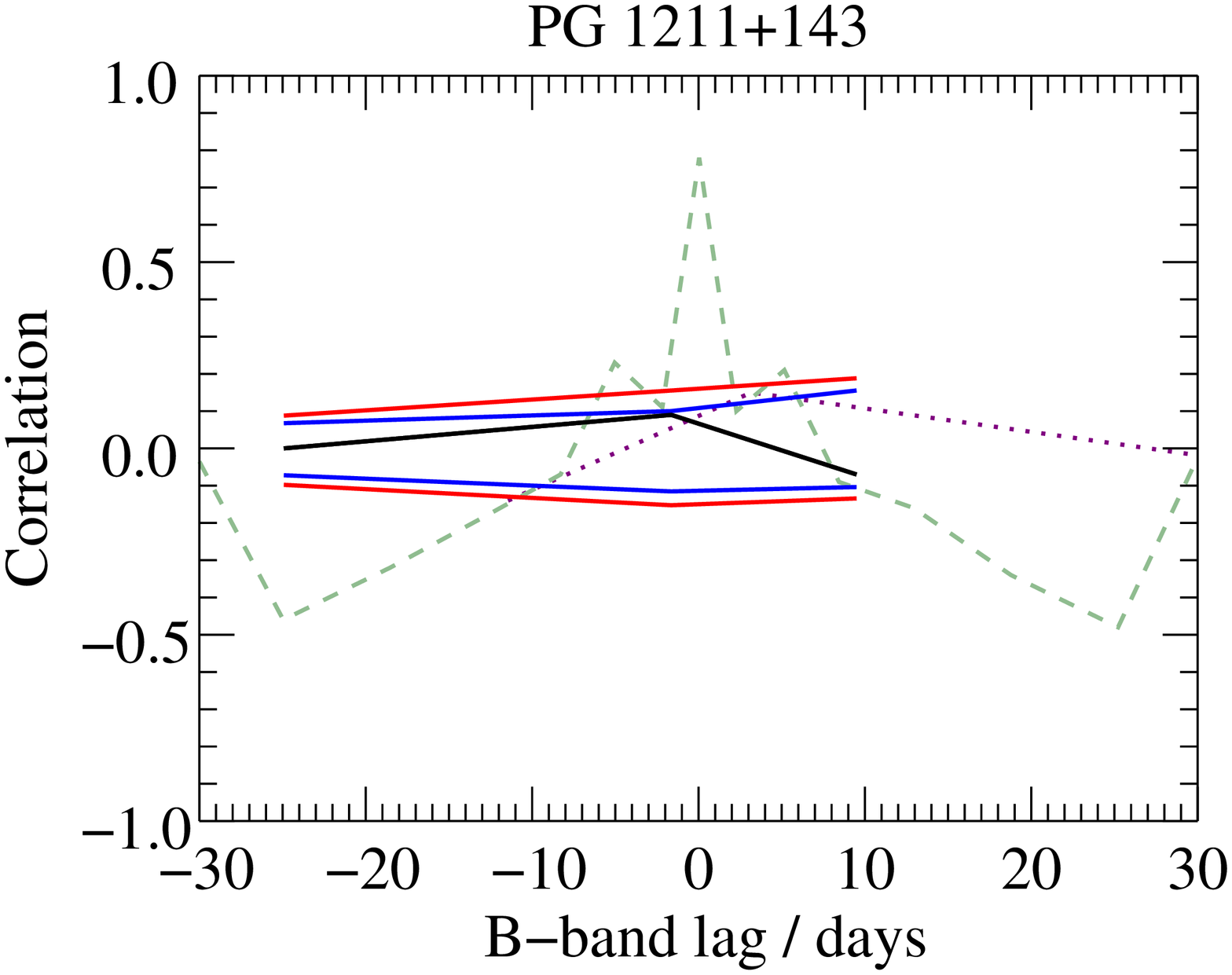}
	\includegraphics[width=5.5cm]{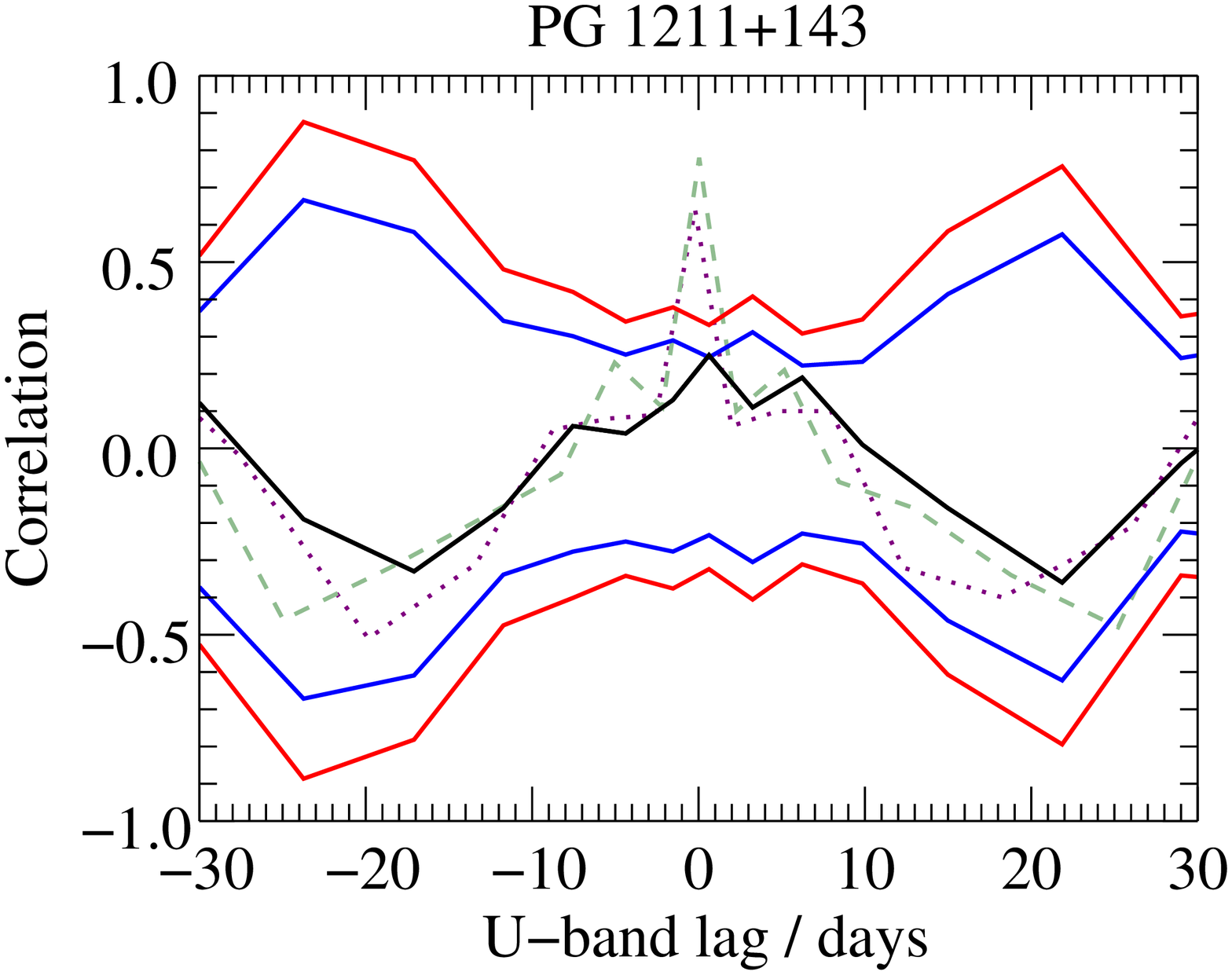}
	\includegraphics[width=5.5cm]{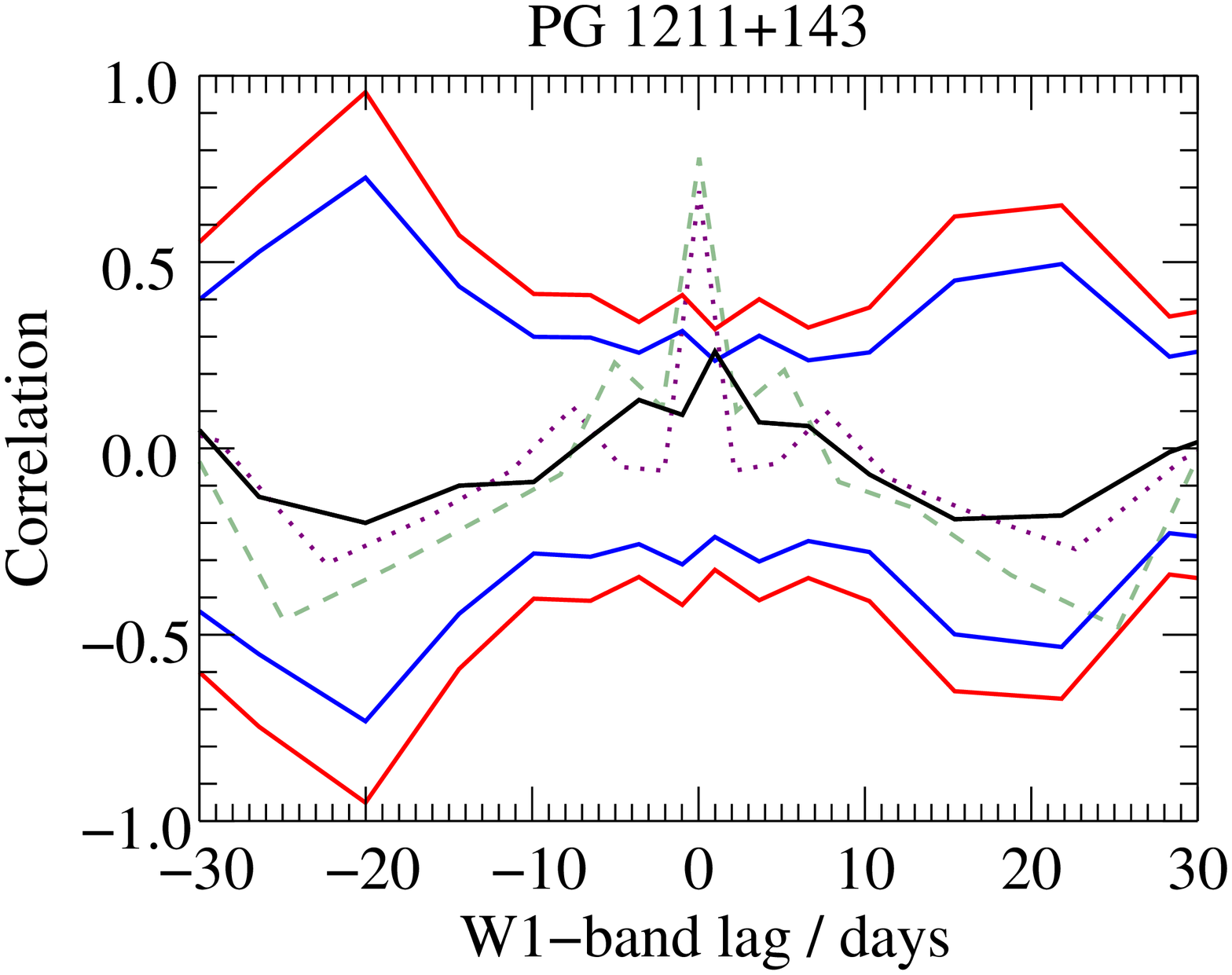}
	\includegraphics[width=5.5cm]{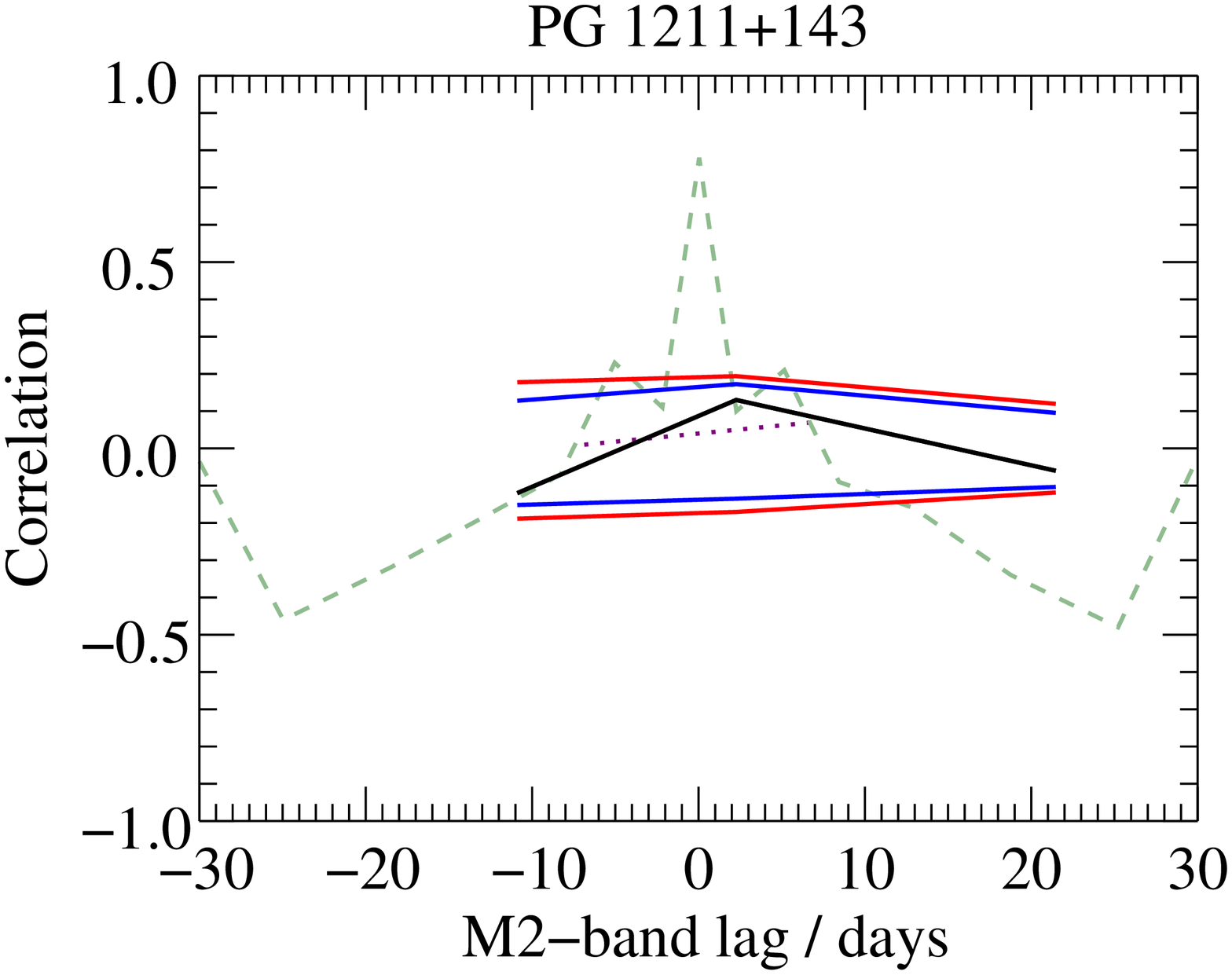}
	\includegraphics[width=5.5cm]{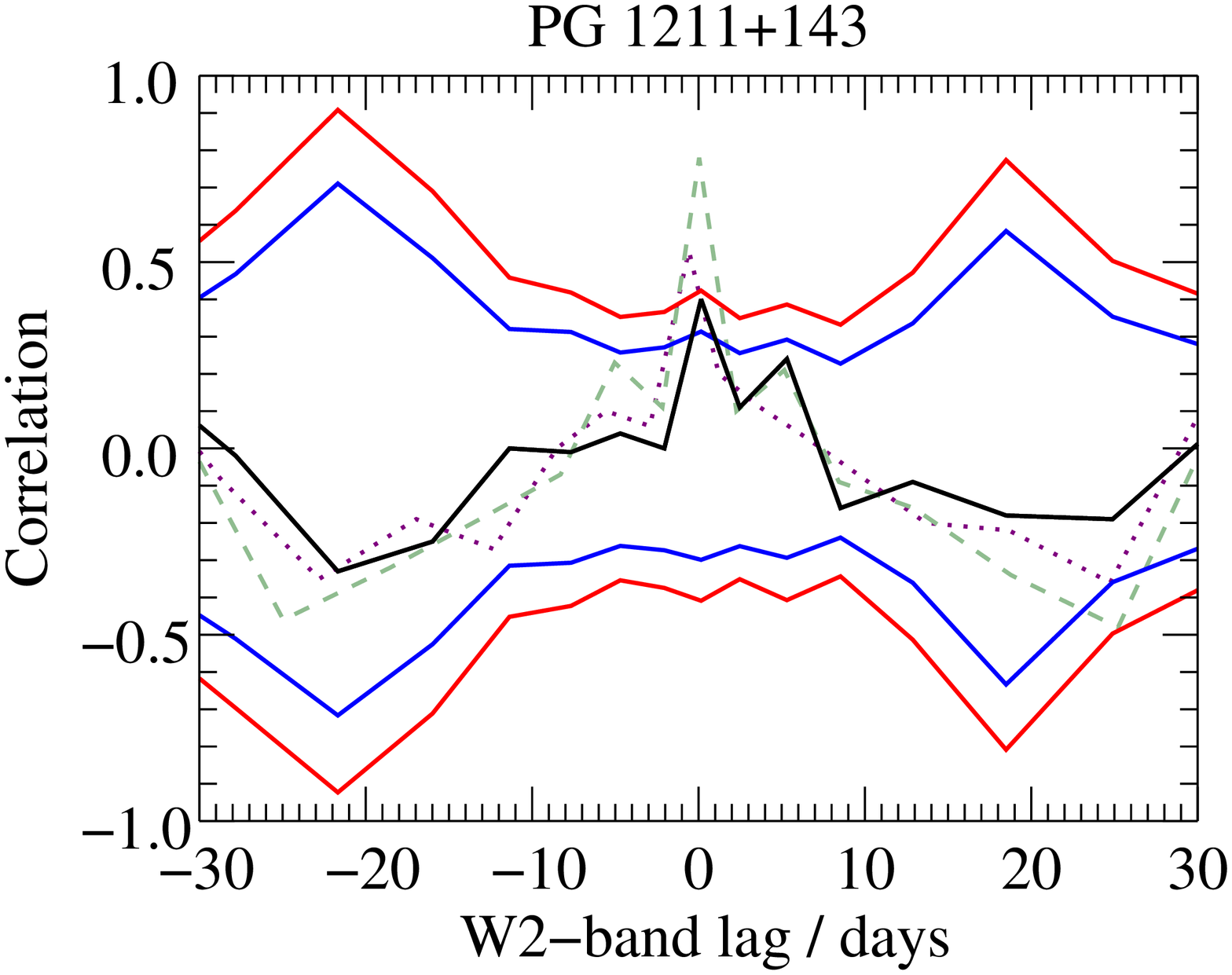}
	\includegraphics[width=5.5cm]{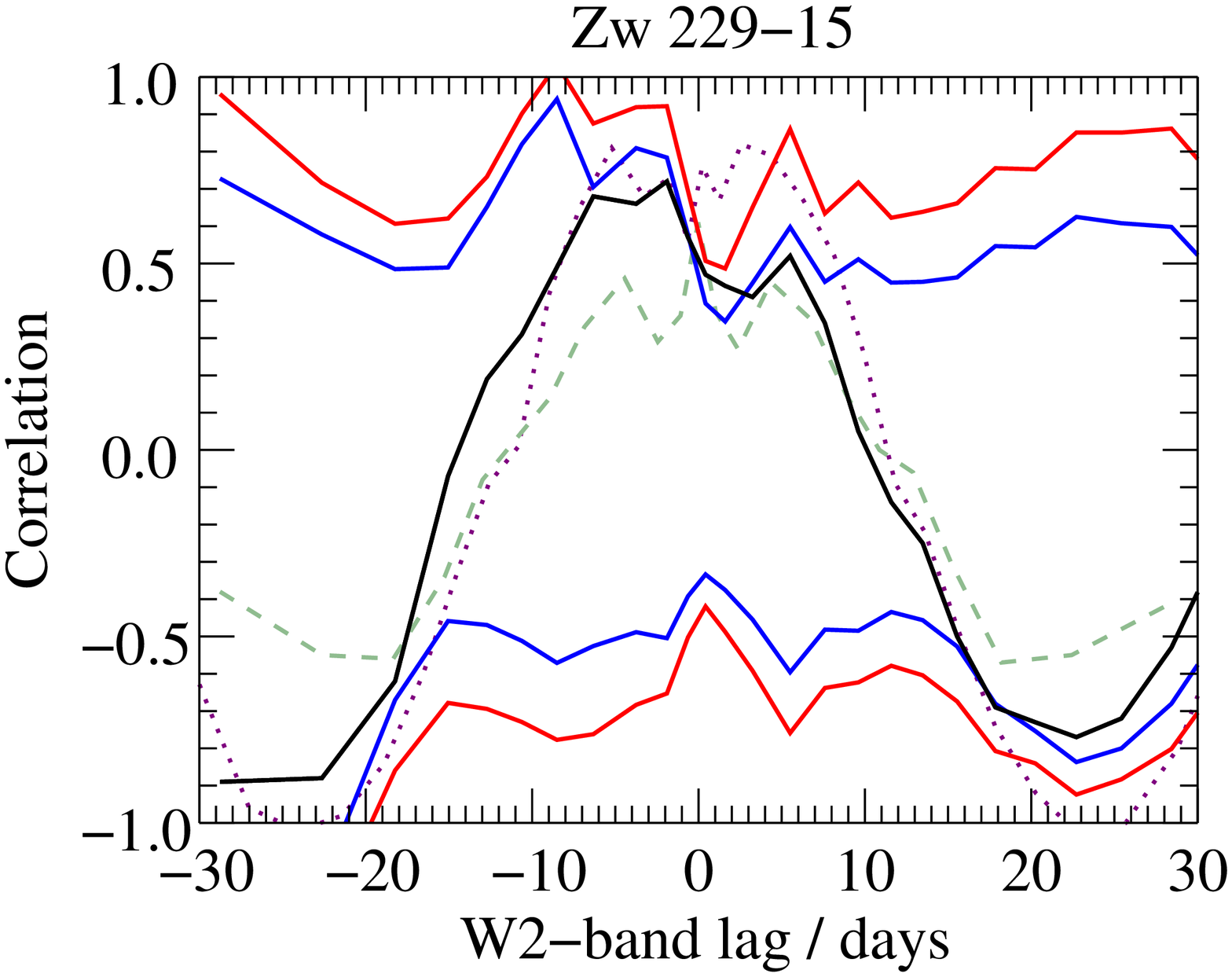}
\end{figure*}

\clearpage

\section{lags}
\label{app:lags}

For each source with UV/X-ray correlations detected at 99\% confidence, the lags are shown as a function of wavelength. These are compared to the predictions for a thin disc (blue) and 1-$\sigma$ errors (red).

\begin{figure}
	\includegraphics[width=\columnwidth]{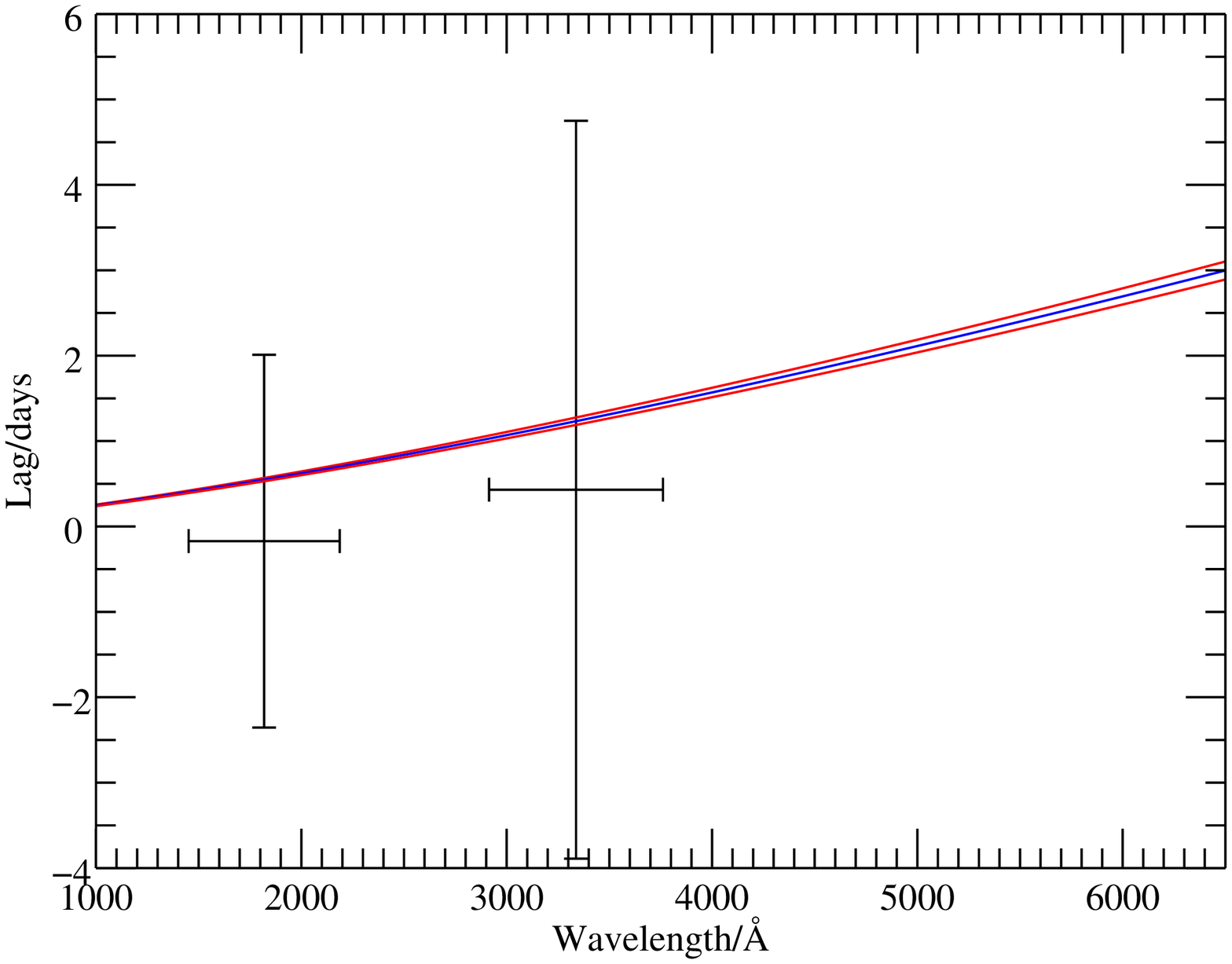}
    \caption{ 3C 120}
    \label{fig:lag3c120}
\end{figure}
\begin{figure}
	\includegraphics[width=\columnwidth]{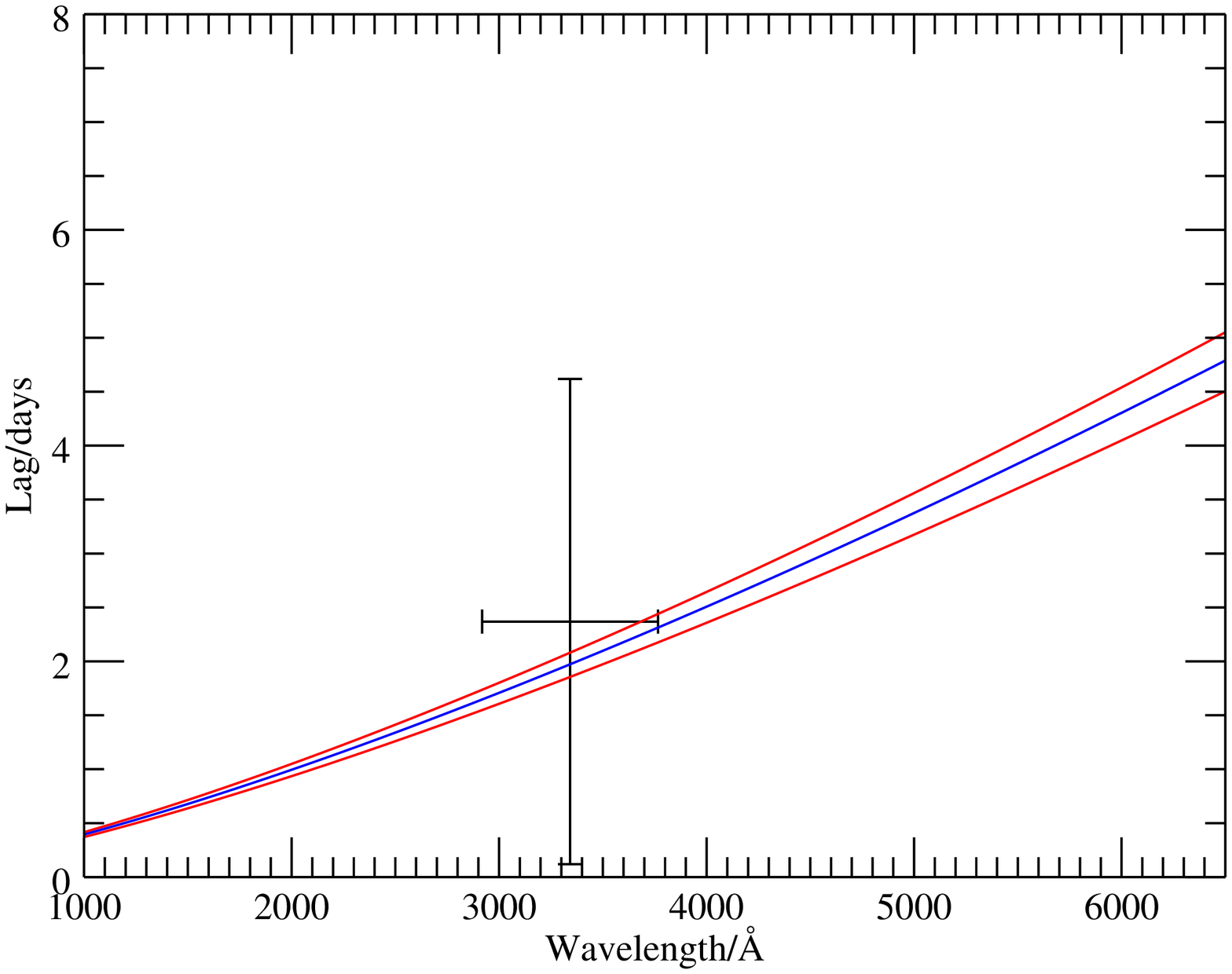}
    \caption{ARK 120}
    \label{fig:lagark120}
\end{figure}
\begin{figure}
	\includegraphics[width=\columnwidth]{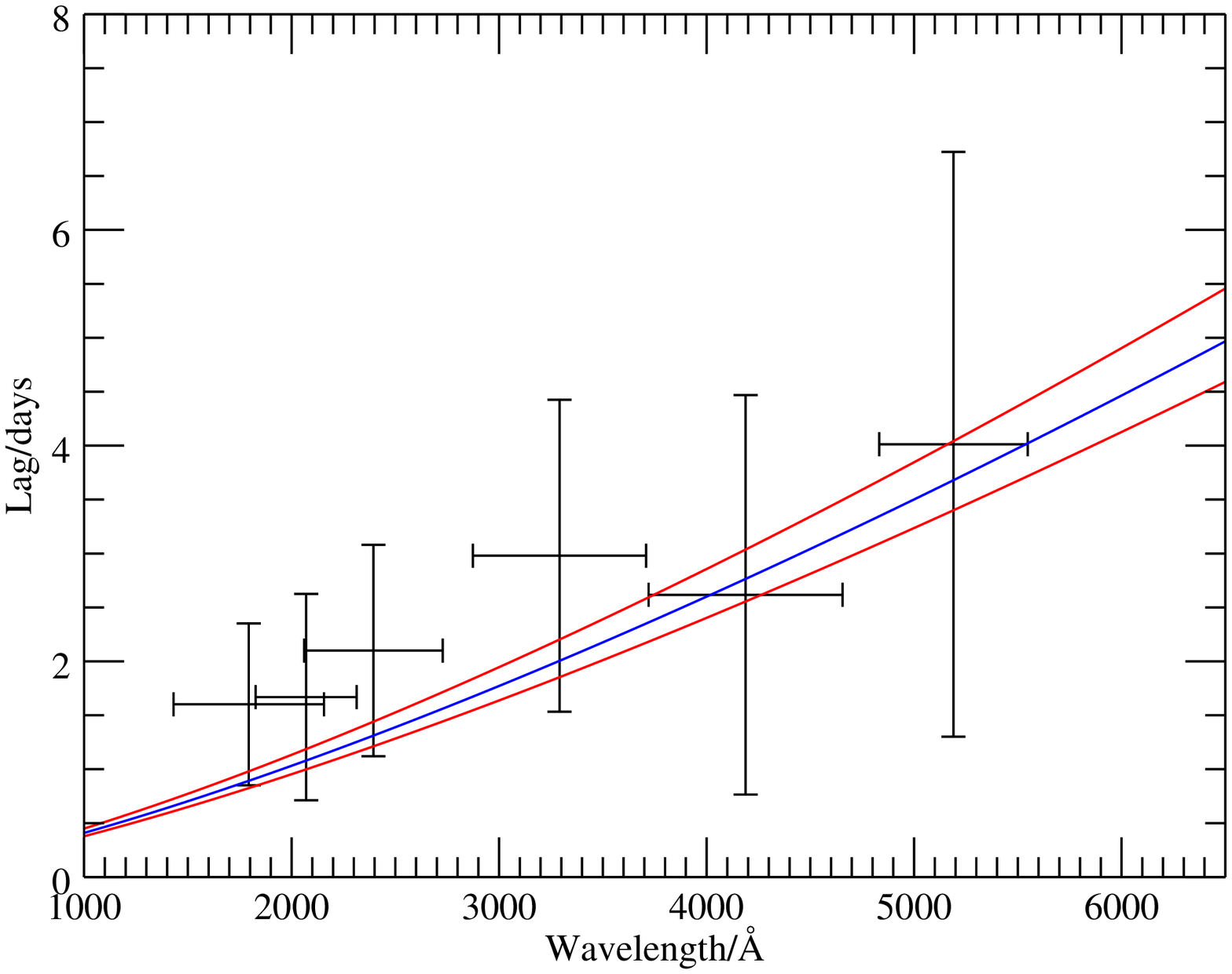}
    \caption{ Fairall 9}
    \label{fig:lagf9}
\end{figure}
\begin{figure}
	\includegraphics[width=\columnwidth]{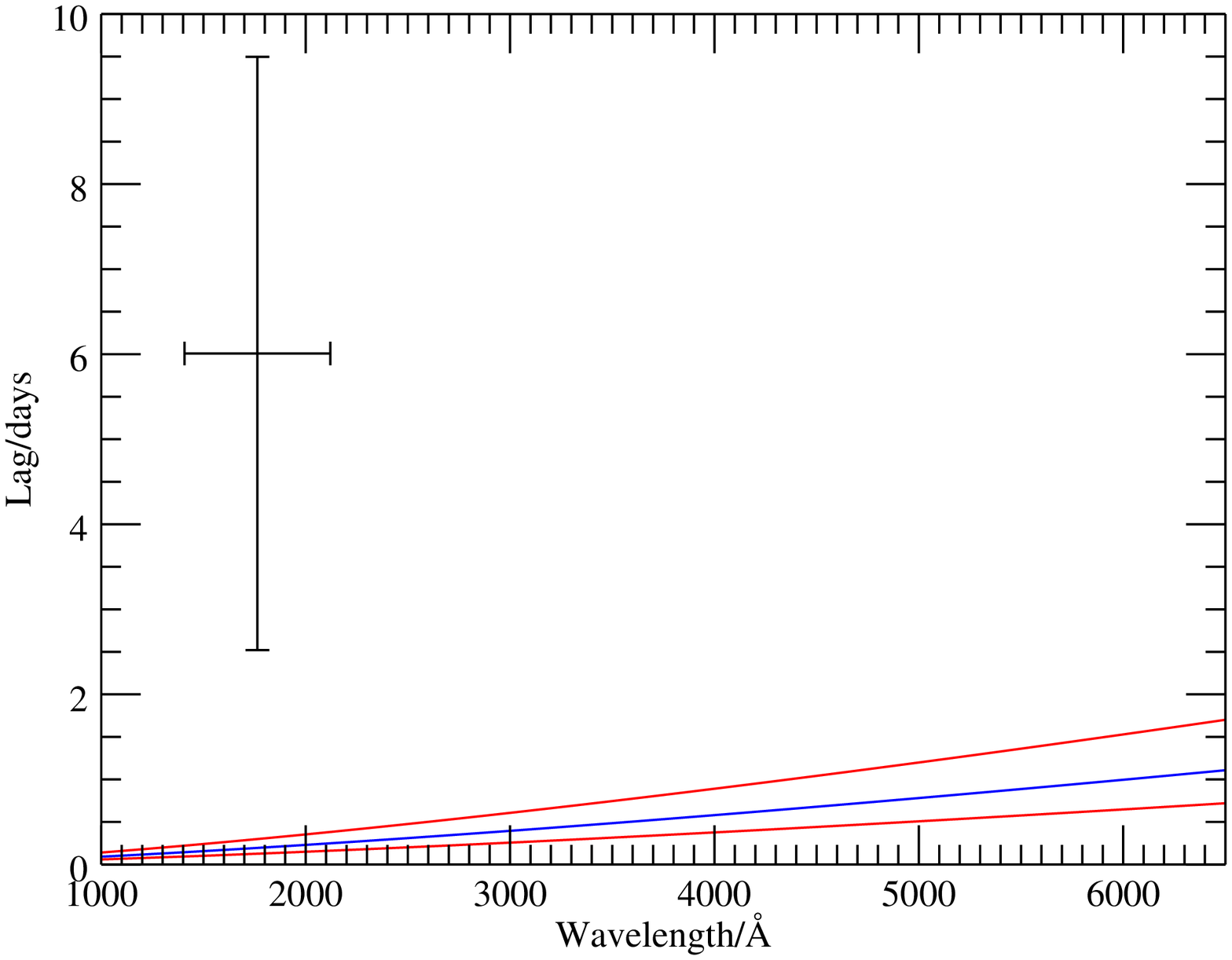}
    \caption{ IRAS 13224-3809}
    \label{fig:lagIRAS13224}
\end{figure}
\begin{figure}
	\includegraphics[width=\columnwidth]{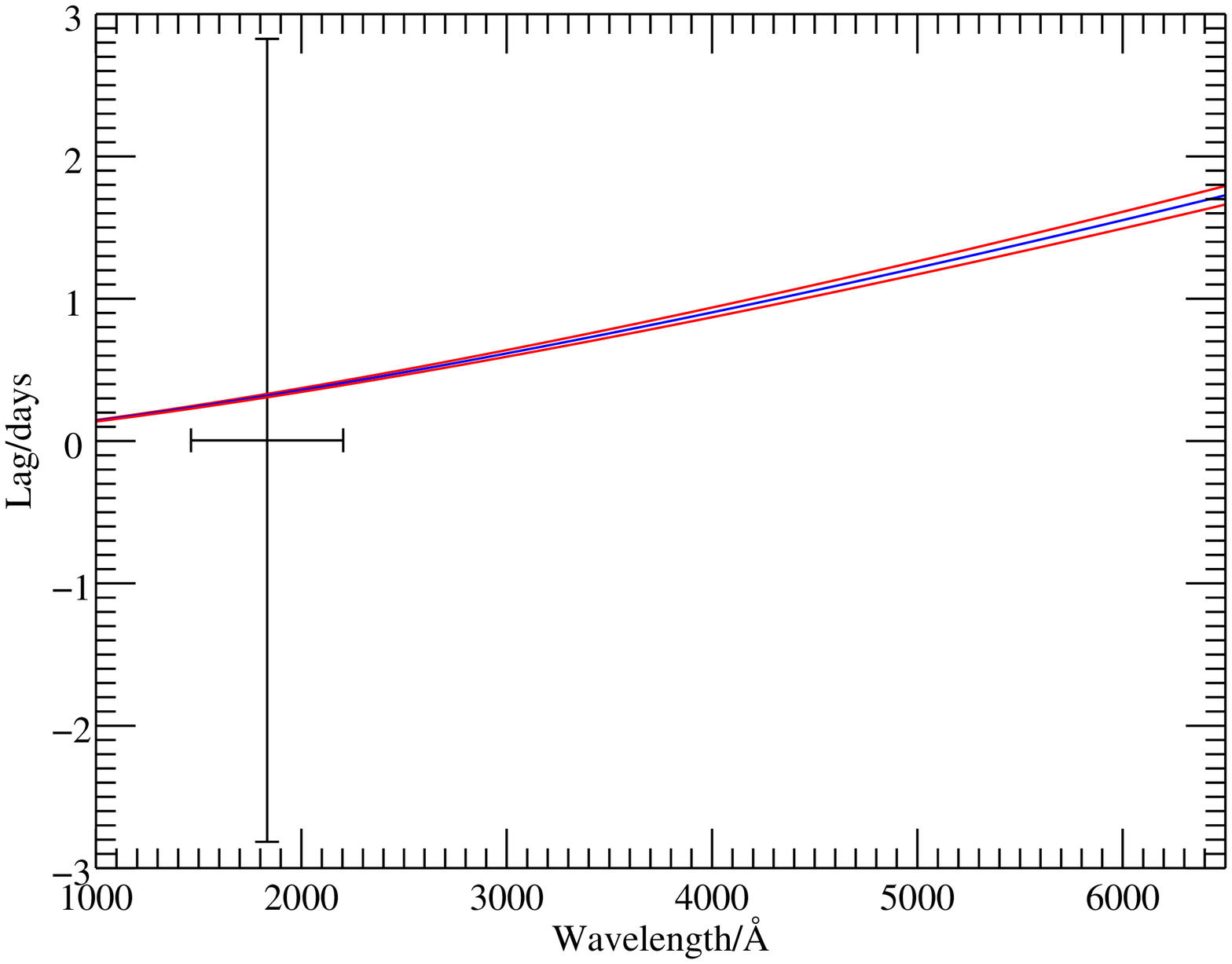}
    \caption{MRK 335}
    \label{fig:lagmrk335}
\end{figure}
\begin{figure}
	\includegraphics[width=\columnwidth]{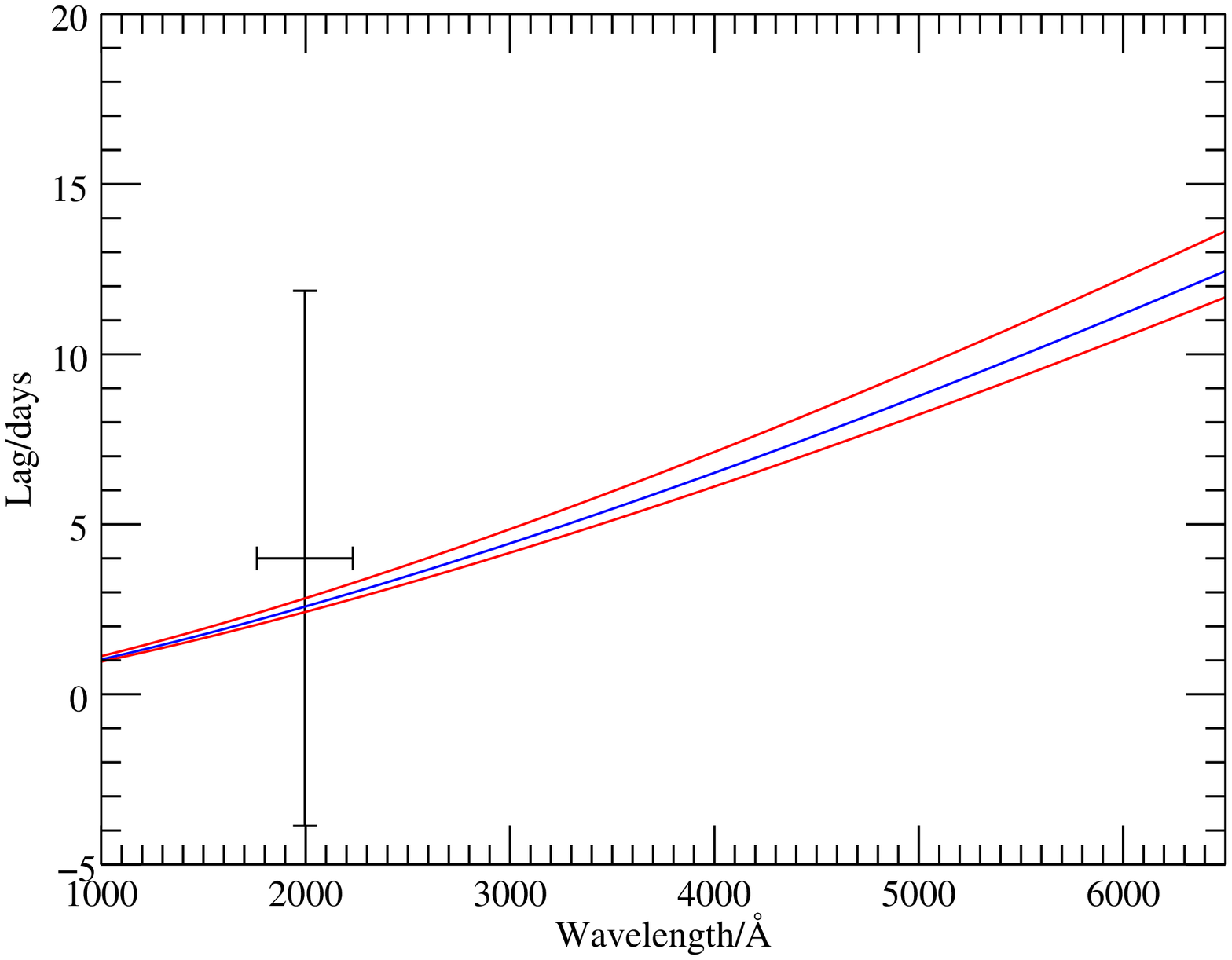}
    \caption{MRK 1383}
    \label{fig:lagmrk1383}
\end{figure}
\begin{figure}
	\includegraphics[width=\columnwidth]{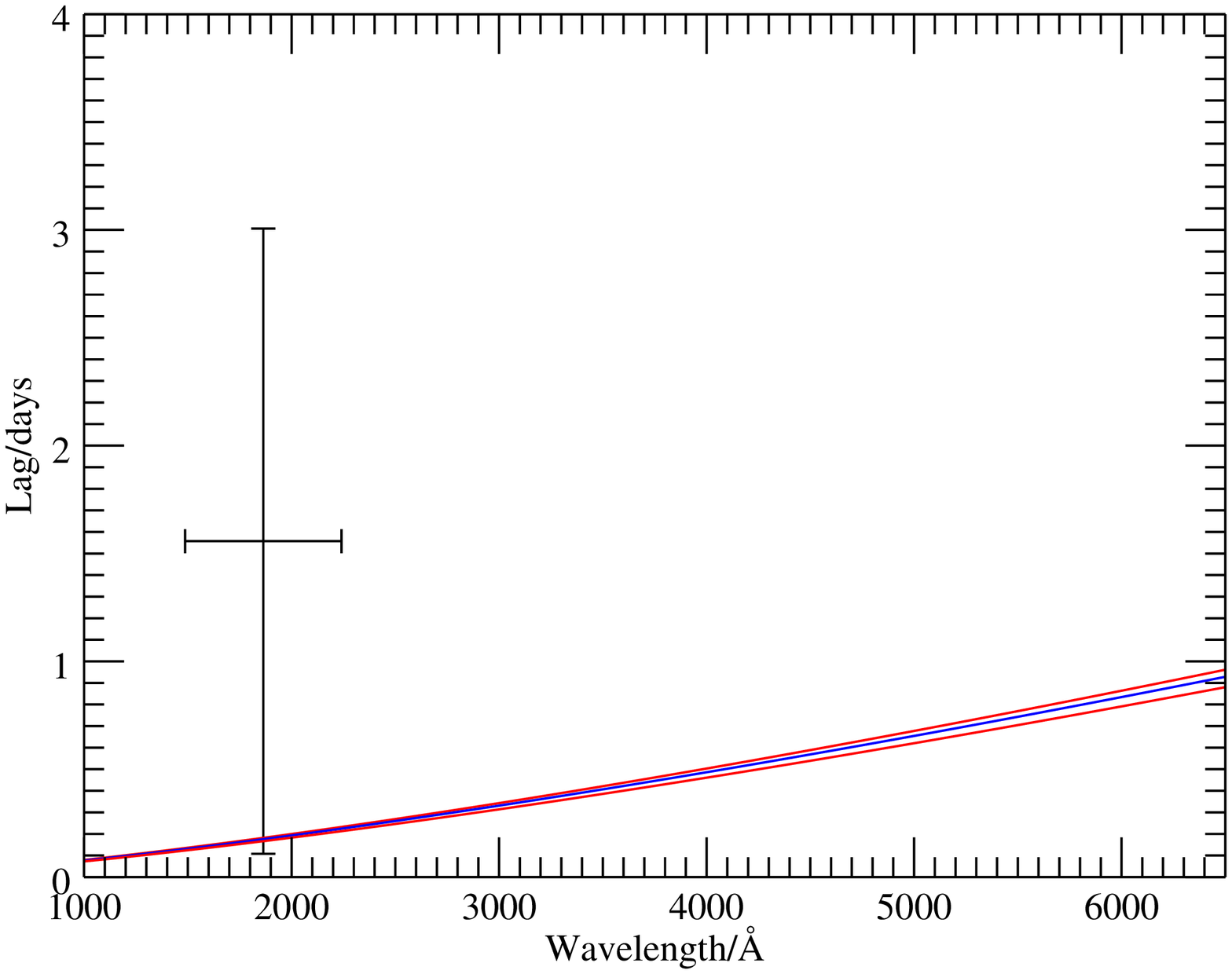}
    \caption{NGC 3516}
    \label{fig:lagngc3516}
\end{figure}
\begin{figure}
	\includegraphics[width=\columnwidth]{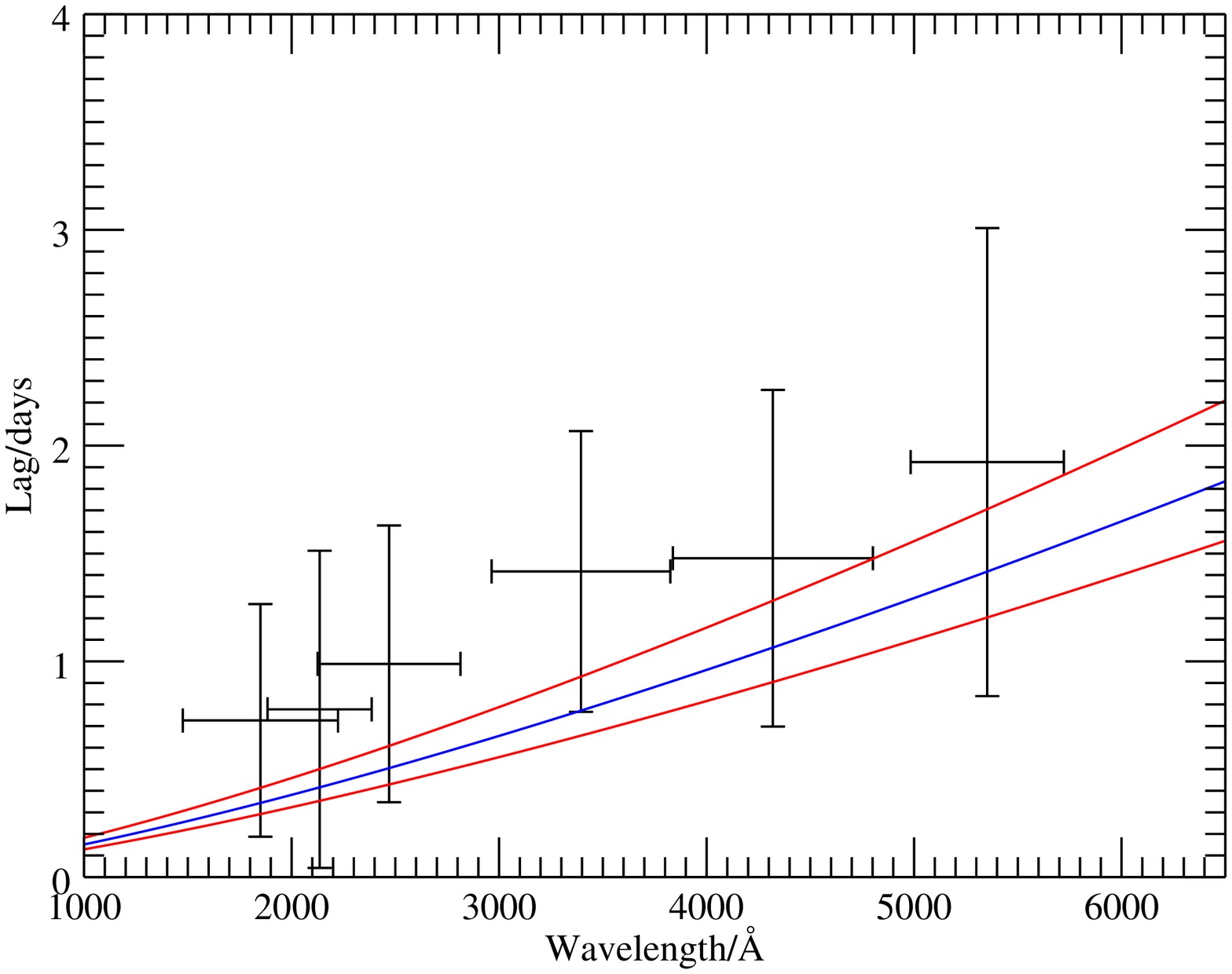}
    \caption{NGC 5548}
    \label{fig:lagngc5548}
\end{figure}
\begin{figure}
	\includegraphics[width=\columnwidth]{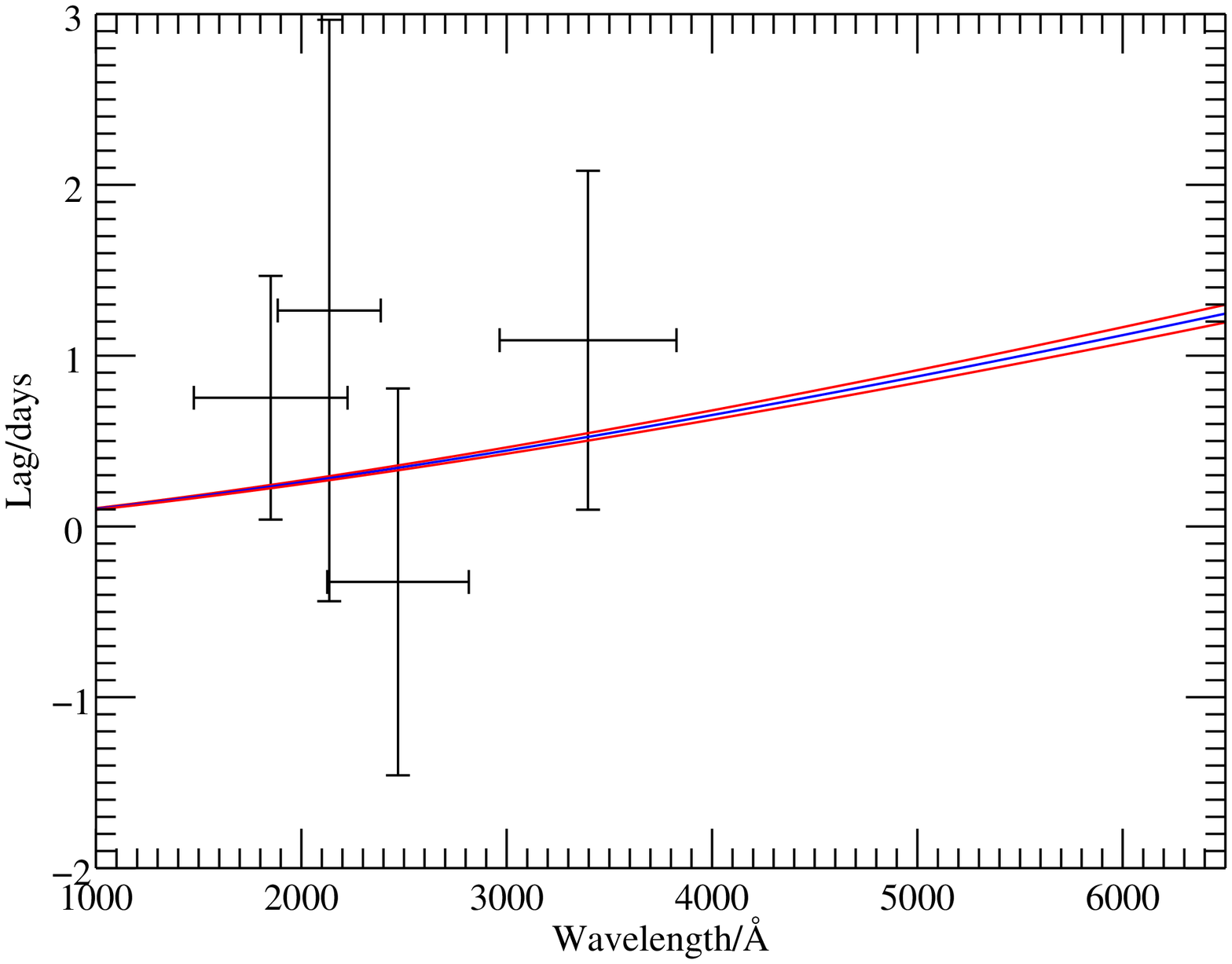}
    \caption{NGC 7469}
    \label{fig:lagngc7469}
\end{figure}

\bsp
\label{lastpage}
\end{document}